\documentclass[12pt]{article}
\usepackage{amsmath,amssymb,mathtools}
\usepackage{color}
\usepackage{xcolor}
\numberwithin{equation}{section}
\definecolor{ired}{RGB}{219 112 147}

\usepackage[font=small]{subfig}
\usepackage[skip=8pt,labelfont={bf,sc},textfont=small]{caption}

\usepackage{graphicx}
\usepackage{hyperref}
\usepackage{cite}
\usepackage{tikz}

\usepackage{mathrsfs}
\usepackage{amsthm}

\newtheoremstyle{thry}
{6pt}
{6pt}
{\itshape}
{}
{\bfseries}
{:}
{.5em}
{}
\theoremstyle{thry}

\newtheoremstyle{remar}
{6pt}
{6pt}
{\upshape}
{}
{\bfseries}
{:}
{.5em}
{}
\theoremstyle{remar}



\usepackage{float} 
\usepackage{ulem}
\usepackage[margin=2cm]{geometry}
\def \be {\begin{equation}}
\def \ee {\end{equation}}
\def \ba {\begin{aligned}}
\def \ea {\end{aligned}}
\def \bea{\begin{eqnarray}}
\def \eea{\end{eqnarray}}

\def \tr{\text{tr}}

\def\CI{{\cal I}}

\def\vev#1{\langle #1\rangle}
\def\p{\partial}

\def\l{\lambda}
\def\m{\mu}
\def\n{\nu}

\def\vev#1{\langle #1\rangle}  
\begin{document}
\begin{titlepage}
\vspace{.5cm}
\begin{center}
{\large $T\overline{T}$ Deformation: Introduction and Some Recent Advances}
\lineskip .75em
\vskip 2.5cm
{ Song He$^{a,b,c,e}$\footnote{hesong@nbu.edu.cn}, Yi Li$^{d,}$\footnote{liyi@fudan.edu.cn}, Hao Ouyang$^{c,}$\footnote{haoouyang@jlu.edu.cn}, Yuan Sun$^{d}$\footnote{sunyuan@csu.edu.cn}}
\vskip 2.5em
{\normalsize\it  $^{a}${Institute of Fundamental Physics and Quantum Technology, \\ Ningbo University, Ningbo, Zhejiang 315211, China }\\
$^{b}${School of Physical Science and Technology, Ningbo University, Ningbo, 315211, China}\\
$^{c}$Center for Theoretical Physics and College of Physics, Jilin University,\\ Changchun 130012, People's Republic of China\\
$^{d}${Institute of Quantum Physics, School of Physics, Central South University, \\Changsha 418003,
China}\\
$^{e}$Max Planck Institute for Gravitational Physics (Albert Einstein Institute),\\
Am M\"uhlenberg 1, 14476 Golm, Germany\\
}
\vskip 3.0em
\end{center}
\begin{abstract}
This review explores recent advances in the theory of $T\bar{T}$ deformation, an irrelevant yet solvable deformation of quantum field theories defined via the quadratic form of the energy-momentum tensor. It addresses classical and quantum aspects, highlighting significant developments across various fields, including field theory, holography, and string theory. Classically, $T\bar{T}$ deformation manifests through multiple geometric interpretations, notably random geometry, Jackiw-Teitelboim-like gravity, and uniform light-cone gauge frameworks. For quantum aspects, the deformation introduces notable features such as non-locality, UV-IR mixing, solvable renormalization structures, and intriguing modifications to correlation functions and entanglement properties. Furthermore, the paper examines the profound relationship between $T\bar{T}$ deformation and holography, particularly within the mixed boundary conditions/cutoff AdS holography proposal and holographic entanglement entropy. Connections to string theory through single-trace deformations and their holographic duals further reveal the deformed structure of the worldsheet. This review synthesizes recent developments and outlines potential directions for future research in the study of $T\bar{T}$-like deformation.

\end{abstract}

\end{titlepage}

\baselineskip=0.7cm

\tableofcontents

\newpage
\section{Introduction}

The study of deformations in quantum field theories (QFTs) has provided crucial insights into the structure and dynamics of physical models. A notable example is the $T\overline{T}$ deformation \cite{Smirnov:2016lqw, Cavaglia:2016oda}, first introduced by Zamolodchikov in 2004 \cite{Zamolodchikov:2004ce}, which has attracted significant attention due to its remarkable properties, such as solvability, integrability, and non-trivial modifications to the system's dynamics. This deformation, which couples the energy-momentum tensor to itself, offers a non-local yet solvable modification of two-dimensional quantum field theories while preserving key physical properties.

One of the primary motivations for reviewing energy-momentum tensor-induced deformations, such as the $T\overline{T}$ deformation, stems from their ability to provide new insights into non-locality and ultraviolet (UV) completion of QFTs. Unlike most irrelevant deformations that typically lead to complex, non-renormalizable theories, the $T\overline{T}$ deformation is solvable (and integrable), suggesting that such deformations can be physically meaningful under certain conditions. As this area of research grows, we expect this review will help identify new connections and directions in the study of quantum gravity\cite{Dubovsky:2017cnj, Dubovsky:2018bmo, Tolley:2019nmm, Iliesiu:2020zld, Okumura:2020dzb, Ebert:2022ehb, Bhattacharyya:2023gvg, Bao:2024ixc}, holography~\cite{McGough:2016lol, Giveon:2017nie, Kraus:2018xrn, Gorbenko:2018oov, Cottrell:2018skz, Taylor:2018xcy, Hartman:2018tkw, Jiang:2019tcq, Caputa:2019pam, Guica:2019nzm, Gross:2019ach, Jafari:2019qns, Li:2020pwa, Griguolo:2021wgy, He:2022bbb, He:2023hoj, He:2023knl, Fichet:2023xbu, Ebert:2024fpc, He:2024xbi}, and string theory\cite{Baggio:2018gct, Dei:2018jyj, Chakraborty:2019mdf, Callebaut:2019omt}. The new variant $\sqrt{T\overline{T}}$ of stress tensor flows inspired by the two-dimensional $T\overline{T}$ deformation has been proposed by \cite{Conti:2022egv, Ferko:2022cix, Ferko:2023ruw, Babaei-Aghbolagh:2022uij, Babaei-Aghbolagh:2022leo, Hadasz:2024pew, Borsato:2022tmu}. {More recently, the generic $T\overline{T}$-like deformation has been reviewed in~\cite{Brizio:2024arr}.}

The study of these deformations spans multiple perspectives beyond standard solvability, integrability, and non-local properties often detailed in early review papers~\cite{Giveon:2019fgr, Jiang:2019epa}, including but not limited to:

{\bf To understand how the graviton emerges from the field theory}

We propose a thought experiment starting from the action of a massless boson in flat spacetime. According to Einstein's equations, this action necessarily induces an energy-momentum tensor, which must result in spacetime curvature. Simultaneously, the curvature alters the background metric of the massless boson's action or, equivalently, modifies the graviton. This modification leads to interactions between the massless boson and the graviton, inducing an interaction that couples the boson and graviton. This process demonstrates the transition from field theory on a fixed background to a gravitational theory coupled with matter field theory.


Studying field theories deformed by $T\overline{T}$-like operators reveals a process remarkably similar to this thought experiment. Beginning with an undeformed field theory, coupling it to a $T\overline{T}$-like deformation operator drives the flow of the field equations. Classically, the $T\overline{T}$ deformation in two dimensions corresponds to coupling undeformed field theories to flat-space Jackiw-Teitelboim-like gravity. This coupling transforms the original seed theory into a gravitational theory, offering a classical framework to explore the emergence of gravity from field theory. One can refer to \cite{Babaei-Aghbolagh:2024hti} for higher dimensional realization.

Although classical, this process provides valuable insights into how gravity may arise from field theory. For example, a free boson subjected to a $T\overline{T}$ deformation evolves into a non-critical string theory in a static gauge\cite{Cavaglia:2016oda,Bonelli:2018kik}. Upon quantization, this theory inevitably produces gravitons, assuming the exclusion of unstable particles such as tachyons. This highlights the importance of studying $T\overline{T}$-like deformations as they offer a mechanism for generating gravitons.

However, understanding this mechanism at the quantum level {(for higher dimensional generalizations as in \cite{Babaei-Aghbolagh:2024hti})} becomes a significant challenge, motivating a comprehensive review of $T\overline{T}$-like deformations. While numerous technical issues remain resolved, this motivation underscores the critical role of energy-momentum tensor deformations in shedding light on the emergence of gravitational interactions.


{\bf Correlation function in the deformed field theories}

Correlation functions are essential for understanding the properties of both undeformed and deformed theories. They are central in holography and black hole physics, particularly in quantum chaos and entanglement. Out-of-time-order correlation functions and K-complexity, key tools for diagnosing chaotic behavior in deformed field theories, highlight their significance. Similarly, the computation of entanglement entropy relies heavily on these functions. Constructing correlation functions is crucial for exploring the core features of deformed theories. Although $T\overline{T}$ is an irrelevant deformation, typically precluding a well-defined renormalized theory with a removable UV cutoff, the $T\overline{T}$ deformation is a notable exception, as shown in~\cite{Cardy:2019qao}. Analyzing these functions in $T\overline{T}$-deformed theories is essential to understanding their non-local behavior. To consider correlation functions in nonlocal theory like $T\bar{T}$
deformed ones, the operator may be best defined in momentum space and thus is a non-local operator {in coordinates space \cite{Aharony:2023dod, Chakraborty:2023wel} }.


{\bf Celestial holographical dual of the $T\overline{T}$ deformed field theory}

At the classical level, a key observation is that the $T\overline{T}$ deformation of a massless free scalar field theory corresponds to the Nambu-Goto action in static gauge. The inherent non-locality of $T\overline{T}$-deformed theories has been identified in studies of the effective dynamics of long relativistic strings~\cite{Dubovsky:2012wk, Dubovsky:2013ira}. The connection between $T\overline{T}$-deformed theories and string theory has been extensively explored in various works~\cite{Frolov:2019nrr, Frolov:2019xzi, Sfondrini:2019smd, Callebaut:2019omt, Apolo:2019zai, Esper:2021hfq}. Celestial holography provides a framework where quantum scattering amplitudes of massless particles in four-dimensional asymptotically Minkowskian spacetimes are reinterpreted as correlation functions on the celestial sphere at null infinity, defining celestial conformal field theories (CFTs)~\cite{Pasterski:2016qvg, Pasterski:2017kqt, Pasterski:2017ylz}. Based on the construction in~\cite{He:2022zcf}, it can be proposed that a two-dimensional $T\overline{T}$ deformation of celestial CFTs may correspond to a four-dimensional gravitational theory, offering potential insights into the construction of a UV-complete formulation of General Relativity. 

{\bf Duality and symmetry of $T\overline{T}$-like deformed theories}

The recent revival of interest in nonlinear electrodynamics (NED) has been significantly driven by the exploration of $T\overline{T}$-like deformations, particularly through the introduction of Modified Maxwell (ModMax) theory \cite{Bandos:2020jsw}. This theory, along with the discovery that both Born-Infeld and ModMax can be derived from the Maxwell theory via Lagrangian flows involving composite fields \cite{Conti:2018jho, Babaei-Aghbolagh:2020kjg, Ferko:2022iru}, has sparked a renewed focus on NED. A central theme in NED is the incorporation of interaction terms into Maxwell's theory while preserving the $SO(2)$ duality symmetry, which deepens our understanding of these systems. Various methods have been proposed to construct self-dual theories of Maxwell's fields with $SO(2)$ duality symmetry, including the Gaillard-Zumino approach \cite{Gaillard:1981rj, Gaillard:1997rt}, the Hamiltonian method by Henneaux and Teitelboim \cite{HT}, and the PST approach by Pasti, Sorokin, and Tonin \cite{Pasti:1995tn, Pasti:1995us, Pasti:1996vs, Pasti:1997gx}. These methods provide different frameworks for incorporating interaction terms while preserving duality. This review paper is motivated by the need to explore duality and symmetry in NED theories through $T\overline{T}$-like deformations, which offer a novel approach for constructing effective Lagrangians that respect these duality symmetries, as demonstrated in previous studies of $T\overline{T}$-like and marginal root-$T\overline{T}$-like deformations in NED theories \cite{Conti:2018jho, Babaei-Aghbolagh:2022uij} and scalar field theory~\cite{Babaei-Aghbolagh:2025lko}.


This review consolidates the knowledge on the $T\overline{T}$ deformation and its generalized versions, spanning from their origins to recent developments. It aims to serve a broad audience, from students to experts. While several reviews already exist in the literature~\cite{Giveon:2019fgr,Jiang:2019epa, Brizio:2024arr,Monica_Guica_TTbar,apolo:review}, this work focuses on recent theoretical advancements associated with the deformations induced by energy-momentum tensor. We highlight their importance in quantum field theory, holography, and string theory, exploring their implications for spacetime structure, non-locality in QFTs, and the interplay between classical and quantum deformations. By synthesizing these developments, this review offers a helpful resource for researchers involved in the ongoing study of energy-momentum tensor-induced deformations.

The structure of this review is organized as follows. Section~\ref{sec:basic} introduces the fundamental concepts of $T\overline{T}$ deformation, beginning with its definition, key properties, and the formulation of the deformed action. We also discuss various geometrical realizations, such as random geometry, Jackiw-Teitelboim gravity, and the uniform light-cone gauge approach. Section~\ref{quantumaspectTT} delves into the quantum aspects of $T\overline{T}$ deformation, examining its impact on the spectrum, partition function, renormalization, correlation functions, and quantum entanglement. 
In Section~\ref{Section TTbarDeformation and Holography}, we investigate the relationship between $T\overline{T}$ deformation and holography, with a focus on the cutoff AdS/$T\overline{T}$ CFT proposal, holographic partition functions, and entanglement entropy.  Section~\ref{string-TTbar} briefly introduces single-trace $T\overline{T}$ deformation in orbifold CFT and its holographic dual, in contrast to the double-trace version discussed in previous sections. Single-trace deformation is described in three aspects: worldsheet theory, bulk geometry, and boundary field theory. Furthermore, the deformed partition function, correlation functions, and non-locality implications from these quantities are reviewed. Finally, Section~\ref{developments-TTbar} reviews recent developments in generalized $T\overline{T}$-like deformations and provides insights into potential future research directions.

\section{Basics of $T\bar T$ deformation}\label{sec:basic}
This section explains the core concepts underlying the $T\overline{T}$-deformation, beginning with its definition and proceeding to the deformed Lagrangians. We then explore various geometric realizations of the $T\bar{T}$ deformation, including random geometry, Jackiw-Teitelboim gravity, field-dependent coordinate transformations, and the uniform light-cone gauge. These geometric realizations collectively lead to a deeper understanding of how the deformation modifies the theory, making it non-local but solvable and preserving key properties, including integrability and symmetry. By analyzing these methods and studying these core concepts, we establish a foundational framework for understanding the effects and applications of the $T\overline{T}$-deformation in quantum field theory, holography, string theory, and other aspects of theoretical physics. 

\subsection{The $T\bar T$ deformation}
The $T\bar T$ deformation is a solvable irrelevant deformation of two-dimensional quantum field theories. It deforms the theory by the $T\bar T$ operator \cite{Zamolodchikov:2004ce} \footnote{Our convention for the stress energy tensor in Euclidean signature is $T_{\mu\nu} = \frac{2}{\sqrt{|\gamma|}}\frac{\delta S}{\delta \gamma^{\mu\nu}}$.}
{ \begin{equation} \label{TTbar operator definition}
 O_{T\bar T}=-\mathrm{det}\left[T^{\mu}{}_\nu\right]=\frac12\left(T^{\mu\nu}T_{\mu\nu}-T^{\mu}{}_{\mu}T^{\nu}{}_{\nu}\right).
 \end{equation}}
For a CFT in a 2D flat space with complex coordinates $z,\bar{z}$ and a flat local metric $ds^2=dzd\bar{z}$, the energy-momentum tensor is traceless and  the $T\bar T$ operator is reduced to $O_{T\bar{T}} = 4 T_{zz}T_{\bar{z}\bar{z}}=\pi^{-2}T \bar{T}$, justifying the nomenclature. The deformed theory depends on a parameter $\lambda$ of mass-dimension -2. When considering an infinitesimal deformation of a theory from $\lambda$ to $\lambda+\delta \lambda$, the action is deformed as
\begin{equation}
S^{(\lambda+\delta\lambda)}= S^{(\lambda)}+\delta \lambda \int d^2 x \sqrt{\pm\gamma}O^{(\lambda)}_{T\bar T},
\end{equation}
where $\gamma_{\mu\nu}$ denotes the background metric with Euclidean or Lorentzian signature.
Equivalently, the flow for theory can be described by the flow equation:
\begin{equation} \label{TTbar flow action}
    \frac{\partial S^{(\lambda)}}{\partial \lambda}=\int d^2 x\sqrt{\pm\gamma}O^{(\lambda)}_{T\bar T}.
\end{equation}
\footnote{The $T\bar{T}$ deformation parameter $\lambda$ we used is related to other conventions or notations by: $\lambda=-t$ for $t$ in \cite{Cavaglia:2016oda} or $\lambda=-\mu$ for $\mu$ in \cite{Guica:2019nzm}, $\lambda=\frac\mu4$ for $\mu$ in \cite{McGough:2016lol}, $\lambda=\pi^2\lambda_{\text{KLM}}$ where $\lambda_{\text{KLM}}$ is the deformation parameter denoted also by $\lambda$ in \cite{Kraus:2018xrn} and $\lambda=16t$ for $t$ in \cite{Aharony:2018vux}.} 
In the following, we will focus on $T\bar T$ deformation in flat spacetime.

The $T\bar{T}$ deformation is well-defined for two-dimensional quantum field theories in flat spacetime.
The $T\bar{T}$ operator in flat spacetime can be defined by point splitting:
\begin{equation}
 -\lim_{x\rightarrow y}\frac{1}{2}\epsilon_{\mu\nu}\epsilon_{\rho\sigma}T^{\mu\rho}(x)T^{\nu\sigma}(y),
\end{equation}
 where we use Euclidean Cartesian coordinates and the Levi-Civita symbol is defined as $\epsilon^{12}=\epsilon_{12}=1$.
Now we show that the $T\bar{T}$ operator is well-defined up to derivatives, following the derivation in \cite{Zamolodchikov:2004ce}. Taking the derivative, we find
\begin{equation}\label{derittbar}
\begin{split}
 &\frac{\partial}{\partial x^{\alpha}}\left(\frac{1}{2}\epsilon_{\mu\nu}\epsilon_{\rho\sigma}T^{\mu\rho}(x)T^{\nu\sigma}(y)  \right)
 \\
 =&\frac{\partial}{\partial x^{\mu}}\left(\frac{1}{2}\epsilon_{\alpha\nu}\epsilon_{\rho\sigma}T^{\mu\rho}(x)T^{\nu\sigma}(y)   \right)+\frac{\partial}{\partial x^{\nu}}\left(\frac{1}{2}\epsilon_{\mu\alpha}\epsilon_{\rho\sigma}T^{\mu\rho}(x)T^{\nu\sigma}(y)    \right)    \\
 =&\left(\frac{\partial}{\partial x^{\nu}}+\frac{\partial}{\partial y^{\nu}}\right)\left(\frac{1}{2}\epsilon_{\mu\alpha}\epsilon_{\rho\sigma}T^{\mu\rho}(x)T^{\nu\sigma}(y)    \right).
\end{split}
\end{equation}
{The second line is antisymmetric in the indices $\mu$ and $\nu$, so it is proportional to $\epsilon_{\mu\nu}$ and can be derived from the first line by computing the coefficient.
To derive the third line, one can use the conservation law $\partial_\mu T^{\mu\nu}=0$.}
The general form of the OPE of the operator in the bracket in the third line can be written as:
\begin{equation}
 \frac{1}{2}\epsilon_{\mu\alpha}\epsilon_{\rho\sigma}T^{\mu\rho}(x)T^{\nu\sigma}(y) =
 \sum_i c_{\alpha}^{i\nu} (x-y)\mathcal{O}_i(y).
\end{equation}
Therefore the OPE of (\ref{derittbar}) takes the form
\begin{equation}\label{TTbarOPE1}
\frac{\partial}{\partial x^{\alpha}}\left(\frac{1}{2}\epsilon_{\mu\nu}\epsilon_{\rho\sigma}T^{\mu\rho}(x)T^{\nu\sigma}(y)  \right)  
=\sum_i c_{\alpha}^{i\nu} (x-y)\frac{\partial}{\partial y^{\nu}}\mathcal{O}_i(y).
\end{equation}
Consider the OPE of the point-splitting operator:
\begin{equation}
\frac{1}{2}\epsilon_{\mu\nu}\epsilon_{\rho\sigma}T^{\mu\rho}(x)T^{\nu\sigma}(y) =   \sum_i C^{i} (x-y)\mathcal{O}_i(y).
\end{equation}
Taking the derivative, we find
\begin{equation}\label{TTbarOPE2}
\frac{\partial}{\partial x^{\alpha}}
\left(
\frac{1}{2}\epsilon_{\mu\nu}\epsilon_{\rho\sigma}T^{\mu\rho}(x)T^{\nu\sigma}(y) \right)=   \sum_i \frac{\partial}{\partial x^{\alpha}}C^{i} (x-y)\mathcal{O}_i(y)
.
\end{equation}
Comparing (\ref{TTbarOPE1}) with (\ref{TTbarOPE2}), one finds that any operator $\mathcal{O}_i$ appearing in the OPE should be a derivative of some other
operators or come with a constant coefficient $C^i$.
Therefore, the $T\bar T$ operator can be defined  up to total derivatives as
\begin{equation}\label{TTderivative}
 -\frac{1}{2}\epsilon_{\mu\nu}\epsilon_{\rho\sigma}T^{\mu\rho}(x)T^{\nu\sigma}(y)=O_{T\bar T}(y)+\mathrm{derivatives}, 
\end{equation}
and the $T\bar T$ deformation is well-defined because the Lagrangian is defined up to total derivatives.
Computing the vacuum expectation value of (\ref{TTderivative}), we find
\begin{equation}
 \langle
 O_{T\bar T}(y)
 \rangle
 =
\langle
T_{12}(x)T_{12}(y)
 \rangle
 -\frac{1}{2}\langle
T_{11}(x)T_{22}(y)
\rangle
-\frac{1}{2}\langle
T_{22}(x)T_{11}(y)
 \rangle.
\end{equation}
The vacuum expectation values of the total derivatives vanish because the theory is invariant under translation. 
Taking $x$ and $y$ infinitely separated from each other, we obtain the factorization formula
\begin{equation}
 \langle
 O_{T\bar T}(y)
 \rangle
 =
\langle
T_{12}(x)
\rangle
\langle
T_{12}(y)
 \rangle
 -\frac{1}{2}\langle
T_{11}(x)
\rangle
\langle
T_{22}(y)
\rangle
-\frac{1}{2}\langle
T_{22}(x)
\rangle
\langle
T_{11}(y)
 \rangle.
\end{equation}
The factorization formula is also valid for non-degenerate eigenstates of the momentum and energy operators \cite{Zamolodchikov:2004ce}.
In the rest of this section, we will focus on the classical aspects of the $T\bar T$ deformation.

$T\bar{T}$ deformation is also considered for curved spacetimes where the existence of the curvature and lack of symmetry can complicate the definition of quantum $T\bar{T}$ operator as what we did for the flat plane. In particular, the expectation value of the $T\bar{T}$ operator was studied in \cite{Jiang:2019tcq}, showing the factorization fails without assuming a large $c$ { (central charge)} limit. Partition function was studied in \cite{Mazenc:2019cfg} by computing path integrals of the 3D gravity theory corresponding to the $T\bar{T}$ deformation. Many geometric realizations of the $T\bar{T}$ deformation to be discussed in this section can be generalized to curved space more or less straightforwardly.

\subsection{Deformed Lagrangians} \label{Subsection Deformed Lagrangians}
In principle (\ref{TTbar flow action}) determines the flow of the action under the $T\bar{T}$ deformation, but in general it's extremely difficult to solve the flow with the stress tensor defined as the functional derivative of the action with respect to the metric. When a Lagrangian describes the theory, the flow equation is reduced to
\begin{equation}\label{TTbar flow Lagrangian}
 \frac{\partial \mathcal{L}}{\partial \lambda}= \frac12\left(T^{\mu\nu}T_{\mu\nu}-T^{\mu}{}_{\mu}T^{\nu}{}_{\nu}\right),
\end{equation}
When the deformed Lagrangian depends algebraically on the metric, we have $T_{\mu\nu}=2\frac{\partial \mathcal{L}}{\partial \gamma^{\mu\nu}}-\gamma_{\mu\nu}\mathcal{L}$ and the flow equation of the Lagrangian takes the form
\begin{align} \label{TTbar flow Lagrangian depends on metric algebraically}
 \frac{\partial \mathcal{L}}{\partial \lambda} = 2 \gamma^{\mu\rho}\gamma^{\nu\sigma}\frac{\partial\mathcal{L}}{\partial \gamma^{\mu\nu}}\frac{\partial\mathcal{L}}{\partial \gamma^{\rho\sigma}} - 2 (\gamma^{\mu\nu}\frac{\partial\mathcal{L}}{\partial \gamma^{\mu\nu}})^2 + 2\mathcal{L}\gamma^{\mu\nu}\frac{\partial\mathcal{L}}{\partial \gamma^{\mu\nu}} - \mathcal{L}^2.
\end{align}
If we further assume the Lagrangian depends on fields through scalar invariants of the fields and their derivatives, then the flow equation can be reduced to a partial differential equation, which in many cases can be solved by the ``method of characteristics'' with the initial condition given by the undeformed theory $\mathcal{L}(\lambda=0)$. We will devote section \ref{sectionchar} to the method of characteristics. Now, for simplicity, we focus on some simple examples in flat spacetimes in the Euclidean signature.

One method to solve the flow equation (\ref{TTbar flow Lagrangian}) is to consider the small $\lambda$ expansion of the Lagrangian:
\begin{equation}\label{expandL}
\mathcal{L}(\lambda)=\sum_{k=0}^{\infty} \lambda^k  \mathcal{L}_k. 
\end{equation}
The energy-momentum tensors can also be expanded as
\begin{equation}\label{expandT}
T^{\mu\nu}(\lambda)=\sum_{k=0}^{\infty} \lambda^k T^{\mu\nu}_k. 
\end{equation}
When $\mathcal{L}(\lambda)$ up to order $\lambda^n$ is known, one can get $T^{\mu\nu}(\lambda)$ up to the same order. Therefore, $\mathcal{L}_k$ can be solved recursively by substituting the expansions (\ref{expandL}) and (\ref{expandT}) into the flow equation. This method has been applied to compute deformed scalar Lagrangians in \cite{Cavaglia:2016oda}.

One can use Lorentz symmetries to simplify the flow equation for relativistic theory. This method has been used in \cite{Kraus:2018xrn, Bonelli:2018kik}. Now, we consider the deformed theory of a free scalar $\phi$ as an example. Lorentz symmetries require that the deformed Lagrangian depend on $\phi$ only through the Lorentz invariant variable $X=\eta^{\mu\nu}\partial_\mu\phi \partial_\nu\phi$:
\begin{equation}\label{ansatzscalar}
\mathcal{L}=\mathcal{L}(\lambda, X).
\end{equation}
In this simple case, it is equivalent to using the canonical energy-momentum tensor
\begin{equation}\label{ansatzT}
T^{\mu\nu}=\frac{\partial\mathcal{L}}{\partial(\partial_\mu\phi)}\partial^\nu\phi-\eta^{\mu\nu}\mathcal{L}=2 \partial_X \mathcal{L} \partial^\mu\phi\partial^\nu\phi-\eta^{\mu\nu}\mathcal{L}.
\end{equation}
Substituting (\ref{ansatzscalar}) and (\ref{ansatzT}) into the flow equation (\ref{TTbar flow Lagrangian}) we get
\begin{equation}
\partial_\lambda\mathcal{L}=X \partial_X \mathcal{L}^2-\mathcal{L}^2 ,
\end{equation}
which is simply the Burgers' equation. The initial condition is given by
\begin{equation}
\mathcal{L}(0, X)=\frac{1}{2} X.
\end{equation}
Solving the Burgers' equation with the  initial condition, we obtained the deformed Lagrangian:
\begin{equation}
\mathcal{L}(\lambda, X)=\frac{1-\sqrt{1-2\lambda X}}{2\lambda}.
\end{equation}
The deformed Lagrangian of a free massless scalar can be interpreted as a Nambu-Goto action in $\mathbb{R}^3$ in the static gauge, which reflects the non-locality of the 
$T \bar T$ deformation. {Geometrization as a String Worldsheet Theory:
The equivalence between the $T \bar T$-deformed theory and the Nambu-Goto action reflects a duality where non-locality in the original field theory is "geometrized" into the dynamics of a string. While the Nambu-Goto action is local on the worldsheet, its interpretation as a string embedding in higher-dimensional space ties the scalar fields to non-local degrees of freedom (e.g., transverse string coordinates). This duality masks the non-locality of the deformation in the original field variables, as the string formulation inherently resums an infinite series of higher-derivative terms into a geometric, locally written action. Typically, the square-root structure of the Nambu-Goto Lagrangian, when expanded, implicitly includes an infinite tower of derivative terms. These terms, if truncated, would represent non-local interactions in the original scalar theory.} In the context of string theory, \cite{Blair:2020ops} interprets the $T\bar T$ flow from free bosons to the Nambu-Goto action as the reverse of a non-relativistic string theory limit, where the target space becomes non-relativistic while retaining a relativistic worldsheet. For a more comprehensive discussion on non-relativistic string theory, see section 7.4.6 in \cite{Demulder:2023bux}. Furthermore, \cite{Blair:2024aqz} extends the findings of \cite{Blair:2020ops}  by exploring non-relativistic brane limits and their application to higher-dimensional generalizations of $T\bar T$ deformation.

The closed form of many $ T\bar T$ deformed theories, including interacting scalar fields, non-linear $\sigma$ models, WZW models, and the massive Thirring model, can be found in \cite{Bonelli:2018kik}. 
Especially in fermion theories, the expansions of deformed Lagrangians in the deformation parameter usually terminate.
$T \bar T$ deformed fermions are further explored in \cite{Lee:2021iut,Dey:2021jyl} and supersymmetric theories in \cite{Baggio:2018rpv,Chang:2018dge,Jiang:2019hux,Chang:2019kiu,Coleman:2019dvf,Jiang:2019trm,Ferko:2019oyv,Ebert:2020tuy,Ferko:2021loo,Lee:2023uxj}.
When there are multiple Lorentz invariants in the theories, the flow equations for the Lagrangians are more complicated, and one often resorts to the method of characteristics, which has also been applied to study stress energy deformations beyond $T\bar{T}$ \cite{Hou:2022csf}.

There are other methods to derive the deformed action based on geometric realizations of the $T\bar{T}$ deformation. One is the field-dependent coordinate transformation \cite{Dubovsky:2017cnj, Conti:2018tca, Coleman:2019dvf, Caputa:2020lpa}, from which solutions \cite{Conti:2018tca,Ceschin:2020jto} and integrals of motion \cite{Conti:2019dxg} can also be derived. Another powerful method is the uniform light-cone gauge approach proposed in \cite{Frolov:2019nrr, Frolov:2019xzi}.

If the seed theory has a traceless energy-momentum tensor, following the flow equation of the Lagrangian (\ref{TTbar flow Lagrangian depends on metric algebraically}), we have the trace relation
\begin{align} \label{Classical trace relation}
    T^\mu_\mu = 2 \lambda O_{T\bar{T}},
\end{align}
or in terms of the Lagrangian
\begin{align} \label{Classical trace relation Lagrangian}
    \lambda\partial_\lambda \mathcal{L} - \gamma^{\mu\nu}\frac{\partial \mathcal{L}}{\partial \gamma^{\mu\nu}} + \mathcal{L} = 0.
\end{align}
To prove the trace relation, we differentiate the left-hand side with respect to $\lambda$, using (\ref{TTbar flow Lagrangian depends on metric algebraically}) we obtain
\begin{align}
  \partial_\lambda \big( \lambda\partial_\lambda \mathcal{L} - \gamma^{\alpha\beta}\frac{\partial \mathcal{L}}{\partial \gamma^{\alpha\beta}} + \mathcal{L} \big) = &4 \gamma^{\mu\rho}\gamma^{\nu\sigma} \frac{\partial}{\partial \gamma^{\mu\nu}} \big( \lambda\partial_\lambda \mathcal{L} - \gamma^{\alpha\beta}\frac{\partial \mathcal{L}}{\partial \gamma^{\alpha\beta}} + \mathcal{L} \big)\frac{\partial\mathcal{L}}{\partial \gamma^{\rho\sigma}} \nonumber\\
  &- 4 \gamma^{\mu\nu}\frac{\partial}{\partial \gamma^{\mu\nu}}\big( \lambda\partial_\lambda \mathcal{L} - \gamma^{\alpha\beta}\frac{\partial \mathcal{L}}{\partial \gamma^{\alpha\beta}} + \mathcal{L} \big) \gamma^{\rho\sigma}\frac{\partial\mathcal{L}}{\partial \gamma^{\rho\sigma}} \nonumber\\
  &+ 2\big( \lambda\partial_\lambda \mathcal{L} - \gamma^{\alpha\beta}\frac{\partial \mathcal{L}}{\partial \gamma^{\alpha\beta}} + \mathcal{L} \big)\gamma^{\mu\nu}\frac{\partial\mathcal{L}}{\partial \gamma^{\mu\nu}} \nonumber\\
  &+ 2\mathcal{L}\gamma^{\mu\nu}\frac{\partial}{\partial \gamma^{\mu\nu}}\big( \lambda\partial_\lambda \mathcal{L} - \gamma^{\alpha\beta}\frac{\partial \mathcal{L}}{\partial \gamma^{\alpha\beta}} + \mathcal{L} \big) \nonumber\\
  &- 2\mathcal{L}\big( \lambda\partial_\lambda \mathcal{L} - \gamma^{\alpha\beta}\frac{\partial \mathcal{L}}{\partial \gamma^{\alpha\beta}} + \mathcal{L} \big). 
\end{align}
(\ref{Classical trace relation Lagrangian}) holds at $\lambda=0$ by the assumption that the seed theory has a traceless energy-momentum tensor, then by the first-order PDE above, the relation holds along the $T\bar{T}$ flow. Our proof assumed the Lagrangian depends on the metric algebraically, a more general proof based on the gravity formulation of $T\bar{T}$ deformation can be found in \cite{Tolley:2019nmm}.


\subsection{The method of characteristics}\label{sectionchar}
This section is devoted to the method of characteristics.\footnote{One can refer to chapter 3 of \cite{evans2010partial}.} Although a bit of a digression, it's important for the derivation of deformed classical actions, deformed energy levels and the large $c$ $T\bar{T}$ flow equations. The method of characteristics is most clearly motivated by the quasilinear partial differential equations
{ \begin{align} \label{Quasilinear first order PDE}
    \vec{a}(\vec{x},u)\cdot \partial_{\vec{x}} u = b(\vec{x},u).
\end{align}}
Suppose the integral surface is given by a graph
\begin{align}
    f(\vec{x},u) = z(\vec{x}) - u = 0.
\end{align}
We have { $df(\vec{a}\cdot\partial_{\vec{x}}+b\partial_u)=0$} on the integral surface. So the vector field { $\vec{a}\cdot\vec{\nabla}+b\partial_u$} is tangent to any integral surface, and integral surfaces are unions of integral curves along the vector field, a.k.a. characteristic curves. Explicitly, characteristic curves are obtained by solving the ODE
{ \be\ba 
\frac{d\vec{x}}{ds} &= \vec{a}(\vec{x},u),\\
    \frac{du}{ds} &= b(\vec{x},u).
\ea\ee}
{  Here $s$ is the parameter of the characteristic curves.} Initial conditions specified as a codimension two surface $\Gamma$ grows into an integral surface by flowing along characteristic curves. If the PDE is nonlinear in the partial derivatives, say, given by a generic function
\begin{align} \label{Nonliear first order PDE}
F(\vec{x},u,\partial_{\vec{x}}u) = 0,
\end{align}
we no longer have the simple geometrical picture described above. However, we can naturally promote $\partial_{\vec{x}}u$ to independent variables denoted by $\vec{p}$ with the constraints $\partial_{\vec{x}}u=\vec{p}$ to hold on an integral surface, now a codimension $n+1$ hypersurface. We want to find a vector field tangent to any integral surface. For a vector field to be tangent to an integral surface, we must have
\be\ba 
\frac{\partial F}{\partial \vec{x}}\cdot \frac{d\vec{x}}{ds} + \frac{\partial F}{\partial u}\frac{du}{ds} + \frac{\partial F}{\partial \vec{p}}\cdot \frac{d\vec{p}}{ds} &= 0 , \\
    \frac{du}{ds} - \vec{p}\cdot \frac{d\vec{x}}{ds} &= 0 ,\\
    \frac{\partial \vec{p}}{\partial \vec{x}} \frac{d\vec{x}}{ds} - \frac{d\vec{p}}{ds} &= 0
\ea\ee
with the first term of the last equation depending on the particular integral surface. To get rid of this term, we differentiate (\ref{Nonliear first order PDE}) with respect to $\vec{x}$ for an integral surface
\begin{align}
    \frac{\partial F}{\partial \vec{x}} + \frac{\partial F}{\partial u} \vec{p} + \frac{\partial F}{\partial \vec{p}} \frac{\partial \vec{p}}{\partial \vec{x}} = 0.
\end{align}
We find the dependence on the particular integral surface can be eliminated if we set $\frac{d\vec{x}}{ds} = \frac{\partial F}{\partial \vec{p}}$. Then, we obtain the equation for characteristic curves of nonlinear first-order PDEs
\be\ba 
\frac{d\vec{x}}{ds} &= \frac{\partial F}{\partial \vec{p}} ,\\
    \frac{du}{ds} &= \vec{p}\cdot\frac{\partial F}{\partial \vec{p}} ,\\
\frac{d\vec{p}}{ds} &= -\frac{\partial F}{\partial\vec{x}} - \frac{\partial F}{\partial u}\vec{p}.
\ea\ee 
In the context of $T\bar{T}$ deformation of classical Lagrangian, the flow equation (\ref{TTbar flow Lagrangian}) is a PDE in which $\frac{\partial\mathcal{L}}{\partial \lambda}$ appears on the left-hand side and nowhere else. So we have $\frac{d\lambda}{ds}=1$ in the characteristic equation. Therefore, the $T\bar{T}$ deformation parameter $\lambda$ can be taken as the parameter of the characteristic curves. Initial conditions are given by $\lambda=0$ hypersurface, determined by the Lagrangian of the seed theory. The $T\bar{T}$ flow is the flow along characteristic curves \cite{Hou:2022csf}.

Taking the interacting scalar for example,
\begin{align}
    \mathcal{L}^{(0)} = \frac{1}{2}g^{\mu\nu}\partial_\mu\phi \partial_\nu\phi + V(\phi).
\end{align}
We assume the deformed Lagrangian depends algebraically on $\phi$, the scalar invariant $X=g^{\mu\nu}\partial_\mu\phi\partial_\nu\phi$ and, of course, the deformation parameter $\lambda$. Then the $T\bar{T}$ flow equation for the Lagrangian (\ref{TTbar flow Lagrangian depends on metric algebraically}) is reduced to
\begin{align} \label{Interacting boson TTbar flow PDE}
    \frac{\partial\mathcal{L}}{\partial\lambda} = 2\mathcal{L} X\frac{\partial\mathcal{L}}{\partial X} - \mathcal{L}^2
\end{align}
with the initial condition being
\begin{align}
    \mathcal{L}(\lambda=0) = \mathcal{L}^{(0)} = \frac{1}{2}X + V(\phi).
\end{align}
Using the method of characteristics, we define
\begin{align}
    p_\lambda = \frac{\partial \mathcal{L}}{\partial \lambda}, \;
    p_X = \frac{\partial \mathcal{L}}{\partial X}, \;
    p_\phi = \frac{\partial \mathcal{L}}{\partial \phi}
\end{align}
and rewrite (\ref{Interacting boson TTbar flow PDE}) as
\begin{align} 
F(\lambda,X,\phi,\mathcal{L},p_\lambda,p_X,p_\phi) = p_\lambda - 2\mathcal{L}X p_X + \mathcal{L}^2 = 0.
\end{align}
The equation for characteristic curves now takes the form
\begin{align}
    \frac{d\lambda}{ds} &= 1,\; \frac{dX}{ds} = -2\mathcal{L}X,\; \frac{d\phi}{ds} = 0 \nonumber\\
    \frac{d\mathcal{L}}{ds} &= p_\lambda - 2\mathcal{L}X p_X = -\mathcal{L}^2 \nonumber\\
    \frac{d p_\lambda}{ds} &= 2(X p_X - \mathcal{L}) p_\lambda,\; \frac{d p_X}{ds} = 2X p_X,\; \frac{d p_\phi}{ds} = 2(X p_X - \mathcal{L}) p_\phi
\end{align}
From the first equation, we set $s=\lambda$. Integrating the fourth equation, we find
\begin{align}
    \mathcal{L} = \frac{\mathcal{L}_0}{1+\lambda\mathcal{L}_0}.
\end{align}
Plugging into the second equation, we obtain
\begin{align}
    X = \frac{X_0}{(1+\lambda \mathcal{L}_0)^2}.
\end{align}
And by the third equation, we have
\begin{align}
    \phi = \phi_0.
\end{align}
For the seed theory we have $\mathcal{L}_0 = \frac{1}{2}X_0 + V(\phi_0)$. Substituting $X_0,\phi_0$ by $X,\phi$, we finally find the deformed Lagrangian
\begin{align}
    \mathcal{L}(\lambda,X,\phi) = \frac{1+2\lambda V(\phi)-\sqrt{1-2\lambda X(1+\lambda V(\phi))}}{2\lambda(1+\lambda V(\phi))}.
\end{align}
This was obtained as a series in \cite{Cavaglia:2016oda} and later in a closed form in \cite{Bonelli:2018kik}, which simplifies to our expression.

\subsection{Geometric realization of $T\bar{T}$ deformation}
\subsubsection*{Random geometry}
Remarkable geometric interpretations exist for the $T\Bar{T}$ deformation. 
It has been shown in \cite{Cardy:2018sdv} that $T\Bar{T}$ deformation can be interpreted as coupling the original theory to a random geometry.
The infinitesimal deformation of the action is:
\begin{equation}
S^{(\lambda+\delta\lambda)}= S^{(\lambda)}-\frac{1}{2}\delta \lambda \int d^2 x   \epsilon_{\mu\nu}\epsilon_{\rho\sigma}T^{\mu\rho}T^{\nu\sigma}.
\end{equation}
In the path integral formulation, one can use the Hubbard-Stratonovich transformation to rewrite the perturbation term as:
\begin{equation}
e^{ \frac{\delta\lambda}{2} \int d^2 x   \epsilon_{\mu\nu}\epsilon_{\rho\sigma}T^{\mu\rho}T^{\nu\sigma}}
=\int \mathcal{D}h\,
e^{-\frac{1}{2\delta\lambda} \int   \epsilon^{\mu\nu}\epsilon^{\rho\sigma}h_{\mu\rho}h_{\nu\sigma }d^2 x -\int h_{\mu\nu} T^{\mu\nu} d^2 x} .
\end{equation}
The saddle point of $h_{\mu\nu}$ is of order $O(\delta \lambda)$:
\begin{equation}\label{saddleh}
  h_{\mu\nu}^*= -\delta \lambda 
  \epsilon_{\mu\rho}\epsilon_{\nu\sigma}
  T^{\rho\sigma}=\delta \lambda (T_{\mu\nu}-\delta_{\mu\nu}T^{\rho}_{\rho}).
\end{equation}
According to the definition of the energy-momentum tensor, the last term can be viewed as an infinitesimal change in the metric $\gamma_{\mu\nu}=\delta_{\mu\nu}+h_{\mu\nu}$.
The path integral of $h_{\mu\nu}$ appears to encompass all possible metrics infinitesimally close to $\delta_{\mu\nu}$.
But we will show that one can restrict to flat metrics.
The infinitesimal variation of the Ricci scalar can be computed as:
\begin{equation}
    \delta R=\partial_\mu\partial_\nu h^{\mu\nu}-\partial^2 h^\mu_\mu.
\end{equation}
At the saddle point (\ref{saddleh}), we find
\begin{equation}
    \delta R=\partial_\mu\partial_\nu h^{*\mu\nu}-\partial^2 h^{*\mu}_{\mu}=0,
\end{equation}
where we use the conservation of energy-momentum tensor $\partial_\mu T^{\mu\nu}$.
Therefore, the metric is flat at the saddle point, and the variation can be restricted to be an infinitesimal diffeomorphism:
\begin{equation} 
  h_{\mu\nu}^*= \partial_\mu \alpha_\nu+\partial_\nu \alpha_\mu.
\end{equation}

Flow equations for partition functions on a torus, a finite cylinder, and connected domains with boundaries can be derived using the random geometry interpretation.

\subsubsection*{{$T\bar{T}$ Deformation as Coupling to Generalized Jackiw--Teitelboim Gravity}}
A particularly insightful approach reinterprets the $T\bar{T}$-deformed theory as a conventional matter theory coupled to a two-dimensional dilaton gravity—namely, generalized Jackiw--Teitelboim (JT) gravity with an arbitrary dilaton potential. This correspondence clarifies the deformation's origin and explains its solvable structure through the dynamics of an auxiliary gravitational sector.

We begin by considering the 2-form curvature $R_{ab}$, expressed in terms of the Riemann curvature tensor $R_{abcd}$ and the zweibein $e^a$. The relationship is given as:
\begin{equation}
\begin{split}
R_{ab} &= \frac{1}{2}R_{abcd} e^c \wedge e^d \\
&= \frac{1}{2} R e_a \wedge e_b \\
&= \frac{1}{2} R e_{a\mu} e_{b\nu} dx^\mu \wedge dx^\nu,
\end{split}
\end{equation}
where we used the 2D identity $R_{abcd} = \frac{R}{2}(\delta_{ac}\delta_{bd} - \delta_{ad}\delta_{bc})$. This formulation shows the curvature expressed in terms of the metric and zwiebein components. 
Comparing the two expressions for $R_{ab}$, we derive the fundamental relation:
\begin{equation}
    R e_{a\mu} e_{b\nu} = \epsilon_{ab} \partial_{[\mu} \omega_{\nu]},
\end{equation}
where $\omega_\mu$ is the spin connection and brackets denote antisymmetrization: $\partial_{[\mu}\omega_{\nu]} \equiv \frac{1}{2}(\partial_\mu\omega_\nu - \partial_\nu\omega_\mu)$. 
Further, we obtain the relation in determinant form:
\begin{equation}
    R \cdot \frac{1}{2} \epsilon^{\mu\nu} \epsilon^{ab} e_{a\mu} e_{b\nu} = \frac{1}{2} \epsilon^{\mu\nu} \epsilon^{ab} \epsilon_{ab} (\partial_\mu \omega_\nu - \partial_\nu \omega_\mu),
\end{equation}
which simplifies to:
\begin{equation}
    R \, \det(e) = \epsilon^{\mu\nu} (\partial_\mu \omega_\nu - \partial_\nu \omega_\mu).
\end{equation}
Finally, we find the expression of $eR$ in terms of the spin connection $\omega_\mu$:
\begin{equation}
\begin{split}
e R &= \epsilon^{\mu\nu} (\partial_\mu \omega_\nu - \partial_\nu \omega_\mu) \\
&= \epsilon^{\mu\nu} \partial_\mu \omega_\nu - \epsilon^{\mu\nu} \partial_\nu \omega_\mu \\
&= 2 \epsilon^{\mu\nu} \partial_\mu \omega_\nu.
\end{split}
\end{equation}

We then consider a coupled system of a conformal matter field and a generalized JT gravity theory~\cite{Kim:2020eqn}. The total action is written as
\begin{equation}
S_{\text{tot}} = S_{\text{GJT}} + S_m(e^a_\mu, \Psi),
\end{equation}
where $S_m$ is the matter action depending on the zweibein $e^a_\mu$ and matter field $\Psi$, and $S_{\text{GJT}}$ is the generalized JT gravity action with arbitrary dilaton potential $W(\phi)$. In the second-order formalism, the gravitational part is given by
\begin{equation}
S_{\text{GJT}} = \int d^2x\, \sqrt{g} \left( \phi R + W(\phi) \right),
\end{equation}
where $\phi$ is the dilaton field and $R$ the Ricci scalar. To make the connection with $T\bar{T}$ deformation explicit, we transition to the first-order formalism and express the metric as $g_{\mu\nu} = e^a_\mu e^b_\nu \delta_{ab}$, so that $\sqrt{g} = \det e \equiv e$.

In this formalism, the curvature scalar becomes $e R = 2 \epsilon^{\mu\nu} \partial_\mu \omega_\nu$, where $\omega_\mu$ is the spin connection. Substituting into the gravitational action yields
\begin{equation}
S_{\text{GJT}} = \int d^2x\, \left( 2 \epsilon^{\mu\nu} \phi \partial_\mu \omega_\nu + e W(\phi) \right).
\end{equation}
To ensure the compatibility condition between the zweibein and the spin connection (metricity), we introduce a Lagrange multiplier $\sigma^a$ that enforces
\begin{equation}
\partial_\mu e^a_\nu + \omega_\mu \epsilon^a_{\ b} e^b_\nu = 0.
\end{equation}
This leads to the total gravitational action in first-order form:
\begin{equation}
S_{\text{GJT}} = \int d^2x\, \epsilon^{\mu\nu} \left( 2\phi \partial_\mu \omega_\nu + \frac{W(\phi)}{2} \epsilon_{ab} e^a_\mu e^b_\nu - \sigma_a (\partial_\mu e^a_\nu + \omega_\mu \epsilon^a_{\ b} e^b_\nu) \right).
\end{equation}

The equation of motion for $\phi$ gives $R + \partial_\phi W(\phi) = 0$, which constrains the scalar curvature. Using this, we rewrite the gravitational action by eliminating the curvature in favor of the dilaton. This leads to a reformulated action containing the function
\begin{equation}
f(\phi) = W(\phi) - \phi \partial_\phi W(\phi),
\end{equation}
which acts as the local deformation parameter in the $T\bar{T}$ language. The full action becomes
\begin{equation}
S_{\text{tot}} = -\frac{1}{16\pi G_N} \int d^2x\, \epsilon^{\mu\nu} \left( \frac{f(\phi)}{2} \epsilon_{ab} e^a_\mu e^b_\nu + D_\mu \sigma^a e^a_\nu \right) + S_m(e^a_\mu, \Psi),
\end{equation}
where $D_\mu \sigma^a = \partial_\mu \sigma^a + \omega_\mu \epsilon^a_{\ b} \sigma^b$ is the covariant derivative.

To proceed further, we define the auxiliary fields $X^a = \epsilon^{ab} \sigma_b$, and introduce a modified zweibein $\tilde{e}^a_\mu$ via
\begin{equation}
\tilde{e}^a_\mu = D_\mu X^a + (\partial_\mu \log f) X^a.
\end{equation}
This new zweibein captures the geometry induced by the deformation. Using it, the gravitational part of the action becomes a quadratic expression in the difference $(e^a_\mu - \tilde{e}^a_\mu)$:
\begin{equation}
S_{\text{GJT}} = -\frac{1}{16\pi G_N} \int d^2x\, \frac{f}{2} \epsilon^{\mu\nu} \epsilon_{ab} (e^a_\mu - \tilde{e}^a_\mu)(e^b_\nu - \tilde{e}^b_\nu) + \frac{f}{2} \epsilon^{\mu\nu} \epsilon_{ab} \tilde{e}^a_\mu \tilde{e}^b_\nu.
\end{equation}

Variation of the action with respect to $\tilde{e}^a_\mu$ defines a stress-energy tensor $\tilde{T}^\mu_{\ a}$ in the tilded geometry:
\begin{equation}
\tilde{T}^\mu_{\ a} = \frac{\delta S_{\text{tot}}}{\delta \tilde{e}^a_\mu} = \frac{f}{16\pi G_N} \epsilon^{\mu\nu} \epsilon_{ab} e^b_\nu.
\end{equation}
Meanwhile, the equation of motion for the original zweibein implies a relation between the deformed and undeformed geometries:
\begin{equation}
e^a_\mu - \tilde{e}^a_\mu = \frac{16\pi G_N}{f} \epsilon^{\mu\nu} \epsilon^a_{\ b} T_\nu^{\ b},
\end{equation}
which shows how the deformation shifts the spacetime background in proportion to the matter stress-energy tensor.

Using this relation, the action can be fully expressed in terms of the tilded fields:
\begin{equation}
S_{\text{tot}} = \int d^2x \left( \frac{16\pi G_N}{2f} \epsilon^{\mu\nu} \epsilon_{ab} \tilde{T}^\mu_{\ a} \tilde{T}^\nu_{\ b} \right) + S_m(\tilde{e}^a_\mu, \Psi).
\end{equation}
Finally, identifying the local deformation parameter as
\begin{equation}
\lambda(x) = \frac{16\pi G_N}{f(\phi(x))},
\end{equation}
we recover the familiar $T\bar{T}$-deformed matter theory in the form:
\begin{equation}
S_{T\bar{T}} = \int d^2x\, \lambda(x) \det \tilde{T} + S_m(\tilde{e}^a_\mu, \Psi),
\end{equation}
with the $T\bar{T}$ operator realized through the geometric coupling.

This construction illustrates that the $T\bar{T}$ deformation of a matter theory corresponds to coupling it to generalized JT gravity, where the spacetime deformation is dynamically encoded in the dilaton potential $W(\phi)$. The deformation parameter is no longer a constant but a local function determined by the gravitational background, giving rise to a geometrically enriched and locally modulated version of the $T\bar{T}$ flow.

\subsubsection*{Field-dependent coordinate transformation} \label{Subsection Field-dependent coordinate transformation}

The deformed theory is related to the original theory by a field-dependent coordinate transformation \cite{Dubovsky:2017cnj, Conti:2018tca, Caputa:2020lpa}.
The deformed theory can be reformulated as
\begin{equation}
    S_\lambda=\int {d^2 w} \mathcal{L}_\lambda
=\int\frac{d^2 w}{\det{\partial_{x^i} w^j}}(\mathcal{L}_0(\phi(x(w)))+\lambda \det((T^{(\lambda=0)})^\mu_\nu)),
\end{equation}
where $x$ and $w$ are coordinates for the original and deformed theory, respectively. They are related by a field-dependent coordinate transformation \begin{equation}
 dw^\mu=dx^\mu-\lambda \epsilon^{\mu\beta}\epsilon_{\alpha\nu}T^{\alpha}_{~\beta}  dx^\nu .
\end{equation}

Now, we show how to use this approach to derive the deformed action using the example of a scalar with an arbitrary potential.
The undeformed action is given by
\begin{equation}
S_0=\int d^2x \mathcal{L}_0
=\int dx^+ dx^-\big(-\partial_{x^+} \phi \partial_{x^-} \phi-V(\phi) \big).
\end{equation} 
where the light-cone coordinates are defined as $x^\pm=(x\pm t)/\sqrt{2}$.
The canonical energy-momentum tensor is
\begin{equation}
T^\mu_{~\nu}=\left(\begin{array}{cc}
V&- \partial_{x^-} \phi \partial_{x^-} \phi \\ 
-\partial_{x^+} \phi \partial_{x^+} \phi&V 
\end{array} \right).
\end{equation}

We defined a field-dependent coordinate transformation with the Jacobian 
\begin{equation}
\left(\begin{array}{cc}
\partial_{x^+}w^+&\partial_{x^+} w^- \\ 
\partial_{x^-}w^+&\partial_{x^-} w^-
\end{array} \right)=\left(\begin{array}{cc}
1- \lambda T^{x^-}_{~x^-}& \lambda T^{x^-}_{~x^+}  \\ 
 \lambda T^{x^+}_{~x^-} &1- \lambda T^{x^+}_{~x^+}
\end{array} \right).
\end{equation}
Solving
\begin{equation}
 \left(\begin{array}{c}
\partial_{x^+}\phi\\ 
\partial_{x^-}\phi
\end{array} \right)=\left(\begin{array}{cc}
\partial_{x^+}w^+&\partial_{x^+} w^- \\ 
\partial_{x^-}w^+&\partial_{x^-} w^-
\end{array} \right)
 \left(\begin{array}{c}
\partial_{w^+}\phi\\ 
\partial_{w^-}\phi
\end{array} \right),
\end{equation}
we get
\begin{equation}
\partial_{x^+}\phi=
\frac{\Omega-1}{2\lambda\partial_{w^-}\phi},~~~
\partial_{x^-}\phi=\frac{\Omega-1}{2\lambda\partial_{w^+}\phi}.
\end{equation}
where
\begin{equation}
\Omega=\sqrt{1+4\lambda \partial_{w^+}\phi\partial_{w^-}\phi (1-\lambda V) }.
\end{equation}
The deformed action is given by 
\begin{equation}\begin{split}\label{defscalar}
S_\lambda=&\int {d^2 w} \mathcal{L}_\lambda
=\int\frac{d^2 w}{\det{\partial_{x^i} w^j}}(\mathcal{L}_0(\phi(x(w)))+\lambda \det(T))\\
=&\int d^2 w \left(  
\frac{1-\Omega-2 \lambda  V}{2 \lambda  (1-\lambda  V)}
\right),
\end{split}\end{equation}
which is consistent with the result in \cite{Bonelli:2018kik}.

\subsection{Uniform light-cone gauge approach}

This subsection reviews the relationship between string theories in the uniform light-cone gauge and $T\bar{T}$ deformation \cite{Frolov:2019nrr, Frolov:2019xzi}.
We consider a string action
\begin{equation}
S = -\frac{1}{2}\int_{-r}^{ r}\, {
d}\sigma{d}\tau\, \gamma^{\alpha\beta}\partial_\alpha X^M\partial_\beta X^N 
G_{MN},
\end{equation}
where the world-sheet coordinate $\sigma$ ranges from $-r$ to $<r$ and the target space metric is given by \footnote{In this subsection, we use $(t,x^\mu,\phi)$ and $(\sigma, \tau)$ to denote target space coordinates and world-sheet coordinates respectively.}
\begin{equation}
ds^2=G_{MN}dX^MdX^N = G_{tt}  dt^2  +  2G_{t\phi}  dtd\phi + G_{\phi\phi}  d\phi^2  + 
G_{\mu\nu} dx^\mu dx^\nu.
\end{equation}
We assume that $G_{MN}$ is independent of $t$ and $\phi$, and therefore, the shifts of $t$ and $\phi$ are two isometries. 
We use $\mu,~\nu=1,...,n$ to denote indices of the transversal coordinates.
It is convenient to use the action in the first-order form
\begin{equation}
S = \int_{- r}^{ r}\, { d}\sigma{ d}\tau\, \left( p_M \dot{X}^M +
\frac{\gamma^{01}}{\gamma^{00}} C_1+\frac{1}{2} { \gamma^{00}}C_2\right)  .
\end{equation}
where $p_M$ are canonical momenta and the constraints $C_{1,2}$ are defined as
\begin{equation}
    C_1=p_MX'^M, ~~~
C_2=G^{MN} p_M p_N + X'^M X'^N G_{MN}.
\end{equation}

The conserved charge corresponds to the invariance under the shift of $t$ is the target space energy. If $\phi$ is an angular variable, the associated shift invariant conserved charge can be viewed as a target space angular momentum. The two conserved charges can be computed as
\begin{equation}
    E = - \int_{-
r}^{ r}  { d}\sigma  p_t,~~~ J= \int_{- r}^{ r}\, { 
d}\sigma  p_\phi. 
\end{equation}
We define the light-cone target space coordinates and momenta as
\begin{equation}
    x^- =\phi -t,~~~
x^+ =\frac{1}{2}(\phi +t)+\alpha  x^-,~~~
p_+ = p_\phi+ p_t,~~~
p_- =\frac{1}{2}(p_\phi- p_t)-\alpha p_+,
\end{equation}
where we introduce an arbitrary parameter $\alpha$ which will turn out to relate to the $T\bar{T}$ deformation $\lambda$. 
In light cone coordinates, the action and constraints can be written as
\begin{align} 
S=&\int_{-r}^{r} d \sigma d \tau\left(p_{+} \dot{x}^{+}+p_{-} \dot{x}^{-}+p_{\mu} \dot{x}^{\mu}+\frac{\gamma^{01}}{\gamma^{00}} C_{1}+\frac{1}{2 \gamma^{00}} C_{2}\right),  \\  
    C_1 =& p_+x'^+
+p_-x'^- + p_\mu x'^\mu,\\
 C_2 =& G^{++} p_+^2  + 2G^{-+}p_-p_+  +
G^{--}p^2_-+
G_{++} (x'^+)^2  +
2G_{-+} x'^-x'^+
+ G_{--}(x'^-)^2  + 
2\mathcal H_x,\\
\mathcal H_x =& \frac{1}{2}(G^{\mu\nu}p_\mu p_\nu + G_{\mu\nu} x'^\mu x'^\nu).    
\end{align}
The target space metric components in light cone coordinates are
\begin{equation}
\begin{split}
G_{++}&=G_{\phi\phi}+2G_{t\phi}+G_{tt} ,\\
G^{--}&=\frac{G_{++}}{\det G_{\mathrm{lc}}} ,\quad\det G_{\mathrm{lc}}\equiv G_{tt}G_{t\phi}-G_{\phi\phi}^{2} , \\
G_{--}&=\left(\frac12-\alpha\right)^{2}G_{\phi\phi}+\left(2\alpha^{2}-\frac12\right)G_{t\phi}+\left(\frac12+\alpha\right)^{2}G_{tt} ,\quad G^{++}=\frac{G_{--}}{\det G_{\mathrm{lc}}} , \\
G_{-+}&=\left(\frac{1}{2}-\alpha\right)G_{\phi\phi}-2\alpha G_{t\phi}-\left(\alpha+\frac{1}{2}\right)G_{tt} ,\quad G^{-+}=-\frac{G_{-+}}{\det G_{\mathrm{lc}}} .    
\end{split}
\end{equation} 
{Now we impose a uniform light-cone gauge by using the worldsheet  diffeomorphism}, which is defined as
\begin{equation}
x^+ = \tau + \pi(\frac{1}{2}+\alpha) \frac{m}{r}R_\phi \sigma ,\quad p_- = 1,
\end{equation}
where $2\pi R_\phi$ is the period of $\phi$ and the integer $m$ is the winding number. In the uniform light-cone gauge, $C_1=0$ can be solved by
\begin{equation}
x^{\prime-} = -\pi(\frac{1}{2}+\alpha)\frac{m}{r}R_{\phi} p_{+} - p_{\mu}x^{\prime\mu} .
\end{equation}
This solution can be substituted into $C_2$, and thus, one can solve $C_2=0$ for $p_+$.
Finally, the  gauge-fixed action can be written as
\begin{equation}
S_\alpha=\int_{-r}^r d\sigma d\tau (p_\mu\dot{x}^\mu + p_+(p_\mu,x^\mu,x^{\prime\mu})).
\end{equation}
The gauge-fixed action can describe $T\bar{T}$ deformed models with scalars, fermions, and chiral bosons by choosing appropriate target space metrics.
For the example of $n$ scalar fields with arbitrary potential, one can take
\begin{equation}
G_{tt}=-(1+\frac V2) ,~~~ G_{\phi\phi}=1-\frac V2 ,~~~ G_{t\phi}=-\frac V2 .
\end{equation}
We also set the winding number $m=0$.
Then, the gauge-fixed action becomes
\begin{equation}
S_\alpha=\int_{-r}^r d\sigma d\tau (p_\mu\dot{x}^\mu - 
\frac{1}{\alpha}+\frac{1}{2 \tilde{\alpha}}+\frac{1}{2 \tilde{\alpha}} \sqrt{1-4 \tilde{\alpha} \mathcal{H}_{x}+4 \tilde{\alpha}^{2}\left(p_{\mu}x^{\prime\mu}\right)^{2}}),
\end{equation}
with $\tilde\alpha=\alpha(1+\alpha V)$.
Integrating out the canonical momenta $p_\mu$, we get
\begin{equation}\label{defscalarlc}
S_\alpha=\int_{-r}^rd\sigma d\tau (-\frac{1}{\alpha}+\frac{1}{2 \tilde{\alpha}}+\frac{1}{2 \tilde{\alpha}} \sqrt{1+2 \tilde{\alpha}\left(\dot{x}^{2}-x^{\prime 2}\right)-4 \tilde{\alpha}^{2}\left(\dot{x}^{2} x^{\prime 2}-\left(\dot{x} x^{\prime}\right)^{2}\right)}),
\end{equation}
which agrees with the $T\bar{T}$ deformation of $n$ scalar fields with arbitrary potential \cite{Conti:2018jho}.
To match the convention in the previous subsections, we take $\alpha=-\lambda$ by comparing (\ref{defscalarlc}) with (\ref{defscalar})
for the $n=1$ case.

\subsection{$T\bar{T}$-deformed conformal symmetry} \label{Subsection TTbar deformed conformal symmetry}
The $T\bar{T}$ deformation is not only solvable but also preserves many important properties of the seed theory. This can usually be understood by the geometric realizations of $T\bar{T}$ deformation. Of particular interest to us, the conformal symmetry is preserved/deformed by $T\bar{T}$ deformation, taking a form of symmetry of field-dependent coordinate transformations as discussed in \ref{Subsection Field-dependent coordinate transformation}. The key observation is, while the energy-momentum tensor is no longer traceless, it still has only two independent components by the trace relation (\ref{Classical trace relation}). Following \cite{Guica:2020uhm}, we work with flat space in light-cone coordinates $U,V$, and the trace relation takes the form
\begin{align}
    T_{UV}=2\lambda(T_{UU}T_{VV}-T_{UV}^2)
\end{align}
It's convenient to parametrize the energy-momentum tensor by two functions $\mathcal{L},\bar{\mathcal{L}}$
\begin{align}
    T_{UU}=\frac{\mathcal{L}}{2(1-\lambda\mathcal{L}\bar{\mathcal{L}})},\; T_{VV}=\frac{\bar{\mathcal{L}}}{2(1-\lambda\mathcal{L}\bar{\mathcal{L}})},\;
    T_{UV}=\frac{\lambda\mathcal{L}\bar{\mathcal{L}}}{2(1-\lambda\mathcal{L}\bar{\mathcal{L}})}.
\end{align}
The conservation equations of the energy-momentum tensor can be written as
\begin{align}
    (\partial_V + \lambda \bar{\mathcal{L}}\partial_U) \mathcal{L}=0,\; (\partial_U + \lambda \mathcal{L}\partial_V) \bar{\mathcal{L}}=0.
\end{align}
It's clear that for $\lambda=0$, $\mathcal{L},\bar{\mathcal{L}}$ are the holomorphic and anti-holomorphic components of the energy-momentum tensor of the seed CFT, and for a CFT the action is invariant under the coordinate transformation
\begin{align} \label{field-independent coordinate transformation CFT}
    U \to U + \epsilon f(U),\; V \to V - \epsilon \bar{f}(V).
\end{align}
Now for a $T\bar{T}$-deformed CFT, we assume the coordinate transformation is
\begin{align} \label{field-dependent coordinate transformation TTbar CFT}
    U \to U + \epsilon f(u),\; V \to V - \epsilon \bar{f}(v).
\end{align}
where $u,v$ is a new set of coordinates. The variation of the action under the coordinate transformation is
\begin{align}
    \delta S = \epsilon \int dUdV \big[-f^{'}(u)(T_{UU}\partial_V u + T_{VU}\partial_U u) + \bar{f}^{'}(v)(T_VV\partial_U v + T_{UV}\partial_V v)\big].
\end{align}
The action is invariant if
\begin{align}
    \partial_V u + \lambda\bar{\mathcal{L}}\partial_U u = 0,\; \partial_U v + \lambda\mathcal{L}\partial_V v = 0.
\end{align}
The solution depends on $\mathcal{L}$ and $\bar{\mathcal{L}}$; that is, the coordinate transformation that leaves the action invariant is field-dependent, in contrast to the field-independent two-dimensional conformal symmetry (\ref{field-independent coordinate transformation CFT}). The conservation equations take a much simpler form in the new coordinates
\begin{align}
    \partial_v \mathcal{L}=0,\;  \partial_u \bar{\mathcal{L}}=0,
\end{align}
so the energy-momentum tensor is parametrized by two functions $\mathcal{L}(u),\bar{\mathcal{L}}(v)$, similar to the case of CFTs. The charge algebra and quantum aspects of the $T\bar{T}$-deformed conformal symmetry were further studied in \cite{Jorjadze:2020ili,Guica:2020uhm,Georgescu:2022iyx,Guica:2022gts,Chakraborty:2023wel}. Given how the conformal symmetry determines CFTs up to conformal dimensions and OPE coefficients subject to bootstrap equations (see \cite{Simmons-Duffin:2016gjk,Poland:2018epd} for example), it's a very interesting question how much we can extract from the deformed conformal symmetry for $T\bar{T}$-deformed CFTs. The first step is to find a set of operators that transform nicely under the deformed conformal symmetry. The definition of a primary operator is given in \cite{Guica:2021fkv} for $J\bar{T}$-deformed CFTs, while it remains an open question for $T\bar{T}$ deformation.


\newpage

\section{$T\overline{T}$ deformation in quantum field theory}\label{quantumaspectTT}

The perturbative and non-perturbative studies of correlation functions in $T\overline{T}$-deformed quantum field theories provide essential insights into the deformation's impact on fundamental properties of the theory. These analysis systematically addresses the induced non-locality and ultraviolet-infrared (UV-IR) mixing effects, clarifying the structure of observables and the renormalization procedures required to maintain the theory's consistency and predictive power. The solvability of $T\overline{T}$-deformed theories further allows exact computations of correlation functions and energy spectra. Beyond perturbation theory, non-perturbative approaches, such as resurgence methods, have explored the partition function and spectral properties. Quantum entanglement entropy calculations in $T\overline{T}$-deformed theories demonstrate modifications from the undeformed case, providing a novel probe of the deformation's non-local nature.
Additionally, the large central charge limit simplifies the flow equations, offering a powerful tool for analyzing quantum corrections and non-perturbative effects. These studies also extend to black hole physics, where the deformation influences thermodynamic quantities, including entropy growth and microstate structure. This section systematically reviews these quantum aspects, emphasizing exact solvability, correlation functions, renormalization, entanglement properties, large central charge behavior, and recent developments, elucidating key characteristics and implications of the $T\overline{T}$ deformation.

\subsection{Deformed spectrum and partition function}
Following the conventions in section 2, Consider a $T\bar{T}$ deformed theory defined on a circle of circumference $L$. The deformed Hamiltonian and its eigenvalues in Euclidean signature  are\cite{McGough:2016lol}
\be\ba 
H_{int}=\int dx L_{int}=\lambda\int dx O_{T\bar{T}},
\ea\ee 
\be\ba 
\frac{\p E_n}{\p \l} =\frac{\p \langle n| H_{int}|n\rangle}{\p \l}= \lambda\int dx \langle n|O_{T\bar{T}}|n \rangle=L\langle n|O_{T\bar{T}}|n \rangle
,\ea\ee 
where 
\be\ba 
\langle n|O_{T\bar{T}}|n\rangle=(\vev{n|T^{x\tau}|n})^2-\vev{n|T^{xx}|n}\vev{n|T^{\tau\tau}|n}.
\ea\ee 
By using the relation 
\be 
\vev{n|T^{\tau\tau}|n}=\frac{E_n}{L},~~\vev{n|T^{xx}|n}=\frac{\p E_n}{\p L},~~\vev{n|T^{\tau x}|n}=\frac{iP_n}{L},
\ee 
one can obtain the so-called Burger's equation
\be \label{Burger}
\frac{\p E_n}{\p \l}+\frac{P_n^2}{L}+E_n\frac{\p E_n}{\p L}=0,
\ee 
whose solution, i.e., the deformed spectrum is \cite{Smirnov:2016lqw,Cavaglia:2016oda,McGough:2016lol}
\be \label{TTspectrum}
E_n(\l)=\frac{L}{2\l}\Big(1-\sqrt{1-\frac{4\l}{L}E_n^{(\text{CFT})}+\frac{4\l^2}{L^2}(P_n^{(\text{CFT})})^2} \Big),~~~~P_n(\l)=P_n^{(\text{CFT})}.
\ee 
For positive $\l$, the highly excited states have complex energies, leading to a non-unitary deformed theory, while for negative spectrum, the spectrum exhibits Hagedorn growth. Notice that
here the undeformed theory was chosen to be a CFT, with the energy and momentum being  
\be 
E_n(0)\equiv E_n^{(\text{CFT})}=\frac{2\pi}{L}(L_0+\bar{L}_0-\frac{c}{12}),~~P_n(0)\equiv P_n^{(\text{CFT})}=\frac{2\pi}{L}(L_0-\bar{L}_0),
\ee 
{ where$L_0$ (and $\bar{L}_0$) are Virasoro zero mode in the expansion of CFT stress tensor, $T(z)=\sum_{n\in \mathbb{Z}} L_n z^{-n-2}$ and $c$ is the CFT central charge}.
Letting $E_L=\frac{1}{2}(E+P),E_R=\frac12 (E-P)$, one have \cite{Apolo:2023aho}
\be \ba \label{despec}
E_R(0)&=E_R(\l)-\frac{2\l}{L}E_R(\l)E_L(\l),\\
E_L(0)&=E_L(\l)-\frac{2\l}{L}E_R(\l)E_L(\l).
\ea\ee  
With periods of torus are $w_1=L,w_2$, and the modular parameter $\tau=w_2/w_1$, the deformed partition function can be expressed as  
\cite{Cardy:2018sdv} \cite{Datta:2018thy}
\be \ba \label{PFTTbar}
Z(\tau,\bar{\tau},\l)=\tr(e^{i\tau_1 L P-\tau_2 L H(\l)})=\tr(e^{i\tau L E_L(\l)-i\bar{\tau}LE_R(\l)}).
\ea\ee 
The CFT partition function is 
\be\ba \label{PFcft}
Z(\tau,\bar{\tau},\l=0)=\tr(q^{L_0-c/24}\bar{q}^{\bar{L}_0-c/24}).
\ea\ee
Since $E_n(\l)L$ is dimensionless, it depends only on $\l/L^2$, one has
\be\ba \label{deriL}
L\frac{\p E}{\p L}+E=-2\l \frac{\p E}{\p \l}.
\ea\ee 
Combining \eqref{Burger} and \eqref{deriL}, one can obtain the flow of partition function defined in \eqref{PFTTbar}
\be 
L^2\p_\l Z+\frac{2}{i}(\tau-\bar{\tau})\p_\tau\p_{\bar{\tau}}Z+2i \l (\p_\tau-\p_{\bar{\tau}})\p_\l Z-\frac{4i\l}{(\tau-\bar{\tau})}\p_\l Z=0.
\ee 
The modular transformation property of the deformed partition function is \cite{Datta:2018thy} (see also \cite{Tian:2024vln} from holographic perspective )
\be 
Z\Big(\frac{a\tau+b}{c\tau +d},\frac{a\bar{\tau}+b}{c\bar{\tau}+d},\frac{\l}{|c\tau +d|^2}\Big)=Z(\tau,\bar{\tau},\l).
\ee 
Notice that $Z(\tau,\bar{\tau},\l)$ does not have a closed form and instead can be expressed as a formal series in $\l$, with expansion coefficients computed via recursion relation. Based on this formal series, non-perturbative contribution to deformed partition function was recently investigated via resurgence method \cite{Gu:2024ogh, Gu:2025tpy}.
It is interesting to note that the resurgence theory was used to study the  $T\bar{T}$-deformed partition function of 2d Maxwell and Yang-Mills theory \cite{Griguolo:2022xcj, Griguolo:2022hek, Griguolo:2022vdx}. 

The $T\bar{T}$-deformed partition function could also be derived from single-trace $T\bar{T}$ deformation  (see section \ref{string-TTbar}) by the argument of universality \cite{Hashimoto:2019wct, Chakraborty:2019mdf}. In \cite{Hashimoto:2019wct} the deformed partition function reads
\begin{equation} \label{inte_trans}
Z(\zeta, \bar{\zeta}, \lambda)=\frac{\zeta_2}{2 \lambda} \int_{\mathcal{H}_{+}} \frac{d^2 \tau}{\tau_2^2} e^{-\frac{\pi}{2 \lambda \tau_2}|\tau-\zeta|^2} Z_{\text {cft }}(\tau, \bar{\tau}).
\end{equation}
Here { $\zeta=\zeta_1+i\zeta_2$ is the modulus of the torus where the deformed theory defined on, and} $\mathcal{H}_+$ is the upper half-plane. \eqref{inte_trans} is consistent with the results obtained via different methods in  \cite{Cardy:2018sdv} and \cite{Dubovsky:2018bmo}. The deformed spectrum \eqref{TTspectrum} follows by substituting CFT partition function  \eqref{PFcft} into the RHS of \eqref{inte_trans}.

The Cardy formula for CFT entropy
\begin{equation}
S=2 \pi \sqrt{\frac{c_R }{6}\left(L_0-\frac{c_R}{24}\right)}+2 \pi \sqrt{\frac{c_L}{6} \left(\bar{L}_0-\frac{c_L}{24}\right)}.
\end{equation}
For simplicity, consider the case $c_L=c_R, L_0=\bar{L}_0$
\be 
S(E(0))=2\pi\sqrt{\frac{c}{3}E(0)}.
\ee 
Since each state in the deformed theory is related to the state in the original CFT via \eqref{TTspectrum}, the entropy for deformed theory
\be \label{HagedornS}
S(E(\l))=2\pi\sqrt{\frac{c}{3}E(0)}=2\pi\sqrt{\frac{c}{3}(E(\l)-\frac{4\l}{L}E(\l)^2)}.
\ee 
For positive $\l$, the entropy exhibit Hagedorn growth $S=\beta_H E$ at high energy, where $\beta_H=2\pi \sqrt{\frac{-4c\l}{3L}}$ is inverse Hagedorn temperature.

\subsection{Renormalization}
Renormalization of $T\bar{T}$ deformed theory is non-trivial since it is non-normalizable from standard quantum field theory. This means there are an infinite number of counterterms. However, since this kind of deformed is integrable, it leads to infinite constraints to fix the counterterms. Finally, it makes the usual perturbation method applicable in the $T\bar{T}$ deformed theories.


It is known that for a  integrable 2D QFT, two-to-two $S$-matrix $S(\theta)$ of a single particle of mass $m$ can {  be determined by unitarity and  crossing symmetry which takes the form 
\be 
S(\theta)=\prod_\alpha S_\alpha (\theta),~~S_\alpha=\frac{\sinh \theta -i\sin \alpha}{\sinh \theta+i\sin \alpha},
\ee
where $\theta=\theta_1-\theta_2$ is the rapidity difference of two particles and $\alpha$ is  a model-dependent real number. Here $S_\alpha$ is called Castillejo-Dalitz-Dyson (CDD) factor.}

In the $T\bar{T}$ deformed theory, the $S$-matrix can be obtained (by integrability) and is modified by picking up a phase factor 
\cite{Smirnov:2016lqw}
\be \label{TTbarSmatrix1}
S_\l(\theta)=S(\theta)S'(\theta),~~S'(\theta)=e^{i\l m^2 \sinh\theta},
\ee 
where $S(\theta)$ is the $S$-matrix of undeformed integrable system. The deformed $S$-matrix grows exponentially at large imaginary momentum, which indicates the deformed theory is not a local QFT.

For example, consider a free massive scalar $S(\theta)=1$, then $S_\l(\theta)=\exp(i\l m^2 \sinh\theta)$. One takes the limit $\theta-\to \infty$ and $ m\to0$ for the massless case while keeping $s=(p_1+p_2)^2$ fixed. Hence  the $T\bar{T}$-deformed $S$-matrix is 
\be \label{TTbarSmatrix}
S_\l(\theta)=\exp(i\l s/2).
\ee 
Alternatively, the $S$-matrix can be computed perturbatively in $\l$ order by order, at least in some leading orders as did in \cite{Rosenhaus:2019utc}. 

To compute the correlation functions perturbatively by bare Lagrangian, one would usually come across UV divergence, which can be removed by introducing counterterms to the Lagrangian. It is ambiguous to determine these counterterms. However, as shown in \cite{Rosenhaus:2019utc}
 for the $T\bar{T}$ deformed free scalar, this ambiguity could be fixed by demanding the renormalized Lagrangian can reproduce $S$-matrix \eqref{TTbarSmatrix1} or \eqref{TTbarSmatrix} order by order. For free massive scalar up to $\l^2$-order, the renormalized Lagrangian turns out to be 
\cite{Rosenhaus:2019utc}
\begin{equation}
\mathcal{L}=2 \partial \phi \bar{\partial} \phi+\frac{1}{2} m^2 \phi^2-4 g(\partial \phi \bar{\partial} \phi)^2+\frac{1}{4} h m^4 \phi^4+\ldots
\end{equation}
with 
\begin{equation}
\begin{aligned}
& g=\lambda+\frac{\lambda^2 m^2}{8 \pi}\left(10 \frac{\Lambda^2}{m^2}+9 \log \frac{m^2}{\Lambda^2}\right), \\
& h=\lambda+\frac{\lambda^2 m^2}{24 \pi}\left(7 \frac{\Lambda^2}{m^2}+39 \log \frac{m^2}{\Lambda^2}\right).
\end{aligned}
\end{equation}

\subsection{Correlation functions} 
\label{sec:corr}
\subsubsection*{Conformal perturbation theory}
One of the most direct ways could be using conformal perturbation theory to study the correlation function in $T\bar{T}$ deformed CFT.
In this method, one usually considers the correlation function of the undeformed operator in the deformed theory, such as the primary fields in CFT, namely
\be\ba\label{correlationO}
\left\langle\mathcal{O}_1 \mathcal{O}_2  \cdots \mathcal{O}_n \right\rangle_\lambda =\int [d\phi] \mathcal{O}_1 \mathcal{O}_2 \cdots \mathcal{O}_n e^{-S_\l[\phi]} 
\ea\ee 
with  $ \mathcal{O}_i \equiv\mathcal{O}_i\left(z_i, \bar{z}_i\right)$ being the primary field in the undeformed CFT. $S_\l[\phi]$ is the deformed action, and $\l=0$ coreesponds to undeformed CFT. Therefore, in principle, one can expand the RHS in power of $\l$ order by order perturbatively. Nevertheless, some operators indeed deformed under $T\bar{T}$ deformation. Such as the conserved current operator and stress tensor. This is because the EOM is deformed, to keep the conservation of the current or stress tensor under deformation, the functional form of these operators must be modified. Furthermore, it would be more important to define new operators that appear more natural in the deformed theory; for example, the symmetry of the deformed theory may be more manifest based on their operators.

The conformal perturbation method was initially employed in \cite{Kraus:2018xrn}\cite{Guica:2019vnb}, where the first-order two and three-point functions of undeformed primary operators were computed.   
In this method, one first 
expands the action  in the power of $\l$ as 
\be \label{actionexp}
S_\l[\phi]=S_{CFT}[\phi]+\l \int d^2z T\bar{T}+\l^2 S_{(2)}[\phi]+...
\ee
where $ S_{(2)}[\phi]$ is the second-order correction of action $S_\l[\phi]$.
It then follows    first-order correction of \eqref{correlationO}, which is
\begin{equation} \label{firston}\ba
\left\langle\mathcal{O}_1  \mathcal{O}_2 \cdots \mathcal{O}_n \right\rangle_\lambda^{(1)}=&\lambda \int d^2 z\left\langle T \bar{T}(z, \bar{z}) \mathcal{O}_1 \mathcal{O}_2 \cdots \mathcal{O}_n \right\rangle,
\ea\end{equation}
where $\langle...\rangle_{\l}^{(n)}$ represents the $n$-th correction of $T\bar{T}$-deformed correlation function. And $\langle...\rangle$ on the RHS denotes the CFT correlation function, which can be computed from conformal Ward identities  
\be\ba \label{CFTwardi}
&  \langle T(z)\bar{T}(\bar{z}) \mathcal{O}_1\mathcal{O}_2...\mathcal{O}_n  \rangle=\sum_{i=1}^n\left(\frac{h_i}{\left(z-z_i\right)^2}+\frac{1}{z-z_i} \partial_{z_i}\right)\left\langle\bar{T} (\bar{z} ) \mathcal{O}_1  \mathcal{O}_2 ... \mathcal{O}_n \right\rangle, \\
& \left\langle\bar{T}\left(\bar{z}\right) \mathcal{O}_1 \mathcal{O}_2 ... \mathcal{O}_n \right\rangle=\sum_{i=1}^3\left(\frac{\bar{h}_1}{\left(\bar{z}-\bar{z}_i\right)^2}+\frac{1}{\bar{z}-\bar{z}_i} \partial_{\bar{z}_i}\right)\left\langle \mathcal{O}_1  \mathcal{O}_2 ... \mathcal{O}_n \right\rangle.
\ea\ee
By making use of explicit expression  for conformal correlation functions, it turns out one can always arrive at the general form
\begin{equation} \label{firston1}\ba 
\left\langle\mathcal{O}_1  \mathcal{O}_2 \cdots \mathcal{O}_n \right\rangle_\lambda^{(1)}=\l\left\langle  \mathcal{O}_1 \mathcal{O}_2 \cdots \mathcal{O}_n \right\rangle \int d^2z f_n(z,\bar{z},\{z_i,\bar{z}_i\}),
\ea\end{equation}
where the function $f_n$ can be worked out for each $n$. The above integral is generally divergent and can usually be regularized by dimensional regularization. 
Consider for example, the cases $n=2$ 
the CFT correlation functions are well-known 
\begin{equation}\label{CFT2pt}
\left\langle\mathcal{O}_1 \mathcal{O}_2\right\rangle=\frac{C_{12}}{z_{12}^{2 h_1} \bar{z}_{12}^{2 \bar{h}_1}},
\ee
where $z_{ij}=z_i-z_j $, and $(h_i,\bar{h}_i)$ the  conformal dimension.
By plugging \eqref{CFT2pt} and \eqref{CFTwardi} into \eqref{firston}, using dimensional regularization, one obtains 
\begin{equation}\label{result1st}
\begin{aligned}
&\left\langle T(z) \bar{T}(z) \mathcal{O}\left(z_1, \bar{z}_1\right) \mathcal{O}\left(z_2, \bar{z}_2\right)\right\rangle_\lambda^{(1)}\\
=&\lambda h \bar{h} \frac{8 \pi}{z_{12}\bar{z}_{12}}\left(\frac{4}{\epsilon}+2 \log \left|z_{12}\right|^2+2 \log \pi+2 \gamma-5\right)\left\langle\mathcal{O}\left(z_1, \bar{z}_1\right) \mathcal{O}\left(z_2, \bar{z}_2\right)\right\rangle\\
=&\lambda h \bar{h} \frac{8 \pi C_{12}}{z_{12}^{2h+1}\bar{z}_{12}^{2\bar{h}+1}}\left(\frac{4}{\epsilon}+2 \log \left|z_{12}\right|^2+2 \log \pi+2 \gamma-5\right),
\end{aligned}
\end{equation}
where $\epsilon\to 0$ is the parameter introduced in dimensional regularization,  $\gamma$ is Euler constant. By redefining $\epsilon$, we can absorb the constants inside the bracket in the last line of  \eqref{result1st}. In addition, the results show that the first-order correction is universal in the sense that it depends only on the dimension and normalization of the operators. 
Notice that there are logarithmic divergences in this order. It was suggested that the logarithmic divergence appears in all order in $\l$ \cite{Cardy:2019qao}, and this point will be discussed in a subsequent section. Also, the logarithmic divergence can be understood as a momentum-dependent correction to the conformal dimension \cite{Cui:2023jrb}. Similar to the two-point case, the three-point function first order correction can also be worked out; one can refer to, for example, \cite{He:2019vzf}, for more details. \footnote{One can also refer to a more recent example for the two-point function of CFT primary operators perturbatively\cite{Menskoy:2024vqv}.}

As for higher points cases $n\geq 4$, even though correlation functions in CFT are model-dependent and we don't have an explicit expression, one can still work out the integral in \eqref{firston} by making use of the following form   
\begin{equation}
\left\langle\mathcal{O}^{\dagger}\left(z_1, \bar{z}_1\right) \mathcal{O}\left(z_2, \bar{z}_2\right) \mathcal{O}^{\dagger}\left(z_3, \bar{z}_3\right) \mathcal{O}\left(z_4, \bar{z}_4\right)\right\rangle=\frac{G(\eta, \bar{\eta})}{z_{13}^{2 h} z_{24}^{2 h} \bar{z}_{13}^{2 \bar{h}} z_{24}^{2 \bar{h}}}
\end{equation}
with the cross ratios
\be 
\eta=\frac{z_{12} z_{34}}{z_{13} z_{24}}, \quad \bar{\eta}=\frac{\bar{z}_{12} \bar{z}_{34}}{\bar{z}_{13} \bar{z}_{24}} .
\ee
Here, $G(\eta,\bar{\eta})$ is a model-dependent cross-ratio function related to conformal blocks.
Then \eqref{firston1} turns out to take the form 
\begin{equation} 
\begin{aligned}
& \left\langle\mathcal{O}^{\dagger}\left(z_1, \bar{z}_1\right) \mathcal{O}\left(z_2, \bar{z}_2\right) \mathcal{O}^{\dagger}\left(z_3, \bar{z}_3\right) \mathcal{O}\left(z_4, \bar{z}_4\right)\right\rangle_\lambda \\
& =\lambda \int d^2 z\left\{\left(\frac{h z_{13}^2}{\left(z-z_1\right)^2\left(z-z_3\right)^2}+\frac{h z_{24}^2}{\left(z-z_2\right)^2\left(z-z_4\right)^2}+\frac{z_{23} z_{14}}{\prod_{j=1}^4\left(z-z_j\right)} \frac{\eta \partial_\eta G(\eta, \bar{\eta})}{G(\eta, \bar{\eta})}\right)\right. \\
& \left(\frac{\bar{h} \bar{z}_{13}^2}{\left(\bar{z}-\bar{z}_1\right)^2\left(\bar{z}-\bar{z}_3\right)^2}+\frac{\bar{h} \bar{z}_{24}^2}{\left(\bar{z}-\bar{z}_2\right)^2\left(\bar{z}-\bar{z}_4\right)^2}+\frac{\bar{z}_{23} \bar{z}_{14}}{\prod_{j=1}^4\left(\bar{z}-\bar{z}_j\right)} \frac{\bar{\eta} \partial_{\bar{\eta}} G(\eta, \bar{\eta})}{G(\eta, \bar{\eta})}\right) \\
& -\eta \bar{\eta} \frac{z_{23} z_{14} \bar{z}_{23} \bar{z}_{14}}{\prod_i^4\left(z-z_i\right)\left(\bar{z}-\bar{z}_i\right)} \frac{\partial_\eta G(\eta, \bar{\eta}) \partial_{\bar{\eta}} G(\eta, \bar{\eta})}{G(\eta, \bar{\eta})^2}+\eta \bar{\eta} \frac{z_{23} z_{14} \bar{z}_{23} \bar{z}_{14}}{\prod_i^4\left(z-z_i\right)\left(\bar{z}-\bar{z}_i\right)} \frac{\partial_\eta \partial_{\bar{\eta}} G(\eta, \bar{\eta})}{G(\eta, \bar{\eta})} \\
&\times \left\langle\mathcal{O}^{\dagger}\left(z_1, \bar{z}_1\right) \mathcal{O}\left(z_2, \bar{z}_2\right) \mathcal{O}^{\dagger}\left(z_3, \bar{z}_3\right) \mathcal{O}\left(z_4, \bar{z}_4\right)\right\rangle .
\end{aligned}
\end{equation}
The integral is worked out in  \cite{He:2019vzf}\cite{He:2020qcs}.

For the general local (non-primary) operator, by the conformal Ward identity, \eqref{firston} can be written as \cite{Cardy:2019qao}
\begin{align}
\langle X \rangle^{(1)}&=-\frac{1}{\pi}\left(\int d^{2}x \sum_{m,n}\sum_{r,s\geq 1}\frac{1}{(z-z_{m}+\varepsilon)^{r}(\bar{z}-\bar{z}_{n})^{s}}\right)\langle L_{r-2,m}\bar{L}_{s-2,n}X \rangle^{(0)},
\end{align}
where $L_{r-2,m}$ is a shorthand notation for $L_{r-2}(z_m)$.
Performing the integration, we obtain
\be\ba \label{X1d}
\langle X\rangle^{(1)}=&\langle\operatorname{d} X\rangle^{(0)} ,~~\operatorname{d}\equiv \sum_{m\neq n}\operatorname{d}_{z_{m},z_{n}}
\ea\ee
with
\be\ba\label{1stgen}
\operatorname{d}_{z_{m},z_{n}}\equiv\sum_{m\not{=}n}\left(\log(|z_{mn}|^2/\varepsilon^2)\partial_{z_{m}}\partial_{\bar{z}_{n}}-\sum_{s\geq 2}\frac{1}{s-1}\frac{\bar{L}_{s-2,n}\partial_{z_{m}}}{\bar{z}_{mn}^{s-1}}-\sum_{r\geq 2}\frac{1}{r-1}\frac{L_{r-2,m}\partial_{\bar{z}_{n}}}{z_{nm}^{r-1}}\right).
\ea\ee 

\subsubsection*{Superconformal case}
The $T\bar{T}$ deformation can be generalized to the case with supersymmetry presented, such as  $\mathcal{N}$ = (0, 1) and extend SUSY with $\mathcal{N}$ = (1, 1),(2, 0),(2, 2) \cite{Baggio:2018rpv,Chang:2018dge,Jiang:2019hux,Chang:2019kiu,Coleman:2019dvf,Ebert:2020tuy}. In
these studies, the supersymmetric versions of $T\bar{T}$ operator was
constructed based on the supercurrent multiplets. It was shown that the deformation constructed in this
way preserves both the solvability and the SUSY. In analogous to the bosonic case discussed above, we can study the correlation functions when the undeformed theory is supersymmetric and conformal. Working in  Euclidean signature with  $\mathcal{N}=(1,1)$ superconformal symmetry,
the variation of action under $T \bar{T}$ deformation can be constructed 
\be 
\delta S=\lambda \int d^2 z T \bar{T}(z)=-\lambda \int d^2 z \int d \theta d \bar{\theta} J(Z) \bar{J}(\bar{Z}),
\ee 
where superspace coordinate is denoted by $Z=(z,\theta)$, the $J(Z)$ is the superfield containing the stress tensor $
J(Z)=\Theta(z)+\theta T(z)
$. \footnote{{ We use "Z"  to denote superspace coordinate only in this subsection, while  it denotes the partition functions elsewhere.}} Here  { $\Theta(z)$ the fermionic component field which is a generator of SUSY transformation}.
In parallel with \eqref{firston} and \eqref{CFTwardi}, consider the first order correlation function of primary superfield  $\Phi(Z, \bar{Z})$, with anomalous dimensions ($\Delta,\bar{\Delta}$)
\be \label{scftfirst}
-\lambda \int d^2 z \int d \theta d \bar{\theta}\left\langle J(Z) \bar{J}(\bar{Z}) \Phi\left(Z_1, \bar{Z}_1\right) \ldots \Phi\left(Z_n, \bar{Z}_n\right)\right\rangle.
\ee
Here, the integrand can be computed via
the $\mathcal{N}=(1,1)$ superconformal Ward identity 
\begin{equation}
\begin{aligned}
& \left\langle J\left(Z_0\right) \Phi_1\left(Z_1, \bar{Z}_1\right) \ldots \Phi_n\left(Z_n, \bar{Z}_n\right)\right\rangle \\
= & \sum_{i=1}^n\left(\frac{\theta_{0 i}}{Z_{0 i}} \partial_{z_i}+\frac{1}{2 Z_{0 i}} D_i+\Delta_i \frac{\theta_{0 i}}{Z_{0 i}^2}\right)\left\langle\Phi_1\left(Z_1, \bar{Z}_1\right) \ldots \Phi_n\left(Z_n, \bar{Z}_n\right)\right\rangle,
\end{aligned}
\end{equation}
and similar expressions for $\bar{J}(\bar{Z})$. Here $D$ is superderivative 
and $Z_{12}=z_1-z_2-\theta_1 \theta_2$ and $\theta_{12}=\theta_1-\theta_2$  the SUSY invariant distance.
Using global osp(2$|$1) superconformal transformation, the undeformed two-point function can be fixed up to a constant $c_{12}$, similar to the bosonic case 
\begin{equation} 
\left\langle\Phi_1\left(Z_1, \bar{Z}_1\right) \Phi_2\left(Z_2, \bar{Z}_2\right)\right\rangle=c_{12} \frac{1}{Z_{12}^{2 \Delta} \bar{Z}_{12}^{2 \bar{\Delta}}}, \quad \Delta \equiv \Delta_1=\Delta_2, \quad \bar{\Delta} \equiv \bar{\Delta}_1=\bar{\Delta}_2. 
\end{equation}
From Ward identity and using dimensional regularization, \eqref{scftfirst} turns out to be \cite{He:2019ahx}
\begin{equation}
\begin{aligned}
& \frac{1}{\left\langle\Phi_1\left(Z_1, \bar{Z}_1\right) \Phi_2\left(Z_2, \bar{Z}_2\right)\right\rangle} \int d^2 z d \theta d \bar{\theta}\left\langle J(Z) \bar{J}(\bar{Z}) \Phi_1\left(Z_1, \bar{Z}_1\right) \Phi_2\left(Z_2, \bar{Z}_2\right)\right\rangle \\
= & -\frac{4 \pi \Delta^2}{Z_{12} \bar{Z}_{12}}\left(-\frac{4}{\epsilon}+2 \ln \left|z_{12}\right|^2+2 \gamma+2 \ln \pi-2\right) .
\end{aligned}
\end{equation}
By setting $\theta_{1,2} \rightarrow 0$
\be \label{scft1stresult}
-\frac{4 \pi \Delta^2}{\left|z_1\right|^2}\left(-\frac{4}{\epsilon}+2 \ln \left|z_{12}\right|^2+2 \gamma+2 \ln \pi-2\right),
\ee 
one recovers results for bosonic CFT \eqref{result1st}. Note that   \eqref{result1st} and \eqref{scft1stresult} equal up to a constant inside the bracket, and this difference can be eliminated by redefining $\epsilon$.
So far, only the 2-point function case has been considered. However, for the general $n$-point function, the first-order correlation functions can also be worked out as the bosonic case. 
\subsubsection*{Torus case}
Consider first-order correlation functions in $T\bar{T}$ deformed CFT on a torus in the next step. In this case, we should instead perform the integral \eqref{firston}  
on a torus $T^2$, which is 
\begin{equation} \label{firsttorus}\ba 
\left\langle\mathcal{O}_1  \mathcal{O}_2 \cdots \mathcal{O}_n \right\rangle_\lambda^{(1)}=&\lambda \int_{T^2} d^2 z\left\langle T \bar{T}(z, \bar{z}) \mathcal{O}_1 \mathcal{O}_2 \cdots \mathcal{O}_n \right\rangle_{T^2}.
\ea\end{equation}
Here, the expectation value on the RHS $\vev{...}_{T^2}$ denotes the CFT correlation functions on a torus \footnote{We will omit subscript $T^2$ in the following for simplicity.}, which can not be evaluated using the Ward identity \eqref{CFTwardi} on a plane. It turns out that the torus CFT correlation function with an arbitrary number of $T$ and $\bar{T}$ insertion satisfies the following  recursion relation \cite{He:2020udl}
\footnote{On a torus, the correlation function is defined as 
\be 
\vev{...}=\frac{1}{Z}\tr(...),\ee
where $Z$ is the partition function.
}
\be\ba \label{nmTTb}
&\tr( T(w)[T(u_1)...T(u_n)\bar{T}(v_1)...\bar{T}(v_m)] Xq^{L_0-c/12})
\\=&2\pi i\frac{\partial}{\partial\tau}\tr(T(u_1)...T(u_n)\bar{T}(v_1)...\bar{T}(v_m)Xq^{L_0-c/24})\\
=&\sum_i h_i \Big(- \zeta'(w_i-w)+2\eta_1 \Big)\tr(T(u_1)...T(u_n)\bar{T}(v_1)...\bar{T}(v_m)Xq^{L_0-c/24})
\\& + \sum_i  \left(- \zeta(w_i-w)+2\eta_1  w_i -2\eta_1 w-\pi i \right) \partial_{w_i}\text{tr}(T(u_1)...T(u_n)\bar{T}(v_1)...\bar{T}(v_m) e^{L_0-c/24}) \\
&+\frac{c}{12}\sum_j P''(u_j-w) \tr( T(u_1)...\hat{T}(u_j)...T(u_n)\bar{T}(v_1)...\bar{T}(v_m) Xq^{L_0-c/24})\\
&+\sum_j 2 \Big(P(w-u_j)+2\eta_1 \Big)\tr( T(u_1)...T(u_n)\bar{T}(v_1)...\bar{T}(v_m)   Xq^{L_0-c/24})\\
&+ \sum_j\Big(\zeta(w-u_j)+2\eta_1 u_j -2\eta_1 w-\pi i \Big)\partial_{u_i}\text{tr}(  T(u_1)...T(u_n)\bar{T}(v_1)...\bar{T}(v_m)  Xq^{L_0-c/24}).
\ea\ee
Here $X=\mathcal{O}_1  \mathcal{O}_2 \cdots \mathcal{O}_n $ is a string of primary operators,  the Weierstrass $P$ elliptic function and the Weierstrass $\zeta$ function is defined in \cite{pfun}. 
From the recursion relation, one has 
\be\ba \label{ttb61}
&\vev{ T(w)\bar{T}(\bar{v}) X}\\
=&2\pi i \partial_\tau\vev{\bar{T}(\bar{v})X }+2\pi i( \partial_\tau\ln Z)\vev{\bar{T}(\bar{v}) X} +\sum_i h_i \Big(- \zeta'(w_i-w)+2\eta_1 \Big)\vev{\bar{T}(\bar{v})X}  \\& + \sum_i  \Big(- \zeta(w_i-w)+2\eta_1 w_i-2\eta_1 w-\pi i  \Big) \partial_{w_i}\vev{\bar{T}(\bar{v})X}.
\ea\ee
where the well-known result for a single $T$ or $\bar{T}$ insertion is given in \cite{DiFrancesco:1997nk}. By substituting \eqref{ttb61} into   \eqref{firsttorus}, and performing the integral by some proper regularization scheme \footnote{The disk regularization method is used.  We excise these singular points $v=w_i$ from the integral domain
\be 
\lambda \int_{T^2-\sum_i D\left(w_i\right)} d^2 v\langle T(v) \bar{T}(\bar{v}) X\rangle,
\ee
where $D\left(w_i\right)$ is a small disk centered at $v=w_i$.  }, one could obtain the first-order correlation function on a torus \cite{He:2020udl}.

\subsubsection*{Partition function: pertubatively}
In the last section, we consider $T\bar{T}$-deformed correlation function on a torus. As a special case, one can consider the zero-point function, i.e., the deformed partition function. The first order correction can be obtained by simply setting $X=1$ in \eqref{ttb61} 
\begin{equation}
\left\langle T\left(w_1\right) \bar{T}\left(\bar{w}_2\right)\right\rangle=-(2 \pi i)^2 \frac{1}{Z} \partial_\tau \partial_{\bar{\tau}} Z.
\end{equation}
Thus  
\begin{equation}
\ba 
Z^{\prime}=&\int D \phi e^{-S+\lambda \int d^2 z T \bar{T}(z)}=Z\left(1+\lambda \int d^2 z\langle T \bar{T}\rangle(z)\right) \ldots=Z+\lambda(2 \pi)^2 \tau_2 \partial_\tau \partial_{\bar{\tau}} Z+...
\ea
\end{equation}
Note the integrand is in the intermediate steps is finite, hence no regularization is needed. This result is in good agreement with \cite{Datta:2018thy}. 

Next, consider the second-order correction of the deformed partition function. To this end, there are two main difficulties. The first one is that the flow of the stress operator in generic theory is necessary for higher-order computation. Fortunately, this problem was solved recently in \cite{Hirano:2024eab}. Second, the second-order perturbation involves more complicated integrals; hence, a more intricate regularization method is needed. As a first attempt, one can start with the free boson 
\be \label{freeb}
S=\frac{g}{2}\int_{T^2} d^2x \partial_\mu\phi\partial^\mu\phi,
\ee
then, the stress tensor up to first-order are 
\be 
T^{(0)}=-2\pi g(\p\phi)^2,\quad\bar{T}^{(0)}=-2\pi g(\bar{\p}\phi)^2,\quad\Theta^{(0)}=0.
\ee
\begin{equation}
T^{(1)}=8 \pi g^2(\partial \phi)^3(\bar{\partial} \phi), \quad \bar{T}^{(1)}=8 \pi g^2(\bar{\partial} \phi)^3(\partial \phi), \quad \Theta^{(1)}=-4 \pi g^2(\partial \phi \bar{\partial} \phi)^2 .
\end{equation}
And up to the second-order, the lagrangian  
\begin{equation}
\mathcal{L}^{(1)}=-\frac{1}{\pi^2} T^{(0)} \bar{T}^{(0)}=-4 g^2(\partial \phi \bar{\partial} \phi)^2,
\end{equation}
\begin{equation}\label{L2boson}
\mathcal{L}^{(2)}=-\frac{1}{\pi^2}\left(T^{(0)} \bar{T}^{(1)}+\bar{T}^{(0)} T^{(1)}\right)=32 g^3(\partial \phi \bar{\partial} \phi)^3.
\end{equation}
It follows that the second-order  partition function is 
\begin{equation}
\begin{aligned}
\mathcal{Z}^{(2)} =&\mathcal{Z}^{(0)} \int_{\mathrm{T}_1^2} \int_{\mathrm{T}_2^2}\left\langle\mathcal{L}^{(1)}\left(z_1, \bar{z}_1\right) \mathcal{L}^{(1)}\left(z_2,\bar{z}_2\right)\right\rangle-\mathcal{Z}^{(0)} \int_{\mathrm{T}^2} \mathcal{L}^{(2)}(z, \bar{z}) \\
=&16 g^4 \mathcal{Z}^{(0)} \int_{\mathrm{T}_1^2} \int_{\mathrm{T}_2^2}\left\langle\left(\partial_1 \phi \bar{\partial}_1 \phi\right)^2\left(\partial_2 \phi \bar{\partial}_2 \phi\right)^2\right\rangle-32 g^3 \mathcal{Z}^{(0)} \int_{\mathrm{T}^2}\left\langle(\partial \phi \bar{\partial} \phi)^3\right\rangle.
\end{aligned}
\end{equation}
Similar to the plane case, the correlation function in the integrand is evaluated in CFT, which can be computed using wick contraction and propagator for torus-free scalar \cite{DiFrancesco:1997nk}
\be\ba \label{boson propagatos} 
\left\langle\phi\left(z_1, \bar{z}_1\right) \phi\left(z_2, \bar{z}_2\right)\right\rangle=&(4 \pi g)^{-1}\left(-\log \left|\frac{\vartheta_1\left(z_{12}\right)}{\eta(\tau)}\right|^2+2 \pi \frac{\left(\operatorname{Im}\left[z_{12}\right]\right)^2}{\tau_2}\right),\\
\left\langle\partial \phi\left(z_1, \bar{z}_1\right) \partial \phi\left(z_2, \bar{z}_2\right)\right\rangle=&(4 \pi g)^{-1}\left(\frac{\pi}{\tau_2}-2 \eta_1-P\left(z_{12}\right)\right),
\\
\left\langle\partial \phi\left(z_1, \bar{z}_1\right) \partial \phi\left(z_1, \bar{z}_1\right)\right\rangle=&\lim _{z_2 \rightarrow z_1}\left(\left\langle\partial \phi\left(z_1, \bar{z}_1\right) \partial \phi\left(z_2, \bar{z}_2\right)\right\rangle+\frac{1}{4 \pi g z_{21}^2}\right)=(4 \pi g)^{-1}\left(\frac{\pi}{\tau_2}-2 \eta_1\right).
\ea\ee 
With the help of this correlation function, the 2nd order partition function can be worked out \cite{He:2020cxp}
\be \label{order2z}
\begin{aligned}
\mathcal{Z}^{(2)}   =16\left(\tau_2^2 \partial_\tau^2 \partial_{\bar{\tau}}^2+i \tau_2\left(\partial_\tau^2 \partial_{\bar{\tau}}-\partial_{\bar{\tau}}^2 \partial_\tau\right)\right) \mathcal{Z}^{(0)},
\end{aligned}
\ee
where minimal subtraction is performed to eliminate the divergent terms. Note \eqref{order2z} is consistent with \cite{Datta:2018thy}. In parallel with the bosonic case, one could also consider the partition function up to two orders for free fermion.   For example, Majorana fermions with spin structure $\nu$, 
Notice that different from the bosonic case, the  deformed Lagrangian for free fermion truncates at the first-order
It turns out that  the second-order correction of the partition
function is also given by
\begin{equation}
\begin{aligned}
\mathcal{Z}_\nu^{(2)}
& =16\left(\tau_2^2 \partial_\tau^2 \partial_{\bar{\tau}}^2+i \tau_2\left(\partial_\tau^2 \partial_{\bar{\tau}}-\partial_{\bar{\tau}}^2 \partial_\tau\right)\right) \mathcal{Z}_\nu^{(0)}.
\end{aligned}
\end{equation}
The correction to all orders of the partition function is universal in the sense that these corrections can be expressed in terms of the undeformed partition function \cite{Datta:2018thy}, as shown here for first and second-order corrections for both free bosons and fermions. Typically, in a recent study \cite{Nakayama:2025lyi}, the author investigated a lattice model constructed solely from fermions that exhibits spontaneously broken supersymmetry. Remarkably, this model connects explicitly to the $T\bar{T}$ deformation of a free fermion theory in the continuum limit. The analysis reveals that the theory transitions from Lorentz-invariant to non-relativistic as the deformation parameter reaches a critical value. This feature positions the model as a particularly intriguing framework for exploring non-perturbative aspects of $T\bar{T}$ deformations beyond the critical threshold.

\subsubsection*{KdV charge one-point function: perturbatively}
{ In CFT, the KdV charges are a set of conserved charges that generate an infinite-dimensional abelian algebra. The corresponding conserved currents are constructed from the power of stress tensor and its derivatives.}
In this subsection, we will proceed with the perturbation method to study the deformation of the one-point function of the first KdV charge  \cite{Bazhanov:1994ft}. There are also studies on deformed KdV charges based on non-perturbation methods \cite{LeFloch:2019wlf, Asrat:2020jsh}. Since the KdV charges are conserved under $T\bar{T}$ deformation \cite{LeFloch:2019wlf}, it will flow
under deformation, similar to the stress tensor, as mentioned at the beginning of this section. The deformation of several conserved charges and currents has been further investigated in the massive integrable models\cite{Travaglino:2024lyf}.

The left-moving KdV charges as $P^\lambda_s$ is a integral of conserved currents ($T_{s+1}^\lambda,\Theta_{s-1}^\lambda)$, with deformation parameter $\l$
\begin{equation}
P_s^\lambda=\frac{1}{2 \pi} \int_0^L\left(\mathrm{~d} z T_{s+1}^\lambda+\mathrm{d} \bar{z} \Theta_{s-1}^\lambda\right),
\end{equation}
{ where the subscript $s$ indicates the charge's spin.}
For the first KdV charge, i.e., $s=1$, is distinguished with others since it relates with Hamiltonian $H^\lambda=-\int_0^L \mathrm{~d} x T_{y y}^\lambda$ and momentum $P^\lambda=-i \int_0^L \mathrm{~d} x T_{x y}^\lambda$ of the system, as follows
\begin{equation}
P_1^\lambda=\frac{1}{2 \pi} \int_0^L\left(\mathrm{~d} z T^\lambda+\mathrm{d} \bar{z} \Theta^\lambda\right) \equiv-\frac{H^\lambda+P^\lambda}{2}.
\end{equation}
The expectation value of $P_1^\lambda$ in the deformed state $|n\rangle^\lambda$ thus reads
\be 
{ }^\lambda\left\langle n\left|P_1^\lambda\right| n\right\rangle^\lambda=-\frac{\mathcal{E}_n^\lambda+P_n^\lambda}{2},
\ee 
where $\mathcal{E}_n^\lambda $ and $P_n^\lambda$  represent the energy and momentum of the state $|n\rangle^\lambda$, which can be read form deformed spectrum on cylinder 
\begin{equation}
\mathcal{E}_n^\lambda=\frac{L}{2 \lambda}\left(\sqrt{1+\frac{4 \lambda E_n}{L}+\frac{4 \lambda^2\left(P_n\right)^2}{L^2}}-1\right), \quad P_n^\lambda=P_n,
\end{equation}
where $E_n$ and $P_n$ are the energy and momentum of the undeformed state, respectively. Therefore 
\begin{equation} \label{non-perkdv}
{ }^\lambda\left\langle n\left|P_1^\lambda\right| n\right\rangle^\lambda=\frac{L}{4 \lambda}\left(1-\sqrt{1+\frac{4 \lambda E_n}{L}+\frac{4 \lambda^2\left(P_n\right)^2}{L^2}}\right)-\frac{P_n}{2}.
\end{equation}
the exact result can be expanded in the power of $\l$ to compare with the results obtained from perturbation theory. In what follows, this comparison will be performed in free CFT with ground state $n=0$ case.

In undeformed case, for free bosons with periodic boundary condition $\phi(z+L)=\phi(z)$, and  fermions with anti-periodic boundary condition $\psi(z+L)=-\psi(z)$, the vacuum energy and momentum are $(n=0)$
\be 
E_0=-\frac{\pi c}{6L},~~P_0=0,
\ee
while for periodic fermions 
\be 
E_0=\frac{\pi c}{3L},~~P_0=0,
\ee
Notice that for ground state $n=0$, cylinder one-point function \eqref{non-perkdv} can be computed from the corresponding one-point functions
on the torus by taking the zero temperature limit
\begin{equation}\label{cy-to}
\lim _{\beta \to  \infty}\left\langle\mathcal{O}^\lambda\right\rangle_{\text {tor. }}^\lambda \equiv \lim _{\beta \to \infty}\left\{\operatorname{Tr}\left[e^{-\beta H^\lambda}\right]^{-1} \cdot \operatorname{Tr}\left[e^{-\beta H^\lambda} \mathcal{O}^\lambda\right]\right\}=\left\langle\mathcal{O}^\lambda\right\rangle_0^\lambda.
\end{equation}
Given \eqref{cy-to}, in what follows, we will reproduce \eqref{non-perkdv} for the ground state up to first-order in $\lambda$ in the context of free boson and fermion perturbatively.

In Lagrangian path integral formalism, the one-point function on the LHS \eqref{cy-to} equals 
\begin{equation}
\begin{aligned}
\left\langle\mathcal{O}^\lambda\right\rangle_{\text {tor. }}^\lambda  =&\frac{1}{\mathcal{Z}^\lambda} \int \mathcal{D} \phi \mathcal{O}^\lambda \exp \left\{-\int_{\mathrm{T}^2} \mathcal{L}^\lambda\right\} 
\end{aligned}
\end{equation}
With the help of free propagators given previously and Wick contraction, one obtains  for free boson case \cite{He:2020cxp}
\begin{equation}
\begin{aligned}
\left\langle T^\lambda\right\rangle_{\text {tor.FB }}^\lambda & =\left(\eta_1-\frac{\pi}{2 \tau_2}\right)+\lambda\cdot\left(\frac{2\left|\eta_1\right|^2}{\pi}-\frac{1}{2 \tau_2}\left(\eta_1+\bar{\eta}_1\right)+\left(\frac{2}{\pi} \tau_2 \bar{\eta}_1-1\right) i \partial_\tau \eta_1\right)+O\left(\lambda^2\right), \\
\left\langle\Theta^\lambda\right\rangle_{\text {tor.FB }}^\lambda & =\lambda \cdot\left(-\frac{\left|\eta_1\right|^2}{\pi}+\frac{2}{\tau_2}\left(\eta_1+\bar{\eta}_1\right)-\frac{3 \pi}{4 \tau_2^2}\right)+O\left(\lambda^2\right),
\end{aligned}
\end{equation}
and for free Majorana fermions
\be 
\begin{aligned}
& \left\langle T^\lambda\right\rangle_{\nu ; \text { tor.MF }}^\lambda=-\frac{e_{\nu-1}}{4}+\lambda \cdot\left(\frac{\left|e_{\nu-1}\right|^2}{8 \pi}+\frac{i \tau_2 \bar{e}_{\nu-1} \partial_\tau e_{\nu-1}}{8 \pi}\right)+O\left(\lambda^2\right), \\
& \left\langle\Theta^\lambda\right\rangle_{\nu ; \text { tor.MF }}^\lambda=\lambda \cdot \frac{-\left|e_{\nu-1}\right|^2}{16 \pi}+O\left(\lambda^2\right), \quad \nu=2,3,4 .
\end{aligned}
\ee 
where $\nu=2,3,4$ correspond to the periodic(space)-
antiperiodic(time), antiperiodic-periodic and antiperiodic-antiperiodic sectors respectively.
Taking zero temperature limit $\tau_2=\beta\to \infty$, \footnote{Here for simplicity the modular parameter of the torus is chosen to be pure imaginary  $\tau=i\tau_2$. }  one obtains the $T \bar{T}$-flow of the $\mathrm{KdV}$ charge $P_1$ up to the first-order for three free theories
\be \ba\label{per1pt}
&\left\langle P_1^\lambda\right\rangle_{o, \mathrm{FB}}^\lambda=\frac{\pi}{12}+\lambda \cdot \frac{\pi^2}{72}+O\left(\lambda^2\right), & \text { (periodic B.C.) } \\
&\left\langle P_1^\lambda\right\rangle_{\mathrm{o}, \mathrm{MF}}^\lambda=\frac{\pi}{24}+\lambda \cdot \frac{\pi^2}{288}+O\left(\lambda^2\right), & \text { (antiperiodic B.C.) } \\ 
&\left\langle P_1^\lambda\right\rangle_{o, \mathrm{MF}}^\lambda=-\frac{\pi}{12}+\lambda \cdot \frac{\pi^2}{72}+O\left(\lambda^2\right). & \text { (periodic B.C.) }
\ea\ee 
It’s easy to check that the perturbative results \eqref{per1pt} match
the results come from the non-perturbative method \eqref{non-perkdv}.
\subsubsection*{Higher-order correlation functions}
In the previous section, we discussed the first-order correlation function \eqref{firston} of undeformed operators on the plane. A natural question is how to compute higher-order corrections. For example, the second order takes the form 
\begin{equation}  \ba
\left\langle\mathcal{O}_1  \mathcal{O}_2 \cdots \mathcal{O}_n \right\rangle_\lambda^{(2)}=&\lambda^2( \int d^2 z_1 d^2z_2\left\langle T \bar{T}(z_1, \bar{z}_1)T \bar{T}(z_2, \bar{z}_2) \mathcal{O}_1 \mathcal{O}_2 \cdots \mathcal{O}_n \right\rangle+...),
\ea\end{equation}
where the dots involves the term containing the second-order action $S_{(2)}$ in \eqref{actionexp}.
Similar to the discussion above \eqref{L2boson},  the explicit expression for $S_{(2)}$ is unknown for general deformed CFT, so the traditional perturbation theory seems not to work for higher-order correlation functions. As an attempt to overcome this problem, a new
method is developed to compute $\left\langle\mathcal{O}_1  \mathcal{O}_2 \cdots \mathcal{O}_n \right\rangle_\lambda^{(2)}$, which based on trace relation of $T\bar{T}$ deformation and conservation of stress tensor \cite{He:2023kgq}. As a consistent check of the universal $\log^2$ behavior at second order of two-point function can be produced by this method, agreed with prior analyses, the works of Cardy \cite{Cardy:2019qao} and also  \cite{Aharony:2023dod, Cui:2023jrb}. 

Non-perturbative two-point functions have been derived in three distinct frameworks: the resummation of leading logarithmic divergences at small momentum in \cite{Cardy:2019qao}, the $JT$-formalism analysis of large-momentum asymptotics in \cite{Aharony:2023dod}, and the worldsheet computation of $\mathrm{TsT}$-deformed vertex operators in \cite{Cui:2023jrb}.\footnote{See \eqref{2ptJT} and \eqref{cardy2pt} { below for the explicit expressions for two-point functions obtained from these methods}.} While the results of \cite{Aharony:2023dod} and \cite{Cui:2023jrb} align precisely, they differ from \cite{Cardy:2019qao} by a momentum-dependent factor. 

This discrepancy is addressed in \cite{Aharony:2023dod}, where a perturbative first-order correction yields agreement in the leading log ($\log^1$) term. The analysis therein emphasizes that, within perturbation theory, only the leading logarithmic term is scheme-independent and thus appropriate for cross-comparison. Consistent with this observation, the result in \cite{He:2023kgq} indicates that the subleading contributions to the first- and second-order corrections depend on the renormalization scale $\mu$. Accordingly, a meaningful consistency check should focus on the leading logarithm.

\subsubsection*{Non-perturbartive method}
For $T\bar{T}$ deformation with general parameter $\l$, it was found that on a 2D plane, the flow of the deformed correlation function can be written as the flow of each field 
\cite{Cardy:2019qao}
\be\ba 
\p_\l\vev{\prod_n\Phi_n(x_n)}_\l=\sum_n\vev{\p_\l\Phi_n^\l(x_n)\prod_{m\neq n}\Phi^\l_m(x_m)}
\ea\ee 
with 
\be \label{flowphi1}
\p_\l\Phi^\l(x)=2\pi \epsilon^{ab}\epsilon^{ij}\int_x^X dx'_jT^\l_{ai}(x'+\varepsilon)\p_{x^b}\Phi^\l(x).
\ee 
The leading divergence on the RHS follows from OPE of $T$ and $\Phi$ \eqref{flowphi1}
\be\ba 
\p_\l\Phi^\l(x)=-(\log|\varepsilon|)\nabla^2_x\Phi^\l(x)+...
\ea\ee 
The all-orders analysis was performed in \cite{Cardy:2019qao}, and it turns out that the logarithmic divergence occurred at all orders, with leading terms of the form $(\l \log \Lambda)^N$ in each order. Resume all such logarithmic terms  divergence and work in momentum space, which may be more convenient than the position space for non-local theory, yielding
\begin{equation}
\Phi^\lambda(k)=e^{-\lambda \log (\Lambda / \mu) k^2} \Phi^0(k).
\end{equation}
It then follows that
the renormalized field can be defined by canceling the logarithmic divergence 
\begin{equation}
\widehat{\Phi}^\lambda(k) \equiv e^{\lambda \log (\Lambda / \mu) k^2} \Phi^\lambda(k),
\end{equation}
which is a non-local renormalization in contrast with the renormalization of local fields.
Since the bare correlation functions of field $\Phi^\l$ are independent of scale $\m$, it follows the  correlation functions satisfy
Callan-Symanzik-like equation. For example the renormalized two-point function in momentum space $\widehat{C}(k ; \lambda, \mu) $ satisfies
\begin{equation}\label{csequ}
\left[\mu \partial_\mu+k \partial_k-2 \lambda \partial_\lambda-2 \Delta\right] \widehat{C}(k ; \lambda, \mu)=0.
\end{equation}
From \eqref{momentum2pt}, for undeformed case, i.e., CFT two-point function in momentum space is $\widehat{C}(k ; \lambda=0)\sim k^{2\Delta-2}$ \footnote{However, the two-point function reads differently  $\widehat{C}(k ; \lambda=0)\sim k^{2\Delta-2}$, as indicated above eq(4.25) in \cite{Cardy:2019qao}.}
At fixed renormalization scale $\mu$, the solution then follows
\be\label{twoptcs}
\widehat{C}(k ; \lambda, \mu)=
k^{2 \Delta-2}(k / \mu)^{2 \lambda k^2}.
\ee 
To match the correlation functions obtained from the dual string theory side, a more general solution to \eqref{csequ} was proposed by multiplying an addition factor $F(\l k^2)$ with $F(0)=1$ to  \eqref{twoptcs} \cite{Cui:2023jrb}.

\subsection{Correlation functions: further developments}
This subsection mainly focuses on the conformal perturbation theory in computing deformed correlation functions. There has been much  progress 
going beyond traditional perturbation methods, including the methods based on dynamical coordinate transformation \cite{Hirano:2024eab} and flat JT gravity perspective \cite{Aharony:2023dod, Barel:2024dgv},
random geometry \cite{Hirano:2020ppu}, and point-splitting methods \cite{Cardy:2019qao,Hirano:2020ppu,Cui:2023jrb}. In this subsection, we will briefly review the first two methods. Let us begin with the former case.
In \cite{Cardy:2019qao}, it was found that the field transformed under $T\bar{T}$ deformation as 
\footnote{
Follow the conventions in \cite{DiFrancesco:1997nk}
\be\ba 
\epsilon^{z\bar{z}}=-2i,~~\epsilon_{z\bar{z}}=\frac{i}{2},~~g_{z\bar{z}}=\frac12,~~g^{z\bar{z}}=2.
\ea\ee 
}
\be \label{flowphi}
\partial_\l\Phi^\l(x)=2\pi \epsilon^{ab}\epsilon^{ij}\int_{x}^X dx'_j T^\l_{ai}(x'+\varepsilon)\p_{x^b}\Phi^\l(x),
\ee
where $X$ is an arbitrary point, the integral path can be anyone, provided it does not intersect with the points where other fields are inserted inside the correlation function. $\varepsilon$ is the point-splitting regulator.
\eqref{flowphi} suggests that instead of deforming the field, one can deform the coordinate, i.e., 
\be\ba 
\partial_\l\Phi^\l(x)=\p_{x^b}\Phi^\l(x)\cdot \p_\l x^b
\ea\ee 
with 
\be 
 \p_\l x^b=2\pi \epsilon^{ab}\epsilon^{ij}\int_{x}^X dx'_j T^\l_{ai}(x'+\varepsilon)=2\pi \epsilon^{ab}\epsilon^{ij}\int_{x+\varepsilon}^X dx'_j T^\l_{ai}(x').
\ee 
In component, this becomes
\be \ba \label{flowx}
\p_\l x^z=&\p_\l z=4\pi\int_{x+\varepsilon}^X\left(dz' T^\l_{\bar{z}z}(z')-d\bar{z}'T^\l_{\bar{z}\bar{z}}(z')\right),\\
\p_\l x^{\bar{z}}=&\p_\l \bar{z}=4\pi\int_{x+\varepsilon}^X\left(d\bar{z}' T^\l_{z\bar{z}}(z')-dz'T^\l_{zz}(z')\right),
\ea\ee 
which can also be derived from a random geometry perspective \cite{Cardy:2018sdv}. Put \eqref{flowx} equivalently into an integral form on a plane that is convenient for computing correlation function below   \cite{Hirano:2024eab}\footnote{Here  $dz\wedge d\bar{z}=-2i dx\wedge dy$, and 
Stokes theorem
\be 
\int dz\wedge d\bar{z} (\partial_z F^z+\partial_{\bar{z}}F^{\bar{z}})=\oint_C (F^zd\bar{z}-F^{\bar{z}}dz).
\ee 
}
\be\ba 
\p_\l x^z=&\p_\l z=4\pi\int_{x+\varepsilon}^X\left(dz' T^\l_{\bar{z}z}(z')-d\bar{z}'T^\l_{\bar{z}\bar{z}}(z')\right)\\
=&-2i\int dz' \wedge d\bar{z}'T^\l_{\bar{z}'\bar{z}'}(z')\left(\frac{1}{x-z'}-\frac{1}{X-z'}\right),
\ea\ee 
or equivalently, 
\begin{equation} \label{infzz}
z \longmapsto Z^{(\mu \mid \delta \mu)}(z, \bar{z})=z+\frac{\delta \mu}{2 \pi^2} \int_{\mathbb{R}^2} d^2 x \frac{\bar{T}^{(\mu)}(x, \bar{x})}{z-x}+\mathcal{O}\left(\delta \mu^2\right).
\end{equation}
This map is called dynamical coordinate transformation. Here the   theory $\mathcal{T}^{(\mu)}$ is defined on the deformed space $Z^{(\mu \mid \delta \mu)}$ and the deformed theory $\mathcal{T}^{(\mu+\delta\mu)}$ is defined on undeformed space $z $ (see also \cite{Dubovsky:2017cnj,Dubovsky:2018bmo,Conti:2018tca}).
It was proposed that under $T\bar{T}$-deformation, the theory deformed by $\l_1$ on space deformed by $\l_2$ is equivalent to the theory deformed by $\m_1$ on deformed space by $\m_2$ \cite{Hirano:2024eab, Cardy:2019qao}.  In the notation of \cite{Hirano:2024eab}, this equivalence can be expressed as 
\begin{equation} \label{equivTT}
\mathcal{T}^{\left(\lambda_1\right)}\left[\mathbb{R}_{\left(\lambda_1 \mid \lambda_2\right)}^2\right]=\mathcal{T}^{\left(\mu_1\right)}\left[\mathbb{R}_{\left(\mu_1 \mid \mu_2\right)}^2\right],~~\text{with}~~\l_1+\l_2=\m_1+\m_2,
\end{equation}
where the coordinates on space $\mathbb{R}_{\left(\lambda_1 \mid \lambda_2\right)}^2$is denoted by $Z^{(\l_1\mid \l_2)}$.
A special case is 
\be  \mathcal{T}^{\left(\m\right)}\left[\mathbb{R}_{\left(\m \mid 0\right)}^2\right]=\mathcal{T}^{\left(0\right)}\left[\mathbb{R}_{\left(0 \mid \mu\right)}^2\right],\ee 
which demonstrates that the $T\bar{T}$-deformed CFT $\mathcal{T}^{\left(\m\right)}$ on undeformed plane $\mathbb{R}_{\left(\m \mid 0\right)}^2\equiv \mathbb{R}^2$ (with coordinate $(z,\bar{z})$) is equivalent to  undeformed CFT $\mathcal{T}^{\left(0\right)}$ on deformed space $\mathbb{R}_{\left(0 \mid \m\right)}^2$ (with coordinate $(Z,\bar{Z})$), the dynamic coordinate transformation in this case is  
\be \label{tran1-dyna}
Z^{(\mu)}(z, \bar{z})=z+\frac{\mu}{2 \pi^2} \int_{\mathbb{R}^2} d^2 x \frac{\bar{T}^{(\mu)}(x, \bar{x})}{z-x},
\ee
Note the direct relation between infinitesimal version \eqref{infzz} and finite version \eqref{tran1-dyna} is checked for some particular case \cite{Hirano:2024eab}.  
This equivalence relation \eqref{equivTT} is proper both technically and conceptually. The flow of stress tensor can be inferred from \eqref{equivTT},
which is inaccessible in traditional perturbation expansion once the Lagrangian is unknown. Furthermore, \eqref{equivTT} plays an important role in computing correlation functions in deformed theory. 
In analogy to quantum mechanics, there are two pictures, namely { "Heisenberg" (where the operator $\mathcal{O}_{\Delta, \bar{\Delta}}^{(\lambda)}\left(z, \bar{z}\right) $ is deformed) and "Schr\"odinger" (where the operator $\mathcal{O}_{\Delta, \bar{\Delta}}^{(0)}\left(z, \bar{z}\right) $ is undeformed ) to describe the deformed correlation function }\cite{Hirano:2024eab, Cardy:2019qao}
\begin{equation}\label{npt-dyna}
\langle\mathcal{O}_{\Delta_1, \bar{\Delta}_1}^{(\lambda)}\left(z_1, \bar{z}_1\right) \cdots \mathcal{O}_{\Delta_n, \bar{\Delta}_n}^{(\lambda)}\left(z_n, \bar{z}_n\right)\rangle_0=\langle\mathcal{O}_{\Delta_1, \bar{\Delta}_1}^{(0)}\left(z_1, \bar{z}_1\right) \cdots \mathcal{O}_{\Delta_n, \bar{\Delta}_n}^{(0)}\left(z_n, \bar{z}_n\right)\rangle_\lambda.
\end{equation}  
Here $\vev{...}_\lambda$ ($\vev{...}_0$) denotes the correlation functions computed in the deformed (undeformed) vacuum state.   
Note the deformed correlation function on the RHS can be converted to a CFT correlation function on the LHS. While the operator on LHS  is related to the undeformed one via dynamic coordinates transformation
\be\ba  \label{tran-dyna}
\mathcal{O}_{\Delta, \bar{\Delta}}^{(\lambda)}(z, \bar{z})=&\left(\operatorname{det} \frac{\partial x}{\partial X}\right)^{-\frac{\Delta+\bar{\Delta}}{2}} \mathcal{O}^{(0 )}_{\Delta, \bar{\Delta}}(X)\\
=&\left(\operatorname{det} \frac{\partial x}{\partial X}\right)^{-\frac{\Delta+\bar{\Delta}}{2}} 
(\mathcal{O}^{(0 )}_{\Delta, \bar{\Delta}}(z)+\delta z \partial_z \mathcal{O}^{(0 )}_{\Delta, \bar{\Delta}}(z)+o(\m^2)),
\ea\ee
where $x=(z,\bar{z})$, and $X=(Z,\bar{Z})$ defined in \eqref{tran1-dyna}. Note the 2nd line is an expansion perturbatively in $\m$ such that operators are defined in CFT and undeformed coordinates. Therefore, using \eqref{tran-dyna},  the LHS of \eqref{npt-dyna} is expressed in terms of the CFT correlation function and can be evaluated by proper regularization. For example, the two-point function  to first  order 
\begin{equation}
\begin{aligned}
&\langle\mathcal{O}_{\Delta_1, \bar{\Delta}_1}^{(\lambda)}\left(z_1, \bar{z}_1\right) \mathcal{O}_{\Delta_n, \bar{\Delta}_n}^{(\lambda)}\left(z_2, \bar{z}_2\right)\rangle_0\\ =&\frac{1}{\left|z_{12}\right|^{4 \Delta}}-\frac{\l \Delta}{\pi^2 z_{12}^{2 \Delta+1}} \int d^2 x \frac{\left\langle\bar{T}(\bar{x}) \mathcal{O}_{\Delta}\left(\bar{z}_1\right) \mathcal{O}_{\Delta}\left(\bar{z}_2\right)\right\rangle_0}{z_1-x}+\left(z_1 \leftrightarrow z_2\right)+\text { c.c. } \\
=&\frac{1}{\left|z_{12}\right|^{4 \Delta}}-\frac{8 \l\Delta^2}{\pi\left|z_{12}\right|^{4 \Delta}}\left(\frac{\ln \left|z_{12} / \epsilon\right|^2}{\left|z_{12}\right|^2}-\frac{1}{\left|z_{12}\right|^2}-\frac{1}{2 \epsilon}\left(\frac{1}{z_{12}}+\frac{1}{\bar{z}_{12}}\right)\right).
\end{aligned}
\end{equation}
For the second order, the leading-log contribution can also be reproduced 
\begin{equation}
\left\langle\mathcal{O}_{\Delta, \Delta}^{(\l)}\left(z_1, \bar{z}_1\right) \mathcal{O}_{\Delta, \Delta}^{(\l)}\left(z_2, \bar{z}_2\right)\right\rangle_0=\cdots+\frac{8 \l^2}{\pi^2} \Delta^2(2 \Delta+1)^2 \frac{\ln ^2\left|z_{12} / \epsilon\right|^2}{\left|z_{12}\right|^{4(\Delta+1)}}+\cdots 
\end{equation}
In summary, the dynamical coordinate transform offers   a  way of computing higher-order correlation function functions, contrasting with the traditional perturbation theory. 

Now, we turn to another method for deformed correlation functions based on flat JT gravity interpretation of $T\bar{T}$ deformation.  
As mentioned in the beginning of sec. \ref{sec:corr}, perturbatively, one usually considers the correlation function of undeformed theory. A question is, how can the appropriate operator be defined in the deformed theory? 
\footnote{For discussion in this regard, see, for example, \cite{Asrat:2017tzd}. In Minkowski signature, two kinds of operators, namely undeformed and dressed ones, and their correlation functions were discussed in \cite{Kruthoff:2020hsi}.
} It was suggested that the flat JT gravity interpretation of $T\bar{T}$ deformation could define the deformed operator as described below.

As introduced in the section \ref{sec:basic}, the action of $T\bar{T}$ deformed theory can be written as 
\begin{equation}
\begin{aligned}
S_{T\bar{T}}\left(\psi, e_\alpha^a, X^a\right) & =S_0\left(\psi, e_\alpha^a\right)+S_{J T}\left(e_\alpha^a, X^a\right)\\
& =S_0\left(\psi, e_\alpha^a\right)+\frac{\Lambda}{2} \int d^2 \sigma \epsilon^{\alpha \beta} \epsilon_{a b}\left(\partial_\alpha X^a-e_\alpha^a\right)\left(\partial_\beta X^b-e_\beta^b\right),
\end{aligned}
\end{equation}
where $X$ is the dynamical coordinates, which is the original space of the undeformed theory. $\sigma^a$ is the coordinates of an auxiliary space.
One can introduce local operators on $X$ as a deformation of undeformed operators 
\footnote{For discussion on correlation functions of undeformed operators in gravity action formalism, see \cite{Tsolakidis:2024wut}.}
\begin{equation}
\mathcal{O}\left(X_0\right)=\int d^2 \sigma \sqrt{g(\sigma)} O(\sigma) \delta\left(X(\sigma)-X_0\right).
\end{equation}
Transforming into momentum space
\begin{equation}
\mathcal{O}\left(q_0\right)=\int d^2 \sigma \sqrt{g(\sigma)} O(\sigma) \exp \left(i q_0 \cdot X(\sigma)\right).
\end{equation}
The corresponding correlation functions are then defined by the path integral
\be 
\begin{aligned}
C\left(q_1, q_2\right) & \equiv\left\langle\mathcal{O}\left(q_1\right) \mathcal{O}\left(q_2\right)\right\rangle=\frac{1}{Z_{T\bar{T}}} \int \frac{\mathcal{D} e \mathcal{D} X \mathcal{D} \psi}{V_{\mathrm{diff}}} \mathcal{O}\left(q_1\right) \mathcal{O}\left(q_2\right) e^{-S_{T\bar{T}}}. 
\end{aligned}
\ee 
In $T\overline{T}$-deformed theories, the two-point correlation function in momentum space in the large momentum limit is given by the expression~\cite{Aharony:2023dod}
\be \label{2ptJT}
\frac{C}{|\Lambda|}\frac{1}{\sin (\frac{q^2}{2|\Lambda|}+\pi \Delta)}\left(\frac{\mu}{2e}\right)^{2\Delta}(\frac{2\pi e|\Lambda|}{\mu|q|})^{\frac{q^2}{\pi|\Lambda|}+2 \Delta)},
\ee 
which  in contrast with Cardy's results \cite{Cardy:2019qao}, which is \eqref{twoptcs} but in a different notation
\begin{equation}\label{cardy2pt}
C_t(q) = C_0(q) \left( \frac{|q|}{\mu} \right)^{\frac{tq^2}{\pi}},
\end{equation}
$C_0(q)$ is the undeformed conformal field theory (CFT) correlation function,  $\mu$ is a renormalization scale, and $t$ is the deformation parameter. 
The correlation function \eqref{2ptJT} reveals the non-local nature of $T\overline{T}$-deformations,
\footnote{In momentum space, power-law behavior $|p|^n$ is a characteristic of local field theory. The two-point correlation function in CFT in momentum space is 
\be\ba \label{momentum2pt}
\int d^2x \frac{e^{i{\bf{p}}\cdot{\bf{x}}}}{|{\bf{x}}|^{2\Delta}}=\int dr d\theta r \frac{e^{ip r\cos\theta }}{r^{2\Delta}}=\int dr \frac{2\pi J_0(pr)}{r^{2\Delta-1}}=2^{2-2\Delta}\pi\frac{\Gamma(1-\Delta)}{\Gamma(\Delta)}p^{2\Delta-2}.
\ea\ee 
} especially at large momenta 
, a key feature of these deformed theories. Additionally, renormalization in these models is momentum-dependent, contrasting with local quantum field theories. The function also exhibits ultraviolet (UV) divergences, which are regulated through techniques like point-splitting, further emphasizing the unique properties of $T\overline{T}$-deformed models in locality and operator renormalization. This result highlights the deformation's impact on short-distance physics, differing significantly from conventional QFTs.

In \cite{Barel:2024dgv}, the two-point function \eqref{2ptJT} on the plane was generalized to the torus, based also on the JT gravity formulation of $T\bar{T}$ deformation. On a torus, at small momenta, $ |q| \ll L/t $, the correlation function behaves similarly to the plane case, where $ C_t(q) \propto |q|^{-\frac{tq^2}{\pi}} $, with $t$ as the deformation parameter, $q$ the momentum, and $L$ the torus length scale. However, for large momenta, $ |q| \gg L/t $, the geometry of the torus modifies the correlation function, yielding the expression
\begin{equation}
C_t(q) \propto \left( 4\sqrt{\frac{tq^2}{\pi e L^3 |T|^2}} \right)^{\frac{tq^2}{\pi}},
\end{equation}
where $T$ is the modular parameter of the torus.
The main properties of the correlation function include UV-IR mixing, where the high-energy (UV) behavior is influenced by the global spacetime structure (IR), a characteristic feature of non-local theories. Furthermore, the deformation introduces momentum smearing, where operators carrying large momentum $q$ are smeared over a distance scale proportional to $t|q|$, highlighting the non-local nature of the deformation. The torus geometry also affects the correlation function significantly. At the same time, small $|q|$ behavior mimics that of the plane; at large $|q|$, the correlation function first decays and then grows, implying that the correlation function diverges at very large momenta. Overall, the correlation function in the $T\overline{T}$-deformed theory on a torus reflects the distinct non-local properties induced by the deformation, deviating from the typical power-law decay found in traditional quantum field theories.

{ Before closing this subsection, let us briefly mention that there are also other ways to investigate the deformed correlation functions, such as holographic method, which will be reviewed in next section, and methods in integrable field theory \cite{Castro-Alvaredo:2023rtl,Castro-Alvaredo:2023wmw,Castro-Alvaredo:2023hap,Castro-Alvaredo:2025nma}. }

\subsection{Quantum entanglement in $T\overline{T}$ deformed field theory}\label{EE-TTbar}

The investigation of entanglement entropy in $T\overline{T}$-deformed quantum field theories has attracted considerable interest for its profound implications in quantum information and non-local field dynamics. As an irrelevant deformation, $T\overline{T}$ induces inherent non-locality and modifies infrared physics, providing a novel framework for probing deformed quantum systems. In two-dimensional spacetime, the classical equivalence between $T\overline{T}$-deformed theories and Jackiw-Teitelboim (JT) gravity-coupled systems has been extended to higher dimensions \cite{Babaei-Aghbolagh:2024hti}, prompting investigations into whether this correspondence persists at the quantum level.

A central challenge emerges from the divergent algebraic structures \cite{Witten:2018zxz, Witten:2021jzq} underlying quantum field theories and gravitational systems, particularly their consistent incorporation within $T\overline{T}$-deformed frameworks. Entanglement entropy is a critical holography probe for analyzing deformation-induced geometric modifications and their dual gravitational manifestations. The intrinsic non-locality of $T\overline{T}$ deformation establishes direct connections to quantum gravity paradigms, where non-local effects play fundamental roles.

The $T\overline{T}$-deformed conformal field theory (CFT) has emerged as a holographic dual to quantum gravity in finite spacetime regions. As proposed in \cite{Donnelly:2018bef} and supported by \cite{McGough:2016lol}, this duality suggests that $T\overline{T}$-deformed CFTs encode gravitational physics within bounded geometries. A key feature of this deformation is its introduction of a finite radial cutoff in AdS spacetimes, which modifies holographic prescriptions like the Ryu-Takayanagi formula \cite{Banerjee:2019ewu}. Notably, the deformation reduces subsystem degrees of freedom, rescales energy scales, and can trigger phase transitions even for fixed subsystem sizes \cite{Banerjee:2019ewu, Ota:2019yfe}.
Recent work extends these insights across diverse holographic settings. In \cite{Chang:2024voo}, the replica sphere method was generalized to derive deformed entanglement entropy for arbitrary single intervals in both AdS and dS holography, revealing non-locality in timelike boundary theories under $T\overline{T}$ deformation. While multi-interval entanglement entropy in holographic CFTs typically sums single-interval contributions at large separations \cite{Jeong:2019ylz}, the deformation introduces novel constraints. For 2D systems, an alternative interpretation frames $T\overline{T}$-deformed CFTs as AdS$_3$ gravity with mixed boundary conditions \cite{Guica:2019nzm}, where coordinate transformations connect Ba\~{n}ados geometries to holographic entanglement entropy calculations \cite{He:2023xnb}, consistent with earlier results \cite{Chen:2018eqk}. Progress in quantifying entanglement includes perturbative expressions for negativity in bipartite mixed states \cite{Basu:2023bov}, mixed state entanglement \cite{Pant:2024eno}, explorations of odd, reflected, { temporal, and timeline entanglement entropies \cite{Grieninger:2023knz,Basu:2023aqz, Basu:2024bal, Basu:2024enr}. }Notably, \cite{Soni:2024aop} recently proposed a holographic covariant entropy bound (HCEB) connecting bulk Cauchy slices to boundary states via codimension-2 surfaces. In 3D AdS gravity with $T\overline{T}$-deformed boundaries, HCEB matches extremal/marginal surface areas but exceeds trapped surface areas (e.g., inside black holes), enabling tensor network constructions for holographic entropy.

The behavior of the entanglement entropy in the deformed CFTs, especially in the context of the $T\bar{T}$-deformations, has been extensively studied in recent years\cite{Chen:2018eqk, He:2023xnb, He:2022xkh, Donnelly:2018bef, Ashkenazi:2023fcn}. Recent studies on entanglement entropy in $T\overline{T}$-deformed CFTs reveal key insights into the quantum structure of these systems. \cite{Chen:2018eqk} explored leading-order corrections in vacuum states using perturbative methods within the twistor operator formalism. In generic 2D CFTs, \cite{He:2019vzf} constructed the perturbative correlation functions to evaluate the entanglement entropy of the locally excited states and further demonstrated that the deformation preserves the maximal chaotic behavior of the holographic CFTs. It has been shown that pseudo entropy, like entanglement entropy, can be used to quantify the topological contributions of entanglement of excited states in two-dimensional rational CFTs\cite{ Nishioka:2021cxe, Guo:2022sfl, He:2023eap}.  More recently, the generalized entanglement entropy called pseudo-entanglement entropy has been proposed via the AdS/CFT correspondence and post-selection \cite{Nakata:2020luh}. The pseudo-entanglement entropy corrected by the deformation has been investigated \cite{He:2023wko}, which shows that the deformation does not change the characteristic behaviors of the quantum chaos \cite{He:2019vzf}. In the massive quantum fields with $T\overline{T}$~\cite{Ashkenazi:2023fcn},  the leading correction to entanglement entropy in $T\overline{T}$-deformed massive quantum field theories arises from the boundary of the entangling surface, with a finite correction for scalars and a log-square divergence for Dirac fermions. By applying a form factor bootstrap approach, {  the paper~\cite{Castro-Alvaredo:2023jbg,He:2023obo} calculates the $T\overline{T}$-deformed }entanglement entropy for integrable quantum field theories, revealing that the UV behavior remains similar to the undeformed case with a modified central charge.

\subsection{$T\bar{T}$ flow at large central charge} \label{Subsection TTbar flow at large central charge}
In this subsection, we briefly switch gears to discuss the ``more classical'' aspects of $T\bar{T}$-deformed quantum field theory. We review the flow equation of partition function and one-point function of stress tensor for $T\bar{T}$-deformed
CFTs in the large central charge limit. This flow equation leads to the important trace relation (\ref{Classical trace relation}) that has been proved for classical field theories in previous subsections. Most importantly, it is closely related to the holographic interpretation of $T\bar{T}$ deformation for holographic CFTs, preparing for the discussion in the next section. Our derivation largely follows the original paper \cite{Guica:2019nzm} with some technical and conceptual tweaks.

The large central charge limit is a limit of large degrees of freedom, similar to the large $N$ limit of gauge theory. It was brought into the context of $T\bar{T}$ deformation in \cite{Aharony:2018vux}
\begin{align} \label{Large c limit}
    c\;\text{large},\; \lambda c\; \text{comparable to the energy scales considered}
\end{align}
and its implication for correlation functions was discussed there. The key simplification of this limit is that the connected correlation functions of the energy-momentum tensor all scale as the central charge $c$, so dominant contributions to a correlation function come from factorizations into the largest number of (nonvanishing) factors (large $c$ factorization). This line of argument was well known for multitrace deformations, for example, in \cite{Gubser:2002vv}. Consider a theory defined by the action $S^{[0]}[\phi,\gamma]$ with $\phi$ being the fundamental field in path integration and $\gamma_{ij}$ the background metric, the partition function and generating functional for energy-momentum tensor are given by
\begin{align}
 Z^{[0]}[\gamma] &= \int {\cal D}\phi e^{-S^{[0]}[\phi,\gamma]}, \nonumber\\
 I^{[0]}[\gamma] &= -\log Z^{[0]}[\gamma].
\end{align}
Connected correlation functions of the energy-momentum tensor are given by functional derivatives of the generating functional with respect to the background metric. If the theory is deformed by the $T\bar{T}$ deformation with infinitesimal deformation parameter $\lambda$, the partition function of the deformed theory is
\begin{align} \label{InfTTbarPathIntegral}
 Z[\gamma] = \int {\cal D}\phi e^{-S^{[0]}[\phi,\gamma] + \frac{1}{2}\lambda \int d^2x \sqrt{\gamma} \epsilon^{\mu\nu}\epsilon^{\rho\sigma}T^{[0]}_{\mu\rho}T^{[0]}_{\nu\sigma}}.
\end{align}
By doing a Hubbard-Stratonovich transformation \footnote{Reviewed in previous sections, this transformation has been used in \cite{Cardy:2018sdv} to derive the flow equation of partition functions in spaces with simple geometries where considerable simplification of the path integral over $h$ is possible.}, that is, inserting the identity
\begin{align}
 1 = \sqrt{\det(\lambda M_{\gamma})} \int {\cal D}h e^{-\frac{1}{2}\lambda \int d^2x\sqrt{\gamma} \epsilon^{\mu\nu}\epsilon^{\rho\sigma}(h_{\mu\rho}+T^{[0]}_{\mu\rho})(h_{\nu\sigma}+T^{[0]}_{\nu\sigma})}.
\end{align}
where $M_{\gamma}$ denotes the operator $M_{\gamma}(f,g) = \int d^2x\sqrt{\gamma} \epsilon^{\alpha\gamma}\epsilon^{\beta\delta} f_{\alpha\beta} g_{\gamma\delta}$, we find
\begin{align} \label{TTbar partition function large c transform}
 Z[\gamma] &= \int {\cal D}\phi \sqrt{\det(\lambda M_{\gamma})} \int {\cal D}h e^{-S^{[0]}[\phi,\gamma_{\mu\nu}] - \frac{1}{2}\lambda \int d^2x \sqrt{\gamma} \epsilon^{\mu\rho}\epsilon^{\nu\sigma}h_{\mu\nu}h_{\rho\sigma} - \lambda \int d^2x \sqrt{\gamma} \epsilon^{\mu\rho}\epsilon^{\nu\sigma}T^{[0]}_{\mu\nu}h_{\rho\sigma}} \nonumber\\
 &=\sqrt{\det(\lambda M_{\gamma})} \int {\cal D}h e^{-\frac{1}{2}\lambda \int d^2x \sqrt{\gamma} \epsilon^{\mu\rho}\epsilon^{\nu\sigma}h_{\mu\nu}h_{\rho\sigma}} \int {\cal D}\phi  e^{-S^{[0]}[\phi,\gamma_{\mu\nu}] - \lambda \int d^2x \sqrt{\gamma} \epsilon^{\mu\rho}\epsilon^{\nu\sigma}T^{[0]}_{\mu\nu}h_{\rho\sigma}} \nonumber\\
 &\approx\sqrt{\det(\lambda M_{\gamma})} \int {\cal D}h e^{-\frac{1}{2}\lambda \int d^2x \sqrt{\gamma} \epsilon^{\mu\rho}\epsilon^{\nu\sigma}h_{\mu\nu}h_{\rho\sigma}} \int {\cal D}\phi  e^{-S^{[0]}[\phi,\gamma^{\mu\nu}+ 2\lambda \epsilon^{\mu\rho}\epsilon^{\nu\sigma}h_{\rho\sigma}]} \nonumber\\
 &=\sqrt{\det(\lambda M_{\gamma})} \int {\cal D}h e^{-\frac{1}{2}\lambda \int d^2x \sqrt{\gamma} \epsilon^{\mu\rho}\epsilon^{\nu\sigma}h_{\mu\nu}h_{\rho\sigma}-I^{[0]}[\gamma^{\mu\nu}+ 2\lambda \epsilon^{\mu\rho}\epsilon^{\nu\sigma}h_{\rho\sigma}]}.
\end{align}
In the third line we only keep terms up to the first order in $\lambda$. In the large $c$ limit, we approximate the path integral by the saddle point, determined by the equation
\begin{align}
 h_{\mu\nu} = -\langle T_{\mu\nu} \rangle^{[0]}[\gamma^{\alpha\beta}+2\lambda \epsilon^{\rho\alpha}\epsilon^{\sigma\beta}h_{\rho\sigma} ].
\end{align}
$h_{\mu\nu}$ appears on the right-hand side in the background metric, making the equation a complicated functional equation, so there is no simple expression for the saddle point. It's thus convenient to introduce the background metric on the right-hand side as a new variable ${\gamma^{[0]}}^{\mu\nu}$, the ``undeformed metric''. Completing the path integral in (\ref{TTbar partition function large c transform}), we find to first order in $\lambda$
\begin{align}
 I[\gamma_{\mu\nu}] = I^{[0]}[{\gamma^{[0]}_{\mu\nu}}] + \frac{1}{2} \lambda \int d^2x \sqrt{\gamma^{[0]}} {\epsilon^{[0]}}^{\mu\rho}{\epsilon^{[0]}}^{\nu\sigma}\langle T_{\mu\nu}\rangle^{[0]} \langle T_{\rho\sigma} \rangle^{[0]}.
\end{align}
In addition we have, by our definition
\begin{align}
    \gamma^{\mu\nu} = {\gamma^{[0]}}^{\mu\nu} + 2\lambda {\epsilon^{[0]}}^{\rho\alpha}{\epsilon^{[0]}}^{\sigma\beta} \langle T_{\rho\sigma} \rangle^{[0]}.
\end{align}
If we consider $T\bar{T}$ deformation with a continuously varying parameter $\lambda$, the infinitesimal equations above become differential equations
\begin{align} \label{TTbar flow large c GeneratingFunctional}
 \partial_\lambda I^{[\lambda]}[\gamma^{[\lambda]}_{\alpha\beta}] &= \frac{1}{2}\int d^2x \sqrt{\gamma^{[\lambda]}} {\epsilon^{[\lambda]}}^{\mu\rho}{\epsilon^{[\lambda]}}^{\nu\sigma}\langle T_{\mu\nu}\rangle^{[\lambda]} \langle T_{\rho\sigma} \rangle^{[\lambda]} \nonumber\\
 &= -\frac{1}{2}\int d^2x \sqrt{\gamma^{[\lambda]}} (\langle T_{\mu\nu}\rangle^{[\lambda]} \langle T^{\mu\nu}\rangle^{[\lambda]} -{\langle T^\rho_\rho \rangle^{[\lambda]}}^2) = -\int d^2x \sqrt{\gamma^{[\lambda]}} \langle O_{T\bar
 T} \rangle^{[\lambda]}
\end{align}
and
\begin{align} \label{TTbar flow large c Metric}
 \partial_\lambda {\gamma^{[\lambda]}}^{\mu\nu} = 2{\epsilon^{[\lambda]}}^{\mu\rho}{\epsilon^{[\lambda]}}^{\nu\sigma} \langle T_{\rho\sigma}\rangle^{[\lambda]} = -2\langle T^{\mu\nu} \rangle^{[\lambda]} + \langle T^\rho_\rho\rangle^{[\lambda]} {\gamma^{[\lambda]}}^{\mu\nu}
\end{align}
To derive the flow equation of the one-point function, we take a variation in the metric of (\ref{TTbar flow large c GeneratingFunctional}) and obtain
\begin{align} \label{TTbar flow large c VariationalPrinciple}
 \partial_\lambda ( \sqrt{\gamma^{[\lambda]}}\langle T_{\mu\nu}\rangle^{[\lambda]} \delta {\gamma^{[\lambda]}}^{\mu\nu}) = -2\delta(\sqrt{\gamma^{[\lambda]}}\langle O_{T\bar
 T}\rangle^{[\lambda]}).
\end{align}
It further simplifies to
\begin{align} \label{TTbar flow large c from VariationalPrinciple}
 \partial_\lambda (\sqrt{\gamma} \langle T_{\mu\nu})\rangle) \delta\gamma^{\mu\nu}  + \sqrt{\gamma} \langle T_{\mu\nu}\rangle \delta(\partial_\lambda \gamma^{\mu\nu}) = &\sqrt{\gamma}\Big[ \big(\langle O_{T\bar{T}}\rangle\gamma_{\mu\nu} + 2\langle T_{\mu\rho}\rangle \langle T^{\rho}_{\;\nu} \rangle - 2 \langle T^\rho_\rho\rangle \langle T_{\mu\nu}\rangle \big) \delta\gamma^{\mu\nu} \nonumber\\
 & - \langle T_{\mu\nu}\rangle \delta \big( 2\langle T^{\mu\nu}\rangle - 2\gamma^{\mu\nu} \langle T^\rho_\rho\rangle \big) \Big]
\end{align}
where for simplicity we have omitted the notation for $\lambda$-dependence. Plugging (\ref{TTbar flow large c Metric}) into the equation, we find
\begin{align} \label{TTbar flow large c OnePointFunction}
 \partial_\lambda(\sqrt{\gamma} \langle T_{\mu\nu} \rangle) = \sqrt{\gamma} \big(\langle O_{T\bar{T}}\rangle\gamma_{\mu\nu} + 2\langle T_{\mu\rho}\rangle \langle T^{\rho}_{\;\nu} \rangle - 2 \langle T^\rho_\rho\rangle \langle T_{\mu\nu}\rangle \big)
\end{align}
Introducing the trace-reversed energy-momentum tensor $\hat{T}_{\mu\nu} = T_{\mu\nu} - T^\rho_\rho \gamma_{\mu\nu}$, (\ref{TTbar flow large c Metric}) and (\ref{TTbar flow large c OnePointFunction}) are further simplified to
\begin{align} \label{TTbar flow large c Trace-Reversed energy-momentumTensor}
 \partial_\lambda \gamma_{\mu\nu} &= 2 \langle \hat{T}_{\mu\nu}\rangle, \nonumber\\
 \partial_\lambda \langle \hat{T}_{\mu\nu} \rangle &= \langle \hat{T}_{\mu\rho}\rangle\langle\hat{T}^\rho_{\;\nu}\rangle.
\end{align}
Leaving the details of the derivation to the original paper \cite{Guica:2019nzm}, the solution to these equations was found to be
\begin{align} \label{TTbar flow large c solution}
 \gamma^{[\lambda]}_{\mu\nu} &= \gamma^{[0]}_{\mu\nu}+ 2\lambda \langle \hat{T}_{\mu\nu} \rangle^{[0]} + \lambda^2 \langle\hat{T}_{\mu\rho}\rangle^{[0]} \gamma^{[0]\rho\sigma} \langle\hat{T}_{\sigma\nu}\rangle^{[0]}, \nonumber\\
 \langle \hat{T}_{\mu\nu}\rangle^{[\lambda]} &= \langle \hat{T}_{\mu\nu}\rangle^{[0]} + \lambda \langle\hat{T}_{\mu\rho}\rangle^{[0]} \gamma^{[0]\rho\sigma} \langle\hat{T}_{\sigma\nu}\rangle^{[0]}.
\end{align}
It was further shown that $\sqrt{\gamma}\langle O_{T\bar{T}}\rangle$ is constant along the flow \cite{Guica:2019nzm}, so for the generating functional we simply have
\begin{align} \label{TTbar large c GeneratingFunctional linear}
 I^{[\lambda]}[\gamma^{[\lambda]}] = I^{[0]}[\gamma^{[0]}] - \lambda \int d^2x \sqrt{\gamma}\langle O_{T\bar{T}}\rangle,
\end{align}
where $\sqrt{\gamma}\langle O_{T\bar{T}}\rangle$ can be taken at any ``time'' along the flow between $0$ and $\lambda$. In addition, if the seed field theory is conformal, the energy-momentum tensor is traceless up to the Weyl anomaly
\begin{align}
    \langle T^\rho_\rho \rangle^{[0]} = \frac{c}{24\pi} R[\gamma^{[0]}]
\end{align}
then we can prove the trace relation follows from the flow equation \cite{Guica:2019nzm}
\begin{align} \label{TTbar flow large c TraceRelation}
 \langle T^\rho_\rho \rangle^{[\lambda]} = \frac{c}{24\pi} R[\gamma^{[\lambda]}] + 2 \lambda \langle O_{T\bar{T}}\rangle^{[\lambda]}.
\end{align}

(\ref{TTbar flow large c VariationalPrinciple}), known as the ``variational principle'', was the starting point of the derivation in \cite{Guica:2019nzm} (also used in \cite{Bzowski:2018pcy} for the $J\bar{T}$ deformation). As the general formalism for multi-trace deformation in \cite{Elitzur:2005kz, Papadimitriou:2007sj}, it encapsulates the effects of multi-trace deformation in a compact form, in analogy to the canonical transformation in the classical field theory. With sources $J^A$ dual to other operators $O_A$ turned on, we generally have
\begin{align} \label{TTbar flow large c AllSourcesAndOperators}
 \partial_\lambda (\frac{1}{2} \sqrt{\gamma^{[\lambda]}}\langle T_{\mu\nu}\rangle^{[\lambda]} \delta {\gamma^{[\lambda]}}^{\mu\nu} + \sum_A \sqrt{\gamma^{[\lambda]}} \langle O_A\rangle^{[\lambda]} \delta{J^A}^{[\lambda]}) = - \delta(\sqrt{\gamma^{[\lambda]}}\langle O_{T\bar{T}}\rangle^{[\lambda]}_{\text{factorized}}).
\end{align}
Since the variation of the action on the right-hand side does not involve $O_A$ or $J^A$, we have
\begin{align}
    {J^A}^{[\lambda]} &= {J^A}^{[0]}. \nonumber\\
    \sqrt{\gamma^{[\lambda]}} \langle O_A\rangle^{[\lambda]} &= \sqrt{\gamma^{[0]}} \langle O_A\rangle^{[0]}.
\end{align}
We would like to remark that, even though the metric is deformed as the $T\bar{T}$ deformation goes in the formulation of multi-trace deformation given above, there is no physical change of the background metric of field theories under the $T\bar{T}$ deformation. The flow of the metric is just a mathematical way to describe the flow of the generating functional as a functional of its argument, the metric. In fact, for double-trace deformations, we get the variation of the generating functional with the ``correct'' sign if we fix the source. That is, for the $T\bar{T}$ deformation  we have
\begin{align} \label{TTbar flow large c GeneratingFunctional FixedMetric}
 \partial_\lambda I^{[\lambda]}[\gamma] = \int d^2x \sqrt{\gamma} \langle O_{T\bar
 T} \rangle^{[\lambda]}
\end{align}
if we fix the metric. The ``variational principle'' and the flow equations (\ref{TTbar flow large c Metric})(\ref{TTbar flow large c GeneratingFunctional})(\ref{TTbar flow large c OnePointFunction}) are solutions to the functional differential equation (\ref{TTbar flow large c GeneratingFunctional FixedMetric}) by the method of characteristics, c.f. our discussion in \ref{sectionchar}. While this flow equation looks simpler, the formulation with varying metrics has its merits for many purposes, for example, $\sqrt{\gamma}\langle O_{T\bar{T}}\rangle$ is constant along the flow, and the flow of metric takes a compelling physical meaning in holography to be discussed in the next section. 

The large $c$ $T\bar{T}$ flow equations derived above enable us to compute the partition function and correlation functions of a $T\bar{T}$-deformed field theory if these were known for the seed theory. For example, the sphere partition function of a $T\bar{T}$-deformed CFT was considered in \cite{Donnelly:2018bef} and \cite{Caputa:2019pam}(generalizing to higher dimensions). Since the sphere is maximally symmetric, the one-point function of the energy-momentum tensor should be proportional to the metric
\begin{align}
    \langle T_{\mu\nu} \rangle = \alpha \gamma_{\mu\nu}
\end{align}
and from the trace relation (\ref{TTbar flow large c TraceRelation}), the constant of proportionality is determined to be
\begin{align}
    \alpha = \frac{\sqrt{1+\frac{\lambda c}{6\pi L^2}}-1}{2\lambda}
\end{align}
where $L$ is the radius of the sphere. The partition function satisfies the differential equation
\begin{align} \label{TTbar SpherePartitionFunction ODE}
    L\frac{d}{dL}\log Z = \int d^2x \sqrt{\gamma} \langle T^\rho_\rho \rangle = 4\pi L^2 \frac{\sqrt{1+\frac{\lambda c}{6\pi L^2}}-1}{\lambda}.
\end{align}
In \cite{Donnelly:2018bef}, the boundary condition $\log Z|_{L=0} = 0$ was chosen, and the sphere partition function was found to be
\begin{align} \label{TTbar SpherePartitionFunction DS}
    \log Z = \frac{c}{3}\log\big(\sqrt{\frac
    {6\pi L^2}{\lambda c}}+\sqrt{1+\frac{6\pi L^2}{\lambda c}}\big) + \frac{2\pi L^2}{\lambda}\big( \sqrt{1+\frac{\lambda c}{6\pi L^2}} -1\big).
\end{align}
Taking the face value, this expression does not have a proper CFT limit as $\lambda\to 0$ \footnote{Some people take the point of view that the deformation parameter $\lambda$ also plays the role of effective UV cutoff, somewhat justifying this expression for the partition function. We think the UV cutoff and the deformation parameter are two different quantities.}. This is not surprising since there is no justification of the boundary condition $\log Z|_{L=0} = 0$ for a CFT, and we don't expect it to hold for a slightly $T\bar{T}$-deformed CFT either. A more serious problem is that (\ref{TTbar SpherePartitionFunction DS}) doesn't satisfy the flow equation (\ref{TTbar flow large c GeneratingFunctional FixedMetric}). To find the partition function with a proper CFT limit, a more straightforward computation can be done using the $T\bar{T}$ flow equations derived in this section \cite{Li:2020zjb}. In particular, (\ref{TTbar large c GeneratingFunctional linear}) is convenient for the computation. For the second term on the right-hand side of (\ref{TTbar large c GeneratingFunctional linear}), we find $\int d^2x \sqrt{\gamma}\langle O_{T\bar{T}}\rangle = -4\pi L^2 \alpha^2 = -\frac{\pi L^2}{\lambda^2}\big(2+\frac{\lambda c}{6\pi L^2}-2\sqrt{1+\frac{\lambda c}{6\pi L^2}} \big)$. For the first term on the right-hand side, the partition function of a CFT on a sphere with radius $L_0$ is given by
\begin{align} \label{CFT sphere partition function}
    \log Z(L_0) = \frac{c}{3} \log \frac{L_0}{\epsilon}
\end{align}
where $\epsilon$ is a UV cutoff. The radius of the ``undeformed metric'' $L_0$ can be determined from the metric flow equation in (\ref{TTbar flow large c solution})
\begin{align}
    L_0 = L \frac{1+\sqrt{1+\frac{\lambda c}{6\pi L^2}}}{2}.
\end{align}
Summing up every term, we find
\begin{align} \label{TTbar SpherePartitionFunction L}
    \log Z = \frac{c}{3}\log \frac{L}{\epsilon} + \frac{c}{3}\log\frac{1+\sqrt{1+\frac{\lambda c}{6\pi L^2}}}{2} + \frac{2\pi L^2}{\lambda} (\sqrt{1+\frac{\lambda c}{6\pi L^2}} - 1) - \frac{c}{6}.
\end{align}
This expression of the sphere partition function satisfies the differential equation (\ref{TTbar SpherePartitionFunction ODE}) as well as the flow equation (\ref{TTbar flow large c GeneratingFunctional FixedMetric}). Still, now it has a proper CFT limit $\log Z = \frac{c}{3}\log\frac{L}{\epsilon}$ as $\lambda\to 0$. In addition, as $\lambda$ goes to $+\infty$, the partition function diverges as $\lambda^\frac{1}{2}$ instead of becoming trivial in (\ref{TTbar SpherePartitionFunction DS}). As far as we know, this expression was first given in \cite{Gorbenko:2018oov}, it was also discussed in \cite{Apolo:2023vnm}.

From the sphere partition function, one can compute the entanglement entropy of two antipodal points \cite{Donnelly:2018bef}. Using the replica trick, the entanglement entropy can be computed by
\begin{align} \label{EntanglementEntropy from Replica}
    S = (1-n\frac{d}{dn})\log Z_n \Big|_{n=1}
\end{align}
where $\log Z_n$ is the partition function of the CFT on the $n$-replica sphere. { The $n$-replica sphere is obtained by gluing $n$ copies of the sphere along a cut from the north pole to the south pole. It has the metric}
\begin{align}
    ds^2 = L^2\big(d\theta^2 + n^2 \sin^2\theta d\phi^2 \big).
\end{align}
The differentiation with respect to $n$ yields the energy-momentum tensor one-point function
\begin{align}
    n\frac{d}{dn} \log Z_n = -\int dV \langle T^\phi_\phi \rangle_n.
\end{align}
Taking the limit $n\to 1$, we have $\lim_{n\to 1}\langle T^\phi_\phi \rangle_n = \frac{1}{2}\langle T^\mu_\mu \rangle$, so we finally obtain
\begin{align}
    S = (1-\frac{L}{2}\frac{d}{dL})\log Z.
\end{align}
Using (\ref{TTbar SpherePartitionFunction DS}), the entanglement entropy was computed to be
\begin{align} \label{TTbar SphereEntanglementEntropy DS}
    S = \frac{c}{3} \log\big(\sqrt{\frac
    {6\pi L^2}{\lambda c}}+\sqrt{1+\frac{6\pi L^2}{\lambda c}}\big)
\end{align}
in \cite{Donnelly:2018bef}. If we use (\ref{TTbar SpherePartitionFunction L}) instead, the entanglement entropy is found to be
\begin{align} \label{TTbar SphereEntanglementEntropy L}
    S= \frac{c}{3}\log\frac{L}{\epsilon} + \frac{c}{3}\log\frac{1+\sqrt{1+\frac{\lambda c}{6\pi L^2}}}{2} - \frac{c}{6}.
\end{align}

\newpage
\section{$T\bar{T}$ deformation and holography}\label{Section TTbarDeformation and Holography}

For a holographic CFT, it's natural to ask what the holographic dual of its $T\bar{T}$ deformation is. It was well-understood that multi-trace deformations correspond to changing boundary conditions in the dual AdS space \cite{Witten:2001ua, Berkooz:2002ug, Elitzur:2005kz, Marolf:2006nd}. A sharp proposal was given in \cite{McGough:2016lol} that for positive $T\bar{T}$ deformation parameter $\lambda$, the holographic dual is a Dirichlet cutoff in the AdS space. Evidences for this holographic proposal and works in this holographic setting include signal propagation speed \cite{McGough:2016lol}, energy levels in a finite volume \cite{McGough:2016lol}, holographic energy-momentum tensor correlation functions \cite{Kraus:2018xrn,Li:2020pwa}, correspondence between the trace relation and Gauss-Codazzi equations \cite{Kraus:2018xrn, Taylor:2018xcy}, partition functions \cite{Donnelly:2018bef, Caputa:2019pam,Apolo:2023aho,Tian:2024vln,FarajiAstaneh:2024fig} and holographic entanglement entropy \cite{Donnelly:2018bef, Chen:2018eqk, Banerjee:2019ewu, Murdia:2019fax,Jeong:2019ylz,Grieninger:2019zts,Lewkowycz:2019xse,Geng:2019ruz,Donnelly:2019pie,Allameh:2021moy,FarajiAstaneh:2022qck,He:2023xnb,Grieninger:2023knz,Pant:2024eno}. Most evidences hinge on the $T\bar{T}$ flow of the partition function and the energy-momentum tensor. The crucial insight into the cutoff AdS holography proposal was given in \cite{Guica:2019nzm}, where the large $c$ $T\bar{T}$ flow equation of the partition function and the one point function of the energy-momentum tensor was derived, as we have reviewed in Section \ref{Subsection TTbar flow at large central charge}. An explicit solution to the flow equation was obtained, translating to a mixed boundary condition for the AdS gravity. Furthermore, when the deformation parameter $\lambda$ is positive, the mixed boundary condition coincides with a Dirichlet boundary condition on a cutoff surface. The cutoff location, determined by $\lambda$, moves into the bulk as $\lambda$ increases. For negative $\lambda$, a glue-on AdS picture was later proposed in \cite{Apolo:2023vnm}. The cutoff AdS or glue-on AdS picture only works in the pure gravity sector or on the field theory side when the physical quantities being considered only involve the partition function or the energy-momentum tensor. A bulk cutoff for all fields requires double-trace deformation involving all operators in the boundary field theory \cite{Heemskerk:2010hk, Faulkner:2010jy, Kraus:2018xrn, Hartman:2018tkw}. Very recently, generalizations of the $T\bar{T}$ deformation that bring the boundary to a finite cutoff with more general boundary conditions  were considered in \cite{Parvizi:2025shq,Parvizi:2025wsg}. We organize this section as follows. The main body of this section, the derivation of the mixed boundary and reduction to the cutoff or glue-on picture, is given as subsection \ref{Subsection MixedBoundaryCondition Cutoff and Glue-on}. We then discuss the holographic partition function and entanglement entropy in \ref{Subsection HolographicPartitionFunction and EntanglementEntropy}. The computation of holographic correlation functions of the energy-momentum tensor is reviewed in \ref{HolographicCorrelationFunctions energy-momentumTensor}, in which the trace relation plays a central role. The asymptotic symmetries of the boundary gravitons that satisfy the mixed boundary condition are discussed in subsection \ref{Subsection AsymptoticSymmetries in the BanadosGeometry}, realizing on the gravity side the $T\bar{T}$-deformed conformal symmetry we discussed in subsection \ref{Subsection TTbar deformed conformal symmetry}. Some generalizations and further developments are briefly mentioned in the last section \ref{Subsection FurtherDevelopments on Holography}, including $T\bar{T}$ in higher dimensions, constructing dS holography with $T\bar{T}$, and $T\bar{T}$ as a quantum theory of boundary gravitons at a finite cutoff (holography beyond large $c$).

\subsection{Mixed boundary condition, cutoff and glue-on} \label{Subsection MixedBoundaryCondition Cutoff and Glue-on}
The AdS/CFT holography is a statement of equivalence of a gravity theory in an AdS space and a conformal field theory on the conformal boundary of the AdS space \cite{Maldacena:1997re,Gubser:1998bc,Witten:1998qj}. Near the conformal boundary, the metric of an asymptotic AdS space takes the form \footnote{Throughout this section, we set the AdS radius $l$ to 1 for simplicity. One can recover the AdS radius in formulae by dimensional analysis.}
\begin{align} \label{AAdS Metric}
    ds^2 = \frac{dr^2}{r^2} + \frac{1}{r^2}g_{ij}(x,r) dx^i dx^j
\end{align}
in the Fefferman-Graham coordinates \cite{Fefferman:1985cfm,Graham:1991jqw}, where $r$ is the holographic radial coordinate, and the conformal boundary is located at $r=0$. The holography is encoded in the Gubser-Polyakov-Klebanov-Witten relation, that the partition function of conformal field theory with operator sources is equal to the gravity partition function with prescribed boundary conditions for bulk fields
\begin{align}
    \langle e^{\int J O} \rangle_{CFT} = Z_{\mathrm{G}}[\phi_0]
\end{align}
with field theory operators corresponding to bulk fields and sources identified as boundary conditions $J\sim \phi_0$. In the semi-classical limit, the gravity on-shell action becomes the generating functional of correlation functions of the conformal field theory. With this holographic prescription, we find an operator $O$ with dimension $\Delta$ corresponds to a bulk field $\phi$ with mass $M$, satisfying the relation
\begin{align}
    \Delta(\Delta-d) = M^2
\end{align}
where $d$ is the dimension of the boundary (and the bulk has dimension $d+1$). The near boundary solution to the on-shell equation of a bulk field, which is a second-order differential equation, takes the form
\begin{align}
    \phi = \phi^{(0)} (r^{d-\Delta}+\ldots) + \phi^{(2\Delta-d)} (r^{\Delta} + \ldots).
\end{align}
The coefficient $\phi^{(0)}$ corresponds to the source $\phi_0$, and $\phi^{(2\Delta-d)}$ corresponds to the one-point function $\langle O \rangle$. Specializing on the bulk metric and its dual operator, the energy-momentum tensor, we have the Fefferman-Graham expansion for the bulk metric
\begin{align}
    g_{ij} = g^{(0)}_{ij} + \ldots + g^{(d)}_{ij} r^d + \ldots
\end{align}
in which $g^{(0)}$ is identified with the field theory background metric $\gamma$ and $g^{(d)}$ corresponds to the one-point function of energy-momentum tensor. For $\text{AdS}_3/\text{CFT}_2$, the expansion truncates as \cite{Skenderis:1999nb}
\begin{equation} \label{FG series in 3D}
    g_{ij}(x,r)=g^{(0)}_{ij}(x) + g^{(2)}_{ij}(x) r^2 + g^{(4)}_{ij}(x) r^4
\end{equation}
and the Einstein's equation is reduced to one equation that determines $g^{(4)}$ in terms of $g^{(0)}$ and $g^{(2)}$:
\begin{equation} \label{g4 by g0 and g2}
    g^{(4)}_{ij}=\frac{1}{4}g^{(2)}_{ik}g^{(0)kl}g^{(2)}_{lj}.
\end{equation}
and another two equations
\begin{align}
    {\nabla^{(0)}}^i g^{(2)}_{ij} &= \nabla^{(0)}_j {g^{(2)}}_i^i, \label{g2 conservation}\\
    {g^{(2)}}_i^i &= -\frac{1}{2}R[g^{(0)}].\label{g2 trace}
\end{align}
We renormalize the gravity action by introducing a cutoff $r=r_c$ and adding the Gibbons-Hawking boundary term and a counter-term \cite{Balasubramanian:1999re,Emparan:1999pm,deHaro:2000vlm}
\begin{align} \label{Gravity renormalized action}
    I = I_{\text{EH}} + I_{\text{GH}} + I_{\text{CT}} = -\frac{1}{16\pi G} \int_{r\geq r_c} dV (R-2\Lambda) - \frac{1}{8\pi G} \int_{r=r_c} d\sigma K + \frac{1}{8\pi G} \int_{r=r_c} d\sigma.
\end{align}
{  Here $K$ is the trace of the extrinsic curvature $K_{ij}$, which is defined as the Lie derivative of the metric in the direction of unit normal of the cutoff surface $K=\frac{1}{2}\mathcal{L}_n g_{ij}$.} As shown in \cite{deHaro:2000vlm}, this renormalized gravity action, when put on-shell, has logarithm divergence as $r_c \to 0$, which is related to the Weyl anomaly \cite{Henningson:1998gx}. The Weyl anomaly counter-term was proposed in \cite{deHaro:2000vlm}
\begin{align} \label{Weyl anomaly counter-term}
    I_{\text{WCT}}= -\frac{1}{16 \pi G} \log r_c \int_{r=r_c} d\sigma \hat{R}
\end{align}
to subtract this logarithm divergence, where $\hat{R}$ is the scalar curvature of the induced metric. It is, in fact, a topological term since
\begin{align}
    \int_{r=r_c} d\sigma \hat{R} = 4\pi \chi = 4\pi(2-2 \mathrm{g})
\end{align}
with $\chi$ being the Euler characteristic of the cutoff surface and $\mathrm{g}$ the genus. As a topological term, it doesn't affect holographic correlation functions. Functionally differentiating the gravity on-shell action with respect to the induced metric on the cutoff surface $h_{ij}$ yields the Brown-York tensor \cite{Brown:1992br} modified by the counter term $I_{CT}$
\begin{align} \label{Modified BY tensor}
    T^{BY}_{ij} = \frac{2}{\sqrt{h}}\frac{\delta I_{\text{on-shell}}}{\delta h^{ij}} = -\frac{1}{8\pi G}(K_{ij} - K h_{ij} + h_{ij}).
\end{align}
Since $\gamma_{ij} = g^{(0)}_{ij} = \lim_{r_c\to 0} h_{ij}r_c^2$, we find the one-point function of energy-momentum tensor
\begin{align} \label{Holographic CFT StressTensor OnePointFunction}
    \langle T_{ij} \rangle = \lim_{r_c\to 0} T^{BY}_{ij} = \frac{1}{8\pi G}(g^{(2)}_{ij} - {g^{(0)}}^{kl}g^{(2)}_{kl} g^{(0)}_{ij}).
\end{align}
The two equations (\ref{g2 conservation},\ref{g2 trace}) translate to the conservation of the energy-momentum tensor and the holographic Weyl anomaly
\begin{align} 
    \nabla^i \langle T_{ij} \rangle &= 0, \label{CFT one-point correlator conservation}\\
    \langle T \rangle &= \frac{R}{16\pi G}. \label{CFT Weyl anomaly}
\end{align}

Now, we consider the $T\bar{T}$ deformation of a holographic conformal field theory. For the seed theory, we have the holographic dictionary
\begin{align}
    \gamma^{[0]}_{ij} &= g^{(0)}_{ij} ,\nonumber\\
    \langle \hat{T}_{ij} \rangle^{[0]} &= \frac{1}{8\pi G} g^{(2)}_{ij}
\end{align}
with the Brown-Henneaux relation \cite{Brown:1986nw}
\begin{align} \label{Brown-Henneaux}
    c=\frac{3}{2G}
\end{align}
and $\hat{T}_{ij}$ denoting the trace-reversed energy-momentum tensor introduced in \ref{Subsection TTbar flow at large central charge}. By the $T\bar{T}$ flow equation (\ref{TTbar flow large c solution}) we find
\begin{align} \label{Metric mixed boundary condition}
    \gamma^{[\lambda]}_{ij} = g^{(0)}_{ij} + \frac{\lambda}{4\pi G} g^{(2)}_{ij} + (\frac{\lambda}{4\pi G})^2 \frac{1}{4} g^{(2)}_{ik} {g^{(0)}}^{kl} g^{(2)}_{lj} = g^{(0)}_{ij} + \frac{\lambda}{4\pi G} g^{(2)}_{ij} + (\frac{\lambda}{4\pi G})^2 g^{(4)}_{ij}.
\end{align}
As expected, the field theory metric corresponds to a mixed boundary condition for gravity. Furthermore, if $\lambda$ is positive, then the mixed boundary can be viewed as a Dirichlet boundary condition on the cutoff surface $r=r_c$
\begin{align} \label{Cutoff AdS Metric}
    \gamma^{[\lambda]}_{ij} = r_c^2 h_{ij}|_{r=r_c}
\end{align}
with the cutoff location $r_c$ determined by $\lambda$
\begin{align} \label{Cutoff and TTbarDeformationParameter}
    r_c^2 = \frac{\lambda}{4 \pi G} = \frac{\lambda c}{6\pi}.
\end{align}
The $\lambda c$ combination in this dictionary makes the large $c$ limit (\ref{Large c limit}) manifest. Furthermore, from (\ref{TTbar flow large c solution}), the deformed one-point function of the trace-reversed energy-momentum tensor is found to be
\begin{align}
    \langle \hat{T}_{ij} \rangle^{[\lambda]} = \frac{1}{8\pi G} (g^{(2)}_{ij} + 2 \frac{\lambda}{4\pi G} g^{(4)}_{ij}) = \frac{1}{8\pi G}(h_{ij}-K_{ij})|_{r=r_c}
\end{align}
with $r_c$ given by (\ref{Cutoff and TTbarDeformationParameter}), and the one-point function of the energy-momentum tensor itself is equal to the modified Brown-York tensor on the cutoff surface
\begin{align} \label{Cutoff AdS OnePointFunction}
    \langle T_{ij} \rangle^{[\lambda]} = -\frac{1}{8\pi G} (K_{ij} - K h_{ij} + h_{ij})|_{r=r_c} = T^{\text{BY}}_{ij}|_{r=r_c}.
\end{align}
Furthermore, from the contracted Gauss and Codazzi equations in an Einstein space with negative curvature (for example, see Appendix D in \cite{Carroll:2004st}) \footnote{Here a hat symbol denotes geometrical quantities pertaining to the intrinsic geometry on the cutoff surface.}
\begin{align}
    K^2 - K_{ij}K^{ij} &= \hat{R} + 2 \label{Contracted Gauss equation}\\
    \hat{\nabla}^{i} K_{ij} - \hat{\nabla}_j K &= 0 \label{Contracted Codazzi equation}
\end{align}
and expressing the extrinsic curvature in terms of the Brown-York tensor
\begin{align} \label{ExtrinsicCurvature in terms of Brown-YorkTensor}
    K_{ij} = -8\pi G(T^{\text{BY}}_{ij}-T^{\text{BY}}h_{ij}) + h_{ij}
\end{align}
we find the Brown-York tensor is conserved
\begin{align} \label{BY Tensor Conservation}
    \hat{\nabla}^i T^{\text{BY}}_{ij} = 0
\end{align}
and satisfies
\begin{align} \label{BY Tensor TraceRelation}
    T^{\text{BY}} = 4\pi G \big( T^{\text{BY}}_{ij}{T^{\text{BY}}}^{ij} - {T^{\text{BY}}}^2 \big) + \frac{1}{16\pi G}\hat{R}[h].
\end{align}
These two equations translate to the conservation of the field theory energy-momentum tensor and the trace relation (\ref{TTbar flow large c TraceRelation}) \cite{Kraus:2018xrn, Taylor:2018xcy}. We see the cutoff AdS picture works both for the metric and the energy-momentum tensor one-point function. In fact, the quickest way to see how $T\bar{T}$ emerges from the cutoff is to consider the flow of the gravity on-shell action as the cutoff moves into the bulk. We find
\begin{align}
    r_c \partial_{r_c} I_{\text{on-shell}} = \frac{1}{2} \int_{r=r_c} d\sigma {T^{\text{BY}}}^{ij} \mathcal{L}_n g_{ij} = \int_{r=r_c} d\sigma {T^{\text{BY}}}^{ij} K_{ij}
\end{align}
where $n=-r\partial_r$ is the unit normal of the cutoff surface. Expressing the extrinsic curvature in terms of the modified Brown-York tensor (\ref{ExtrinsicCurvature in terms of Brown-YorkTensor}) and using the trace relation (\ref{BY Tensor TraceRelation}), we find
\begin{align}
    r_c \partial_{r_c} I_{\text{on-shell}} = -8\pi G \int_{r=r_c} d\sigma O_{T\bar{T}}^{\text{BY}} + \frac{1}{16\pi G} \int_{r=r_c} d\sigma \hat{R}[h]
\end{align}
where $O^{\text{BY}}_{T\bar{T}}$ means the combination of the $T\bar{T}$ form of the Brown-York tensor. With the Weyl anomaly counter-term (\ref{Weyl anomaly counter-term}), we rewrite the equation as
\begin{align} \label{Cutoff GravityOnShellAction FlowEquation}
    r_c\partial_{r_c} I^{'}_{\text{on-shell}} = -8\pi G \int_{r=r_c} d\sigma O_{T\bar{T}}^{\text{BY}}
\end{align}
with
\begin{align} \label{GravityAction with WeylAnomalyCounter-term}
    I^{'} = I + I_{\text{WCT}}.
\end{align}
This flow equation of the gravity on-shell action as the cutoff moving into the bulk agrees with the $T\bar{T}$ flow equation of the generating functional (\ref{TTbar flow large c GeneratingFunctional}) given the relation (\ref{Cutoff and TTbarDeformationParameter}) between the AdS cutoff and the $T\bar{T}$ deformation parameter. A version of the statement for a finite displacement of the cutoff was also given in \cite{Caputa:2020lpa}. We see that even if we omit the Weyl anomaly counter-term for holographic CFT to subtract the logarithm divergence, this term emerges as the theory is $T\bar{T}$-deformed, which can be written as
\begin{align}
    I_{\text{WCT}}= -\frac{1}{16 \pi G} \log \frac{r_c}{\epsilon} \int_{r=r_c} d\sigma \hat{R}
\end{align}
where $\epsilon$ is the CFT UV cutoff. Inclusion of this term is essential to match the $T\bar{T}$ flow equation as discussed in \cite{Caputa:2020lpa,Li:2020zjb,Tian:2023fgf}. We can obtain a holographic sphere partition function that agrees with (\ref{TTbar SpherePartitionFunction L}), which has a proper CFT limit and satisfies the differential equation from the $T\bar{T}$ flow. The Weyl anomaly counter-term also gives us a generalization of the Ryu-Takayanagi formula \cite{Ryu:2006bv} for the holographic entanglement entropy, as we will see in (\ref{Subsection HolographicPartitionFunction and EntanglementEntropy}).

The cutoff AdS picture only works for positive deformation parameter $\lambda$, as seen from (\ref{Cutoff and TTbarDeformationParameter}). A glue-on AdS picture was proposed in \cite{Apolo:2023vnm}, that for negative deformation parameter, the boundary where the field theory lives moves beyond $r=0$ to an AdS glued on the original AdS space\footnote{ As a powerful tool, the glue-on picture has been applied to construct the multiple AdS spacetime \cite{Kawamoto:2023wzj} and traversable wormholes wormhole\cite{Kawamoto:2025oko}. Specifically, a generalized $T_1T_2$ deformation was introduced by coupling already $T\bar{T}$-deformed theories together \cite{Ferko:2022dpg}. Notably, this approach resolves issues associated with negative-sign $T\bar{T}$ deformations by combining two negatively deformed theories and performing a joint deformation with a positive sign. Within certain parameter regimes, this method effectively prevents the appearance of problematic complex energy states, illustrating a broader class of sequential flow generalizations within $T\bar{T}$-deformed theories.}. More precisely, we can use the holographic radial coordinate $\rho=r^2$ in which the bulk metric takes the form
\begin{align}
    \label{AAdS Metric in rho coordinate}
    ds^2 = \frac{d\rho^2}{4\rho^2} + \frac{1}{\rho}g_{ij}(x,\rho) dx^i dx^j
\end{align}
Now, we allow $\rho$ to take values in the whole real line, with $\rho>0$ corresponding to the original AdS space and $\rho<0$ corresponding to the AdS space glued on. The relation between the cutoff and the $T\bar{T}$ deformation parameter now reads
\begin{align}
    \rho_c = \frac{\lambda}{4\pi G} = \frac{\lambda c}{6\pi}
\end{align}
so a negative $\lambda$ corresponds to cutoff at a negative $\rho_c$. This glue-on AdS proposal can be understood as a geometric realization of analytic continuation of the cutoff AdS holography to the region $\lambda<0$, and various physics quantities, including the energy levels, the trace relation,  partition functions and the entanglement entropy, were studied in \cite{Apolo:2023vnm,Apolo:2023ckr} to support this picture.

One must keep in mind that the cutoff or glue-on picture only works for the pure gravity sector. In particular, (\ref{Cutoff AdS Metric}) does not hold in the presence of matter. The solution to Einstein's equation is much more complicated with matter, and the deformed metric will have no simple geometric meaning. In \cite{Guica:2019nzm}, a bulk geometry with a matter shell, BTZ outside, and thermal AdS inside was considered. If the perceived cutoff surface by (\ref{Cutoff and TTbarDeformationParameter}) was inside the shell, the Brown-York tensor on the cutoff surface cannot reproduce the energy level of the $T\bar{T}$-deformed field theory. The universal deformation of energy levels of the field theory should come from the universal asymptotics of the bulk metric near the conformal boundary. Therefore, in holography, we can't remove the part of AdS space outside the perceived cutoff. Holographic correlation functions of scalar operators were also computed in \cite{Kraus:2018xrn} with the boundary put at a finite cutoff. The results do not agree with the field theory unless the double-trace deformation of scalar operators is turned on. All in all, as is known before \cite{Heemskerk:2010hk, Faulkner:2010jy} and discussed in the context of $T\bar{T}$ deformation in \cite{Hartman:2018tkw}, we need double-trace deformation involving all operators to move the conformal boundary into the bulk in holography.

\subsection{Holographic partition function and entanglement entropy} \label{Subsection HolographicPartitionFunction and EntanglementEntropy}
In this subsection, we consider the holographic partition function and entanglement entropy, which directly follow the holography analysis in the previous subsection. First, we compute the holographic sphere partition function. Working in the ``global patch'' of AdS space
\begin{align} \label{AdS GlobalPatch}
    ds^2 = d\varrho^2 + \sinh^2\varrho d\Omega_2^2
\end{align}
the radial Fefferman-Graham coordinate $r$, which must be proportional to $e^{-\varrho}$, needs to be chosen such that at the cutoff $r=r_c$, the induced metric corresponds to a field theory metric of a sphere with radius $L$. Let the cutoff in terms of $\varrho$ coordinate be $\varrho_c$, we must have
\begin{align}
    r_c \sinh\varrho_c = L.
\end{align}
The gravity on-shell action, with the Weyl anomaly counter-term, is evaluated to be 
\begin{align} \label{Holographic sphere partition function}
    I^{'}_{\text{on-shell}} &= -\frac{1}{2G}e^{-\varrho_c}\sinh\varrho_c - \frac{1}{2G}\varrho_c - \frac{1}{2G} \log\frac{r_c}{\epsilon} \nonumber\\
    &= -\frac{1}{2G} \frac{L^2}{r_c^2}\big( \sqrt{1+\frac{r_c^2}{L^2}} - 1 \big) - \frac{1}{2G} \log (1+\sqrt{1+\frac{r_c^2}{L^2}}) - \frac{1}{2G} \log\frac{L}{\epsilon}.
\end{align}
It agrees with the $T\bar{T}$-deformed CFT sphere partition function (\ref{TTbar SpherePartitionFunction L}) up to a constant by the parameter map (\ref{Brown-Henneaux}) and (\ref{Cutoff and TTbarDeformationParameter}). Without the Weyl anomaly counter-term, the on-shell action will agree with (\ref{CFT sphere partition function}) \cite{Donnelly:2018bef, Caputa:2019pam}. The same can be said for the holographic entanglement entropy, (\ref{TTbar SphereEntanglementEntropy DS}) agrees with the Ryu-Takayanagi formula \cite{Ryu:2006bv} with the geodesic endpoints on the cutoff surface \cite{Donnelly:2018bef}, as expected given that the Ryu-Takayanagi formula can be derived from the Einstein's gravity \cite{Lewkowycz:2013nqa, Dong:2016hjy}. With the Weyl anomaly counter-term, a modification to the Ryu-Takayanagi formula has to be made. 

The modification turns out to be simple and intuitive. Suppose we consider the entanglement entropy of $m$ intervals, the $n$-replica manifold $\mathcal{M}_n$ has genus $\text{g}_n = (n-1)(m-1)$. By the replica trick (\ref{EntanglementEntropy from Replica}), the contribution of the Weyl anomaly counter-term to the entanglement entropy is $\frac{1}{2G} m \log\frac{r_c}{\epsilon}$, that is, each endpoint contributes $\frac{1}{4G}\log\frac{r_c}{\epsilon}$, which is exactly $\frac{1}{4G}$ times the distance from the endpoint to the cutoff $\epsilon$ near the conformal boundary. Therefore, we have the modified Ryu-Takayanagi formula
\begin{align} \label{RTFormula TTbar}
    S = \frac{A}{4G} + \frac{2m}{4G} \log\frac{r_c}{\epsilon} = \frac{A}{4G} + \frac{A^{'}}{4G}.
\end{align}
The first term $\frac{A}{4G}$ is the Ryu-Takayanagi formula adapted to the cutoff AdS picture advocated in \cite{Donnelly:2018bef} and many subsequent works, with $A$ being the length of the Ryu-Takayanagi surface, namely the geodesics connecting the points of the entangling surface located on the cutoff surface in the bulk. In the correction term, $2m$ is the area of the entangling surface, namely the number of points, explicitly retaining the area law of UV-divergence of the entanglement entropy \cite{Srednicki:1993im}. $A^{'}$ is the length of the lines connecting the entangling surface to the conformal boundary. This picture of extending the Ryu-Takayanagi surface from the cutoff surface to the conformal boundary supports the perspective that the AdS space outside the perceived cutoff must be retained, see Figure \ref{Extended RT surface}. In our case, the geometry of the extension of the Ryu-Takayanagi surface is simple since the cutoff surface is a surface of constant Fefferman-Graham radial coordinates, and all points in the entangling surface are equidistant from the conformal boundary. In general, we may consider a state represented by a cutoff surface that is not necessarily at a constant radial cutoff, which can be prepared by an inhomogeneous $T\bar{T}$ deformation. The extension of the Ryu-Takayanagi surface will be ``less trivial''.
\begin{figure}
\center
\begin{tikzpicture}
 \draw[black, thick] (-2,4) -- (-2,-4);
 \draw[black, dashed] (0,4) -- (0,-4);
 \node[] at (-3.5,-4.5) {Conformal boundary};
 \node[] at (0.5,-4.5) {Perceived cutoff};
 \draw[red, thick] (0,3) arc (60:-60:3.46);
 \draw[blue, thick] (-2,3) -- (0,3);
 \draw[blue, thick] (-2,-3) -- (0,-3);
 \node[] at (-1,0) {$A^{'}$};
 \draw[black, thick,->] (-1,0.5) -- (-1,2.8);
 \draw[black, thick,->] (-1,-0.5) -- (-1,-2.8);
 \node[] at (2.5,0) {$A$};
\end{tikzpicture}
\caption{Ryu-Takayanagi surface extended to the conformal boundary}\label{Extended RT surface}
\end{figure}
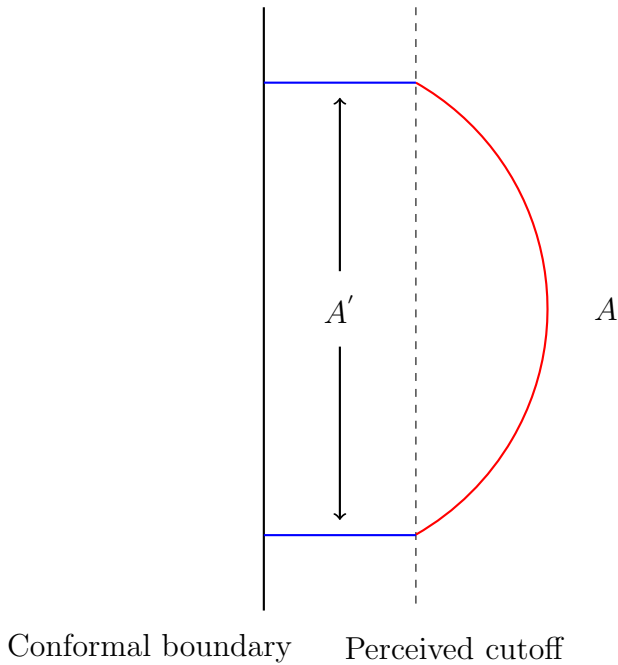

We can check our modified Ryu-Takayanagi formula (\ref{RTFormula TTbar}) for the perturbative computation of entanglement entropy to the first order in the deformation parameter $\lambda$ as was done in \cite{Chen:2018eqk, Jeong:2019ylz}. Consider an interval of length $L$ in a thermal state with inverse temperature $\beta$. The dual of the thermal state in the high-temperature limit is the BTZ black hole
\begin{align}
  ds^2 = (\varrho^2 - \varrho_+^2) dt^2 + \frac{d\varrho^2}{\varrho^2-\varrho_+^2} + \varrho^2 dx^2 = \frac{dr^2}{r^2} + (\frac{1}{r}-\frac{\varrho_+^2 r}{4})^2 dt^2 + (\frac{1}{r}+\frac{\varrho_+^2 r}{4})^2 dx^2
\end{align}
with $r=\frac{2}{\varrho_+^2}(\varrho-\sqrt{\varrho^2-\varrho_+^2})$ (or $\varrho = \frac{\varrho_+}{2}\big(\frac{\varrho_+ r}{2}+(\frac{\varrho_+ r}{2})^{-1}\big) = \frac{1}{r} + \frac{\varrho_+^2 r}{4}$). The geodesic distance between two points in the Euclidean AdS space takes a particularly simple form in the coordinates of flat space $\mathbb{R}_{1,3}$ where the Euclidean AdS space is embedded as a hyperboloid (as an ``analytic continuation'' of the apparent formula for the sphere)
\begin{align}
    d(X_1,X_2) = \cosh^{-1} (-X_1 \cdot X_2).
\end{align}
The map from the usual coordinates of the BTZ black hole to the embedding coordinates is
\begin{align}
    X^0 &= \frac{\varrho}{\varrho_+} \cosh(\varrho_+ x) \nonumber\\
    X^1 &= \frac{\varrho}{\varrho_+} \sinh(\varrho_+ x) \nonumber\\
    X^2 &= \sqrt{\frac{\varrho^2}{\varrho_+^2}-1} \cos(\varrho_+ t) \nonumber\\
    X^3 &= \sqrt{\frac{\varrho^2}{\varrho_+^2}-1} \sin(\varrho_+ t).
\end{align}
Then the length of the Ryu-Takayanagi surface for a single interval with coordinate difference $\Delta x$ at cutoff $\varrho_c$ is found to be
\begin{align}
    A_{\text{BTZ}} = \cosh^{-1}\Big( 1+2\frac{\varrho_c^2}{\varrho_+^2}\sinh^2\frac{\varrho_+\Delta x}{2} \Big) = \cosh^{-1}\Big( 1+\frac{1}{2}(\frac{\varrho_+ r_c}{2}+\frac{2}{\varrho_+ r_c})^2 \sinh^2\frac{\varrho_+\Delta x}{2} \Big).
\end{align}
Plugging in the length of the interval in the field theory metric $L = (1+\frac{\varrho_+^2 r_c^2}{4})\Delta x$, and the inverse temperature $\beta=\frac{2\pi}{\varrho_+}(1-\frac{\varrho_+^2 r_c^2}{4})$, we find
\begin{align}
    A_{\text{BTZ}} = \cosh^{-1}\Big( \frac{(4\pi^2 r_c^2 + \beta^2)\cosh\frac{2\pi L}{\sqrt{4\pi^2 r_c^2+\beta^2}}-\beta^2}{4\pi^2 r_c^2} \Big).
\end{align}
As $r_c\to 0$ we find
\begin{align} \label{ThermalState RTSurface}
    A_{\text{BTZ}} = 2\log\frac{\beta\sinh\frac{\pi L}{\beta}}{\pi r_c} - \frac{4\pi^2 r_c^2}{\beta^2} \frac{\pi L}{\beta}\coth\frac{\pi L}{\beta} + \frac{2\pi^2 r_c^2}{\beta^2} (\coth^2\frac{\pi L}{\beta}+1) + o(r_c^2).
\end{align}
The $\log r_c$ is canceled exactly by the correction term in (\ref{RTFormula TTbar}), we find
\begin{align} \label{ThermalState ExtendedRTSurface}
    A_{\text{BTZ}}+A^{'}_{\text{BTZ}} = 2\log\frac{\beta\sinh\frac{\pi L}{\beta}}{\pi \epsilon} - \frac{4\pi^2 r_c^2}{\beta^2} \frac{\pi L}{\beta}\coth\frac{\pi L}{\beta} + \frac{2\pi^2 r_c^2}{\beta^2} (\coth^2\frac{\pi L}{\beta}+1) + o(r_c^2).
\end{align}
In the high-temperature limit, the leading $T\bar{T}$ deformation of the holographic entanglement entropy is
\begin{align}
    \delta S = - \frac{\pi^2 r_c^2}{G \beta^2} \frac{\pi L}{\beta}\coth\frac{\pi L}{\beta}.
\end{align}
It was found to agree with the field theory perturbative computation in \cite{Chen:2018eqk}
\begin{align}
    \delta S = -\frac{\lambda \pi^2 c^2 L}{9\beta^3} \coth\frac{\pi L}{\beta}.
\end{align}
by the parameter map (\ref{Brown-Henneaux}) and (\ref{Cutoff and TTbarDeformationParameter}). The same result was obtained in \cite{He:2023xnb} using the Chern-Simons formulation of three-dimensional gravity. We remark that without the correction term in the Ryu-Takayanagi formula, $r_c$ plays a dual role of the UV cutoff and the $T\bar{T}$ deformation parameter in the original paper \cite{Chen:2018eqk}, it's thus ambiguous in which part of the expression (\ref{ThermalState RTSurface}) $r_c$ should be identified as the UV cutoff, and in which part identified as the $T\bar{T}$ deformation parameter. {  Our standpoint is that the UV cutoff (denoted by $\epsilon$ in (\ref{ThermalState ExtendedRTSurface})) and the $T\bar{T}$ deformation parameter (represented by $r_c$ in (\ref{ThermalState ExtendedRTSurface})) are different quantities.} We can repeat this computation for the entanglement entropy of the vacuum state in a circle as in \cite{Chen:2018eqk}, but we can't find agreement between both sides of holography for reasons not yet clear.

\subsection{Holographic correlation functions of the energy-momentum tensor} \label{HolographicCorrelationFunctions energy-momentumTensor}
Energy-momentum tensor correlation functions were computed as early evidence for the proposed cutoff AdS holography for $T\bar{T}$-deformed CFTs \cite{Kraus:2018xrn}. It has been further studied on the field theory side in \cite{Aharony:2018vux, Li:2020pwa, Hirano:2020ppu} and on the gravity side of the holography in \cite{Li:2020pwa, He:2023hoj, He:2023knl, He:2024xbi, He:2024fdm, Li:2025sfn}. Since three-dimensional gravity has no propagating degrees of freedom, the holographic computation can be reduced to the conservation and the trace relation of the Brown-York tensor on a finite cutoff surface. The Brown-York tensor, with three independent components, is determined by the three equations from the conservation and the trace relation up to constants that, in principle, can be further determined by the global geometry, physical considerations, or some mathematical tricks. In the previous subsection, we have shown that the one-point function corresponds to the Brown-York tensor (\ref{Cutoff AdS OnePointFunction}) and the conservation and trace relation of the one-point function correspond to those of the Brown-York tensor, respectively. As a result, the holographic energy-momentum tensor correlation functions agree with the large $c$ field theory computation based on the trace relation \cite{Kraus:2018xrn, Li:2020pwa}. The holographic correlation functions are computed by perturbatively solving for the one-point function from the conservation equation
\begin{align} \label{one-point function conservation}
    \nabla^i \langle T_{ij} \rangle = 0,
\end{align}
and the trace relation
\begin{align} \label{one-point function trace relation}
    \langle T \rangle = \frac{R}{16\pi G} + 4\pi G r_c^2 (\langle T^{ij} \rangle \langle T_{ij} \rangle - \langle T \rangle^2).
\end{align}
for a given metric variation. Consider a $T\bar{T}$-deformed CFT on a compactified plane with the metric
\begin{align}
    ds^2 = dz d\bar{z}
\end{align}
and a variation of the metric $\epsilon f_{ij} dx^idx^j$, where $\epsilon$ is an infinitesimal parameter. The one-point function can be solve as a series in $\epsilon$ $\langle T_{ij} \rangle = \epsilon \langle T_{ij} \rangle_1 + \epsilon^2 \langle T_{ij} \rangle_2 + \ldots$ from (\ref{one-point function conservation}) and (\ref{one-point function trace relation}). For the first order, we find
\begin{align}
    \langle T_{z\bar{z}} \rangle_1 = \frac{1}{16\pi 
    G} \big( \partial_{\bar{z}}^2 f_{zz} - 2\partial_z\partial_{\bar{z}} f_{z\bar{z}} + \partial_z^2 f_{\bar{z}\bar{z}} \big)
\end{align}
and
\begin{align}
    \partial_{\bar{z}} \langle T_{zz} \rangle_1 + \partial_z \langle T_{z\bar{z}} \rangle_1 &= 0 \nonumber\\
    \partial_{z} \langle T_{\bar{z}\bar{z}} \rangle_1 + \partial_{\bar{z}} \langle T_{z\bar{z}} \rangle_1 &= 0.
\end{align}
The Green's function on the compactified plane is $\frac{1}{z-w}$ ($\frac{1}{\bar{z}-\bar{w}}$), that is
\begin{align}
    \frac{1}{\pi}\partial_{\bar{z}} \frac{1}{z-w} = \delta^{(2)}(z-w) \;\text{and c.c.}.
\end{align}
Using the Green's function, we find
\begin{align}
    \langle T_{zz}(z) \rangle_1 = \frac{1}{16\pi G} \big( -\partial_z\partial_{\bar{z}} f_{zz}(z) + 2\partial_z^2 f_{z\bar{z}}(z) + \frac{6}{\pi} \int d^2w \frac{f_{\bar{z}\bar{z}}(w)}{(z-w)^4} \big) \;\text{and c.c.}
\end{align}
where we have ruled out a possible constant by the regularity at the infinity point of the compactified plane. Therefore, we obtain the non-vanishing two-point functions for distinct points 
\begin{align}
    \langle T_{zz}(z) T_{zz}(w)\rangle^c = \frac{3}{16\pi^2 G} \frac{1}{(z-w)^4} \; \text{and c.c.}.
\end{align}
We see the two-point functions are unaffected by the cutoff (or the $T\bar{T}$ deformation on the field theory side). For the second order, we find
\begin{align}
    \langle T_{z\bar{z}} \rangle_2 =& 8\pi G r_c^2 (\langle T_{zz} \rangle_1 \langle T_{\bar{z}\bar{z}} \rangle_1 - \langle T_{z\bar{z}} \rangle_1^2) - (f_{zz}\langle T_{\bar{z}\bar{z}}\rangle_1 + f_{\bar{z}\bar
    {z}}\langle T_{zz} \rangle_1) \nonumber\\
    &+ \text{local quadratic functions of} \;f\; \text{and its derivatives}.
\end{align}
Plugging in $\langle T_{ij} \rangle_1$ we have obtained, the only non-vanishing three-point function involving $T_{z\bar{z}}$ is found to be
\begin{align}
    \langle T_{z\bar{z}}(z) T_{zz}(w) T_{\bar{z}\bar{z}}(v) \rangle^c = \frac{9 r_c^2}{32 \pi^3 G} \frac{1}{(z-w)^4(\bar{z}-\bar{v})^4}.
\end{align}
Furthermore, from the conservation equation to the second order, we find
\begin{align}
    \partial_{\bar{z}} \langle T_{zz} \rangle_2 + \partial_z \langle T_{z\bar{z}} \rangle_2 = &2 \langle T_{z\bar{z}} \rangle_1 (\partial_{\bar{z}} f_{zz} + \partial_z f_{z\bar{z}}) + 3 \langle T_{zz} \rangle_1 \partial_z f_{\bar{z}\bar{z}} + \langle T_{\bar{z}\bar{z}}\rangle_1 \partial_z f_{zz}  \nonumber\\
    &+ f_{z\bar{z}} (\partial_{\bar{z}}\langle T_{zz}\rangle_1 + \partial_z \langle T_{z\bar{z}} \rangle_1) + 2f_{zz} \partial_{\bar{z}} \langle T_{z\bar{z}} \rangle_1 + 2f_{\bar{z}\bar{z}} \partial_z \langle T_{zz} \rangle_1 \;\text{and c.c.}.
\end{align}
Other non-vanishing three-point functions are found to be
\begin{align}
    \langle T_{zz}(z) T_{zz}(w) T_{zz}(v) \rangle^c &= -\frac{3}{16\pi^3 G} \frac{1}{(z-w)^2(w-v)^2(v-z)^2} \;\text{and c.c.}, \nonumber\\
    \langle T_{zz}(z) T_{\bar{z}\bar{z}}(w) T_{\bar{z}\bar{z}}(v) \rangle^c &= \frac{3 r_c^2}{8\pi^3 G} (\frac{1}{(z-w)^3(\bar{w}-\bar{v})^5} + \frac{1}{(z-v)^3(\bar{v}-\bar{w})^5}) \;\text{and c.c.}.
\end{align}
Following this spirit, one can proceed to compute higher-point functions \cite{Li:2020pwa} or holographic correlation functions on other boundary spacetimes (with nontrivial topology), for example, the sphere \cite{Li:2020pwa}, torus \cite{He:2023hoj, He:2024fdm}, higher-genus surfaces \cite{He:2024xbi}, and non-orientable manifolds \cite{Li:2025sfn}.

\subsection{Symmetries of boundary gravitons at a finite cutoff}\label{Subsection AsymptoticSymmetries in the BanadosGeometry}

An important aspect of the $\text{AdS}_3/\text{CFT}_2$ correspondence is the symmetry, the Virasoro algebra emerges as the algebra of the boundary charges of the boundary gravitons of $\text{AdS}_3$ gravity \cite{Brown:1986nw}. It's thus a natural and important question how the symmetry evolves as the $T\bar{T}$ deformation is turned on. This question has been studied on the CFT side in \cite{Guica:2020uhm,Georgescu:2022iyx,Guica:2022gts,Chakraborty:2023wel}, and on the gravity side in \cite{Guica:2019nzm,Kraus:2021cwf,He:2021bhj}. Restricting to pure gravity, now the boundary gravitons are on a finite cutoff surface (or equivalently with mixed boundary conditions on the conformal boundary). We will study the symmetries, boundary charges, and the algebra of the boundary gravitons in this subsection, mainly following \cite{Kraus:2021cwf}. To this end, it's helpful first to recapitulate the canonical formalism for gravity.

Hamilton's equation and Darboux's theorem inspire the symplectic-geometrical description of classical mechanics \cite{Arnold:1989who, Abraham1994Foundations}. The phase space is a $2n$-dimensional space with a symplectic form $\Omega$, a non-degenerate closed 2-form. A symmetry is represented by a symplectomorphism, infinitesimally given by a vector field $V$ that preserves the symplectic form
\begin{align}
    \mathcal{L}_V \Omega = 0.
\end{align}
By the Cartan formula $\mathcal{L}_V = di_V + i_Vd$ and the Poincare lemma, the vector is locally generated by a ``Hamiltonian'' $F$
\begin{align}
    i_V \Omega = - dF
\end{align}
or in components
\begin{align}
    V^i = \Omega^{ij}\partial_j F.
\end{align}
Conversely, Hamiltonians generate symplectomorphisms. The Poisson bracket is defined as
\begin{align}
    \{F,G\} = \Omega^{ij}\partial_i F \partial_j G = -V_F(G) = V_G(F)
\end{align}
and the equation of motion
\begin{align}
    \frac{d F}{dt} = \{F,H\} = V_H (F)
\end{align}
is geometrically a flow along the vector field generated by the Hamiltonian $H$. The canonical formalism can be generalized to field theories in a covariant way \cite{Crnkovic:1986ex}. In this case, the phase space is defined as the space of all on-shell solutions $\phi$. A typical function is $\phi(x)$, the value of the on-shell solution $\phi$ at the point $x$, or more generally a functional $F[\phi]$. A tangent vector is defined as an infinitesimal variation of an on-shell solution $\delta\phi$ (in fact, a solution to the linearized equation); it acts on the functions naturally by computing the variation of the function by this variation of the on-shell solution. One forms are linear functionals on the tangent space, a typical one is $\delta \phi(x)$, namely the value of $\delta\phi$ at $x$. A $k$-form can be represented as a wedge product of 1-forms
\begin{align}
    A = \int dx_1\ldots dx_k a_{x_1,\ldots,x_k}[\phi]\delta\phi(x_1)\ldots\delta\phi(x_k).
\end{align}
The exterior derivative is defined by its action on a function
\begin{align}
    \delta F[\phi] = \int dx \frac{\delta F}{\delta \phi(x)} \delta\phi(x).
\end{align}
and extended to arbitrary $k$-forms by the usual algebraic properties of an exterior derivative. The symplectic form of the gravity theory was given in \cite{Crnkovic:1986ex}
\begin{align} \label{Gravity Symplecticform}
    \Omega = i\int d\Sigma_i \sqrt{g} J^i
\end{align}
where the symplectic current is
\begin{align}
    J^i = \frac{1}{16\pi G}\big[ \delta\Gamma^i_{jk}(\delta g^{jk}+\frac{1}{2}g^{jk}\delta\log g) - \delta\Gamma^j_{kj}(\delta g^{ik}+\frac{1}{2}g^{ik}\delta\log g) \big].
\end{align}
It can be shown that for a diffeomorphism $\xi$, the charge
\begin{align}
    Q[\xi] = i \int_{\partial\Sigma} d\sigma T_{ij}n^i\xi^j
\end{align}
generates the diffeomorphism in the sense that
\begin{align} \label{GravityBoundaryCharges Generate Diffeomorphisms}
    i_{\delta_\xi\phi}\Omega = -\delta Q[\xi]
\end{align}
where $\delta_\xi\phi$ is a tangent vector, the variation of an on-shell solution by the diffeomorphism $\xi$.

Now we consider a boundary at a finite cutoff $\rho_c$ with a cylinder topology and the metric
\begin{align}
    ds^2 = dt^2 + d\phi^2 = dwd\bar{w}.
\end{align}
The general Banados geometry that satisfies this boundary condition is
\begin{align} \label{BanadosGeometry MixedBoundaryCondition}
    ds^2 = \frac{d\rho^2}{4\rho^2}+\frac{\big[(1-\rho\rho_c\mathcal{L}\bar{\mathcal{L}})dw + (\rho-\rho_c)\bar{\mathcal{L}}d\bar{w}\big]\big[(1-\rho\rho_c\mathcal{L}\bar{\mathcal{L}})d\bar{w} + (\rho-\rho_c)\mathcal{L}dw\big]}{\rho(1-\rho_c^2\mathcal{L}\bar{\mathcal{L}})}
\end{align}
with $\mathcal{L},\bar{\mathcal{L}}$ satisfying
\begin{align}
    \partial_{\bar{w}}\mathcal{L} + \rho_c\bar{\mathcal{L}}\partial_w\mathcal{L} = 0, \; \partial_w\bar{\mathcal{L}} + \rho_c\mathcal{L}\partial_{\bar{w}}\bar{\mathcal{L}} = 0.
\end{align}
Symmetries are given by the boundary-preserving diffeomorphisms, which, to the order that can be used to compute the boundary charges, are found to be
\begin{align}
    \xi^w &= f^w - \frac{1}{2}\partial_w f^\rho +O((\rho-\rho_c)^2) \nonumber\\
    \xi^{\bar{w}} &= f^{\bar{w}} - \frac{1}{2}\partial_{\bar{w}} f^\rho + O((\rho-\rho_c)^2) \nonumber\\
    \xi^\rho &=\rho f^\rho
\end{align}
with
\begin{align}
    f^\rho = \frac{1-\rho_c^2\mathcal{L}\bar{\mathcal{L}}}{1+\rho_c^2\mathcal{L}\bar{\mathcal{L}}}(\partial_w f^w + \partial_{\bar{w}} f^{\bar{w}}).
\end{align}
These diffeomorphisms depend on $\mathcal{L}$ and $\bar{\mathcal{L}}$. In other words, the symmetries are ``state-dependent'' as compared to the case without $T\bar{T}$-deformation. On a time slice, we decompose the diffeomorphism in Fourier modes
\begin{align}
    \xi_n = e^{-in\phi}\partial_w,\; \bar{\xi}_n = -e^{-in\phi}\partial_{\bar{w}}
\end{align}
and we define
\begin{align}
    Q_n = Q[\xi_n]=\frac{1}{8\pi G}\int d\phi\frac{(1-\rho_c\bar{\mathcal{L}})\mathcal{L}}{1-\rho_c^2\mathcal{L}\bar{\mathcal{L}}}e^{-in\phi},\; \bar{Q}_n = Q[\bar{\xi}_n]=\frac{1}{8\pi G}\int d\phi\frac{(1-\rho_c\mathcal{L})\bar{\mathcal{L}}}{1-\rho_c^2\mathcal{L}\bar{\mathcal{L}}}e^{in\phi}.
\end{align}
From (\ref{GravityBoundaryCharges Generate Diffeomorphisms}) we have the relation
\begin{align}
    \{Q[\xi],Q[\chi]\} = - \delta_{\xi}Q[\chi] = \delta_{\chi}Q[\xi]
\end{align}
and using this relation, we can find the charge algebra \cite{Kraus:2021cwf}
\begin{align}
    i\{Q_m,Q_n\} &= \frac{1}{8\pi G}\int_0^{2\pi} d\phi e^{-i(m+n)\phi} \big[ \frac{1}{2}m^3 + (m-n)\frac{(1-\rho_cq)q}{1-\rho_c(q+\bar{q})} - \frac{1}{2}mn(m-n)\rho_c\frac{\bar{q}-\frac{1}{2}\rho_c(q+\bar{q})^2}{(1-\rho_c(q+\bar{q}))^2} \big] \nonumber\\
    \text{and} &\; \text{c.c.} \nonumber\\
    i\{Q_m,\bar{Q}_n\} &= \frac{1}{8\pi G}\int_0^{2\pi} d\phi e^{-i(m-n)\phi}\big[ \frac{\rho_c(m-n)q\bar{q}}{1-\rho_c(q+\bar{q})} +\frac{1}{2}\rho_c mn\frac{nq-m\bar{q}+\frac{\rho_c}{2}(m-n)(q+\bar{q})^2}{(1-\rho_c(q+\bar{q}))^2} \big]
\end{align}
where
\begin{align}
    q=4G\sum_n Q_n e^{-in\phi}, \; \bar{q}=4G\sum_n \bar{Q}_n e^{in\phi}.
\end{align}
This is a nonlinear generalization of the Virasoro algebra that reduces to it when $\rho_c=0$.

\subsection{Further developments on holography} \label{Subsection FurtherDevelopments on Holography}
There are many generalizations and further developments of the relation between $T\bar{T}$ deformation and holography, which we can't discuss in detail in this review. We mention some of them in this last subsection and refer the readers to the original paper.
\subsubsection*{Quantum boundary gravitons: holography beyond large $c$}
Following the symmetry analysis, boundary gravitons at a finite cutoff were further studied in \cite{Kraus:2021cwf,Ebert:2022cle,Kraus:2022mnu}, focusing on quantization which can be interpreted as holography for $T\bar{T}$-deformed CFTs beyond large $c$. The theory of boundary gravitons on the conformal boundary is essentially a constrained free scalar. By $T\bar{T}$ deformation, it's turned into a Nambu-Goto action in the static gauge. One can quantize in the light-cone gauge, with some ambiguities settled by the trace relation. The energy levels \cite{Smirnov:2016lqw} and energy-momentum tensor correlation functions \cite{Kraus:2018xrn} are reproduced to the first few orders of $\frac{1}{c}$ expansion.
\subsubsection*{$T\bar{T} + \Lambda^2$ and dS holography}
The picture of a flow with a moving cutoff can be generalized to space with a positive cosmological constant, particularly for dS holography. For dS space, the contracted Gauss equation is
\begin{align}
    K^2 - K_{ij}K^{ij} &= \hat{R} - 2
\end{align}
while the { 
Codazzi equation (\ref{Contracted Codazzi equation})} is unchanged. The trace relation of the modified Brown-York tensor now has an extra term, and as a consequence, the flow equation of the on-shell action as the cutoff moves also takes an extra term
\begin{align}
    r_c \partial_{r_c} I_{\text{on-shell}} = -8\pi G \int_{r=r_c} d\sigma O_{T\bar{T}}^{\text{BY}} + \frac{1}{16\pi G} \int_{r=r_c} d\sigma \hat{R}[h] - \frac{1}{4\pi G} \int_{r=r_c} d\sigma.
\end{align}
Keeping the same holographic dictionary, this translates to the flow equation of the field theory generating functional
\begin{align}
    \partial_\lambda I^{[\lambda]}[\gamma^{[\lambda]}] = -\int d^2x \sqrt{\gamma^{[\lambda]}} \langle O_{T\bar
 T} \rangle^{[\lambda]} - \frac{1}{2\lambda^2} \int d^2x \sqrt{\gamma^{[\lambda]}}.
\end{align}
With the extra term on the right-hand side, a $\lambda$-dependent two-dimensional cosmological constant term, the deformation was introduced by the name $T\bar{T} + \Lambda_2$ in \cite{Gorbenko:2018oov} and subsequently studied in \cite{Lewkowycz:2019xse,Shyam:2021ciy,Coleman:2021nor,Torroba:2022jrk,Batra:2024kjl,Aguilar-Gutierrez:2024nst}.
\subsubsection*{Higher dimensional $T\bar{T}$ deformation from cutoff AdS}
There have been proposals for generalizing $T\bar{T}$ deformation to higher dimensions from the cutoff AdS holography. They are based on the trace relation or the flow of the on-shell action \cite{Taylor:2018xcy,Hartman:2018tkw}. In general, the Brown-York tensor modified by counter-terms takes the form
\begin{align}
    T^{\text{BY}}_{ij} = -\frac{1}{8\pi G}(K_{ij}-K h_{ij} + C_{ij})
\end{align}
where $C_{ij}$ is the contribution from the local counter-term \cite{deHaro:2000vlm}. The contracted Gauss equation (\ref{Contracted Gauss equation}) in an Einstein space with negative curvature in general dimensions is
\begin{align} \label{ContractedGaussEquation NegativeCurvature in GeneralDimensions}
    K^2 - K_{ij}K^{ij} &= \hat{R} + d(d-1).
\end{align}
In terms of the Brown-York tensor, it reads
\begin{align}
    &(8\pi G)^2(T^{\text{BY}}_{ij}{T^{\text{BY}}}^{ij} - \frac{1}{d-1}{T^{\text{BY}}}^2) + (C_{ij}C^{ij}-\frac{1}{d-1}C^2) + 16\pi G(T^{\text{BY}}_{ij}C^{ij} - \frac{1}{d-1}T^{\text{BY}}C) \nonumber\\
    &+ \hat{R}[h] + d(d-1) = 0.
\end{align}
Using this relation, the flow equation for the on-shell action can be written as
\begin{align}
    r_c \partial_{r_c} I_{\text{on-shell}} &= \frac{1}{2} \int_{r=r_c} d\sigma {T^{\text{BY}}}^{ij} \mathcal{L}_n g_{ij} = \int_{r=r_c} d\sigma {T^{\text{BY}}}^{ij} K_{ij} \nonumber\\
    &= \int_{r=r_c} d\sigma \Big[ (-8\pi G)(T^{\text{BY}}_{ij}{T^{\text{BY}}}^{ij} - \frac{1}{d-1}{T^{\text{BY}}}^2) - (T^{\text{BY}}_{ij}C^{ij} - \frac{1}{d-1}T^{\text{BY}}C) \Big] \nonumber\\
    &= \int_{r=r_c} d\sigma \Big[ (-4\pi G)(T^{\text{BY}}_{ij}{T^{\text{BY}}}^{ij} - \frac{1}{d-1}{T^{\text{BY}}}^2) - \frac{1}{16\pi G} (C_{ij}C^{ij}-\frac{1}{d-1}C^2 + \hat{R}[h] + d(d-1)) \Big].
\end{align}
Now we can see the natural higher dimensional $T\bar{T}$ operator inspired by the cutoff AdS holography is
\begin{align}
    O_{T\bar{T}} = \frac{1}{2}(T_{ij}T^{ij} - \frac{1}{d-1}T^2)
\end{align}
Taking the leading counter-term to be $(d-1)h_{ij}$ as in \cite{deHaro:2000vlm} and using the dictionary
\begin{align}
    \langle T_{ij} \rangle = r_c^{2-d} T^{\text{BY}}_{ij},\; \gamma_{ij} = r_c^2 h_{ij}|_{r=r_c},
\end{align}
we can map the Gauss equation in terms of the Brown-York tensor to the trace relation and the flow of the gravity on-shell action to the flow of the generating functional, up to contribution from counter-terms, with the relation between the deformation parameter and the cutoff
\begin{align}
    \lambda = \frac{8\pi G r_c^d}{d}.
\end{align}



\newpage
\section{$T\bar{T}$ and string theory}\label{string-TTbar}
This section will briefly introduce the application of $T\bar{T}$ deformation in string holography, i.e., single trace  $T\bar{T}$ deformation. 
In generic translational and rotational invariant CFT, one can define the $T\bar{T}$ deformation. In addition, we can define another deformation also induced by the bilinear form of energy-momentum tensors at the same time if the CFT is a symmetric product orbifold $\mathcal{M}^p/S_p$, where $\mathcal{M}$ represents the seed CFT and $S_p$ is permutation group. The deformed theory remains an orbifold with a different seed theory $(\mathcal{M}_\l)^p/S_p$. $\mathcal{M}^i_\l$ is $T\bar{T}$ deformation of $\mathcal{M}^i$ induced by $T_\l^i \bar{T}^i_\l$. Here $i$ is $i-$th copy of the orbifold. Therefore, the deformation of $\mathcal{M}^p/S_p$ is triggered by 
\be 
\sum_{i=1}^p (T_\l^i \bar{T}^i_\l- \Theta_\l^i \Theta^i_\l).
\ee 
This is the so-called single-trace $T\bar{T}$ deformation of $\mathcal{M}^p/S_p$, in contrast with the usual  $T\bar{T}$ deformation (also called double-trace deformation). The latter is induced by 
\be 
\sum_{i=1}^p T_\l^i \sum_{j=1}^p\bar{T}^j_\l-\sum_{i=1}^p \Theta_\l^i \sum_{j=1}^p\Theta^j_\l.
\ee 
The entropy for single-trace $T\bar{T}$ deformed orbifold CFT $\mathcal{M}^p/S_p$ can be obtained from  \eqref{HagedornS}, which turns out to be \cite{Giveon:2017nie}
\be \label{singleS}
S(\mathcal{M})=p S_{\mathcal{M}}(E/p),
\ee 
where the total energy and entropy is a sum of each copy $E=\sum_i E_i, S(\mathcal{M})=\sum_i S_{\mathcal{M}}(E_i)$, $S_{\mathcal{M}}$ is the entropy of seed theory, taking the form \eqref{HagedornS}. \eqref{singleS} also exhibit Hagedorn behavior for negative
$\l$ in \eqref{HagedornS}.

We are interested in single-trace deformation because the above orbifold is supposed to be holographically dual to long string sector of type IIB string theory in AdS$_3$. 
Thus, it is natural to ask about the behavior of this kind of holographical duality when turning on the single-trace deformation. 

For string on AdS$_3\times \mathcal{N}_7$, a single long string, which is a string configuration that can wrap around the conformal boundary of AdS$_3$, is described by a sigma model on the following background \cite{Seiberg:1999xz} (see also \cite{Chakraborty:2020yka, Chakraborty:2019mdf}, and recently \cite{Knighton:2024pqh})
\be \label{syorbi}
\mathcal{M} =R_\phi\times \mathcal{N}_7,
\ee
where $\phi$ is the radius coordinate of AdS$_3$. The factor $R_\phi$ denotes the  linear dilaton along coordinate $\phi$,
with slope 
\be 
Q=(k-1)\sqrt{\frac{2}{k}}.
\ee { By the analysis of Nambu-Goto action near the boundary of AdS$_3$, one can read off the action of this sigma model, which turns out to be the action of a linear dilaton CFT. From the background charge of the linear dilaton CFT, the central charge follows as $c_{\mathcal{M}}=6k$, and $k$ is related to AdS radius (see below \eqref{wzwL}).} 
The long string coupling 
\be 
g\sim e^{Q\phi}.
\ee
Thus, the string is strongly coupled at the boundary $\phi\to \infty$ ($\phi$ can be considered a radial direction). The complete boundary theory dual to string on AdS$_3$ is unknown for generic $k$. 
For $p$ long strings, it is described by the symmetric orbifold
\begin{equation}
\left(\mathcal{M}\right)^p/S_p.
\end{equation}

\subsubsection*{String on AdS$_3$}

String propagate on  AdS$_3$ with NS-NS fields have a tractable world-sheet description of the WZW model. The AdS$_3$ metric is   \footnote{Euclidean AdS$_3$: $H_3^+$=SL(2,C)/SU(2).}
\be 
ds^2=l^2(d\phi^2+e^{2\phi}d\gamma d\bar{\gamma})
\ee
and NS-NS $B$ field
\be\ba 
B=&-\frac12 l^2e^{2\phi}d\gamma\wedge d\bar{\gamma}.
\ea\ee  
The corresponding   WZW model is based on group $SL(2,R)$ with element $g $   parametrized as 
\be 
g=\left(\begin{array}{cc}
e^{\phi} & 0 \\ 
0 & e^{-\phi}
\end{array} \right)\left(\begin{array}{cc}
1 & 0 \\ 
\gamma & 1
\end{array} \right)\left(\begin{array}{cc}
1 & \bar{\gamma} \\ 
0 & 1
\end{array} \right)=
\left(\begin{array}{cc}
e^\phi & \bar{\gamma}e^\phi \\ 
\gamma e^\phi & e^{-\phi}+\gamma\bar{\gamma}e^\phi
\end{array} \right).
\ee 
The world-sheet action is written as
\be\ba \label{wzwL}
&\mathcal{L}=\frac{k}{\pi}\int d^2z(\p\phi\bar{\p}\phi+e^{2\phi}\bar{\p}\gamma\p\bar{\gamma}),~~k=\frac{l^2}{\alpha'}.
\ea\ee
which is SL(2,R) WZW model at level $l^2/\alpha'$.  
\eqref{wzwL} can be equivalently expressed in Wakamoto representation
\be\ba 
\mathcal{L}=&\frac{2 l^2}{l_s^2}\left(\partial \phi \bar{\partial} \phi+e^{2 \phi} \bar{\partial} \gamma \partial \bar{\gamma}\right)=\frac{2 l^2}{l_s^2}\left(\partial \phi \bar{\partial} \phi+\beta \bar{\partial} \gamma+\bar{\beta} \partial \bar{\gamma}-e^{-2 \phi} \beta \bar{\beta}\right),
\ea\ee
where action is described by a free field $\phi$ and a holomorphic bosonic $\beta$, $\gamma$ system with conformal dimension
\begin{equation}
h(\beta)=1, h(\gamma)=0,
\end{equation}
and two-point function
\begin{equation}
\langle\phi(z) \phi(0)\rangle=-\log |z|^2,\langle\beta(z) \gamma(0)\rangle=1 / z.
\end{equation}

Taking quantum effect into account, the renormalized Lagrangian is \cite{Giveon:1998ns} 
\begin{equation}\label{lagwa}
\mathcal{L}_{\mathrm{ws}}=\partial \phi \bar{\partial} \phi-\frac{2}{\alpha_{+}} \widehat{R} \phi+\beta \bar{\partial} \gamma+\bar{\beta} \partial \bar{\gamma}-\beta \bar{\beta} \exp \left(-\frac{2}{\alpha_{+}} \phi\right),~~\alpha_+^2=2k-4.
\end{equation}
The corresponding SL(2,R) current 
\be \label{Jwa}
\begin{aligned}
J^3 & =\beta \gamma+\frac{\alpha_{+}}{2} \partial \phi, \\
J^{+} & =\beta \gamma^2+\alpha_{+} \gamma \partial \phi+k \partial \gamma, \\
J^{-} & =\beta,
\end{aligned}
\ee 
whose modes expansion $J^i(z)=\sum_{n=-\infty}^\infty z^{-n-1}J^i_n$ obey the SL(2,R) current algebra
\be \ba \label{mocom}
&[J_n^3,J_m^3]=-\frac{1}{2}kn\delta(n+m),\\
&[J_n^3,J_m^\pm]=\pm J^\pm_{n+m},\\
&[J^+_n,J^-_m]=k n \delta(n+m)-2J^3_{m+n}.
\ea\ee
The representation of the current algebra can be found in \cite{Maldacena:2000hw}, and the long string sector corresponds to vertex operators belonging to the continuous representation of the SL(2, R). 
To figure out which operator in the world-sheet theory duals to single-trace $T\bar{T}$ operator in the spacetime CFT, we need to investigate the vertex operator in the world-sheet WZW model. Two relevant kinds of vertex operators will be relevant: the current operator and the energy-momentum tensor. 
\subsubsection*{Current operator}
From the  current \eqref{Jwa}, one introduces \cite{Giveon:2017nie} \footnote{
For example, from the algebra \eqref{mocom}, one can obtain \cite{Kutasov:1999xu}
\begin{equation}
\begin{aligned}
& e^{-x J_0^{-}} J^{+}(z) e^{x J_0^{-}}\\
=&J^{+}(z)+\left[-x J_0^{-} , J^{+}(z)\right]+\frac{1}{2}\left[-x J_0^{-} ,\left[-x J_0^{-}, J^{+}(z)\right]\right]+\cdots\\
=&J^{+}(z)-2 x J^3(z)+x^2 J^{-}(z).
\end{aligned}
\end{equation}
}
\begin{equation}
\begin{aligned}
J^{+}(x ; z) & =e^{-x J_0^{-}} J^{+}(z) e^{x J_0^{-}}=J^{+}(z)-2 x J^3(z)+x^2 J^{-}(z), \\
J^3(x ; z) & =e^{-x J_0^{-}} J^3(z) e^{x J_0^{-}}=J^3(z)-x J^{-}(z)=-\frac{1}{2} \partial_x J^{+}(x ; z), \\
J^{-}(x ; z) & =e^{-x J_0^{-}} J^{-}(z) e^{x J_0^{-}}=J^{-}(z)=\frac{1}{2} \partial_x^2 J^{+}(x ; z),
\end{aligned}
\end{equation}
where $z$ is world-sheet coordinates and $x$ can be regarded as spacetime coordinates where the dual CFT lives. It is convenient  to compact the currents together as follows
\begin{equation}\label{Jdef}
J(x ; z) \equiv-J^{+}(x ; z)=2 x J^3(z)-J^{+}(z)-x^2 J^{-}(z),
\end{equation}
whose zero mode (on the worldsheet) is 
\begin{equation}
J_0(x)=\int d z J(x, z).
\end{equation}

Another useful operator is 
\begin{equation}
\Phi_h(x ; z)=\frac{1}{\pi}\left(\frac{1}{|\gamma-x|^2 e^\phi+e^{-\phi}}\right)^{2 h},
\end{equation}
which is an eigenfunction of Laplacian on AdS$_3$. $\Phi_h(x ; z)$ can be viewed as a primary of spacetime with dimension $(h,h)$, and also primary of world-sheet with dimension $(-h(h-1)/(k-2),-h(h-1)/(k-2))$. The transformation  generated by the  current for $\Phi_h(x ; z)$ is 
\begin{equation}
\begin{aligned}
{\left[J_0(x), \Phi_h(y , \bar{y})\right] } & =\int d z\left[J(x, z), \Phi_h(y, \bar{y})\right] \\
& =\int d z\left(2 x\left[J^3(z), \Phi_h\right]-\left[J^{+}(z), \Phi_h\right]-x^2\left[J(z), \Phi_h\right]\right) \\
& =2 x\left[J_0^3, \Phi_h\right]-\left[J_0^{+}, \Phi_h\right]-x^2\left[J_0^{-}, \Phi_h\right] \\
& =2 x(-1)\left(y \partial_y+h\right) \Phi_h-(-1)\left(y^2 \partial_y+2 h y\right) \Phi_h-x^2(-1) \partial_y \Phi_h \\
& =(x-y)^2 \partial_y \Phi_h+2(y-x) h \Phi_h.
\end{aligned}
\end{equation}

\subsubsection*{Spacetime energy-momentum tensor }
From Sugawara construction, the world-sheet energy-momentum tensor can be expressed as 
\footnote{
From \eqref{Jdef}, for $J(x;z)$
\begin{equation}
\begin{aligned}
& J \p_x^2 J-\frac{1}{2} (\partial_x J)^2  
=  -2J J^{-}-\frac{1}{2}\left(2 J^3-2 x J^{-}\right)^2  
= 2J^+J^--2(J^3)^2.
\end{aligned}
\end{equation}
}
\begin{equation}
T^{\mathrm{ws}}=\frac{1}{2(k-2)}\left[J \partial_x^2 J-\frac{1}{2}\left(\partial_x J\right)^2\right] .
\end{equation}
Here $J(x;z)$ has spacetime dimension (-1,0),  $x$ has spacetime dimension (-1,0), thus $T^{ws}$ has spacetime dimension $(0,0)$.
$\Phi_{h=1}$ as spacetime dimension $(h,\bar{h})=(1,1)$.

Thus, to construct $(2, 0)$ spacetime tensor, which is the right dimension for spacetime energy-momentum tensor, one needs to combine ($\p_x^2, J, \bar{J},\Phi_1$), taking BRST condition into account, it turns out that the spacetime stress tension could be constructed as 
\begin{equation}\ba 
T(x)=&\frac{1}{2 k} \int d^2 z\left[J(x ; z) \partial_x^2 \Phi_1+3 \partial_x J \partial_x \Phi_1+3 \partial_x^2 J \Phi_1\right] \bar{J}(\bar{x} ; \bar{z})\\
=&\frac{1}{2 k} \int d^2 z\left(\partial_x J \partial_x \Phi_1+2 \partial_x^2 J \Phi_1\right) \bar{J}(\bar{x} ; \bar{z}).
\ea
\end{equation}
In parallel, one can also obtain a similar expression for $\bar{T}(x)$. By taking the product of $T$ and $\bar{T}$, one can obtain the $T\bar{T}$ operator with spacetime dimension $(2,2)$ that triggers double-trace $T\bar{T}$ deformation. What interests us here is the operator corresponds to the single-trace $T\bar{T}$ operator, which turns out to be 
the operator of spacetime dimension $(2,2)$,
the vertex operator of the massive dilaton \cite{Giveon:2017nie}
\begin{equation}
D(x)=\int d^2 z\left(\partial_x J \partial_x+2 \partial_x^2 J\right)\left(\partial_{\bar{x}} \bar{J} \partial_{\bar{x}}+2 \partial_{\bar{x}}^2 \bar{J}\right) \Phi_1 .
\end{equation}
This is a quasi-primary operator. 
This operator is supposed to trigger single-trace $T\bar{T}$ deformation  in the spacetime theory 
\begin{equation}
D(x)=A \sum_{i=1}^p T_i(x) \bar{T}_i(\bar{x}).
\end{equation}
Integrating over spacetime, one has \footnote{
Note that $\p_x$ commute with $\left(\partial_{\bar{x}} \bar{J} \partial_{\bar{x}}+2 \partial_{\bar{x}}^2 \bar{J}\right)$.
And for example
\begin{equation}
\begin{aligned}
& \int \partial_x J \partial_x \cdot\left[\partial_{\bar{x}} \bar{J} \partial_{\bar{x}} \Phi_1\right] d z=\int \partial_x J \cdot \partial_{\bar{x}} \bar{J} \partial_x \partial_{\bar{x}} \Phi_1 d^2 z = \int \partial_x^2 J \cdot \partial_{\bar{x}}^2\bar{J} \Phi_1 d^2 z.
\end{aligned}
\end{equation}
}
\begin{equation}\label{intD}\ba 
\int d^2 x D(x, \bar{x})=&\int d^2 x \int d^2 z \partial_x^2 J \partial_{\bar{x}}^2 \bar{J} \Phi_1=4 \int d^2 x \int d^2 z J^{-}(z) \bar{J}^{-}(\bar{z}) \Phi_1(x ; z) \\
\propto & \int d^2 z J^{-}(z) \bar{J}^{-}(\bar{z}), 
\ea\end{equation}
which implies that the single-trace $T\bar{T}$ deformation corresponds to current-current $J^-\bar{J}^-$ deformation in the world-sheet, while the latter is a marginal deformation on the world-sheet classically.

\subsubsection*{ Deformation of world-sheet WZW model}
In the Wakamoto representation 
\eqref{lagwa}, the Lagrangian deformed by \eqref{intD} can be expressed as $-\l \int d^2 z J^{-}(z) \bar{J}^{-}(\bar{z})=-\l\int d^2z \beta(z)\bar{\beta}(\bar{z})$. Thus the deformed world-sheet Lagrangian is 
\begin{equation}
\mathcal{L}_{\mathrm{ws}}=\partial \phi \bar{\partial} \phi-\frac{2}{\alpha_{+}} \widehat{R} \phi+\beta \bar{\partial} \gamma+\bar{\beta} \partial \bar{\gamma}-\beta\bar{\beta}\left[\l+\exp \left(-\frac{2}{\alpha_{+}} \phi\right)\right].
\end{equation}
After integrating over $\beta,\bar{\beta}$, yields
\begin{equation}\label{deL}
\mathcal{L}_{\mathrm{ws}}=\partial \phi \bar{\partial} \phi-\frac{2}{\alpha_{+}} \widehat{R} \phi +\frac{\bar{\p}\gamma \p \bar{\gamma}}{\l+\exp \left(-\frac{2}{\alpha_{+}} \phi\right)},
\end{equation}
which is $\sigma$-model on a space interpolating between { flat spacetime with a linear dilaton } in the UV region $\phi\to \infty$ and AdS$_3$ in the IR $\phi\to -\infty$. Note that $\l>0$ is assumed in the above discussion. 
When $\l<0$, 
the IR behavior remains AdS$_3$, while there is a curvature singularity located at 
certain value of $\phi$. It was argued in \cite{Chakraborty:2020swe}
that to study single-trace $T\bar{T}$, both side of the singular wall should be included. 
\subsubsection*{Deformation of symmetric orbifold CFT}
An asymmetric orbifold like \eqref{syorbi}, based on the symmetric group $S_N$, is a particular kind of orbifold CFT.
It is related to holographic CFT in many aspects \cite{Argurio:2000tb}. As the same with generic orbifold CFT,
the Hilbert space of symmetric orbifold 2D CFT contains untwist and twist sectors, where the latter can be derived by imposing modular invariance of the partition function. Following this spirit, the authors of \cite{Apolo:2023aho}
(see also \cite{Chakraborty:2023wel}) figure out the modular invariant partition functions under single-trace $T\bar{T}$ in terms of generalized Heck operator denoted as $T'_n$, 
\begin{equation}\label{depfsingle}
Z_N(\tau, \bar{\tau} ; \l)=\sum_{\left\{k_1, \ldots, k_N\right\}} \frac{1}{\prod_{n=1}^N n^{k_n} k_{n}!} \prod_{n=1}^N\left(T_n^{\prime} Z\right)(\tau, \bar{\tau} ; \l)^{k_n},~~\sum_{n=1}^N n k_n=N
\end{equation}
{ 
with 
\begin{equation}
\left(T_n^{\prime} Z\right)(\tau, \bar{\tau} ; \l)=\sum_{\gamma \mid n} \sum_{\alpha=0}^{\gamma-1} Z\left(\frac{n \tau+\alpha \gamma}{\gamma^2}, \frac{n \bar{\tau}+\alpha \gamma}{\gamma^2} ; \frac{\l}{\gamma^2}\right) ,
\end{equation}}
which leads to the appearance of twisted sectors in the deformed theory. The deformed spectrum of single particle twisted states also follows from the above-deformed partition function, which (up to different conventions) matches the result  \eqref{detwspectrum} below derived in the bulk side.  The generating function for partition function $Z_N$ can be defined as 
\begin{equation}
\mathcal{Z}(\tau, \bar{\tau} ; \l ; p):=\sum_{N=0}^{\infty} p^N Z_N(\tau, \bar{\tau} ; \l)=\exp \left(\sum_{n=1}^{\infty} \frac{p^n}{n}\left(T_n^{\prime} Z\right)(\tau, \bar{\tau} ; \l)\right).
\end{equation}
The deformed partition function \eqref{depfsingle} matches the holographic calculation from bulk string theory side \cite{Pozsgay:2019ekd}, suggested the single-trace $T\bar{T}$  deformed CFT is holographic dual to string in linear dilation background. Recently, an exact holographic duality was proposed, i.e., tensionless string/2D symmetric orbifold duality, with the seed CFT of the orbifold being a supersymmetric sigma model on $T^4$ \cite{Eberhardt:2018ouy, Eberhardt:2019ywk}. This framework further investigated the single trace $T\bar{T}$ deformation in \cite{Dei:2024sct}. The deformed partition function of the orbifold is computed explicitly, including the spin structure for fermions, which matches the string partition function with current-current deformation on the background with the torus boundary. Therefore, it provides strong evidence that current-current deformation of tensionless strings on AdS$_3\times S^3\times T^4$ is holographically dual to single-trace $T\bar{T}$ deformed $T^4$ orbifold.


\subsection{Little string theory perspective}
As mentioned, the background implied by the lagrangian \eqref{deL} interpolates between AdS$_3$ and { flat spacetime with a linear dilaton}. The latter are also known as little string theory (LST) background. LST
is the decoupled theory of NS5 branes that are magnetically charged under the NS B field. 
In type II string theory, $k$  NS5-branes extending in $\mathbb{R}^{5,1}$ with coordinate $x^\m$, is described in string frame as  
\be 
\begin{aligned}
d s^2 & =d x_\mu d x^\mu+\left(1+\frac{N \alpha^{\prime}}{r^2}\right) d x^i d x^i,~~\mu=0,1,...,5.\\
e^{2 \Phi} & =g_s^2\left(1+\frac{N \alpha^{\prime}}{r^2}\right)
\end{aligned}
\ee
with transverse direction
$
d x^i d x^i=d r^2+r^2 d \Omega_3^2.
$
To decouple the dynamics on the
fivebranes from the bulk, consider the limit $g_s\to 0, r/g_s$ fixed, one obtain the background 
\begin{equation}\label{dec}
\begin{aligned}
d s^2 =d x_\mu d x^\mu+N \alpha^{\prime}\left(d \sigma^2+d \Omega_3^2\right), ~~
e^{2\Phi}=&\frac{g_s^2N\alpha'}{r^2}  \equiv e^{-2\sigma}.
\end{aligned}
\end{equation}
This background is holographically dual to LST on the fivebrane.
The topology of the background \eqref{dec} becomes
\begin{equation}\label{geo1}
\mathbb{R}^{5,1} \times \mathbb{R}_\phi \times S^3.
\end{equation}
Note the string coupling vanishes far away from brane $\sigma\to +\infty$ and divergent near the brane $\sigma\to -\infty$. 
 
So far NS5-branes wrapped around $\mathbb{R}^{5,1}$ is considered, one can also let the $k$ NS5-branes wrapped on compact $T^4\times S^1$, then the geometry \eqref{geo1}  becomes \cite{Giveon:2017nie}
\be  \label{geo2}
\mathbb{R}_t\times \mathbb{R}_\phi \times S^1\times \mathcal{N}~~\text{with}~~\mathcal{N}=T^4\times S^3.
\ee 
String theory in this background 
is holographically dual to non-local, non-gravitational LST theory \cite{Aharony:1998ub}. 
To make contact \eqref{geo2} with geometry induced by  \eqref{deL}, one can
add $p$ F1 string around $S^1$ in \eqref{geo2}, which will modify the IR geometry (near the brane region) while leaving the UV geometry intact. Meanwhile, the string coupling will grow when approaching the IR and saturate $g_s^2\sim 1/p.$ Thus, the perturbation theory applies to large $p$.
Now the IR geometry becomes
\be \label{geo3}
\text{AdS}_3\times T^4\times S^3.
\ee 
Therefore, we have obtained a geometry indicated by \eqref{deL} starting from LST background \cite{Aharony:1998ub}. Notice that the background with a black hole in the linear dilaton in the UV and BTZ in the IR  arises in the SL$(2,\mathbb{R})\times$U(1)/U(1) coset model, which can be thought as finite temperature generalization of background implied by \eqref{deL} \cite{Chakraborty:2020yka}. In addition, the above geometry can also arise by null gauging of the CFT on $R^{1,1}\times$AdS$_3\times\mathcal{N}$\cite{Giveon:2017myj, Asrat:2017tzd}.

As an example for the background, namely, local AdS in the IR and { background with} linear dilaton in the UV,
let us consider 
type IIB string theory on $\mathbb{R}^{4,1}\times T^4\times S^1$, $k$ NS5-branes on $T^4\times S^1$ and $p$ fundamental strings on $S^1$ (smeared over the $T^4$), with $n$ units of momentum along $S^1$, on string frame \cite{Chakraborty:2020yka,Chakraborty:2020swe}
\begin{equation}\label{NS5F1}
\begin{aligned}
d s^2 & =\frac{1}{f_1}\left[-\frac{f}{f_n} d t^2+f_n\left(d x+\frac{r_0^2 \sinh 2 \alpha_n}{2 f_n r^2} d t\right)^2\right]+f_5\left(\frac{1}{f} d r^2+r^2 d \Omega_3^2\right)+\sum_{i=1}^4 d x_i^2, \\
e^{2 \Phi} & =g^2 \frac{f_5}{f_1}, ~~~H  =d x \wedge d t \wedge d\left(\frac{r_0^2 \sinh 2 \alpha_1}{2 f_1 r^2}\right)+r_0^2 \sinh 2 \alpha_5 d \Omega_3
\end{aligned}
\end{equation}
with 
\be 
f=1-\frac{r_0^2}{r^2}, \quad f_{1,5, n}=1+\frac{r_{1,5, n}^2}{r^2}, \quad r_{1,5, n}^2=r_0^2 \sinh ^2 \alpha_{1,5, n}.
\ee\be 
\sinh 2 \alpha_1= \frac{2 l_s^2 p}{v X}, \quad \sinh 2 \alpha_5=\frac{2 l_s^2 k}{r_0^2}, \quad \sinh 2 \alpha_n=\frac{2 l_s^4 n}{R^2 v X},~~X=\frac{r_0^2}{g^2} .
 \ee
Note that for $p=0,n=0$, \eqref{NS5F1} reduces to (non-extremal) NS5-branes
\begin{equation}
\begin{aligned}
d s^2 & = \left[-f d t^2+dx^2\right]+f_5\left(\frac{1}{f} d r^2+r^2 d \Omega_3^2\right)+\sum_{i=1}^4 d x_i^2, \\
e^{2 \Phi} & =g^2 f_5, ~~H= r_0^2 \sinh 2 \alpha_5 d \Omega_3,
\end{aligned}
\end{equation}
which further reduces to extremal NS5 when $r_0=0$.


To investigate the thermodynamics of the dual field, transform 
\eqref{NS5F1} to  Einstein frame  
\be 
ds^2_E=e^{-\Phi/2}ds^2,
\ee 
in which frame we can compute ADM mass from tress tensor $T_{00}$ as 
\be \ba 
M=&\frac{V_{S_1}V_{T^4}}{8\pi G_N^{(10)}}\int d^4 y T_{00}=\frac{V_{S_1}V_{T^4}}{8\pi G_N^{(10)}}\int d^4 y (R_{00}^{(1)}-\frac12 \eta_{00}R^{(1)})\\
=& \frac{R v X}{2 l_s^4}\left(\cosh 2 \alpha_1+\cosh 2 \alpha_5+\cosh 2 \alpha_n\right)
\ea\ee 
with the flat transverse coordinates $y^i,i=1,...,4$
satisfying
\be 
dy^idy^i=dr^2+r^2 d\Omega_3^2.
\ee 
{ Note that the mass can be computed with covariant phase space formalism \cite{Chang:2023kkq}}.
By evaluating the area of the horizon at $r=r_0$, the entropy turns out to be
\be 
S=\frac{A}{4G_{N}^{(10)}}=\frac{2 \pi R v X r_0}{l_s^4} \cosh \alpha_1 \cosh \alpha_5 \cosh \alpha_n,
\ee 
Next, taking  the decoupling limit, the energy and entropy become
\be 
S=\frac{2\pi Rv X r_0}{l_s^3}\sqrt{k}\cosh \alpha_1
,~~
E=\frac{RvX}{2l_s^4}(e^{-2\alpha_1}+1),
\ee 
which satisfies
\be \label{entropyde}
S=2\pi\sqrt{\frac{kl_s^2 \mathcal{E}^2}{R^2}+2kp \mathcal{E}},~~\mathcal{E}=ER,
\ee 
which is consistent with the entropy \eqref{singleS} of single-trace  $T\bar{T}$ deformed theory, provided one identifies the deformation parameter as  $\l=l^2_s/R^2$.   Therefore, it is suggested that the gravity in the above background is dual to single-trace $T\bar{T}$ deformed CFT. As one varies the $\l$, the energy $\mathcal{E}$ will be changed, but not the number of states 
\begin{equation}
S(\l)=2 \pi \sqrt{k \lambda \mathcal{E}(\lambda)^2+2 k p \mathcal{E}(\lambda)}=S(0)=2 \pi \sqrt{2 k p \mathcal{E}(0)}.
\end{equation}
It then follows the deformed spectrum
\begin{equation}
\frac{1}{p} \mathcal{E}(\lambda)=\frac{1}{\lambda}(-1+\sqrt{1+2 \lambda \mathcal{E}(0) / p}) .
\end{equation}
\subsection{TsT and single trace $T\bar{T}$}\label{subsectionsingletrace}

As introduced in the last subsection, the geometry dual to the single-trace $T\bar{T}$ deformation interpolated AdS$_3$ and { background with} linear dilaton, which is relevant in studying little string theory. More recently, it was suggested that the single-trace $T\bar{T}$ deformed CFT dual to long string sector on a background generated by TsT transformation of $AdS_3\times T^4\times S^3$. \footnote{{ A related interesting work showed that the deformation can be solved using O($d,d$) duality of string theory \cite{Araujo:2018rho}.}} The latter duality is supported by matching the spectrum and thermodynamics \cite{Apolo:2019zai}.  Firstly, let us review the background of the TsT transformation. Following the convention in \cite{Apolo:2019zai}, the world-sheet action 
is 
\begin{equation}
\tilde{S}=-\ell_s^{-2} \int d^2 z\left(\sqrt{-\eta} \eta^{a b} \tilde{G}_{\mu \nu}+\epsilon^{a b} \tilde{B}_{\mu \nu}\right) \partial_a \tilde{X}^\mu \partial_b \tilde{X}^\nu=\ell_s^{-2} \int d^2 z \tilde{M}_{\mu \nu} \partial \tilde{X}^\mu \bar{\partial} \tilde{X}^\nu,
\end{equation}
where $\tilde{M}_{\mu \nu}=\tilde{G}_{\mu \nu}+\tilde{B}_{\mu \nu}$ is the background fields \cite{Giveon:1994fu}. The field $\tilde{G}_{\mu \nu},\tilde{B}_{\mu \nu}$ and dilaton $\tilde{\Phi }$ is assumed to be independent of two coordinates denoted as 
$\tilde{X}^1,\tilde{X}^{\bar{2}}$. The associated Neother currents 
\be \label{Neocurrent}
j^a_{(n)}=\frac{\p L}{\p (\p_a X^n)},~~n=1,\bar{2},
\ee 
which will be useful in a moment.

Consider the TsT transformation of the background and coordinates, i.e., a T transformation along $\tilde{X}^1$, followed by a shift of $\tilde{x}^2$
\be \label{shift}
\tilde{x}^1=x^1,~~\tilde{x}^2=x^2-2\tilde{\l}x^1/l_s^2,
\ee
and another T transformation along $x^1$. { Varies coordinates in  each step of transformation are summarized as follows  }
\footnote{The coordinate transformation corresponds T duality \cite{Frolov:2005dj}, taking the first T duality as an example,  can be written \begin{equation}\label{Tcoordi}
\begin{aligned}
\partial_a \tilde{X}^{1} & =-(\sqrt{-\eta} \varepsilon_{a b} \eta^{c b} \tilde{g}_{1 \mu} \partial_c \tilde{x}^\mu+\tilde{b}_{1 \mu} \partial_a \tilde{x}^\mu )=-\epsilon_{ab}\tilde{J}^b_{(1)}\\
\tilde{X}^i & =\tilde{x}^i,~~i\neq 1
\end{aligned}
\end{equation}
}
\be \label{TsT}
(\tilde{G}_{\mu \nu},\tilde{B}_{\mu \nu},\tilde{\Phi},\tilde{X}^\m)\xrightarrow{T} (\tilde{g}_{\mu \nu},\tilde{b}_{\mu \nu},\tilde{\phi},\tilde{x}^\m)\xrightarrow{s}(g_{\mu \nu},b_{\mu \nu},\phi,x^\m)\xrightarrow{T}(G_{\mu \nu},B_{\mu \nu},\Phi,X^\m).
\ee 
The T-duality and shift can be embedded in O($d,d$) transformation (see, for example, \cite{Orlando:2019rjg}). The action of $g\in$ O($d,d$) on background  organized as 
\begin{equation}
H(G, B)_{\hat{I} \hat{J}}=\left(\begin{array}{cc}
G-B G^{-1} B & B G^{-1} \\
-G^{-1} B & G^{-1}
\end{array}\right)_{\hat{I} \hat{J}}, \quad \hat{I}, \hat{J}=1, \ldots, 2 d.
\end{equation}
takes the form
\be \label{ghg}
H(G',B')=g^TH(G,B)g.
\ee
Here O($d,d$) element $g$ is defined by obeying 
\be 
g^T L g=L,~~\text{with}~~
L=\left(\begin{array}{ll}
0 & \mathbb{I} \\
\mathbb{I} & 0
\end{array}\right),~~
g=\left(\begin{array}{ll}
a & b \\
c& d
\end{array}\right).
\ee 
Beside \eqref{ghg}, the background field $M$ is transformed as \cite{Giveon:1994fu}
\be 
M\to M'=[aM+b][cM+d]^{-1}.
\ee 
It turns out that a coordinate shift, or more generally, a diffeomorphism transformation of background, can be realized by 
\be 
g=\left(\begin{array}{cc}
A & 0\\
0& \left(A^{-1}\right)^t
\end{array}\right),
\ee 
such that
\begin{equation}
\begin{aligned}
& g^t H(G,B) g  =H(A^T G A,A^T B A). 
\end{aligned}
\end{equation}
Note that for the shift in  \eqref{TsT}, one has
\be 
A_{\m\n}=\frac{\p \tilde{X}^\m}{\p \tilde{x}^\n}=\left(
    \begin{array}{cc}
        1 & 0\\
        -2\tilde{\l}/l_s^2 & 1\\
    \end{array}\right).
\ee 
As for $T$-duality along the $k$-th direction , the corresponding O($d,d$) element is  given by 
\be 
g_{T_k}=\left(\begin{array}{cc}
\mathbb{I}-E_k & E_k \\
E_k & \mathbb{I}-E_k
\end{array}\right) .
\ee 
$E_k$ the $d \times d$-matrix with 1 in the $(k, k)$-entry and zero everywhere else. Then
the metric and $B$-field transform according to the standard Buscher rules
($k=1$)
\be\label{buscher}
\ba 
\tilde G_{11}=& \frac{1}{G_{11}}  ~~G_{1i}=\frac{B_{1i}}{G_{11}},~~
\tilde G_{ij}=  G_{ij}-\frac{G_{1i}G_{1j}-B_{1i}B_{1j}}{G_{11}}.\cr
\tilde B_{1i}=& \frac{G_{1i}}{G_{11}},i\neq 1;~~~~
\tilde B_{ij}= B_{ij}-\frac{G_{1i}B_{1j}-B_{1i}G_{1j}}{G_{11}},i\neq 1.\\
\ea\ee    

Based on the above results, the background and after TsT transformation \eqref{TsT} is given by   \cite{Apolo:2019zai}\cite{Giveon:1994fu}\cite{Lunin:2005jy}
\begin{equation} \label{Mtrans}
M=\tilde{M}\left(I+\frac{2 \tilde{\lambda}}{\ell_s^2} \Gamma \tilde{M}\right)^{-1}, \quad \Phi_{\mathrm{TsT}}=\tilde{\Phi}+\frac{1}{4} \log \left(\frac{\operatorname{det} G_{\mu \nu}}{\operatorname{det} \tilde{G}_{\mu \nu}}\right)
\end{equation}
with $
\Gamma_{\mu \nu}=\delta_\mu^1 \delta_\nu^{\overline{2}}-\delta_\mu^{\overline{2}} \delta_\nu^1
$. Here, the TsT transformation of the dilaton field is included. The action after TsT is 
\begin{equation}\label{defsx}
S =\ell_s^{-2} \int d^2 z M_{\mu \nu} \partial X^\mu \bar{\partial} X^\nu,
\end{equation}
Making use of \eqref{Mtrans}, one has 
\begin{equation}
\frac{\partial M}{\partial \tilde{\lambda}}=-2 \ell_s^{-2} M \Gamma M.
\end{equation}
Thus, one obtains the flow of deformed world-sheet action
\begin{equation}\label{defS}
\frac{\partial S}{\partial \tilde{\lambda}}=2 \int \boldsymbol{j}_{(1)} \wedge \boldsymbol{j}_{(\overline{2})},
\end{equation}
where $\boldsymbol{j}_{(1)}$ is one-form current related to two Noether currents \eqref{Neocurrent} \cite{Apolo:2019zai}.
\eqref{defS} implies that world-sheet action with TsT transformed background field corresponds to current-current deformation, which is reminiscent of \eqref{intD}, which dual to single-trace $T\bar{T}$ deformation. TsT transformed action is related to single-trace $T\bar{T}$ deformation on the boundary once proper coordinate $X^1, X^{(2)}$ is chosen. This holographic duality is further supported by matching the spectrum on world-sheet \eqref{defsx} and the deformed CFT side. The spectrum on the world sheet can be computed by relating the energy-momentum tensor before and after TsT.
By repeated use coordinates transformation for T-duality \eqref{Tcoordi} and shift \eqref{shift}, one can relate coordinates in TsT transformation \cite{Apolo:2019zai} (see \eqref{TsT}) \footnote{\eqref{coordTsT} can be obtained by the following  coordinate transformations
\begin{equation}
\begin{aligned}
\partial_a \tilde{X}^{1}  =&-(\sqrt{-\eta} \varepsilon_{a b} \eta^{c b} \tilde{g}_{1 \mu} \partial_c \tilde{x}^\mu+\tilde{b}_{1 \mu} \partial_a \tilde{x}^\mu )=-\epsilon_{ab}\tilde{j}^b_{(1)}(\tilde{x}),~~\tilde{X}^2 =\tilde{x}^2,\\
\tilde{g}_{1\m}\p \tilde{x}^\m=&\frac{\p x^\rho}{\p \tilde{x}^1}g_{\rho\m}\partial x^\m=g_{1\m}\p x^\m+2(\tilde{\l}/l_s^2) g_{2\m}\p x^\m,
\\
\partial_a X^{1}  =&-(\sqrt{-\eta} \varepsilon_{a b} \eta^{c b} g_{1 \mu} \partial_c x^\mu+b_{1 \mu} \partial_a x^\mu )=-\epsilon_{ab}j^b_{(1)},~~~X^2 =x^2,
\end{aligned}
\end{equation}
which corresponds to T-duality, shift, and another T-duality in \eqref{TsT}, respectively. Also, the Buscher rules \eqref{buscher} for T-duality is used
\be\ba 
m_{12}=&\frac{M_{12}}{G_{11}},~~m_{22}=
M_{22}-\frac{M_{12}M_{21}}{G_{11}},~~m_{i2}=
M_{i2}-\frac{M_{i1}M_{12}}{G_{11}}.
\ea\ee 

}
\be \label{coordTsT}
\p_a\tilde{X}^n=\p_a X^n-2(\tilde{\l}/l_s^2)\epsilon^{nn'}\epsilon_{ab}J^b_{(n')}(X),~~\epsilon^{12}=-1.
\ee 
In \cite{Apolo:2019zai}, $\tilde{X}$ is replaced by auxiliary fields $\hat{X}^n$ to distinguish the non-trivial boundary condition for $\hat{X}^n$.
With \eqref{coordTsT} and boundary conditions for auxiliary fields $\hat{X}^n$, one can relate the Virasoro constraint before and after TsT so that the deformed spectrum follows 
\begin{equation}\label{detwspectrum}
E_L(0)=E_L(\lambda)-\frac{2 \lambda \ell}{w k} E_L(\lambda) E_R(\lambda), \quad E_R(0)=E_R(\lambda)-\frac{2 \lambda \ell}{w k} E_L(\lambda) E_R(\lambda),
\end{equation}
{ It is conjectured that the states with $w<-1$ in the long string sector of undeformed theory dual to twisted sector of the symmetric product orbifold CFT. $|w|$ counts the number of times the string winds along spatial direction.} Notice \eqref{detwspectrum} with $w=-1$ is a reminiscent of deformed spectrum \eqref{despec} of double trace $T\bar{T}$ deformation.  
For $w=-1$, the deformed spectrum matches the spectrum of single-particle states in the untwisted sector of single-trace deformed orbifold theory. While for $w<-1$, it corresponds to $Z_{-w}$ twisted sector of the deformed orbifold. Therefore, the spectrum on both sides of the duality match each other.

In addition to the spectrum, a further piece of evidence is provided by matching the deformed correlation functions \cite{Cui:2023jrb}\cite{Chakraborty:2023wel}\cite{Giveon:2023gzh}, where the two-point function of the vertex operators on worldsheet after TsT is computed, both perturbative and non-perturbative results are presented.  Meanwhile, in the dual single-trace $T\bar{T}$ orbifold, the two-point function of the twist operator, which is dual to vertex operator with a specific quantum number in the worldsheet, is calculated by conformal perturbation theory, as well as by the nonperturbative method (for untwisted sector correlation functions) developed in \cite{Cardy:2019qao}. Consistent results from both sides were established. However, the matching of correlation functions of general primary operators in the twisted sector remains unsolved.
In addition to the long string sector, the deformed correlation functions were also considered in the short string sector, which can not be described in symmetric orbifold \cite{Giribet:2017imm, Asrat:2017tzd, Chakraborty:2020yka}.



\newpage

\section{Recent Developments and Perspectives}\label{developments-TTbar}
In recent years, significant progress has been made in the study of $T\bar{T}$-like deformation and its generalizations, providing new insights and establishing connections across field theory, holography, and string theory. Generalized $T\bar{T}$-like operators broaden the original scope of the investigation. These developments enhance our understanding of non-locality and integrability and offer new perspectives on the relationship between classical geometric formulations and deformed field theories. This section briefly reviews recent advances and outlines potential directions for future research.
\subsection{Generalized $T\bar{T}$-like deformations}
The generalization of the $T\bar{T}$ deformation can be explored from various perspectives, including spacetime symmetries, conserved currents, and dimensionality. Extensions to different settings often adapt the two-dimensional $T\bar{T}$ deformation framework to accommodate variations in symmetry, composite operators, and spacetime dimensions.

{\bf Beyond relativistic QFT}

The $T\bar{T}$ deformation has been generalized to various systems beyond relativistic QFT and string theory. There are generalizations to non-relativistic QFTs \cite{Cardy:2018jho,Hansen:2020hrs,Ceschin:2020jto, Chen:2020jdi}, non-relativistic string theory~\cite{Blair:2020ops,Demulder:2023bux}, and many body systems such as interacting particles\cite{Cardy:2020olv,Jiang:2020nnb,Pavshinkin:2021jpy}. Classically, the $T\bar{T}$-like deformations of free particles lead to novel classical integrable many-body systems\cite{Doyon:2023bvo,  Doyon:2023hiq}.
Additionally, a lot of works have been focused on the $T\bar{T}$ deformation of integrable spin chains\cite{Bargheer:2008jt, Pozsgay:2019ekd, Marchetto:2019yyt, Pozsgay:2021rwc}, where the $T\bar{T}$ deformation preserves integrability and the S-matrix is deformed by a CDD factor. Recently, $T\bar{T}$ deformation has also been studied in lattice theories in \cite{Jiang:2023rxa}.

The analog of the $T\bar{T}$ deformation in one dimension has been introduced in \cite{Gross:2019ach, Gross:2019uxi}, which can be defined through the flow equation of the Hamiltonian:
\be \label{TTbar1D}
\partial_\lambda H=\frac{H^2}{1/2-2\lambda H},
\ee
which can be defined via the dimensional reduction of {$T\bar{T}$ deformation} in two dimensions. The $T\bar{T}$ deformation in one dimension can be viewed as a special case, a broad class solvable of deformations that maps the Hamiltonian $H$ to a function $f(H)$ of $H$. The deformed spectrum and partition function can be obtained from the undeformed theory. The geometrical realization of the above flow equation (\ref{TTbar1D}) has been explored in~\cite{He:2021dhr}.

The one-dimensional analog of $T\bar{T}$ deformation also has interesting holographic interpretations. The two-dimensional Jackiw-Teitelboim gravity is
holographically dual to a solvable one-dimensional quantum theory known as the Schwarzian theory \cite{Jensen:2016pah, Maldacena:2016upp, Engelsoy:2016xyb}. 
 The holographic duality between two-dimensional Jackiw-Teitelboim gravity at finite cutoff and $T\bar{T}$ deformed Schwarzian theory was studied in \cite{Gross:2019ach, Gross:2019uxi, Iliesiu:2020zld}.
 The genus expansion over two-dimensional surfaces of the partition function of the Jackiw-Teitelboim gravity can be computed by a particular double-scaled matrix integral \cite{Saad:2019lba}. $T\bar{T}$ deformation of matrices models are investigated in \cite{Rosso:2020wir, Ebert:2022gyn, He:2022bbb}.

{\bf Other double current deformations}

The $T\bar{T}$ deformation is a special case of double current deformations driven by bilinears of Noether currents of the original theory.
Double current deformations play an important role in integrable models and string theory.

When there is a $U(1)$ current in a two-dimensional CFT, one can define a solvable Lorentz-breaking $J\bar{T}$ deformation \cite{Guica:2017lia}.
The holographic dual of the $J\bar{T}$ deformed CFT is related to mixed boundary conditions between the metric and a $U(1)$ gauge field \cite{Bzowski:2018pcy}.
There are interesting relations between $J\bar{T}$ deformation and string theory\cite{Apolo:2018qpq,Chakraborty:2018vja,Chakraborty:2019mdf}.
Correlation functions and entanglement entropy of $J\bar{T}$ deformed CFTs were computed in \cite{He:2019vzf}.
Modular covariance of $J\bar{T}$ deformed CFTs was discussed in
\cite{Aharony:2018ics}.
For theories with higher spin conserved currents, such as two-dimensional integrable field theories or CFTs, one can construct deformations using these currents \cite{Smirnov:2016lqw, LeFloch:2019rut, Conti:2019dxg, Hernandez-Chifflet:2019sua}.

{\bf Generic energy-momentum tensor induced deformation in various spacetime}

Besides $T\bar{T}$, one can construct other Lorentz invariant operators from the energy-momentum tensor. One interesting example is the root-$T\bar{T}$ operator \cite{Conti:2022egv,Babaei-Aghbolagh:2022leo,Ferko:2022cix}:
\begin{equation}
 R=\sqrt{\frac12 T^{\mu\nu}T_{\mu\nu}-\frac14 T^{\mu}{}_{\mu}T^{\nu}{}_{\nu}}.
 \end{equation}
The rigorous definition of the  root-$T\bar{T}$ operator at the quantum level remains uncertain, and a recent proposal has been presented in \cite{Hadasz:2024pew}. Nevertheless, the deformation triggered by the root-$T\bar{T}$ operator has increased interest because it exhibits some nice properties at the classical level. 
The root-$T\bar{T}$ deformation commutes with the $T\bar{T}$ deformation. 
The classical integrability \cite{Borsato:2022tmu} of specific integrable models is preserved by the root-$T\bar{T}$ deformation.
The root-$T\bar{T}$ deformed conformal field theories are related to the ultra-relativistic (BMS$_{3}$)  field theories \cite{Rodriguez:2021tcz}\cite{He:2024yzx}. Further results \cite{Tempo:2022ndz} prove that the $\sqrt{T\bar{T}}$ deformation admits a geometrical interpretation and preserves key structural features of the original CFT, such as integrability and classical conformal symmetry. The role of the root-$T\bar{T}$ deformation in holography has been explored in \cite{Ebert:2023tih, Ebert:2024zwv}. The modular properties of root-$T\bar{T}$ deformed conformal field theories can be derived from the holographic description \cite{Tian:2024vln}.
The geometric formulation of the combination of the $T\bar{T}$ and root-$T\bar{T}$ deformation has been presented in \cite{Babaei-Aghbolagh:2024hti, Tsolakidis:2024wut}. 

We review some interesting details about the root-$T\bar{T}$ deformation.
The root-$T\bar{T}$ deformed classical action of $N$ real scalars can be computed exactly \cite{Conti:2022egv,Babaei-Aghbolagh:2022leo,Ferko:2022cix}. In the following, we mainly follow the derivation in \cite{Ferko:2022cix}.
To obtain the deformed Lagrangian, one needs to solve the flow equation
\begin{equation}
    \frac{\partial \mathcal{L}_\gamma}{\partial \gamma}=R_\gamma=  \sqrt{\frac12 T^{\mu\nu}T_{\mu\nu}-\frac14 T^{\mu}{}_{\mu}T^{\nu}{}_{\nu}},
\end{equation}
with the initial condition
\begin{equation}\label{GBscalar}
\mathcal{L}_0=\frac{1}{2}\Big(G_{ij}(\phi)g^{\mu\nu}+B_{ij}(\phi)\varepsilon^{\mu\nu}\Big)\partial_{\mu}\phi^i\partial_{\nu}\phi^j-V(\phi).
\end{equation}
We use $\gamma$ to denote the dimensionless deformation parameter. $G_{ij}$, $B_{ij}$, and $V$ represent the target space metric, Kalb-Ramond fields, and the potential, respectively.
One can use the Hilbert energy-momentum tensor
\begin{equation}
T_{\mu\nu}=-2\frac{\partial \mathcal{L}}{\partial \gamma^{\mu\nu}}+\gamma_{\mu\nu}\mathcal{L} .
\end{equation}
Now, we simplify the flow equation by using Lorentz invariant variables.
We define matrices 
\begin{equation}
(X_1)_{\mu}^{~\nu}=G_{ij}\partial_\mu\phi^i\partial^\nu\phi^j,~~~(X_n)_{\mu}^{~\nu}=(X_1^n)_{\mu}^{~\nu},
\end{equation}
and traces $x_n=\mathrm{tr}X_n$. In two dimension, $x_{n>2}$ algebraically depends on $x_1$ and $x_2$. For example, we have
\begin{equation}
2x_3=3x_1x_2-x_1^3,~~~ 2x_4=2x^2_1x_2-x_1^4+x_2^2.
\end{equation}
Then the energy-momentum tensor is given by
\begin{equation}
T_{\mu\nu}=-2(X_1)_{\mu\nu}\frac{\partial \mathcal{L}_\gamma}{\partial x_1}-4(X_2)_{\mu\nu}\frac{\partial \mathcal{L}_\gamma}{\partial x_2}+\gamma_{\mu\nu}\mathcal{L}_\gamma .
\end{equation}
The square of the root-$T\bar{T}$ operator is
\begin{equation}
R^2=\left(2x_2-x_1^2\right)\left(\frac{\partial \mathcal{L}_\gamma}{\partial x_1}+2x_1\frac{\partial \mathcal{L}_\gamma}{\partial x_2}\right)^2.
\end{equation}
Remarkably, it does not explicitly depend on $\mathcal{L}_\gamma$ and depends on the derivatives of $\mathcal{L}_\gamma$ through a perfect square. The flow equation can be simplified as
\begin{equation}
\frac{\partial\mathcal{L}_\gamma}{\partial\gamma}=\pm\sqrt{2x_2-x^2_1}\left(\frac{\partial \mathcal{L}_\gamma}{\partial x_1}+2x_1\frac{\partial \mathcal{L}_\gamma}{\partial x_2}\right) ,
\end{equation}
which can be solved by using the method of characteristics. The solution is
\begin{equation}
\mathcal{L}_\gamma\left(x_0,x_1,x_2\right)=\mathcal{L}_{\gamma=0}\left(x_0,x_1^{(\gamma)},x_2^{(\gamma)}\right) ,
\end{equation}
with
\begin{align}
x_1^{(\gamma)}&=\cosh (\gamma) x_1\pm
\sinh (\gamma)
\sqrt{2x_2-x_1^2} ,\\x_2^{(\gamma)}&=\cosh(2\gamma) x_2\pm\sinh(2\gamma) x_1\sqrt{2x_2-x_1^2} ,\end{align}
and $x_0$ denotes the terms independent of $x_{1,2}$. For the initial condition (\ref{GBscalar}), we have
\begin{equation}
x_0=\frac{1}{2}B_{ij}(\phi)\varepsilon^{\mu\nu}\partial_{\mu}\phi^i\partial_{\nu}\phi^j-V(\phi).    
\end{equation}
In the special case of free scalar with $x_0=0$ and $G_{ij}=\delta_{ij}$, the deformed Lagrangian is given by
\begin{equation}
\mathcal{L}_\gamma=\frac{\cosh\gamma}2\Phi_\mu{}^\mu\pm\frac{\sinh\gamma}2\sqrt{2\Phi_\mu{}^\nu\Phi_\nu{}^\mu-(\Phi_\mu{}^\mu)^2}.
\end{equation}
When there are two scalars, it can be obtained from the four-dimensional  ModMax theory by dimensional reduction \cite{Conti:2022egv}.

Like the $T\bar{T}$ deformation, the combination of the root-$T\bar{T}$ and $T\bar{T}$ deformation can be reformulated as coupling the undeformed theory to a gravity action\cite{Babaei-Aghbolagh:2024hti, Tsolakidis:2024wut}:
\begin{equation}\label{gravityformulation}
    S_{\gamma,\lambda}[\phi,e^a_\mu,f^a_\mu]=S_0[\phi,e_\mu^a]+S_{\mathrm{grav}}[e_\mu^a,f^a_\mu],
\end{equation}
where $S_0$ denotes the action of the undeformed theory, and $e^a_\mu$ denotes an auxiliary dynamical zweibein. The physical background zweibein $f^a_\mu$ couples to $e^a_\mu$ through the gravity action $S_{\mathrm{grav}}$ 
\begin{equation}
 S_{\mathrm{grav}}[e^a_\mu,f^a_\mu]
 = \frac{1}{2\lambda}\int d^2x \det e \left(
 2+y_1^2-y_2-2y_1 \cosh \frac{\gamma }{2}
 +2\sqrt{2y_2-y_1^2}
 \sinh \frac{\gamma }{2}
 \right),    
\end{equation}
where
\begin{equation}
    y_1=f^a_\mu e_a^\mu ,~~~
    y_2=f^a_\mu e_b^\mu f^b_\nu e_a^\nu,    
\end{equation}
are two Lorentz invariant variables.

The explicit form is
\begin{equation}
\begin{split}
   S_{\mathrm{grav}}&=  \frac{1}{\lambda} \int d^2x \Bigg({\left(e_1^2 f_2^1-e_1^1 f_2^2+e_2^1 f_1^2-e_2^2 f_1^1\right) \cosh \frac{\gamma }{2}-e_2^1 e_1^2+e_2^2 e_1^1-f_2^1 f_1^2+f_1^1 f_2^2}\Bigg. \\
    &+ {\sqrt{\left(e_1^1 e_2^2-e_2^1 e_1^2\right) \left(\frac{\left(-e_1^2 f_2^1+e_1^1 f_2^2-e_2^1 f_1^2+e_2^2 f_1^1\right){}^2}{e_1^1 e_2^2-e_2^1 e_1^2}+4 f_2^1 f_1^2-4 f_2^2 f_1^1\right)} \sinh \frac{\gamma }{2}}\Bigg) .      
\end{split}
\end{equation}

The deformed action can be obtained by solving 
the equation of motion for $e^a_\nu$ 
\begin{equation}
    \det e\, T^\nu_{0a}\equiv \frac{\delta S_{0}}{\delta e^a_\nu}=-\frac{\delta S_{\mathrm{grav}}}{\delta e^a_\nu},
\end{equation}
and plugging the solution back to (\ref{gravityformulation}).

The energy-momentum tensor of the deformed theory can be computed as
\begin{equation}
    T^\nu_{~\mu}\equiv \frac{1}{\det f}\frac{\delta S_{\gamma,\lambda}}{\delta f^a_\nu} f_\mu^a
    =\frac{1}{\det f}\frac{\delta S_{\mathrm{grav}}}{\delta f^a_\nu}f^a_\mu.
\end{equation}
{Note that the result is the same whether we perform the extermination of the action with respect to $e^a_\nu$  before or after we compute the variation of $f^a_\nu$.}  One can verify the action satisfies the flow equations
\begin{align}
    \frac{\partial S_{\gamma,\lambda}}{\partial \lambda}&=-\int d^2x \det f \det(T^\nu_{~\mu})\nonumber\\&
    =-\int d^2x \det f
    \left(
    \frac{2 \sqrt{2 y_2-y_1^2} \sinh \frac{\gamma }{2}-2 y_1 \cosh \frac{\gamma }{2}+y_1^2-y_2+2}{\lambda ^2 \left(y_1^2-y_2\right)}
    \right),\\
    \frac{\partial S_{\gamma,\lambda}}{\partial \gamma}&=\int d^2x \det f \sqrt{\frac{1}{2}T^\mu_\nu T^\nu_\mu-\frac{1}{4}(T^\nu_\nu)^2}
    \nonumber\\&
    =\int d^2x \det f
    \left(\frac{\sqrt{2 y_2-y_1^2} \cosh \frac{\gamma }{2}- y_1 \sinh \frac{\gamma }{2}}{ \lambda \left(y_1^2-y_2\right)}
    \right).
\end{align}
Therefore the action (\ref{gravityformulation}) gives a gravitational
description of the combination of the root-$T\bar{T}$ and $T\bar{T}$ deformations.
Using a change of variables $y_1=\alpha_1+\alpha_2$ and $y_2=\alpha_1^2+\alpha_2^2$, the gravity action can be simplified as
\begin{equation}
  S_{\mathrm{grav}}[e^a_\mu,f^a_\mu]
 = \frac{1}{\lambda}\int d^2x \,\det e
(\alpha_1-e^{\frac{\gamma}{2}})(\alpha_2-e^{-\frac{\gamma}{2}}).
\end{equation}
The $\alpha$ parameters can be viewed as the eigenvalues of the matrix $e^{\mu}_af^a_\nu$. This expression is useful for generalizing root-$T\bar{T}$ deformation to higher dimensions\cite{Babaei-Aghbolagh:2024hti}.


Energy-momentum tensor deformations in higher dimensions which generalize the $T\bar{T}$ deformation have been proposed in\cite{Cardy:2018sdv,Bonelli:2018kik,Taylor:2018xcy, Conti:2018tca, Babaei-Aghbolagh:2020kjg, Conti:2022egv,Hou:2022csf, Ferko:2023sps,Ferko:2024yua, Blair:2024aqz}. 
Specific models in four dimensions include
the Modified Maxwell theory\cite{Bandos:2020jsw} and its Born-Infeld-like extension  \cite{Bandos:2020hgy} which have been studied  as energy-momentum tensor flows \cite{Conti:2018jho,Babaei-Aghbolagh:2022uij,Ferko:2022iru}.
There are also non-linear gauge theories as energy-momentum tensor flows in three \cite{Ferko:2023sps}, six and higher dimensions \cite{Ferko:2024zth}. Another nonlinear extension of electrodynamics emerges naturally from the $T\bar{T}$ deformation in two dimensions and can be generalized to the non-Abelian case \cite{Brennan:2019azg}. This theory is analogous to Dirac–Born–Infeld (DBI) electrodynamics and notably preserves maximal supersymmetry.

Inquiring whether general energy-momentum tensor deformations exhibit rich and interesting properties in various physical models is natural. Recent investigations have uncovered several significant features related to such deformations. It has been demonstrated in \cite{Ferko:2024ali} that energy-momentum tensor deformations of the two-dimensional principal chiral models can be expressed through an auxiliary field formulation, which facilitates the proof of the classical integrability of these models. This result highlights the robustness of integrable structures in the presence of such deformations. Moreover, it has been shown that the non-Abelian T-duality of the principal chiral models commutes with arbitrary energy-momentum tensor deformations. This intriguing result is derived using the auxiliary field approach \cite{Bielli:2024khq}, indicating that the integrability properties are preserved under both operations. The applications of the auxiliary field formulation extend beyond the classical integrability framework and have been further explored in various contexts. For instance, \cite{Fukushima:2024nxm, Bielli:2024ach, Bielli:2024fnp} delve into more intricate properties of the auxiliary field formulation, providing deeper insights into its utility in both integrable and non-integrable systems. For four-dimensional electrodynamics, general energy-momentum tensor deformations preserve the electric-magnetic duality of Maxwell's theory. This preservation of duality under deformations is a remarkable result, as demonstrated in a series of works \cite{Babaei-Aghbolagh:2020kjg, Babaei-Aghbolagh:2022uij, Ferko:2022iru, Babaei-Aghbolagh:2022itg, Ferko:2023ruw, Ferko:2023wyi}.
Furthermore, the relationship between duality symmetries and energy-momentum tensor deformations in $p$-form theories across various dimensions has been systematically investigated in \cite{Ferko:2024zth}, revealing the broad applicability of these deformations in different physical settings. Also, energy-momentum tensor deformations have a natural connection with modified gravity models. These deformations are pivotal in various modified gravity theories, providing a fertile ground for exploring non-trivial gravitational dynamics. Such connections have been extensively studied in \cite{Floss:2023nod, Morone:2024ffm, Babaei-Aghbolagh:2024hti, Tsolakidis:2024wut, Brizio:2024arr, Morone:2024sdg, Hao:2024stt}, shedding light on how energy-momentum tensor deformations can lead to modifications in gravitational theories, including the coupling of matter fields to modified geometries.

In summary, the study of general energy-momentum tensor deformations is an active area of research with implications across a wide range of physical models, including integrable systems, duality symmetries, and gravitational theories. Recent results suggest that these deformations provide a versatile framework for probing the deeper structures of classical and quantum field theories.

\subsection{Some perspectives}

Several promising directions can be pursued to understand various aspects of $T\overline{T}$-like deformed theories regarding holography, UV completion, duality, etc.

{\bf Holography and $T\overline{T}$-like deformation}

The holographic interpretation of the $T\overline{T}$ deformation corresponds to asymptotically anti-de Sitter (AdS) geometries with modified boundary conditions or radial cutoffs~\cite{McGough:2016lol, Kraus:2018xrn, Guica:2019nzm}. Within this framework, bulk correlation functions are computed via holographic duality and mapped to boundary observables. In pure Einstein gravity, the boundary metric—determined by the Ricci flow equation~\cite{Brizio:2024arr, Morone:2024sdg}—serves as the foundation for bulk reconstruction. Consistency with the AdS/CFT correspondence is ensured by verifying that the boundary energy-momentum tensor’s correlation functions align with the standard holographic dictionary. Extending this to theories with matter fields involves imposing $T\overline{T}$-compatible boundary conditions, after which the Gubser-Klebanov-Polyakov-Witten prescription enables systematic analysis of deformed holographic correlators. Recent developments further enhance this duality by emulating $T\overline{T}$-like deformations on hyperlattices using quantum circuits~\cite{Shi:2021nkx, Basteiro:2022pyp, Chen:2023cad}.  

A discrete formulation of holography offers a powerful paradigm for studying $T\overline{T}$-like deformations. Two key frameworks emerge:  
\begin{itemize}  
    \item \textbf{Holographic Tensor Networks}: By discretizing bulk spacetime into interconnected tensors encoding local geometry and entanglement, this approach \cite{Swingle:2009bg} provides a tractable representation of bulk fields. The resulting discrete structure simplifies computations of holographic correlation functions while preserving essential features of continuum physics.  
      
    \item \textbf{$p$-adic Conformal Field Theories}: As a non-archimedean generalization of standard CFTs, $p$-adic frameworks directly relate deformation parameters to bulk geometric modifications~\cite{Gubser:2016guj}. This connection facilitates exploration of non-perturbative effects and insights into strongly coupled regimes~\cite{Chen:2021ipv}.  
\end{itemize}  
These discrete frameworks not only enable controlled studies of $T\overline{T}$-like deformations but also reveal novel deformation classes. For instance, holographic tensor networks~\cite{Chen:2022wvy,Cheng:2023kxh,Hung:2024gma,Bao:2024ixc} reinterpret bulk geometry as entanglement structures, while $p$-adic CFTs establish geometric dualities inaccessible to conventional continuum methods. The interplay between discrete bulk structures and boundary theories opens new avenues for understanding quantum field theories and their holographic duals, particularly in regimes where traditional continuum approaches face computational or conceptual challenges.

{\bf UV Completion and Nonlocality}

Using bootstrap constraints to explore the possibility of UV completion or the nature of nonlocal interactions induced by the deformation~\cite{Taylor:2018xcy, Giveon:2017nie}. At the classical level, the deformed theory can be regarded as seed theory coupled with flat JT-like gravity. One can expect to check whether the equivalent relation can be held at the quantum level. The one direct check is to obtain finite entanglement entropy in the deformed theory due to UV completion and nonlocality. Whether the entanglement entropy can be regarded as the quantum extremal surface in the UV region, the Hilbert space of the deformed theory should satisfy the type II von Neumann algebra~\cite{Witten:2018zxz}.  However, the entanglement entropy in the deformed theory is divergent regarding the perturbation theory approach. From the IR to UV, how to understand the $T\bar{T}$ deformation that induces the dramatic changes in Hilbert space is an interesting open question.

    
{\bf Dualities and Symmetries}

By leveraging deformation methods tied to the energy-momentum tensor, we can identify deformation theories that satisfy both duality constraints and integrability requirements. This approach enables the systematic construction of theories that preserve these dual principles. One can refer to some interesting examples ~\cite{Babaei-Aghbolagh:2024uqp, Babaei-Aghbolagh:2025lko}. The investigation focuses on uncovering potential dualities or hidden symmetries in deformed theories, which may be constrained or revealed through bootstrap methods~\cite{Chakraborty:2020swe} or classical duality equations, such as those in~\cite{Bialynicki-Birula:1981, Bialynicki-Birula:1992rcm, Gaillard:1981rj, Gaillard:1997rt, Gibbons:1995ap}. For instance, three principal frameworks exist for constructing self-dual theories of Maxwell fields with $SO(2)$ duality symmetry: Gaillard-Zumino approach~\cite{Gaillard:1981rj, Gaillard:1997rt}, refined by Gibbons and Rasheed~\cite{Gibbons:1995ap}; non-covariant Hamiltonian formalism introduced by Henneaux and Teitelboim~\cite{HT}, extended by Deser et al.~\cite{Deser:1997mz, Deser:1997se}; PST method by Pasti, Sorokin, and Tonin~\cite{Pasti:1995tn, Pasti:1995us, Pasti:1996vs, Pasti:1997gx}.  
These frameworks provide distinct pathways to incorporate interactions while retaining duality symmetry. Recent work in~\cite{Bielli:2024khq} further demonstrates that T-duality transformations commute with energy-momentum tensor deformations via an auxiliary field formulation~\cite{Ferko:2024ali, Bielli:2024ach}, offering a unified perspective on duality-preserving deformations.

{\bf Bootstrap $T\bar{T}$}

The $T\overline{T}$ deformation leads to intriguing modifications of the spectrum and dynamics, often resulting in nonlocal theories with characteristics such as finite energy spectra and Hagedorn-like behavior~\cite{Cavaglia:2016oda}. Despite its nontrivial nature, the solvability of $T\overline{T}$-deformed theories provides a valuable framework for analytical studies, offering fertile ground to explore nonperturbative phenomena in quantum field theory. 

A powerful tool in studying such theories has been the bootstrap approach, particularly the conformal bootstrap, which leverages symmetries and consistency conditions like crossing symmetry and unitarity to solve for operator dimensions and correlation functions in conformal field theories. However, extending these techniques to $T\overline{T}$-deformed theories presents several challenges. First, the original conformal symmetry is broken, and we are left with a $T\bar{T}$-deformed conformal symmetry consisting of field-dependent transformations. It's an open question of utilizing this complicated deformed symmetry to find constraints imposed on the theory. In particular, we don't know what operators can make the symmetry manifest. Second, the deformation introduces nonlocal interactions, complicating the definition and calculation of correlation functions. Traditional bootstrap methods rely on local operator product expansions (OPEs), but these may not hold in the context of nonlocal theories~\cite{Datta:2018thy}. 
Finally, as an irrelevant deformation, the $T\overline{T}$ operator introduces ultraviolet divergences, making the theory sensitive to high-energy details. This sensitivity poses a significant challenge to the bootstrap approach, which often depends on infrared (IR) data and universal behavior~\cite{Taylor:2018xcy}. Thus, while the $T\overline{T}$ deformation opens up intriguing avenues for nonperturbative study, its nonlocality, loss of conformal symmetry, and UV sensitivity introduce substantial challenges to applying traditional bootstrap techniques.

{\bf Acknowledgements}

We would like to thank H. Babaei-Aghbolagh, Ronggen Cai, Bin Chen, Jingcheng Chang, Yi-Hong Gao, Miao He, Jue Hou, Ling-Yan Hung, Yunfeng Jiang, Yun-Ze Li, Yu-Xiao Liu, Zhangcheng Liu,  Xin-Cheng Mao, Pujian Mao, Tommaso Morone, Hongfei Shu, Jia Tian, Roberto Tateo, Huajia Wang, Jun-Bao Wu, Wei Song, Hao-Yu Sun, Jia-Rui Sun, Stefan Theisen, Zhuo-Yu Xian, Yunfei Xie, Jiashi Yin, Yu-Xuan Zhang, Long Zhao and Zi-Xuan Zhao for valuable discussions related to this work at various stages. SH would like to appreciate the financial support from Ningbo University, the Max Planck Partner group, and the Natural Science Foundation of China Grants (No. 12475053, No. 12235016).
HO is supported by the National Natural Science Foundation of China, Grant No. 12205115, and by the Science and Technology Development Plan Project of Jilin Province of China, Grant No. 20240101326JC. YS is supported by the National Natural Science Foundation of China, Grant No. 12105113.

\appendix
\section{Appendix: Useful integrals}
\subsection{Dimensional regularization }
In this appendix, we present the detailed computations of integrals appearing in the conformal perturbation theory in section~\ref{sec:corr}, where the deformed correlation is investigated. 

We will consider the plane case, namely, the deformed theory is defined on a 2D plane, while the torus case can be found in \cite{He:2020udl, He:2020cxp}. 
The general 
integral on a plane can be expressed as \cite{He:2019vzf}
\be
\label{eq:integrals}
{\cal I}_{a_1,\cdots, a_m, b_1,\cdots, b_n}(z_{i_1},\cdots,z_{i_m}, \bar{z}_{j_1},\cdots, \bar{z}_{j_n})\equiv\int \frac{ d^2z}{(z-z_{i_1})^{a_1}\cdots (z-z_{i_m})^{a_m} (\bar{z}-\bar{z}_{j_1})^{b_1}\cdots (\bar{z}-\bar{z}_{j_n})^{b_n}}.
\ee
Using Feynman parametrization and dimensional regularization, one can obtain the following basic integral \cite{Guica:2019vnb}
\be \label{CI11}
 \CI_{11}(z_1,\bar{z}_2) =\int d^2z\frac{1}{z_{01}\bar{z}_{02}}=-\pi\Big(-\frac{2}{\epsilon}+\ln|z_{12}|^2+\gamma+\ln\pi\Big)+O(\epsilon),
\ee
with $\epsilon$ being an infinitesimal constant.
Next example is  $\CI_{12}(z_1,\bar{z}_2)$ 
\be\ba\label{CI12}
\int d^2z\frac{1}{z_{01}\bar{z}_{02}^2}&=\int d^2z \frac{\bar{z}_{01}z_{02}^2}{|z_{01}|^2|z_{02}|^4}\\
&=2\int_0^1 du (1-u) \int d^2z\frac{\bar{z}_{01}z_{02}^2 }{(u|z_{01}|^2+(1-u)|z_{02}|^2)^3}\\
&=2\int_0^1 du (1-u) \int d^2y\frac{2uz_{12}|y|^2-(1-u)u^2z^2_{12}\bar{z}_{12} }{(|y|^2+(1-u)u|z_{12}|^2)^3}\\
&=2z_{12}\int_0^1 du u(1-u) \int d^2y\frac{2|y|^2-A^2 }{(|y|^2+A^2)^3}\\
&=2z_{12}\int_0^1 du u(1-u) V_d\int d\rho \rho^{d-1}\frac{2\rho^2-A^2 }{(\rho^2+A^2)^3}\\
&=2z_{12}\int_0^1 du u(1-u) V_dA^{-2}\frac{1}{4}=\frac{\pi}{\bar{z}_{12}},
\ea\ee
with $d=2$ and $
V_d=2\pi^{d/2}/\Gamma(d/2)$ is the area of $(d-1)$-sphere with unit radius, also
we denote $A^2=(1-u)u|z_{12}|^2$ and use the coordinates transformation
\be
z=y+uz_1+(1-u)z_2,~~z_{01}=y-(1-u)z_{12},~~z_{02}=y+uz_{12}.
\ee
Let us mention that the result in eq.(\ref{CI12}) is consistent with eq.(\ref{CI11}), i.e. they satisfy $\p_{\bar{z}_2} \CI_{11}(z_i,\bar{z}_j)=\CI_{12}(z_i,\bar{z}_j)$.

For $\CI_{22}(z_1,\bar{z}_2)$ with $z_1\neq z_2$, similarly we can obtain
\be\ba
\int d^2z\frac{1}{z_{01}^2\bar{z}_{02}^2}&=\int d^2z \frac{\bar{z}_{01}^2z_{02}^2}{|z_{01}|^4|z_{02}|^4}\\
&=6\int_0^1 du u(1-u) \int d^2y\frac{(\bar{y}-(1-u)\bar{z}_{12})^2(y+uz_{12})^2 }{(|y|^2+(1-u)u|z_{12}|^2)^4}\\
&=6\int_0^1 duu (1-u) \int d^2y\frac{ |y|^4-4|y|^2u(1-u)|z_{12}|^2+(1-u)^2u^2|z_{12}|^4 }{(|y|^2+(1-u)u|z_{12}|^2)^4}\\
&=6\int_0^1 duu (1-u) \int d^2y\frac{ |y|^4-4|y|^2A^2+A^4 }{(|y|^2+A^2)^4}=0.
\ea\ee
In summary, by using dimensional regularization, we can obtain all the integrals needed, which are \cite{He:2019ahx}
\be \ba\label{DR1}
&\CI_{11}(z_i,\bar{z}_j)=-\pi(-\frac{2}{\epsilon}+\ln|z_{ij}|^2+\gamma+\ln\pi+\mathcal{O}(\epsilon)),\\
&
\CI_{12}(z_i,\bar{z}_j)=\frac{ \pi}{\bar{z}_{ij}} ,~~\CI_{21}(z_i,\bar{z}_j)=-\frac{ \pi}{z_{ij}} ,~~\CI_{22}(z_i,\bar{z}_j)=0,\\
&\CI_{13}(z_i,\bar{z}_j)=\frac{\pi}{(\bar{z}_{ij})^2},~~
\CI_{31}(z_i,\bar{z}_j)=\frac{\pi}{(z_{ij})^2},\\
&  \CI_{23}(z_i,\bar{z}_j)=\CI_{32}(z_i,\bar{z}_j)=\CI_{33}(z_i,\bar{z}_j)=0.
\ea\ee

\providecommand{\href}[2]{#2}\begingroup\raggedright\endgroup


\begin{thebibliography}{100}

\bibitem{Smirnov:2016lqw}
F.~A. Smirnov and A.~B. Zamolodchikov, {On space of integrable quantum field theories}, \href{http://dx.doi.org/10.1016/j.nuclphysb.2016.12.014}{{ Nucl. Phys. B} {915} (2017) 363--383}, \href{http://arxiv.org/abs/1608.05499}{{\ttfamily arXiv:1608.05499}}.

\bibitem{Cavaglia:2016oda}
A.~Cavagli\`a, S.~Negro, I.~M. Sz\'ecs\'enyi, and R.~Tateo, {$T \bar{T}$-deformed 2D Quantum Field Theories}, \href{http://dx.doi.org/10.1007/JHEP10(2016)112}{{ JHEP} {10} (2016) 112}, \href{http://arxiv.org/abs/1608.05534}{{\ttfamily arXiv:1608.05534}}.

\bibitem{Zamolodchikov:2004ce}
A.~B. Zamolodchikov, {Expectation value of composite field T anti-T in two-dimensional quantum field theory}, \href{http://arxiv.org/abs/hep-th/0401146}{{\ttfamily arXiv:hep-th/0401146}}.

\bibitem{Dubovsky:2017cnj}
S.~Dubovsky, V.~Gorbenko, and M.~Mirbabayi, {Asymptotic fragility, near AdS$_{2}$ holography and $ T\overline{T} $}, \href{http://dx.doi.org/10.1007/JHEP09(2017)136}{{ JHEP} {09} (2017) 136}, \href{http://arxiv.org/abs/1706.06604}{{\ttfamily arXiv:1706.06604}}.

\bibitem{Dubovsky:2018bmo}
S.~Dubovsky, V.~Gorbenko, and G.~Hern\'andez-Chifflet, {$ T\overline{T} $ partition function from topological gravity}, \href{http://dx.doi.org/10.1007/JHEP09(2018)158}{{ JHEP} {09} (2018) 158}, \href{http://arxiv.org/abs/1805.07386}{{\ttfamily arXiv:1805.07386}}.

\bibitem{Tolley:2019nmm}
A.~J. Tolley, {$ T\overline{T} $ deformations, massive gravity and non-critical strings}, \href{http://dx.doi.org/10.1007/JHEP06(2020)050}{{ JHEP} {06} (2020) 050}, \href{http://arxiv.org/abs/1911.06142}{{\ttfamily arXiv:1911.06142}}.

\bibitem{Iliesiu:2020zld}
L.~V. Iliesiu, J.~Kruthoff, G.~J. Turiaci, and H.~Verlinde, {JT gravity at finite cutoff}, \href{http://dx.doi.org/10.21468/SciPostPhys.9.2.023}{{ SciPost Phys.} {9} (2020) 023}, \href{http://arxiv.org/abs/2004.07242}{{\ttfamily arXiv:2004.07242}}.

\bibitem{Okumura:2020dzb}
S.~Okumura and K.~Yoshida, {$T\bar{T}$-deformation and Liouville gravity}, \href{http://dx.doi.org/10.1016/j.nuclphysb.2020.115083}{{ Nucl. Phys. B} {957} (2020) 115083}, \href{http://arxiv.org/abs/2003.14148}{{\ttfamily arXiv:2003.14148}}.

\bibitem{Ebert:2022ehb}
S.~Ebert, C.~Ferko, H.-Y. Sun, and Z.~Sun, {$T\bar{T}$ in JT Gravity and BF Gauge Theory}, \href{http://dx.doi.org/10.21468/SciPostPhys.13.4.096}{{ SciPost Phys.} {13} no.~4, (2022) 096}, \href{http://arxiv.org/abs/2205.07817}{{\ttfamily arXiv:2205.07817}}.

\bibitem{Bhattacharyya:2023gvg}
A.~Bhattacharyya, S.~Ghosh, and S.~Pal, {Aspects of $T\bar{T}+J\bar{T }$ deformed 2D topological gravity : from partition function to late-time SFF}, \href{http://arxiv.org/abs/2309.16658}{{\ttfamily arXiv:2309.16658}}.

\bibitem{Bao:2024ixc}
N.~Bao, L.-Y. Hung, Y.~Jiang, and Z.~Liu, {QG from SymQRG: AdS$_3$/CFT$_2$ Correspondence as Topological Symmetry-Preserving Quantum RG Flow}, \href{http://arxiv.org/abs/2412.12045}{{\ttfamily arXiv:2412.12045}}.

\bibitem{McGough:2016lol}
L.~McGough, M.~Mezei, and H.~Verlinde, {Moving the CFT into the bulk with $ T\overline{T} $}, \href{http://dx.doi.org/10.1007/JHEP04(2018)010}{{ JHEP} {04} (2018) 010}, \href{http://arxiv.org/abs/1611.03470}{{\ttfamily arXiv:1611.03470}}.

\bibitem{Giveon:2017nie}
A.~Giveon, N.~Itzhaki, and D.~Kutasov, {$ \mathrm{T}\overline{\mathrm{T}} $ and LST}, \href{http://dx.doi.org/10.1007/JHEP07(2017)122}{{ JHEP} {07} (2017) 122}, \href{http://arxiv.org/abs/1701.05576}{{\ttfamily arXiv:1701.05576}}.

\bibitem{Kraus:2018xrn}
P.~Kraus, J.~Liu, and D.~Marolf, {Cutoff AdS$_{3}$ versus the $ T\overline{T} $ deformation}, \href{http://dx.doi.org/10.1007/JHEP07(2018)027}{{ JHEP} {07} (2018) 027}, \href{http://arxiv.org/abs/1801.02714}{{\ttfamily arXiv:1801.02714}}.

\bibitem{Gorbenko:2018oov}
V.~Gorbenko, E.~Silverstein, and G.~Torroba, {dS/dS and $ T\overline{T} $}, \href{http://dx.doi.org/10.1007/JHEP03(2019)085}{{ JHEP} {03} (2019) 085}, \href{http://arxiv.org/abs/1811.07965}{{\ttfamily arXiv:1811.07965}}.

\bibitem{Cottrell:2018skz}
W.~Cottrell and A.~Hashimoto, {Comments on $T \bar T$ double trace deformations and boundary conditions}, \href{http://dx.doi.org/10.1016/j.physletb.2018.09.068}{{ Phys. Lett. B} {789} (2019) 251--255}, \href{http://arxiv.org/abs/1801.09708}{{\ttfamily arXiv:1801.09708}}.

\bibitem{Taylor:2018xcy}
M.~Taylor, {$T \bar{T}$ deformations in general dimensions}, \href{http://dx.doi.org/10.4310/ATMP.2023.v27.n1.a2}{{ Adv. Theor. Math. Phys.} {27} no.~1, (2023) 37--63}, \href{http://arxiv.org/abs/1805.10287}{{\ttfamily arXiv:1805.10287}}.

\bibitem{Hartman:2018tkw}
T.~Hartman, J.~Kruthoff, E.~Shaghoulian, and A.~Tajdini, {Holography at finite cutoff with a $T^2$ deformation}, \href{http://dx.doi.org/10.1007/JHEP03(2019)004}{{ JHEP} {03} (2019) 004}, \href{http://arxiv.org/abs/1807.11401}{{\ttfamily arXiv:1807.11401}}.

\bibitem{Jiang:2019tcq}
Y.~Jiang, {Expectation value of $\mathrm{T}\overline{\mathrm{T}}$ operator in curved spacetimes}, \href{http://dx.doi.org/10.1007/JHEP02(2020)094}{{ JHEP} {02} (2020) 094}, \href{http://arxiv.org/abs/1903.07561}{{\ttfamily arXiv:1903.07561}}.

\bibitem{Caputa:2019pam}
P.~Caputa, S.~Datta, and V.~Shyam, {Sphere partition functions \textbackslash{}\& cut-off AdS}, \href{http://dx.doi.org/10.1007/JHEP05(2019)112}{{ JHEP} {05} (2019) 112}, \href{http://arxiv.org/abs/1902.10893}{{\ttfamily arXiv:1902.10893}}.

\bibitem{Guica:2019nzm}
M.~Guica and R.~Monten, {$T\bar T$ and the mirage of a bulk cutoff}, \href{http://dx.doi.org/10.21468/SciPostPhys.10.2.024}{{ SciPost Phys.} {10} no.~2, (2021) 024}, \href{http://arxiv.org/abs/1906.11251}{{\ttfamily arXiv:1906.11251}}.

\bibitem{Gross:2019ach}
D.~J. Gross, J.~Kruthoff, A.~Rolph, and E.~Shaghoulian, {$T\overline{T}$ in AdS$_2$ and Quantum Mechanics}, \href{http://dx.doi.org/10.1103/PhysRevD.101.026011}{{ Phys. Rev. D} {101} no.~2, (2020) 026011}, \href{http://arxiv.org/abs/1907.04873}{{\ttfamily arXiv:1907.04873}}.

\bibitem{Jafari:2019qns}
G.~Jafari, A.~Naseh, and H.~Zolfi, {Path Integral Optimization for $T\bar{T}$ Deformation}, \href{http://dx.doi.org/10.1103/PhysRevD.101.026007}{{ Phys. Rev. D} {101} no.~2, (2020) 026007}, \href{http://arxiv.org/abs/1909.02357}{{\ttfamily arXiv:1909.02357}}.

\bibitem{Li:2020pwa}
Y.~Li and Y.~Zhou, {Cutoff AdS$_{3}$ versus $ T\overline{T} $ CFT$_{2}$ in the large central charge sector: correlators of energy-momentum tensor}, \href{http://dx.doi.org/10.1007/JHEP12(2020)168}{{ JHEP} {12} (2020) 168}, \href{http://arxiv.org/abs/2005.01693}{{\ttfamily arXiv:2005.01693}}.

\bibitem{Griguolo:2021wgy}
L.~Griguolo, R.~Panerai, J.~Papalini, and D.~Seminara, {Nonperturbative effects and resurgence in Jackiw-Teitelboim gravity at finite cutoff}, \href{http://dx.doi.org/10.1103/PhysRevD.105.046015}{{ Phys. Rev. D} {105} no.~4, (2022) 046015}, \href{http://arxiv.org/abs/2106.01375}{{\ttfamily arXiv:2106.01375}}.

\bibitem{He:2022bbb}
S.~He, H.~Ouyang, and Y.~Sun, {Note on $T{\bar{T}}$ deformed matrix models and JT supergravity duals}, \href{http://dx.doi.org/10.1140/epjc/s10052-023-12019-3}{{ Eur. Phys. J. C} {83} no.~10, (2023) 885}, \href{http://arxiv.org/abs/2204.13636}{{\ttfamily arXiv:2204.13636}}.

\bibitem{He:2023hoj}
S.~He, Y.~Li, Y.-Z. Li, and Y.~Zhang, {Holographic torus correlators of stress tensor in AdS$_{3}$/CFT$_{2}$}, \href{http://dx.doi.org/10.1007/JHEP06(2023)116}{{ JHEP} {06} (2023) 116}, \href{http://arxiv.org/abs/2303.13280}{{\ttfamily arXiv:2303.13280}}.

\bibitem{He:2023knl}
S.~He, Y.-Z. Li, and Y.~Zhang, {Holographic torus correlators in AdS$_{3}$ gravity coupled to scalar field}, \href{http://dx.doi.org/10.1007/JHEP05(2024)254}{{ JHEP} {05} (2024) 254}, \href{http://arxiv.org/abs/2311.09636}{{\ttfamily arXiv:2311.09636}}.

\bibitem{Fichet:2023xbu}
S.~Fichet, E.~Megias, and M.~Quiros, {Holography of linear dilaton spacetimes from the bottom up}, \href{http://dx.doi.org/10.1103/PhysRevD.109.106011}{{ Phys. Rev. D} {109} no.~10, (2024) 106011}, \href{http://arxiv.org/abs/2309.02489}{{\ttfamily arXiv:2309.02489}}.

\bibitem{Ebert:2024fpc}
S.~Ebert, { {Holographic Renormalization Group and Stress Tensor Operators}}.
\newblock PhD thesis, UCLA, Los Angeles (main), UCLA, 2024.
\newblock \href{http://arxiv.org/abs/2404.10190}{{\ttfamily arXiv:2404.10190}}.

\bibitem{He:2024xbi}
S.~He, Y.-Z. Li, and Y.~Xie, {Holographic stress tensor correlators on higher genus Riemann surfaces}, \href{http://arxiv.org/abs/2406.04042}{{\ttfamily arXiv:2406.04042}}.

\bibitem{Baggio:2018gct}
M.~Baggio and A.~Sfondrini, {Strings on NS-NS Backgrounds as Integrable Deformations}, \href{http://dx.doi.org/10.1103/PhysRevD.98.021902}{{ Phys. Rev. D} {98} no.~2, (2018) 021902}, \href{http://arxiv.org/abs/1804.01998}{{\ttfamily arXiv:1804.01998}}.

\bibitem{Dei:2018jyj}
A.~Dei and A.~Sfondrini, {Integrable S matrix, mirror TBA and spectrum for the stringy AdS$_{3}$ \texttimes{} S$^{3}$ \texttimes{} S$^{3}$ \texttimes{} S$^{1}$ WZW model}, \href{http://dx.doi.org/10.1007/JHEP02(2019)072}{{ JHEP} {02} (2019) 072}, \href{http://arxiv.org/abs/1812.08195}{{\ttfamily arXiv:1812.08195}}.

\bibitem{Chakraborty:2019mdf}
S.~Chakraborty, A.~Giveon, and D.~Kutasov, {$T\bar{T}$, $J\bar{T}$, $T\bar{J}$ and String Theory}, \href{http://dx.doi.org/10.1088/1751-8121/ab3710}{{ J. Phys. A} {52} no.~38, (2019) 384003}, \href{http://arxiv.org/abs/1905.00051}{{\ttfamily arXiv:1905.00051}}.

\bibitem{Callebaut:2019omt}
N.~Callebaut, J.~Kruthoff, and H.~Verlinde, {$ T\overline{T} $ deformed CFT as a non-critical string}, \href{http://dx.doi.org/10.1007/JHEP04(2020)084}{{ JHEP} {04} (2020) 084}, \href{http://arxiv.org/abs/1910.13578}{{\ttfamily arXiv:1910.13578}}.

\bibitem{Conti:2022egv}
R.~Conti, J.~Romano, and R.~Tateo, {Metric approach to a $ \mathrm{T}\overline{\mathrm{T}} $-like deformation in arbitrary dimensions}, \href{http://dx.doi.org/10.1007/JHEP09(2022)085}{{ JHEP} {09} (2022) 085}, \href{http://arxiv.org/abs/2206.03415}{{\ttfamily arXiv:2206.03415}}.

\bibitem{Ferko:2022cix}
C.~Ferko, A.~Sfondrini, L.~Smith, and G.~Tartaglino-Mazzucchelli, {Root-$T \bar T$ Deformations in Two-Dimensional Quantum Field Theories}, \href{http://dx.doi.org/10.1103/PhysRevLett.129.201604}{{ Phys. Rev. Lett.} {129} no.~20, (2022) 201604}, \href{http://arxiv.org/abs/2206.10515}{{\ttfamily arXiv:2206.10515}}.

\bibitem{Ferko:2023ruw}
C.~Ferko, L.~Smith, and G.~Tartaglino-Mazzucchelli, {Stress Tensor flows, birefringence in non-linear electrodynamics and supersymmetry}, \href{http://dx.doi.org/10.21468/SciPostPhys.15.5.198}{{ SciPost Phys.} {15} no.~5, (2023) 198}, \href{http://arxiv.org/abs/2301.10411}{{\ttfamily arXiv:2301.10411}}.

\bibitem{Babaei-Aghbolagh:2022uij}
H.~Babaei-Aghbolagh, K.~B. Velni, D.~M. Yekta, and H.~Mohammadzadeh, {Emergence of non-linear electrodynamic theories from TT\textasciimacron{}-like deformations}, \href{http://dx.doi.org/10.1016/j.physletb.2022.137079}{{ Phys. Lett. B} {829} (2022) 137079}, \href{http://arxiv.org/abs/2202.11156}{{\ttfamily arXiv:2202.11156}}.

\bibitem{Babaei-Aghbolagh:2022leo}
H.~Babaei-Aghbolagh, K.~Babaei~Velni, D.~Mahdavian~Yekta, and H.~Mohammadzadeh, {Marginal TT\textasciimacron{}-like deformation and modified Maxwell theories in two dimensions}, \href{http://dx.doi.org/10.1103/PhysRevD.106.086022}{{ Phys. Rev. D} {106} no.~8, (2022) 086022}, \href{http://arxiv.org/abs/2206.12677}{{\ttfamily arXiv:2206.12677}}.

\bibitem{Hadasz:2024pew}
L.~Hadasz and R.~von Unge, {Defining the root-TT\textasciimacron{} operator}, \href{http://dx.doi.org/10.1103/PhysRevD.111.045024}{{ Phys. Rev. D} {111} no.~4, (2025) 045024}, \href{http://arxiv.org/abs/2405.17945}{{\ttfamily arXiv:2405.17945}}.

\bibitem{Borsato:2022tmu}
R.~Borsato, C.~Ferko, and A.~Sfondrini, {Classical integrability of root-TT\textasciimacron{} flows}, \href{http://dx.doi.org/10.1103/PhysRevD.107.086011}{{ Phys. Rev. D} {107} no.~8, (2023) 086011}, \href{http://arxiv.org/abs/2209.14274}{{\ttfamily arXiv:2209.14274}}.

\bibitem{Brizio:2024arr}
N.~Brizio, T.~Morone, and R.~Tateo, {Stress-energy tensor deformations, Ricci flows and black holes}, \href{http://arxiv.org/abs/2408.06031}{{\ttfamily arXiv:2408.06031}}.

\bibitem{Giveon:2019fgr}
A.~Giveon, {Comments on $T\bar T$, $J\bar{T}$ and String Theory}, \href{http://arxiv.org/abs/1903.06883}{{\ttfamily arXiv:1903.06883}}.

\bibitem{Jiang:2019epa}
Y.~Jiang, {A pedagogical review on solvable irrelevant deformations of 2D quantum field theory}, \href{http://dx.doi.org/10.1088/1572-9494/abe4c9}{{ Commun. Theor. Phys.} {73} no.~5, (2021) 057201}, \href{http://arxiv.org/abs/1904.13376}{{\ttfamily arXiv:1904.13376}}.

\bibitem{Babaei-Aghbolagh:2024hti}
H.~Babaei-Aghbolagh, S.~He, T.~Morone, H.~Ouyang, and R.~Tateo, {Geometric formulation of generalized root-$T\bar{T}$ deformations}, \href{http://arxiv.org/abs/2405.03465}{{\ttfamily arXiv:2405.03465}}.

\bibitem{Bonelli:2018kik}
G.~Bonelli, N.~Doroud, and M.~Zhu, {$T \bar{T}$-deformations in closed form}, \href{http://dx.doi.org/10.1007/JHEP06(2018)149}{{ JHEP} {06} (2018) 149}, \href{http://arxiv.org/abs/1804.10967}{{\ttfamily arXiv:1804.10967}}.

\bibitem{Cardy:2019qao}
J.~Cardy, {$T\bar T$ deformation of correlation functions}, \href{http://dx.doi.org/10.1007/JHEP12(2019)160}{{ JHEP} {12} (2019) 160}, \href{http://arxiv.org/abs/1907.03394}{{\ttfamily arXiv:1907.03394}}.

\bibitem{Aharony:2023dod}
O.~Aharony and N.~Barel, {Correlation functions in $ \textrm{T}\overline{\textrm{T}} $-deformed Conformal Field Theories}, \href{http://dx.doi.org/10.1007/JHEP08(2023)035}{{ JHEP} {08} (2023) 035}, \href{http://arxiv.org/abs/2304.14091}{{\ttfamily arXiv:2304.14091}}.

\bibitem{Chakraborty:2023wel}
S.~B. Chakraborty, S.~Georgescu, and M.~Guica, {States, symmetries and correlators of $T\bar{T}$ and $ J\bar{T} $ symmetric orbifolds}, \href{http://dx.doi.org/10.21468/SciPostPhys.16.1.011}{{ SciPost Phys.} {16} no.~1, (2024) 011}, \href{http://arxiv.org/abs/2306.16454}{{\ttfamily arXiv:2306.16454}}.

\bibitem{Dubovsky:2012wk}
S.~Dubovsky, R.~Flauger, and V.~Gorbenko, {Solving the Simplest Theory of Quantum Gravity}, \href{http://dx.doi.org/10.1007/JHEP09(2012)133}{{ JHEP} {09} (2012) 133}, \href{http://arxiv.org/abs/1205.6805}{{\ttfamily arXiv:1205.6805}}.

\bibitem{Dubovsky:2013ira}
S.~Dubovsky, V.~Gorbenko, and M.~Mirbabayi, {Natural Tuning: Towards A Proof of Concept}, \href{http://dx.doi.org/10.1007/JHEP09(2013)045}{{ JHEP} {09} (2013) 045}, \href{http://arxiv.org/abs/1305.6939}{{\ttfamily arXiv:1305.6939}}.

\bibitem{Frolov:2019nrr}
S.~Frolov, {$T\overline{T}$ Deformation and the Light-Cone Gauge}, \href{http://dx.doi.org/10.1134/S0081543820030098}{{ Proc. Steklov Inst. Math.} {309} (2020) 107--126}, \href{http://arxiv.org/abs/1905.07946}{{\ttfamily arXiv:1905.07946}}.

\bibitem{Frolov:2019xzi}
S.~Frolov, {$T{\overline T}$, $\widetilde JJ$, $JT$ and $\widetilde JT$ deformations}, \href{http://dx.doi.org/10.1088/1751-8121/ab581b}{{ J. Phys. A} {53} no.~2, 025401}, \href{http://arxiv.org/abs/1907.12117}{{\ttfamily arXiv:1907.12117}}.

\bibitem{Sfondrini:2019smd}
A.~Sfondrini and S.~J. van Tongeren, {$T\bar{T}$ deformations as $TsT$ transformations}, \href{http://dx.doi.org/10.1103/PhysRevD.101.066022}{{ Phys. Rev. D} {101} no.~6, (2020) 066022}, \href{http://arxiv.org/abs/1908.09299}{{\ttfamily arXiv:1908.09299}}.

\bibitem{Apolo:2019zai}
L.~Apolo, S.~Detournay, and W.~Song, {TsT, $T\bar{T}$ and black strings}, \href{http://dx.doi.org/10.1007/JHEP06(2020)109}{{ JHEP} {06} (2020) 109}, \href{http://arxiv.org/abs/1911.12359}{{\ttfamily arXiv:1911.12359}}.

\bibitem{Esper:2021hfq}
C.~Esper and S.~Frolov, {$ T\overline{T} $ deformations of non-relativistic models}, \href{http://dx.doi.org/10.1007/JHEP06(2021)101}{{ JHEP} {06} (2021) 101}, \href{http://arxiv.org/abs/2102.12435}{{\ttfamily arXiv:2102.12435}}.

\bibitem{Pasterski:2016qvg}
S.~Pasterski, S.-H. Shao, and A.~Strominger, {Flat Space Amplitudes and Conformal Symmetry of the Celestial Sphere}, \href{http://dx.doi.org/10.1103/PhysRevD.96.065026}{{ Phys. Rev. D} {96} no.~6, (2017) 065026}, \href{http://arxiv.org/abs/1701.00049}{{\ttfamily arXiv:1701.00049}}.

\bibitem{Pasterski:2017kqt}
S.~Pasterski and S.-H. Shao, {Conformal basis for flat space amplitudes}, \href{http://dx.doi.org/10.1103/PhysRevD.96.065022}{{ Phys. Rev. D} {96} no.~6, (2017) 065022}, \href{http://arxiv.org/abs/1705.01027}{{\ttfamily arXiv:1705.01027}}.

\bibitem{Pasterski:2017ylz}
S.~Pasterski, S.-H. Shao, and A.~Strominger, {Gluon Amplitudes as 2d Conformal Correlators}, \href{http://dx.doi.org/10.1103/PhysRevD.96.085006}{{ Phys. Rev. D} {96} no.~8, (2017) 085006}, \href{http://arxiv.org/abs/1706.03917}{{\ttfamily arXiv:1706.03917}}.

\bibitem{He:2022zcf}
S.~He, P.~Mao, and X.-C. Mao, {TT\textasciimacron{} deformed soft theorem}, \href{http://dx.doi.org/10.1103/PhysRevD.107.L101901}{{ Phys. Rev. D} {107} no.~10, (2023) L101901}, \href{http://arxiv.org/abs/2209.01953}{{\ttfamily arXiv:2209.01953}}.

\bibitem{Bandos:2020jsw}
I.~Bandos, K.~Lechner, D.~Sorokin, and P.~K. Townsend, {A non-linear duality-invariant conformal extension of Maxwell's equations}, \href{http://dx.doi.org/10.1103/PhysRevD.102.121703}{{ Phys. Rev. D} {102} (2020) 121703}, \href{http://arxiv.org/abs/2007.09092}{{\ttfamily arXiv:2007.09092}}.

\bibitem{Conti:2018jho}
R.~Conti, L.~Iannella, S.~Negro, and R.~Tateo, {Generalised Born-Infeld models, Lax operators and the $ \mathrm{T}\overline{\mathrm{T}} $ perturbation}, \href{http://dx.doi.org/10.1007/JHEP11(2018)007}{{ JHEP} {11} (2018) 007}, \href{http://arxiv.org/abs/1806.11515}{{\ttfamily arXiv:1806.11515}}.

\bibitem{Babaei-Aghbolagh:2020kjg}
H.~Babaei-Aghbolagh, K.~Babaei~Velni, D.~M. Yekta, and H.~Mohammadzadeh, {$ T\overline{T} $-like flows in non-linear electrodynamic theories and S-duality}, \href{http://dx.doi.org/10.1007/JHEP04(2021)187}{{ JHEP} {04} (2021) 187}, \href{http://arxiv.org/abs/2012.13636}{{\ttfamily arXiv:2012.13636}}.

\bibitem{Ferko:2022iru}
C.~Ferko, L.~Smith, and G.~Tartaglino-Mazzucchelli, {On Current-Squared Flows and ModMax Theories}, \href{http://dx.doi.org/10.21468/SciPostPhys.13.2.012}{{ SciPost Phys.} {13} no.~2, (2022) 012}, \href{http://arxiv.org/abs/2203.01085}{{\ttfamily arXiv:2203.01085}}.

\bibitem{Gaillard:1981rj}
M.~K. Gaillard and B.~Zumino, Duality rotations for interacting fields, \href{http://dx.doi.org/10.1016/0550-3213(81)90527-7}{{ Nucl. Phys. B} {193} (1981) 221--244}.

\bibitem{Gaillard:1997rt}
M.~K. Gaillard and B.~Zumino, Nonlinear electromagnetic selfduality and legendre transformations,. [arXiv:hep-th/9712103 [hep-th]].

\bibitem{HT}
M.~Henneaux and C.~Teitelboim, Dynamics of chiral (self-dual) p-forms, { Phys. Lett. B} {206} (1988) 650.

\bibitem{Pasti:1995tn}
P.~Pasti, D.~P. Sorokin, and M.~Tonin, Duality symmetric actions with manifest space-time symmetries, \href{http://dx.doi.org/10.1103/PhysRevD.52.R4277}{{ Phys. Rev. D} {52} (1995) R4277--R4281}. [arXiv:hep-th/9506109 [hep-th]].

\bibitem{Pasti:1995us}
P.~Pasti, D.~P. Sorokin, and M.~Tonin, Space-time symmetries in duality symmetric models,. [arXiv:hep-th/9509052 [hep-th]].

\bibitem{Pasti:1996vs}
P.~Pasti, D.~P. Sorokin, and M.~Tonin, On lorentz invariant actions for chiral p-forms, \href{http://dx.doi.org/10.1103/PhysRevD.55.6292}{{ Phys. Rev. D} {55} (1997) 6292--6298}. [arXiv:hep-th/9611100 [hep-th]].

\bibitem{Pasti:1997gx}
P.~Pasti, D.~P. Sorokin, and M.~Tonin, Covariant action for a d = 11 five-brane with the chiral field, \href{http://dx.doi.org/10.1016/S0370-2693(97)00188-3}{{ Phys. Lett. B} {398} (1997) 41--46}. [arXiv:hep-th/9701037 [hep-th]].

\bibitem{Babaei-Aghbolagh:2025lko}
H.~Babaei-Aghbolagh, S.~He, and H.~Ouyang, {Generalized $T\bar{T}$-like flows for scalar theories in two dimensions}, \href{http://arxiv.org/abs/2501.14583}{{\ttfamily arXiv:2501.14583}}.

\bibitem{Monica_Guica_TTbar}
M.~Guica, { $T\overline{T}$ deformations and holography}, { \url{https://indico.cern.ch/event/857396/contributions/3706292/attachments/2036750/3410352/ttbar_cern_v1s.pdf}} .

\bibitem{apolo:review}
L.~Apolo, {Single-trace $T\overline{T}$ deformations and string theory}, { \url{https://lui-apolo.github.io/}} .

\bibitem{Aharony:2018vux}
O.~Aharony and T.~Vaknin, {The TT* deformation at large central charge}, \href{http://dx.doi.org/10.1007/JHEP05(2018)166}{{ JHEP} {05} (2018) 166}, \href{http://arxiv.org/abs/1803.00100}{{\ttfamily arXiv:1803.00100}}.

\bibitem{Mazenc:2019cfg}
E.~A. Mazenc, V.~Shyam, and R.~M. Soni, {A $T \bar{T}$ Deformation for Curved Spacetimes from 3d Gravity}, \href{http://arxiv.org/abs/1912.09179}{{\ttfamily arXiv:1912.09179}}.

\bibitem{Blair:2020ops}
C.~D.~A. Blair, {Non-relativistic duality and $T \bar T$ deformations}, \href{http://dx.doi.org/10.1007/JHEP07(2020)069}{{ JHEP} {07} (2020) 069}, \href{http://arxiv.org/abs/2002.12413}{{\ttfamily arXiv:2002.12413}}.

\bibitem{Demulder:2023bux}
S.~Demulder, S.~Driezen, B.~Knighton, G.~Oling, A.~L. Retore, F.~K. Seibold, A.~Sfondrini, and Z.~Yan, {Exact approaches on the string worldsheet}, \href{http://arxiv.org/abs/2312.12930}{{\ttfamily arXiv:2312.12930}}.

\bibitem{Blair:2024aqz}
C.~D.~A. Blair, J.~Lahnsteiner, N.~A. Obers, and Z.~Yan, {Matrix theory reloaded: a BPS road to holography}, \href{http://dx.doi.org/10.1007/JHEP02(2025)024}{{ JHEP} {02} (2025) 024}, \href{http://arxiv.org/abs/2410.03591}{{\ttfamily arXiv:2410.03591}}.

\bibitem{Lee:2021iut}
K.-S. Lee, P.~Yi, and J.~Yoon, {$ T\overline{T} $-deformed fermionic theories revisited}, \href{http://dx.doi.org/10.1007/JHEP07(2021)217}{{ JHEP} {07} (2021) 217}, \href{http://arxiv.org/abs/2104.09529}{{\ttfamily arXiv:2104.09529}}.

\bibitem{Dey:2021jyl}
A.~Dey and A.~Fortinsky, {Perturbative renormalization of the $ \mathrm{T}\overline{\mathrm{T}} $-deformed free massive Dirac fermion}, \href{http://dx.doi.org/10.1007/JHEP12(2021)200}{{ JHEP} {12} (2021) 200}, \href{http://arxiv.org/abs/2109.10525}{{\ttfamily arXiv:2109.10525}}.

\bibitem{Baggio:2018rpv}
M.~Baggio, A.~Sfondrini, G.~Tartaglino-Mazzucchelli, and H.~Walsh, {On $ T\overline{T} $ deformations and supersymmetry}, \href{http://dx.doi.org/10.1007/JHEP06(2019)063}{{ JHEP} {06} (2019) 063}, \href{http://arxiv.org/abs/1811.00533}{{\ttfamily arXiv:1811.00533}}.

\bibitem{Chang:2018dge}
C.-K. Chang, C.~Ferko, and S.~Sethi, {Supersymmetry and $ T\overline{T} $ deformations}, \href{http://dx.doi.org/10.1007/JHEP04(2019)131}{{ JHEP} {04} (2019) 131}, \href{http://arxiv.org/abs/1811.01895}{{\ttfamily arXiv:1811.01895}}.

\bibitem{Jiang:2019hux}
H.~Jiang, A.~Sfondrini, and G.~Tartaglino-Mazzucchelli, {$T\bar{T}$ deformations with $\mathcal{N}=(0,2)$ supersymmetry}, \href{http://dx.doi.org/10.1103/PhysRevD.100.046017}{{ Phys. Rev. D} {100} no.~4, (2019) 046017}, \href{http://arxiv.org/abs/1904.04760}{{\ttfamily arXiv:1904.04760}}.

\bibitem{Chang:2019kiu}
C.-K. Chang, C.~Ferko, S.~Sethi, A.~Sfondrini, and G.~Tartaglino-Mazzucchelli, {$T\bar{T}$ flows and (2,2) supersymmetry}, \href{http://dx.doi.org/10.1103/PhysRevD.101.026008}{{ Phys. Rev. D} {101} no.~2, (2020) 026008}, \href{http://arxiv.org/abs/1906.00467}{{\ttfamily arXiv:1906.00467}}.

\bibitem{Coleman:2019dvf}
E.~A. Coleman, J.~Aguilera-Damia, D.~Z. Freedman, and R.~M. Soni, {$ T\overline{T} $ -deformed actions and (1,1) supersymmetry}, \href{http://dx.doi.org/10.1007/JHEP10(2019)080}{{ JHEP} {10} (2019) 080}, \href{http://arxiv.org/abs/1906.05439}{{\ttfamily arXiv:1906.05439}}.

\bibitem{Jiang:2019trm}
H.~Jiang and G.~Tartaglino-Mazzucchelli, {Supersymmetric J$ \overline{T} $ and T$ \overline{J} $ deformations}, \href{http://dx.doi.org/10.1007/JHEP05(2020)140}{{ JHEP} {05} (2020) 140}, \href{http://arxiv.org/abs/1911.05631}{{\ttfamily arXiv:1911.05631}}.

\bibitem{Ferko:2019oyv}
C.~Ferko, H.~Jiang, S.~Sethi, and G.~Tartaglino-Mazzucchelli, {Non-linear supersymmetry and $ T\overline{T} $-like flows}, \href{http://dx.doi.org/10.1007/JHEP02(2020)016}{{ JHEP} {02} (2020) 016}, \href{http://arxiv.org/abs/1910.01599}{{\ttfamily arXiv:1910.01599}}.

\bibitem{Ebert:2020tuy}
S.~Ebert, H.-Y. Sun, and Z.~Sun, {T$ \overline{T} $ deformation in SCFTs and integrable supersymmetric theories}, \href{http://dx.doi.org/10.1007/JHEP09(2021)082}{{ JHEP} {09} (2021) 082}, \href{http://arxiv.org/abs/2011.07618}{{\ttfamily arXiv:2011.07618}}.

\bibitem{Ferko:2021loo}
C.~Ferko, { {Supersymmetry and Irrelevant Deformations}}.
\newblock PhD thesis, Chicago U., Chicago U., 2021.
\newblock \href{http://arxiv.org/abs/2112.14647}{{\ttfamily arXiv:2112.14647}}.

\bibitem{Lee:2023uxj}
K.-S. Lee and J.~Yoon, {TT\textasciimacron{} deformation of N=(1,1) off-shell supersymmetry and partially broken supersymmetry}, \href{http://dx.doi.org/10.1103/PhysRevD.110.025001}{{ Phys. Rev. D} {110} no.~2, (2024) 025001}, \href{http://arxiv.org/abs/2306.08030}{{\ttfamily arXiv:2306.08030}}.

\bibitem{Hou:2022csf}
J.~Hou, {$ T\overline{T} $ flow as characteristic flows}, \href{http://dx.doi.org/10.1007/JHEP03(2023)243}{{ JHEP} {03} (2023) 243}, \href{http://arxiv.org/abs/2208.05391}{{\ttfamily arXiv:2208.05391}}.

\bibitem{Conti:2018tca}
R.~Conti, S.~Negro, and R.~Tateo, {The $ \mathrm{T}\overline{\mathrm{T}} $ perturbation and its geometric interpretation}, \href{http://dx.doi.org/10.1007/JHEP02(2019)085}{{ JHEP} {02} (2019) 085}, \href{http://arxiv.org/abs/1809.09593}{{\ttfamily arXiv:1809.09593}}.

\bibitem{Caputa:2020lpa}
P.~Caputa, S.~Datta, Y.~Jiang, and P.~Kraus, {Geometrizing $ T\overline{T} $}, \href{http://dx.doi.org/10.1007/JHEP03(2021)140}{{ JHEP} {03} (2021) 140}, \href{http://arxiv.org/abs/2011.04664}{{\ttfamily arXiv:2011.04664}}. [Erratum: JHEP 09, 110 (2022)].

\bibitem{Ceschin:2020jto}
P.~Ceschin, R.~Conti, and R.~Tateo, {$ \mathrm{T}\overline{\mathrm{T}} $-deformed nonlinear Schr\"odinger}, \href{http://dx.doi.org/10.1007/JHEP04(2021)121}{{ JHEP} {04} (2021) 121}, \href{http://arxiv.org/abs/2012.12760}{{\ttfamily arXiv:2012.12760}}.

\bibitem{Conti:2019dxg}
R.~Conti, S.~Negro, and R.~Tateo, {Conserved currents and $\text{T}\bar{\text{T}}_s$ irrelevant deformations of 2D integrable field theories}, \href{http://dx.doi.org/10.1007/JHEP11(2019)120}{{ JHEP} {11} (2019) 120}, \href{http://arxiv.org/abs/1904.09141}{{\ttfamily arXiv:1904.09141}}.

\bibitem{evans2010partial}
L.~C. Evans, { Partial Differential Equations}.
\newblock American Mathematical Society, 2010.

\bibitem{Cardy:2018sdv}
J.~Cardy, {The $ T\overline{T} $ deformation of quantum field theory as random geometry}, \href{http://dx.doi.org/10.1007/JHEP10(2018)186}{{ JHEP} {10} (2018) 186}, \href{http://arxiv.org/abs/1801.06895}{{\ttfamily arXiv:1801.06895}}.

\bibitem{Kim:2020eqn}
K.~K. Kim, J.-H. Baek, and Y.~Seo, {Phase transition in JT gravity and $ T\overline{T} $ deformation}, \href{http://dx.doi.org/10.1007/JHEP02(2021)224}{{ JHEP} {02} (2021) 224}, \href{http://arxiv.org/abs/2012.11871}{{\ttfamily arXiv:2012.11871}}.

\bibitem{Guica:2020uhm}
M.~Guica and R.~Monten, {Infinite pseudo-conformal symmetries of classical $T \bar T$, $J \bar T $ and $J T_a$ - deformed CFTs}, \href{http://dx.doi.org/10.21468/SciPostPhys.11.4.078}{{ SciPost Phys.} {11} no.~4, (2021) 078}, \href{http://arxiv.org/abs/2011.05445}{{\ttfamily arXiv:2011.05445}}.

\bibitem{Jorjadze:2020ili}
G.~Jorjadze and S.~Theisen, {Canonical maps and integrability in $T\bar T$ deformed 2d CFTs}, \href{http://arxiv.org/abs/2001.03563}{{\ttfamily arXiv:2001.03563}}.

\bibitem{Georgescu:2022iyx}
S.~Georgescu and M.~Guica, {Infinite $\mathrm{T\bar T}$-like symmetries of compactified LST}, \href{http://dx.doi.org/10.21468/SciPostPhys.16.1.006}{{ SciPost Phys.} {16} no.~1, (2024) 006}, \href{http://arxiv.org/abs/2212.09768}{{\ttfamily arXiv:2212.09768}}.

\bibitem{Guica:2022gts}
M.~Guica, R.~Monten, and I.~Tsiares, {Classical and quantum symmetries of $T\bar T$-deformed CFTs}, \href{http://arxiv.org/abs/2212.14014}{{\ttfamily arXiv:2212.14014}}.

\bibitem{Simmons-Duffin:2016gjk}
D.~Simmons-Duffin, \href{http://dx.doi.org/10.1142/9789813149441_0001}{{The Conformal Bootstrap},} in { {Theoretical Advanced Study Institute in Elementary Particle Physics}: {New Frontiers in Fields and Strings}}, pp.~1--74.
\newblock 2017.
\newblock \href{http://arxiv.org/abs/1602.07982}{{\ttfamily arXiv:1602.07982}}.

\bibitem{Poland:2018epd}
D.~Poland, S.~Rychkov, and A.~Vichi, {The Conformal Bootstrap: Theory, Numerical Techniques, and Applications}, \href{http://dx.doi.org/10.1103/RevModPhys.91.015002}{{ Rev. Mod. Phys.} {91} (2019) 015002}, \href{http://arxiv.org/abs/1805.04405}{{\ttfamily arXiv:1805.04405}}.

\bibitem{Guica:2021fkv}
M.~Guica, {A definition of primary operators in $J\bar T$-deformed CFTs}, \href{http://dx.doi.org/10.21468/SciPostPhys.13.3.045}{{ SciPost Phys.} {13} no.~3, (2022) 045}, \href{http://arxiv.org/abs/2112.14736}{{\ttfamily arXiv:2112.14736}}.

\bibitem{Apolo:2023aho}
L.~Apolo, W.~Song, and B.~Yu, {On the universal behavior of $ T\overline{T} $-deformed CFTs: single and double-trace partition functions at large c}, \href{http://dx.doi.org/10.1007/JHEP05(2023)210}{{ JHEP} {05} (2023) 210}, \href{http://arxiv.org/abs/2301.04153}{{\ttfamily arXiv:2301.04153}}.

\bibitem{Datta:2018thy}
S.~Datta and Y.~Jiang, {$T\bar{T}$ deformed partition functions}, \href{http://dx.doi.org/10.1007/JHEP08(2018)106}{{ JHEP} {08} (2018) 106}, \href{http://arxiv.org/abs/1806.07426}{{\ttfamily arXiv:1806.07426}}.

\bibitem{Tian:2024vln}
J.~Tian, T.~Lai, and F.~Omidi, {Modular transformations of on-shell actions of (root-)$\text{T}\overline{\text{T}}$ deformed holographic CFTs}, \href{http://arxiv.org/abs/2404.16354}{{\ttfamily arXiv:2404.16354}}.

\bibitem{Gu:2024ogh}
J.~Gu, Y.~Jiang, and H.~Wang, {Resurgence of $T\bar{T}$-deformed Partition Function}, \href{http://arxiv.org/abs/2410.19633}{{\ttfamily arXiv:2410.19633}}.

\bibitem{Gu:2025tpy}
J.~Gu, Y.~Jiang, and H.~Wang, {Resurgent properties of $T\overline{T}$-deformed conformal field theories}, \href{http://arxiv.org/abs/2503.19350}{{\ttfamily arXiv:2503.19350}}.

\bibitem{Griguolo:2022xcj}
L.~Griguolo, R.~Panerai, J.~Papalini, and D.~Seminara, {Exact TT\textasciimacron{} Deformation of Two-Dimensional Maxwell Theory}, \href{http://dx.doi.org/10.1103/PhysRevLett.128.221601}{{ Phys. Rev. Lett.} {128} no.~22, (2022) 221601}, \href{http://arxiv.org/abs/2203.09683}{{\ttfamily arXiv:2203.09683}}.

\bibitem{Griguolo:2022hek}
L.~Griguolo, R.~Panerai, J.~Papalini, and D.~Seminara, {Exact $ T\overline{T} $ deformation of two-dimensional Yang-Mills theory on the sphere}, \href{http://dx.doi.org/10.1007/JHEP10(2022)134}{{ JHEP} {10} (2022) 134}, \href{http://arxiv.org/abs/2207.05095}{{\ttfamily arXiv:2207.05095}}.

\bibitem{Griguolo:2022vdx}
L.~Griguolo, R.~Panerai, J.~Papalini, and D.~Seminara, {The phase diagram of $ T\overline{T} $-deformed Yang-Mills theory on the sphere}, \href{http://dx.doi.org/10.1007/JHEP11(2022)078}{{ JHEP} {11} (2022) 078}, \href{http://arxiv.org/abs/2209.06222}{{\ttfamily arXiv:2209.06222}}.

\bibitem{Hashimoto:2019wct}
A.~Hashimoto and D.~Kutasov, {$ T\overline{T},J\overline{T},T\overline{J} $ partition sums from string theory}, \href{http://dx.doi.org/10.1007/JHEP02(2020)080}{{ JHEP} {02} (2020) 080}, \href{http://arxiv.org/abs/1907.07221}{{\ttfamily arXiv:1907.07221}}.

\bibitem{Rosenhaus:2019utc}
V.~Rosenhaus and M.~Smolkin, {Integrability and renormalization under $T \bar T$}, \href{http://dx.doi.org/10.1103/PhysRevD.102.065009}{{ Phys. Rev. D} {102} no.~6, (2020) 065009}, \href{http://arxiv.org/abs/1909.02640}{{\ttfamily arXiv:1909.02640}}.

\bibitem{Guica:2019vnb}
M.~Guica, {On correlation functions in $J\bar T$-deformed CFTs}, \href{http://dx.doi.org/10.1088/1751-8121/ab0ef3}{{ J. Phys. A} {52} no.~18, (2019) 184003}, \href{http://arxiv.org/abs/1902.01434}{{\ttfamily arXiv:1902.01434}}.

\bibitem{Cui:2023jrb}
W.~Cui, H.~Shu, W.~Song, and J.~Wang, {Correlation functions in the ${\text{TsT}}/T\overline{T }$ correspondence}, \href{http://dx.doi.org/10.1007/JHEP04(2024)017}{{ JHEP} {04} (2024) 017}, \href{http://arxiv.org/abs/2304.04684}{{\ttfamily arXiv:2304.04684}}.

\bibitem{He:2019vzf}
S.~He and H.~Shu, {Correlation functions, entanglement and chaos in the $ T\overline{T}/J\overline{T} $-deformed CFTs}, \href{http://dx.doi.org/10.1007/JHEP02(2020)088}{{ JHEP} {02} (2020) 088}, \href{http://arxiv.org/abs/1907.12603}{{\ttfamily arXiv:1907.12603}}.

\bibitem{Menskoy:2024vqv}
D.~Menskoy, {On correlators in $T\bar{T}$-deformed conformal field theories}, \href{http://arxiv.org/abs/2407.20774}{{\ttfamily arXiv:2407.20774}}.

\bibitem{He:2020qcs}
S.~He, {Note on higher-point correlation functions of the $T\bar T$ or $J\bar T$ deformed CFTs}, \href{http://dx.doi.org/10.1007/s11433-021-1741-1}{{ Sci. China Phys. Mech. Astron.} {64} no.~9, (2021) 291011}, \href{http://arxiv.org/abs/2012.06202}{{\ttfamily arXiv:2012.06202}}.

\bibitem{He:2019ahx}
S.~He, J.-R. Sun, and Y.~Sun, {The correlation function of (1,1) and (2,2) supersymmetric theories with $T\bar{T}$ deformation}, \href{http://dx.doi.org/10.1007/JHEP04(2020)100}{{ JHEP} {04} (2020) 100}, \href{http://arxiv.org/abs/1912.11461}{{\ttfamily arXiv:1912.11461}}.

\bibitem{He:2020udl}
S.~He and Y.~Sun, {Correlation functions of CFTs on a torus with a $T\overline{T}$ deformation}, \href{http://dx.doi.org/10.1103/PhysRevD.102.026023}{{ Phys. Rev. D} {102} no.~2, (2020) 026023}, \href{http://arxiv.org/abs/2004.07486}{{\ttfamily arXiv:2004.07486}}.

\bibitem{pfun}
N.~I. Akhiezer, { {Elements of the Theory of Elliptic Functions}}.
\newblock American Mathematical Society, Providence, 1990.

\bibitem{DiFrancesco:1997nk}
P.~Di~Francesco, P.~Mathieu, and D.~Senechal, \href{http://dx.doi.org/10.1007/978-1-4612-2256-9}{{ {Conformal Field Theory}}}.
\newblock Graduate Texts in Contemporary Physics. Springer-Verlag, New York, 1997.

\bibitem{Hirano:2024eab}
S.~Hirano and M.~Shigemori, {Conformal field theory on $ T\overline{T} $-deformed space and correlators from dynamical coordinate transformations}, \href{http://dx.doi.org/10.1007/JHEP07(2024)190}{{ JHEP} {07} (2024) 190}, \href{http://arxiv.org/abs/2402.08278}{{\ttfamily arXiv:2402.08278}}.

\bibitem{He:2020cxp}
S.~He, Y.~Sun, and Y.-X. Zhang, {$T \overline{T} $-flow effects on torus partition functions}, \href{http://dx.doi.org/10.1007/JHEP09(2021)061}{{ JHEP} {09} (2021) 061}, \href{http://arxiv.org/abs/2011.02902}{{\ttfamily arXiv:2011.02902}}.

\bibitem{Nakayama:2025lyi}
Y.~Nakayama, {Is chiral supersymmetry emanant or emergent?}, \href{http://arxiv.org/abs/2502.20815}{{\ttfamily arXiv:2502.20815}}.

\bibitem{Bazhanov:1994ft}
V.~V. Bazhanov, S.~L. Lukyanov, and A.~B. Zamolodchikov, {Integrable structure of conformal field theory, quantum KdV theory and thermodynamic Bethe ansatz}, \href{http://dx.doi.org/10.1007/BF02101898}{{ Commun. Math. Phys.} {177} (1996) 381--398}, \href{http://arxiv.org/abs/hep-th/9412229}{{\ttfamily arXiv:hep-th/9412229}}.

\bibitem{LeFloch:2019wlf}
B.~Le~Floch and M.~Mezei, {KdV charges in $T\bar{T}$ theories and new models with super-Hagedorn behavior}, \href{http://dx.doi.org/10.21468/SciPostPhys.7.4.043}{{ SciPost Phys.} {7} no.~4, (2019) 043}, \href{http://arxiv.org/abs/1907.02516}{{\ttfamily arXiv:1907.02516}}.

\bibitem{Asrat:2020jsh}
M.~Asrat, {KdV charges and the generalized torus partition sum in $T \bar T$ deformation}, \href{http://dx.doi.org/10.1016/j.nuclphysb.2020.115119}{{ Nucl. Phys. B} {958} (2020) 115119}, \href{http://arxiv.org/abs/2002.04824}{{\ttfamily arXiv:2002.04824}}.

\bibitem{Travaglino:2024lyf}
R.~Travaglino, M.~Mazzoni, and O.~A. Castro-Alvaredor, {Generalised Hydrodynamics of $\mathrm{T\bar{T}}$-Deformed Integrable Quantum Field Theories}, \href{http://arxiv.org/abs/2405.06564}{{\ttfamily arXiv:2405.06564}}.

\bibitem{He:2023kgq}
S.~He, Y.~Sun, and J.~Yin, {A systematic approach to correlators in $T\bar{T}$ deformed CFTs}, \href{http://arxiv.org/abs/2310.20516}{{\ttfamily arXiv:2310.20516}}.

\bibitem{Barel:2024dgv}
N.~Barel, {Correlation Functions in $\textrm{T}\bar{\textrm{T}}$-deformed Theories on the Torus}, \href{http://arxiv.org/abs/2407.15090}{{\ttfamily arXiv:2407.15090}}.

\bibitem{Hirano:2020ppu}
S.~Hirano, T.~Nakajima, and M.~Shigemori, {$ T\overline{T} $ Deformation of stress-tensor correlators from random geometry}, \href{http://dx.doi.org/10.1007/JHEP04(2021)270}{{ JHEP} {04} (2021) 270}, \href{http://arxiv.org/abs/2012.03972}{{\ttfamily arXiv:2012.03972}}.

\bibitem{Asrat:2017tzd}
M.~Asrat, A.~Giveon, N.~Itzhaki, and D.~Kutasov, {Holography Beyond AdS}, \href{http://dx.doi.org/10.1016/j.nuclphysb.2018.05.005}{{ Nucl. Phys. B} {932} (2018) 241--253}, \href{http://arxiv.org/abs/1711.02690}{{\ttfamily arXiv:1711.02690}}.

\bibitem{Kruthoff:2020hsi}
J.~Kruthoff and O.~Parrikar, {On the flow of states under $T\overline{T}$}, \href{http://arxiv.org/abs/2006.03054}{{\ttfamily arXiv:2006.03054}}.

\bibitem{Tsolakidis:2024wut}
E.~Tsolakidis, {Massive gravity generalization of $T\overline{T}$ deformations}, \href{http://arxiv.org/abs/2405.07967}{{\ttfamily arXiv:2405.07967}}.

\bibitem{Castro-Alvaredo:2023rtl}
O.~A. Castro-Alvaredo, S.~Negro, and F.~Sailis, {Completing the bootstrap program for -deformed massive integrable quantum field theories}, \href{http://dx.doi.org/10.1088/1751-8121/ad5395}{{ J. Phys. A} {57} no.~26, (2024) 265401}, \href{http://arxiv.org/abs/2305.17068}{{\ttfamily arXiv:2305.17068}}.

\bibitem{Castro-Alvaredo:2023wmw}
O.~A. Castro-Alvaredo, S.~Negro, and F.~Sailis, {Form factors and correlation functions of $ \textrm{T}\overline{\textrm{T}} $-deformed integrable quantum field theories}, \href{http://dx.doi.org/10.1007/JHEP09(2023)048}{{ JHEP} {09} (2023) 048}, \href{http://arxiv.org/abs/2306.01640}{{\ttfamily arXiv:2306.01640}}.

\bibitem{Castro-Alvaredo:2023hap}
O.~A. Castro-Alvaredo, S.~Negro, and I.~M. Sz\'ecs\'enyi, {On the representation of minimal form factors in integrable quantum field theory}, \href{http://dx.doi.org/10.1016/j.nuclphysb.2024.116459}{{ Nucl. Phys. B} {1000} (2024) 116459}, \href{http://arxiv.org/abs/2311.16955}{{\ttfamily arXiv:2311.16955}}.

\bibitem{Castro-Alvaredo:2025nma}
O.~A. Castro-Alvaredo, S.~Negro, and F.~Sailis, {Boundary Quantum Field Theories Perturbed by ${\rm T}\bar{\rm T}$: Towards a Form Factor Program}, \href{http://arxiv.org/abs/2501.11647}{{\ttfamily arXiv:2501.11647}}.

\bibitem{Witten:2018zxz}
E.~Witten, Aps medal for exceptional achievement in research: Invited article on entanglement properties of quantum field theory, \href{http://dx.doi.org/10.1103/RevModPhys.90.045003}{{ Rev. Mod. Phys.} {90} no.~4, (2018) 045003}, \href{http://arxiv.org/abs/1803.04993}{{\ttfamily arXiv:1803.04993}}.

\bibitem{Witten:2021jzq}
E.~Witten, { Why Does Quantum Field Theory in Curved Spacetime Make Sense? And What Happens to the Algebra of Observables in the Thermodynamic Limit?}, \href{http://dx.doi.org/10.1007/978-3-031-17523-7$\_$11}{pp.~241--284}.
\newblock Springer International Publishing, Cham, 2022.
\newblock \url{https://doi.org/10.1007/978-3-031-17523-7_11}.

\bibitem{Donnelly:2018bef}
W.~Donnelly and V.~Shyam, {Entanglement entropy and $T \overline{T}$ deformation}, \href{http://dx.doi.org/10.1103/PhysRevLett.121.131602}{{ Phys. Rev. Lett.} {121} no.~13, (2018) 131602}, \href{http://arxiv.org/abs/1806.07444}{{\ttfamily arXiv:1806.07444}}.

\bibitem{Banerjee:2019ewu}
A.~Banerjee, A.~Bhattacharyya, and S.~Chakraborty, {Entanglement Entropy for $TT$ deformed CFT in general dimensions}, \href{http://dx.doi.org/10.1016/j.nuclphysb.2019.114775}{{ Nucl. Phys. B} {948} (2019) 114775}, \href{http://arxiv.org/abs/1904.00716}{{\ttfamily arXiv:1904.00716}}.

\bibitem{Ota:2019yfe}
T.~Ota, {Comments on holographic entanglements in cutoff AdS}, \href{http://arxiv.org/abs/1904.06930}{{\ttfamily arXiv:1904.06930}}.

\bibitem{Chang:2024voo}
J.-C. Chang, S.~He, Y.-X. Liu, and L.~Zhao, {The holographic $T\bar{T}$ deformation of the entanglement entropy in (A)dS$_3$/CFT$_2$}, \href{http://arxiv.org/abs/2409.08198}{{\ttfamily arXiv:2409.08198}}.

\bibitem{Jeong:2019ylz}
H.-S. Jeong, K.-Y. Kim, and M.~Nishida, {Entanglement and R\'enyi entropy of multiple intervals in $T\overline{T}$-deformed CFT and holography}, \href{http://dx.doi.org/10.1103/PhysRevD.100.106015}{{ Phys. Rev. D} {100} no.~10, (2019) 106015}, \href{http://arxiv.org/abs/1906.03894}{{\ttfamily arXiv:1906.03894}}.

\bibitem{He:2023xnb}
M.~He and Y.~Sun, {Holographic entanglement entropy in $T \bar{T}$-deformed AdS3}, \href{http://dx.doi.org/10.1016/j.nuclphysb.2023.116190}{{ Nucl. Phys. B} {990} (2023) 116190}, \href{http://arxiv.org/abs/2301.04435}{{\ttfamily arXiv:2301.04435}}.

\bibitem{Chen:2018eqk}
B.~Chen, L.~Chen, and P.-X. Hao, {Entanglement entropy in $T\overline{T}$-deformed CFT}, \href{http://dx.doi.org/10.1103/PhysRevD.98.086025}{{ Phys. Rev. D} {98} no.~8, (2018) 086025}, \href{http://arxiv.org/abs/1807.08293}{{\ttfamily arXiv:1807.08293}}.

\bibitem{Basu:2023bov}
D.~Basu, Lavish, and B.~Paul, {Entanglement negativity in TT\textasciimacron{}-deformed CFT2s}, \href{http://dx.doi.org/10.1103/PhysRevD.107.126026}{{ Phys. Rev. D} {107} no.~12, (2023) 126026}, \href{http://arxiv.org/abs/2302.11435}{{\ttfamily arXiv:2302.11435}}.

\bibitem{Pant:2024eno}
S.~Pant and H.~Parihar, {Mixed state entanglement in deformed field theory at finite temperature}, \href{http://arxiv.org/abs/2412.19680}{{\ttfamily arXiv:2412.19680}}.

\bibitem{Grieninger:2023knz}
S.~Grieninger, K.~Ikeda, and D.~E. Kharzeev, {Temporal entanglement entropy as a probe of renormalization group flow}, \href{http://dx.doi.org/10.1007/JHEP05(2024)030}{{ JHEP} {05} (2024) 030}, \href{http://arxiv.org/abs/2312.08534}{{\ttfamily arXiv:2312.08534}}.

\bibitem{Basu:2023aqz}
D.~Basu, S.~Biswas, A.~Dey, B.~Paul, and G.~Sengupta, {Odd entanglement entropy in TT\textasciimacron{} deformed CFT2s and holography}, \href{http://dx.doi.org/10.1103/PhysRevD.108.126013}{{ Phys. Rev. D} {108} no.~12, (2023) 126013}, \href{http://arxiv.org/abs/2307.04832}{{\ttfamily arXiv:2307.04832}}.

\bibitem{Basu:2024bal}
D.~Basu and V.~Raj, {Reflected entropy and timelike entanglement in TT\textasciimacron{}-deformed CFT2s}, \href{http://dx.doi.org/10.1103/PhysRevD.110.046009}{{ Phys. Rev. D} {110} no.~4, (2024) 046009}, \href{http://arxiv.org/abs/2402.07253}{{\ttfamily arXiv:2402.07253}}.

\bibitem{Basu:2024enr}
D.~Basu and S.~Biswas, {Entanglement, $\textrm{T}\bar{\textrm{T}}$ and rotating black holes}, \href{http://arxiv.org/abs/2410.06363}{{\ttfamily arXiv:2410.06363}}.

\bibitem{Soni:2024aop}
R.~M. Soni and A.~C. Wall, {A New Covariant Entropy Bound from Cauchy Slice Holography}, \href{http://arxiv.org/abs/2407.16769}{{\ttfamily arXiv:2407.16769}}.

\bibitem{He:2022xkh}
S.~He, Z.-C. Liu, and Y.~Sun, {Entanglement entropy and modular Hamiltonian of free fermion with deformations on a torus}, \href{http://dx.doi.org/10.1007/JHEP09(2022)247}{{ JHEP} {09} (2022) 247}, \href{http://arxiv.org/abs/2207.06308}{{\ttfamily arXiv:2207.06308}}.

\bibitem{Ashkenazi:2023fcn}
S.~Ashkenazi, S.~Chakraborty, Z.~Ma, and T.~Shachar, {Linear response of entanglement entropy to $ T\overline{T} $ in massive QFTs}, \href{http://dx.doi.org/10.1007/JHEP04(2023)077}{{ JHEP} {04} (2023) 077}, \href{http://arxiv.org/abs/2302.06688}{{\ttfamily arXiv:2302.06688}}.

\bibitem{Nishioka:2021cxe}
T.~Nishioka, T.~Takayanagi, and Y.~Taki, {Topological pseudo entropy}, \href{http://dx.doi.org/10.1007/JHEP09(2021)015}{{ JHEP} {09} (2021) 015}, \href{http://arxiv.org/abs/2107.01797}{{\ttfamily arXiv:2107.01797}}.

\bibitem{Guo:2022sfl}
W.-z. Guo, S.~He, and Y.-X. Zhang, {On the real-time evolution of pseudo-entropy in 2d CFTs}, \href{http://dx.doi.org/10.1007/JHEP09(2022)094}{{ JHEP} {09} (2022) 094}, \href{http://arxiv.org/abs/2206.11818}{{\ttfamily arXiv:2206.11818}}.

\bibitem{He:2023eap}
S.~He, J.~Yang, Y.-X. Zhang, and Z.-X. Zhao, {Pseudoentropy for descendant operators in two-dimensional conformal field theories}, \href{http://dx.doi.org/10.1103/PhysRevD.109.025014}{{ Phys. Rev. D} {109} no.~2, (2024) 025014}, \href{http://arxiv.org/abs/2301.04891}{{\ttfamily arXiv:2301.04891}}.

\bibitem{Nakata:2020luh}
Y.~Nakata, T.~Takayanagi, Y.~Taki, K.~Tamaoka, and Z.~Wei, {New holographic generalization of entanglement entropy}, \href{http://dx.doi.org/10.1103/PhysRevD.103.026005}{{ Phys. Rev. D} {103} no.~2, (2021) 026005}, \href{http://arxiv.org/abs/2005.13801}{{\ttfamily arXiv:2005.13801}}.

\bibitem{He:2023wko}
S.~He, J.~Yang, Y.-X. Zhang, and Z.-X. Zhao, {Pseudo entropy of primary operators in $ T\overline{T}/J\overline{T} $-deformed CFTs}, \href{http://dx.doi.org/10.1007/JHEP09(2023)025}{{ JHEP} {09} (2023) 025}, \href{http://arxiv.org/abs/2305.10984}{{\ttfamily arXiv:2305.10984}}.

\bibitem{Castro-Alvaredo:2023jbg}
O.~A. Castro-Alvaredo, S.~Negro, and F.~Sailis, {Entanglement entropy from form factors in $ \textrm{T}\overline{\textrm{T}} $-deformed integrable quantum field theories}, \href{http://dx.doi.org/10.1007/JHEP11(2023)129}{{ JHEP} {11} (2023) 129}, \href{http://arxiv.org/abs/2306.11064}{{\ttfamily arXiv:2306.11064}}.

\bibitem{He:2023obo}
M.~He, J.~Hou, and Y.~Jiang, {$ T\overline{T} $-deformed entanglement entropy for IQFT}, \href{http://dx.doi.org/10.1007/JHEP03(2024)056}{{ JHEP} {03} (2024) 056}, \href{http://arxiv.org/abs/2306.07784}{{\ttfamily arXiv:2306.07784}}.

\bibitem{Gubser:2002vv}
S.~S. Gubser and I.~R. Klebanov, {A Universal result on central charges in the presence of double trace deformations}, \href{http://dx.doi.org/10.1016/S0550-3213(03)00056-7}{{ Nucl. Phys. B} {656} (2003) 23--36}, \href{http://arxiv.org/abs/hep-th/0212138}{{\ttfamily arXiv:hep-th/0212138}}.

\bibitem{Bzowski:2018pcy}
A.~Bzowski and M.~Guica, {The holographic interpretation of $J \bar T$-deformed CFTs}, \href{http://dx.doi.org/10.1007/JHEP01(2019)198}{{ JHEP} {01} (2019) 198}, \href{http://arxiv.org/abs/1803.09753}{{\ttfamily arXiv:1803.09753}}.

\bibitem{Elitzur:2005kz}
S.~Elitzur, A.~Giveon, M.~Porrati, and E.~Rabinovici, {Multitrace deformations of vector and adjoint theories and their holographic duals}, \href{http://dx.doi.org/10.1088/1126-6708/2006/02/006}{{ JHEP} {02} (2006) 006}, \href{http://arxiv.org/abs/hep-th/0511061}{{\ttfamily arXiv:hep-th/0511061}}.

\bibitem{Papadimitriou:2007sj}
I.~Papadimitriou, {Multi-Trace Deformations in AdS/CFT: Exploring the Vacuum Structure of the Deformed CFT}, \href{http://dx.doi.org/10.1088/1126-6708/2007/05/075}{{ JHEP} {05} (2007) 075}, \href{http://arxiv.org/abs/hep-th/0703152}{{\ttfamily arXiv:hep-th/0703152}}.

\bibitem{Li:2020zjb}
Y.~Li, {Comments on large central charge $T\bar{T}$ deformed conformal field theory and cutoff AdS holography}, \href{http://arxiv.org/abs/2012.14414}{{\ttfamily arXiv:2012.14414}}.

\bibitem{Apolo:2023vnm}
L.~Apolo, P.-X. Hao, W.-X. Lai, and W.~Song, {Glue-on AdS holography for $ T\overline{T} $-deformed CFTs}, \href{http://dx.doi.org/10.1007/JHEP06(2023)117}{{ JHEP} {06} (2023) 117}, \href{http://arxiv.org/abs/2303.04836}{{\ttfamily arXiv:2303.04836}}.

\bibitem{Witten:2001ua}
E.~Witten, {Multitrace operators, boundary conditions, and AdS / CFT correspondence}, \href{http://arxiv.org/abs/hep-th/0112258}{{\ttfamily arXiv:hep-th/0112258}}.

\bibitem{Berkooz:2002ug}
M.~Berkooz, A.~Sever, and A.~Shomer, {'Double trace' deformations, boundary conditions and space-time singularities}, \href{http://dx.doi.org/10.1088/1126-6708/2002/05/034}{{ JHEP} {05} (2002) 034}, \href{http://arxiv.org/abs/hep-th/0112264}{{\ttfamily arXiv:hep-th/0112264}}.

\bibitem{Marolf:2006nd}
D.~Marolf and S.~F. Ross, {Boundary Conditions and New Dualities: Vector Fields in AdS/CFT}, \href{http://dx.doi.org/10.1088/1126-6708/2006/11/085}{{ JHEP} {11} (2006) 085}, \href{http://arxiv.org/abs/hep-th/0606113}{{\ttfamily arXiv:hep-th/0606113}}.

\bibitem{FarajiAstaneh:2024fig}
A.~Faraji~Astaneh, {Holographic action principle for TT\textasciimacron{}-deformation}, \href{http://dx.doi.org/10.1016/j.physletb.2024.139227}{{ Phys. Lett. B} {860} (2025) 139227}, \href{http://arxiv.org/abs/2407.16391}{{\ttfamily arXiv:2407.16391}}.

\bibitem{Murdia:2019fax}
C.~Murdia, Y.~Nomura, P.~Rath, and N.~Salzetta, {Comments on holographic entanglement entropy in $TT$ deformed conformal field theories}, \href{http://dx.doi.org/10.1103/PhysRevD.100.026011}{{ Phys. Rev. D} {100} no.~2, (2019) 026011}, \href{http://arxiv.org/abs/1904.04408}{{\ttfamily arXiv:1904.04408}}.

\bibitem{Grieninger:2019zts}
S.~Grieninger, {Entanglement entropy and $ T\overline{T} $ deformations beyond antipodal points from holography}, \href{http://dx.doi.org/10.1007/JHEP11(2019)171}{{ JHEP} {11} (2019) 171}, \href{http://arxiv.org/abs/1908.10372}{{\ttfamily arXiv:1908.10372}}.

\bibitem{Lewkowycz:2019xse}
A.~Lewkowycz, J.~Liu, E.~Silverstein, and G.~Torroba, {$ T\overline{T} $ and EE, with implications for (A)dS subregion encodings}, \href{http://dx.doi.org/10.1007/JHEP04(2020)152}{{ JHEP} {04} (2020) 152}, \href{http://arxiv.org/abs/1909.13808}{{\ttfamily arXiv:1909.13808}}.

\bibitem{Geng:2019ruz}
H.~Geng, {Some Information Theoretic Aspects of De-Sitter Holography}, \href{http://dx.doi.org/10.1007/JHEP02(2020)005}{{ JHEP} {02} (2020) 005}, \href{http://arxiv.org/abs/1911.02644}{{\ttfamily arXiv:1911.02644}}.

\bibitem{Donnelly:2019pie}
W.~Donnelly, E.~LePage, Y.-Y. Li, A.~Pereira, and V.~Shyam, {Quantum corrections to finite radius holography and holographic entanglement entropy}, \href{http://dx.doi.org/10.1007/JHEP05(2020)006}{{ JHEP} {05} (2020) 006}, \href{http://arxiv.org/abs/1909.11402}{{\ttfamily arXiv:1909.11402}}.

\bibitem{Allameh:2021moy}
K.~Allameh, A.~F. Astaneh, and A.~Hassanzadeh, {Aspects of holographic entanglement entropy for TT\textasciimacron{}-deformed CFTs}, \href{http://dx.doi.org/10.1016/j.physletb.2022.136914}{{ Phys. Lett. B} {826} (2022) 136914}, \href{http://arxiv.org/abs/2111.11338}{{\ttfamily arXiv:2111.11338}}.

\bibitem{FarajiAstaneh:2022qck}
A.~Faraji~Astaneh and K.~Allameh, {Energy of decomposition and entanglement thermodynamics for T2-deformation}, \href{http://dx.doi.org/10.1016/j.physletb.2023.137772}{{ Phys. Lett. B} {839} (2023) 137772}, \href{http://arxiv.org/abs/2212.02816}{{\ttfamily arXiv:2212.02816}}.

\bibitem{Heemskerk:2010hk}
I.~Heemskerk and J.~Polchinski, {Holographic and Wilsonian Renormalization Groups}, \href{http://dx.doi.org/10.1007/JHEP06(2011)031}{{ JHEP} {06} (2011) 031}, \href{http://arxiv.org/abs/1010.1264}{{\ttfamily arXiv:1010.1264}}.

\bibitem{Faulkner:2010jy}
T.~Faulkner, H.~Liu, and M.~Rangamani, {Integrating out geometry: Holographic Wilsonian RG and the membrane paradigm}, \href{http://dx.doi.org/10.1007/JHEP08(2011)051}{{ JHEP} {08} (2011) 051}, \href{http://arxiv.org/abs/1010.4036}{{\ttfamily arXiv:1010.4036}}.

\bibitem{Parvizi:2025shq}
A.~Parvizi, M.~M. Sheikh-Jabbari, and V.~Taghiloo, {Freelance Holography, Part I: Setting Boundary Conditions Free in Gauge/Gravity Correspondence}, \href{http://arxiv.org/abs/2503.09371}{{\ttfamily arXiv:2503.09371}}.

\bibitem{Parvizi:2025wsg}
A.~Parvizi, M.~M. Sheikh-Jabbari, and V.~Taghiloo, {Freelance Holography, Part II: Moving Boundary in Gauge/Gravity Correspondence}, \href{http://arxiv.org/abs/2503.09372}{{\ttfamily arXiv:2503.09372}}.

\bibitem{Maldacena:1997re}
J.~M. Maldacena, {The Large N limit of superconformal field theories and supergravity}, \href{http://dx.doi.org/10.4310/ATMP.1998.v2.n2.a1}{{ Adv. Theor. Math. Phys.} {2} (1998) 231--252}, \href{http://arxiv.org/abs/hep-th/9711200}{{\ttfamily arXiv:hep-th/9711200}}.

\bibitem{Gubser:1998bc}
S.~S. Gubser, I.~R. Klebanov, and A.~M. Polyakov, {Gauge theory correlators from noncritical string theory}, \href{http://dx.doi.org/10.1016/S0370-2693(98)00377-3}{{ Phys. Lett. B} {428} (1998) 105--114}, \href{http://arxiv.org/abs/hep-th/9802109}{{\ttfamily arXiv:hep-th/9802109}}.

\bibitem{Witten:1998qj}
E.~Witten, {Anti-de Sitter space and holography}, \href{http://dx.doi.org/10.4310/ATMP.1998.v2.n2.a2}{{ Adv. Theor. Math. Phys.} {2} (1998) 253--291}, \href{http://arxiv.org/abs/hep-th/9802150}{{\ttfamily arXiv:hep-th/9802150}}.

\bibitem{Fefferman:1985cfm}
C.~Fefferman and C.~R. Graham, {Conformal invariants}, { Ast{\'e}risque} {S131} (1985) 95--116.

\bibitem{Graham:1991jqw}
C.~R. Graham and J.~M. Lee, {Einstein metrics with prescribed conformal infinity on the ball}, \href{http://dx.doi.org/10.1016/0001-8708(91)90071-E}{{ Adv. Math.} {87} no.~2, (1991) 186--225}.

\bibitem{Skenderis:1999nb}
K.~Skenderis and S.~N. Solodukhin, {Quantum effective action from the AdS / CFT correspondence}, \href{http://dx.doi.org/10.1016/S0370-2693(99)01467-7}{{ Phys. Lett. B} {472} (2000) 316--322}, \href{http://arxiv.org/abs/hep-th/9910023}{{\ttfamily arXiv:hep-th/9910023}}.

\bibitem{Balasubramanian:1999re}
V.~Balasubramanian and P.~Kraus, {A Stress tensor for Anti-de Sitter gravity}, \href{http://dx.doi.org/10.1007/s002200050764}{{ Commun. Math. Phys.} {208} (1999) 413--428}, \href{http://arxiv.org/abs/hep-th/9902121}{{\ttfamily arXiv:hep-th/9902121}}.

\bibitem{Emparan:1999pm}
R.~Emparan, C.~V. Johnson, and R.~C. Myers, {Surface terms as counterterms in the AdS / CFT correspondence}, \href{http://dx.doi.org/10.1103/PhysRevD.60.104001}{{ Phys. Rev. D} {60} (1999) 104001}, \href{http://arxiv.org/abs/hep-th/9903238}{{\ttfamily arXiv:hep-th/9903238}}.

\bibitem{deHaro:2000vlm}
S.~de~Haro, S.~N. Solodukhin, and K.~Skenderis, {Holographic reconstruction of space-time and renormalization in the AdS / CFT correspondence}, \href{http://dx.doi.org/10.1007/s002200100381}{{ Commun. Math. Phys.} {217} (2001) 595--622}, \href{http://arxiv.org/abs/hep-th/0002230}{{\ttfamily arXiv:hep-th/0002230}}.

\bibitem{Henningson:1998gx}
M.~Henningson and K.~Skenderis, {The Holographic Weyl anomaly}, \href{http://dx.doi.org/10.1088/1126-6708/1998/07/023}{{ JHEP} {07} (1998) 023}, \href{http://arxiv.org/abs/hep-th/9806087}{{\ttfamily arXiv:hep-th/9806087}}.

\bibitem{Brown:1992br}
J.~D. Brown and J.~W. York, Jr., {Quasilocal energy and conserved charges derived from the gravitational action}, \href{http://dx.doi.org/10.1103/PhysRevD.47.1407}{{ Phys. Rev. D} {47} (1993) 1407--1419}, \href{http://arxiv.org/abs/gr-qc/9209012}{{\ttfamily arXiv:gr-qc/9209012}}.

\bibitem{Brown:1986nw}
J.~D. Brown and M.~Henneaux, {Central Charges in the Canonical Realization of Asymptotic Symmetries: An Example from Three-Dimensional Gravity}, \href{http://dx.doi.org/10.1007/BF01211590}{{ Commun. Math. Phys.} {104} (1986) 207--226}.

\bibitem{Carroll:2004st}
S.~M. Carroll, \href{http://dx.doi.org/10.1017/9781108770385}{{ {Spacetime and Geometry}: {An Introduction to General Relativity}}}.
\newblock Cambridge University Press, 7, 2019.

\bibitem{Tian:2023fgf}
J.~Tian, {On-shell action of $\text{T}\bar{\text{T}}$-deformed Holographic CFTs}, \href{http://arxiv.org/abs/2306.01258}{{\ttfamily arXiv:2306.01258}}.

\bibitem{Ryu:2006bv}
S.~Ryu and T.~Takayanagi, {Holographic derivation of entanglement entropy from AdS/CFT}, \href{http://dx.doi.org/10.1103/PhysRevLett.96.181602}{{ Phys. Rev. Lett.} {96} (2006) 181602}, \href{http://arxiv.org/abs/hep-th/0603001}{{\ttfamily arXiv:hep-th/0603001}}.

\bibitem{Kawamoto:2023wzj}
T.~Kawamoto, S.-M. Ruan, and T.~Takayanagia, {Gluing AdS/CFT}, \href{http://dx.doi.org/10.1007/JHEP07(2023)080}{{ JHEP} {07} (2023) 080}, \href{http://arxiv.org/abs/2303.01247}{{\ttfamily arXiv:2303.01247}}.

\bibitem{Kawamoto:2025oko}
T.~Kawamoto, R.~Maeda, N.~Nakamura, and T.~Takayanagi, {Traversable AdS Wormhole via Non-local Double Trace or Janus Deformation}, \href{http://arxiv.org/abs/2502.03531}{{\ttfamily arXiv:2502.03531}}.

\bibitem{Ferko:2022dpg}
C.~Ferko and S.~Sethi, {Sequential flows by irrelevant operators}, \href{http://dx.doi.org/10.21468/SciPostPhys.14.5.098}{{ SciPost Phys.} {14} no.~5, (2023) 098}, \href{http://arxiv.org/abs/2206.04787}{{\ttfamily arXiv:2206.04787}}.

\bibitem{Apolo:2023ckr}
L.~Apolo, P.-X. Hao, W.-X. Lai, and W.~Song, {Extremal surfaces in glue-on AdS/$ T\overline{T} $ holography}, \href{http://dx.doi.org/10.1007/JHEP01(2024)054}{{ JHEP} {01} (2024) 054}, \href{http://arxiv.org/abs/2311.04883}{{\ttfamily arXiv:2311.04883}}.

\bibitem{Lewkowycz:2013nqa}
A.~Lewkowycz and J.~Maldacena, {Generalized gravitational entropy}, \href{http://dx.doi.org/10.1007/JHEP08(2013)090}{{ JHEP} {08} (2013) 090}, \href{http://arxiv.org/abs/1304.4926}{{\ttfamily arXiv:1304.4926}}.

\bibitem{Dong:2016hjy}
X.~Dong, A.~Lewkowycz, and M.~Rangamani, {Deriving covariant holographic entanglement}, \href{http://dx.doi.org/10.1007/JHEP11(2016)028}{{ JHEP} {11} (2016) 028}, \href{http://arxiv.org/abs/1607.07506}{{\ttfamily arXiv:1607.07506}}.

\bibitem{Srednicki:1993im}
M.~Srednicki, {Entropy and area}, \href{http://dx.doi.org/10.1103/PhysRevLett.71.666}{{ Phys. Rev. Lett.} {71} (1993) 666--669}, \href{http://arxiv.org/abs/hep-th/9303048}{{\ttfamily arXiv:hep-th/9303048}}.

\bibitem{He:2024fdm}
S.~He, Y.~Li, Y.-Z. Li, and Y.~Zhang, {Note on holographic torus stress tensor correlators in $AdS_3$ gravity}, \href{http://arxiv.org/abs/2405.01255}{{\ttfamily arXiv:2405.01255}}.

\bibitem{Li:2025sfn}
Y.-Z. Li, Y.~Xie, and S.~He, {Holographic Correlators of Boundary/Crosscap CFTs in Two Dimensions}, \href{http://arxiv.org/abs/2501.18386}{{\ttfamily arXiv:2501.18386}}.

\bibitem{Kraus:2021cwf}
P.~Kraus, R.~Monten, and R.~M. Myers, {3D Gravity in a Box}, \href{http://dx.doi.org/10.21468/SciPostPhys.11.3.070}{{ SciPost Phys.} {11} (2021) 070}, \href{http://arxiv.org/abs/2103.13398}{{\ttfamily arXiv:2103.13398}}.

\bibitem{He:2021bhj}
M.~He, S.~He, and Y.-h. Gao, {Surface charges in Chern-Simons gravity with $ T\overline{T} $ deformation}, \href{http://dx.doi.org/10.1007/JHEP03(2022)044}{{ JHEP} {03} (2022) 044}, \href{http://arxiv.org/abs/2109.12885}{{\ttfamily arXiv:2109.12885}}.

\bibitem{Arnold:1989who}
V.~I. Arnold, \href{http://dx.doi.org/10.1007/978-1-4757-2063-1}{{ {Mathematical Methods of Classical Mechanics}}}.
\newblock Graduate Texts in Mathematics. Springer, 1989.

\bibitem{Abraham1994Foundations}
R.~Abraham and J.~E. Marsden, { {Foundations of Mechanics}}.
\newblock Addison-Wesley, second~ed., 1994.

\bibitem{Crnkovic:1986ex}
C.~Crnkovic and E.~Witten, {Covariant description of canonical formalism in geometrical theories.}, in { Three Hundred Years of Gravitation}, S.~W. {Hawking} and W.~{Israel}, eds., pp.~676--684.
\newblock 9, 1986.

\bibitem{Ebert:2022cle}
S.~Ebert, E.~Hijano, P.~Kraus, R.~Monten, and R.~M. Myers, {Field Theory of Interacting Boundary Gravitons}, \href{http://dx.doi.org/10.21468/SciPostPhys.13.2.038}{{ SciPost Phys.} {13} no.~2, (2022) 038}, \href{http://arxiv.org/abs/2201.01780}{{\ttfamily arXiv:2201.01780}}.

\bibitem{Kraus:2022mnu}
P.~Kraus, R.~Monten, and K.~Roumpedakis, {Refining the cutoff 3d gravity/$ T\overline{T} $ correspondence}, \href{http://dx.doi.org/10.1007/JHEP10(2022)094}{{ JHEP} {10} (2022) 094}, \href{http://arxiv.org/abs/2206.00674}{{\ttfamily arXiv:2206.00674}}.

\bibitem{Shyam:2021ciy}
V.~Shyam, {$ \mathrm{T}\overline{\mathrm{T}} $ + \ensuremath{\Lambda}$_{2}$ deformed CFT on the stretched dS$_{3}$ horizon}, \href{http://dx.doi.org/10.1007/JHEP04(2022)052}{{ JHEP} {04} (2022) 052}, \href{http://arxiv.org/abs/2106.10227}{{\ttfamily arXiv:2106.10227}}.

\bibitem{Coleman:2021nor}
E.~Coleman, E.~A. Mazenc, V.~Shyam, E.~Silverstein, R.~M. Soni, G.~Torroba, and S.~Yang, {De Sitter microstates from T$ \overline{T} $ + \ensuremath{\Lambda}$_{2}$ and the Hawking-Page transition}, \href{http://dx.doi.org/10.1007/JHEP07(2022)140}{{ JHEP} {07} (2022) 140}, \href{http://arxiv.org/abs/2110.14670}{{\ttfamily arXiv:2110.14670}}.

\bibitem{Torroba:2022jrk}
G.~Torroba, {$ T\overline{T} $ + \ensuremath{\Lambda}$_{2}$ from a 2d gravity path integral}, \href{http://dx.doi.org/10.1007/JHEP01(2023)163}{{ JHEP} {01} (2023) 163}, \href{http://arxiv.org/abs/2212.04512}{{\ttfamily arXiv:2212.04512}}.

\bibitem{Batra:2024kjl}
G.~Batra, G.~B. De~Luca, E.~Silverstein, G.~Torroba, and S.~Yang, {Bulk-local dS$_{3}$ holography: the matter with $ T\overline{T} $ + \ensuremath{\Lambda}$_{2}$}, \href{http://dx.doi.org/10.1007/JHEP10(2024)072}{{ JHEP} {10} (2024) 072}, \href{http://arxiv.org/abs/2403.01040}{{\ttfamily arXiv:2403.01040}}.

\bibitem{Aguilar-Gutierrez:2024nst}
S.~E. Aguilar-Gutierrez, A.~Svesko, and M.~R. Visser, {$ \textrm{T}\overline{\textrm{T}} $ deformations from AdS$_{2}$ to dS$_{2}$}, \href{http://dx.doi.org/10.1007/JHEP01(2025)120}{{ JHEP} {01} (2025) 120}, \href{http://arxiv.org/abs/2410.18257}{{\ttfamily arXiv:2410.18257}}.

\bibitem{Seiberg:1999xz}
N.~Seiberg and E.~Witten, {The D1 / D5 system and singular CFT}, \href{http://dx.doi.org/10.1088/1126-6708/1999/04/017}{{ JHEP} {04} (1999) 017}, \href{http://arxiv.org/abs/hep-th/9903224}{{\ttfamily arXiv:hep-th/9903224}}.

\bibitem{Chakraborty:2020yka}
S.~Chakraborty, {$ \frac{\mathrm{SL}\left(2,\mathrm{\mathbb{R}}\right)\times \mathrm{U}(1)}{\mathrm{U}(1)} $ CFT, NS5+F1 system and single trace $ T\overline{T} $}, \href{http://dx.doi.org/10.1007/JHEP03(2021)113}{{ JHEP} {03} (2021) 113}, \href{http://arxiv.org/abs/2012.03995}{{\ttfamily arXiv:2012.03995}}.

\bibitem{Knighton:2024pqh}
B.~Knighton, {Deriving the long-string CFT in AdS$_3$}, \href{http://arxiv.org/abs/2410.16904}{{\ttfamily arXiv:2410.16904}}.

\bibitem{Giveon:1998ns}
A.~Giveon, D.~Kutasov, and N.~Seiberg, {Comments on string theory on AdS(3)}, \href{http://dx.doi.org/10.4310/ATMP.1998.v2.n4.a3}{{ Adv. Theor. Math. Phys.} {2} (1998) 733--782}, \href{http://arxiv.org/abs/hep-th/9806194}{{\ttfamily arXiv:hep-th/9806194}}.

\bibitem{Maldacena:2000hw}
J.~M. Maldacena and H.~Ooguri, {Strings in AdS(3) and SL(2,R) WZW model 1.: The Spectrum}, \href{http://dx.doi.org/10.1063/1.1377273}{{ J. Math. Phys.} {42} (2001) 2929--2960}, \href{http://arxiv.org/abs/hep-th/0001053}{{\ttfamily arXiv:hep-th/0001053}}.

\bibitem{Kutasov:1999xu}
D.~Kutasov and N.~Seiberg, {More comments on string theory on AdS(3)}, \href{http://dx.doi.org/10.1088/1126-6708/1999/04/008}{{ JHEP} {04} (1999) 008}, \href{http://arxiv.org/abs/hep-th/9903219}{{\ttfamily arXiv:hep-th/9903219}}.

\bibitem{Chakraborty:2020swe}
S.~Chakraborty, A.~Giveon, and D.~Kutasov, {$ T\overline{T} $, black holes and negative strings}, \href{http://dx.doi.org/10.1007/JHEP09(2020)057}{{ JHEP} {09} (2020) 057}, \href{http://arxiv.org/abs/2006.13249}{{\ttfamily arXiv:2006.13249}}.

\bibitem{Argurio:2000tb}
R.~Argurio, A.~Giveon, and A.~Shomer, {Superstrings on AdS(3) and symmetric products}, \href{http://dx.doi.org/10.1088/1126-6708/2000/12/003}{{ JHEP} {12} (2000) 003}, \href{http://arxiv.org/abs/hep-th/0009242}{{\ttfamily arXiv:hep-th/0009242}}.

\bibitem{Pozsgay:2019ekd}
B.~Pozsgay, Y.~Jiang, and G.~Tak\'acs, {$T\bar T$-deformation and long range spin chains}, \href{http://dx.doi.org/10.1007/JHEP03(2020)092}{{ JHEP} {03} (2020) 092}, \href{http://arxiv.org/abs/1911.11118}{{\ttfamily arXiv:1911.11118}}.

\bibitem{Eberhardt:2018ouy}
L.~Eberhardt, M.~R. Gaberdiel, and R.~Gopakumar, {The Worldsheet Dual of the Symmetric Product CFT}, \href{http://dx.doi.org/10.1007/JHEP04(2019)103}{{ JHEP} {04} (2019) 103}, \href{http://arxiv.org/abs/1812.01007}{{\ttfamily arXiv:1812.01007}}.

\bibitem{Eberhardt:2019ywk}
L.~Eberhardt, M.~R. Gaberdiel, and R.~Gopakumar, {Deriving the AdS$_{3}$/CFT$_{2}$ correspondence}, \href{http://dx.doi.org/10.1007/JHEP02(2020)136}{{ JHEP} {02} (2020) 136}, \href{http://arxiv.org/abs/1911.00378}{{\ttfamily arXiv:1911.00378}}.

\bibitem{Dei:2024sct}
A.~Dei, B.~Knighton, K.~Naderi, and S.~Sethi, {Tensionless AdS$_3$/CFT$_2$ and Single Trace $T\overline{T}$}, \href{http://arxiv.org/abs/2408.00823}{{\ttfamily arXiv:2408.00823}}.

\bibitem{Aharony:1998ub}
O.~Aharony, M.~Berkooz, D.~Kutasov, and N.~Seiberg, {Linear dilatons, NS five-branes and holography}, \href{http://dx.doi.org/10.1088/1126-6708/1998/10/004}{{ JHEP} {10} (1998) 004}, \href{http://arxiv.org/abs/hep-th/9808149}{{\ttfamily arXiv:hep-th/9808149}}.

\bibitem{Giveon:2017myj}
A.~Giveon, N.~Itzhaki, and D.~Kutasov, {A solvable irrelevant deformation of AdS$_{3}$/CFT$_{2}$}, \href{http://dx.doi.org/10.1007/JHEP12(2017)155}{{ JHEP} {12} (2017) 155}, \href{http://arxiv.org/abs/1707.05800}{{\ttfamily arXiv:1707.05800}}.

\bibitem{Chang:2023kkq}
C.-K. Chang, C.~Ferko, and S.~Sethi, {Holography and irrelevant operators}, \href{http://dx.doi.org/10.1103/PhysRevD.107.126021}{{ Phys. Rev. D} {107} no.~12, (2023) 126021}, \href{http://arxiv.org/abs/2302.03041}{{\ttfamily arXiv:2302.03041}}.

\bibitem{Araujo:2018rho}
T.~Araujo, E.~O. Colg\'ain, Y.~Sakatani, M.~M. Sheikh-Jabbari, and H.~Yavartanoo, {Holographic integration of $T \bar{T}$ \textbackslash{}\& $J \bar{T}$ via $O(d,d)$}, \href{http://dx.doi.org/10.1007/JHEP03(2019)168}{{ JHEP} {03} (2019) 168}, \href{http://arxiv.org/abs/1811.03050}{{\ttfamily arXiv:1811.03050}}.

\bibitem{Giveon:1994fu}
A.~Giveon, M.~Porrati, and E.~Rabinovici, {Target space duality in string theory}, \href{http://dx.doi.org/10.1016/0370-1573(94)90070-1}{{ Phys. Rept.} {244} (1994) 77--202}, \href{http://arxiv.org/abs/hep-th/9401139}{{\ttfamily arXiv:hep-th/9401139}}.

\bibitem{Frolov:2005dj}
S.~Frolov, {Lax pair for strings in Lunin-Maldacena background}, \href{http://dx.doi.org/10.1088/1126-6708/2005/05/069}{{ JHEP} {05} (2005) 069}, \href{http://arxiv.org/abs/hep-th/0503201}{{\ttfamily arXiv:hep-th/0503201}}.

\bibitem{Orlando:2019rjg}
D.~Orlando, S.~Reffert, Y.~Sekiguchi, and K.~Yoshida, {O(d,d) transformations preserve classical integrability}, \href{http://dx.doi.org/10.1016/j.nuclphysb.2019.114880}{{ Nucl. Phys. B} {950} (2020) 114880}, \href{http://arxiv.org/abs/1907.03759}{{\ttfamily arXiv:1907.03759}}.

\bibitem{Lunin:2005jy}
O.~Lunin and J.~M. Maldacena, {Deforming field theories with U(1) x U(1) global symmetry and their gravity duals}, \href{http://dx.doi.org/10.1088/1126-6708/2005/05/033}{{ JHEP} {05} (2005) 033}, \href{http://arxiv.org/abs/hep-th/0502086}{{\ttfamily arXiv:hep-th/0502086}}.

\bibitem{Giveon:2023gzh}
A.~Giveon, {2pf in single-trace $ T\overline{T} $ holography}, \href{http://dx.doi.org/10.1007/JHEP10(2023)112}{{ JHEP} {10} (2023) 112}, \href{http://arxiv.org/abs/2309.15629}{{\ttfamily arXiv:2309.15629}}.

\bibitem{Giribet:2017imm}
G.~Giribet, {$T\bar{T}$-deformations, AdS/CFT and correlation functions}, \href{http://dx.doi.org/10.1007/JHEP02(2018)114}{{ JHEP} {02} (2018) 114}, \href{http://arxiv.org/abs/1711.02716}{{\ttfamily arXiv:1711.02716}}.

\bibitem{Cardy:2018jho}
J.~Cardy, {$T\overline T$ deformations of non-Lorentz invariant field theories}, \href{http://arxiv.org/abs/1809.07849}{{\ttfamily arXiv:1809.07849}}.

\bibitem{Hansen:2020hrs}
D.~Hansen, Y.~Jiang, and J.~Xu, {Geometrizing non-relativistic bilinear deformations}, \href{http://dx.doi.org/10.1007/JHEP04(2021)186}{{ JHEP} {04} (2021) 186}, \href{http://arxiv.org/abs/2012.12290}{{\ttfamily arXiv:2012.12290}}.

\bibitem{Chen:2020jdi}
B.~Chen, J.~Hou, and J.~Tian, {Note on the nonrelativistic $ T\overline{T} $ deformation}, \href{http://dx.doi.org/10.1103/PhysRevD.104.025004}{{ Phys. Rev. D} {104} no.~2, (2021) 025004}, \href{http://arxiv.org/abs/2012.14091}{{\ttfamily arXiv:2012.14091}}.

\bibitem{Cardy:2020olv}
J.~Cardy and B.~Doyon, {$ T\overline{T} $ deformations and the width of fundamental particles}, \href{http://dx.doi.org/10.1007/JHEP04(2022)136}{{ JHEP} {04} (2022) 136}, \href{http://arxiv.org/abs/2010.15733}{{\ttfamily arXiv:2010.15733}}.

\bibitem{Jiang:2020nnb}
Y.~Jiang, {$\mathrm{T}\overline{\mathrm{T}}$-deformed 1d Bose gas}, \href{http://dx.doi.org/10.21468/SciPostPhys.12.6.191}{{ SciPost Phys.} {12} no.~6, (2022) 191}, \href{http://arxiv.org/abs/2011.00637}{{\ttfamily arXiv:2011.00637}}.

\bibitem{Pavshinkin:2021jpy}
D.~Pavshinkin, {$T\overline T$ deformation of the Calogero\textendash{}Sutherland model via dimensional reduction}, \href{http://dx.doi.org/10.1134/S0040577923110089}{{ Theor. Math. Phys.} {217} no.~2, (2023) 1726--1742}, \href{http://arxiv.org/abs/2111.12080}{{\ttfamily arXiv:2111.12080}}.

\bibitem{Doyon:2023bvo}
B.~Doyon, F.~H\"ubner, and T.~Yoshimura, {New Classical Integrable Systems from Generalized TT\textasciimacron{}-Deformations}, \href{http://dx.doi.org/10.1103/PhysRevLett.132.251602}{{ Phys. Rev. Lett.} {132} no.~25, (2024) 251602}, \href{http://arxiv.org/abs/2311.06369}{{\ttfamily arXiv:2311.06369}}.

\bibitem{Doyon:2023hiq}
B.~Doyon, F.~H\"ubner, and T.~Yoshimura, {Generalised $T\bar{T}$-deformations of classical free particles}, \href{http://arxiv.org/abs/2312.14855}{{\ttfamily arXiv:2312.14855}}.

\bibitem{Bargheer:2008jt}
T.~Bargheer, N.~Beisert, and F.~Loebbert, {Boosting Nearest-Neighbour to Long-Range Integrable Spin Chains}, \href{http://dx.doi.org/10.1088/1742-5468/2008/11/L11001}{{ J. Stat. Mech.} {0811} (2008) L11001}, \href{http://arxiv.org/abs/0807.5081}{{\ttfamily arXiv:0807.5081}}.

\bibitem{Marchetto:2019yyt}
E.~Marchetto, A.~Sfondrini, and Z.~Yang, {$T\bar{T}$ Deformations and Integrable Spin Chains}, \href{http://dx.doi.org/10.1103/PhysRevLett.124.100601}{{ Phys. Rev. Lett.} {124} no.~10, (2020) 100601}, \href{http://arxiv.org/abs/1911.12315}{{\ttfamily arXiv:1911.12315}}.

\bibitem{Pozsgay:2021rwc}
B.~Pozsgay, T.~Gombor, and A.~Hutsalyuk, {Integrable hard-rod deformation of the Heisenberg spin chains}, \href{http://dx.doi.org/10.1103/PhysRevE.104.064124}{{ Phys. Rev. E} {104} no.~6, (2021) 064124}, \href{http://arxiv.org/abs/2108.13724}{{\ttfamily arXiv:2108.13724}}.

\bibitem{Jiang:2023rxa}
Y.~Jiang, {$T\bar{T}$~Deformation: A Lattice Approach}, \href{http://dx.doi.org/10.3390/sym15122212}{{ Symmetry} {15} no.~12, (2023) 2212}, \href{http://arxiv.org/abs/2312.12078}{{\ttfamily arXiv:2312.12078}}.

\bibitem{Gross:2019uxi}
D.~J. Gross, J.~Kruthoff, A.~Rolph, and E.~Shaghoulian, {Hamiltonian deformations in quantum mechanics, $T\bar T$, and the SYK model}, \href{http://dx.doi.org/10.1103/PhysRevD.102.046019}{{ Phys. Rev. D} {102} no.~4, (2020) 046019}, \href{http://arxiv.org/abs/1912.06132}{{\ttfamily arXiv:1912.06132}}.

\bibitem{He:2021dhr}
S.~He and Z.-Y. Xian, {TT\textasciimacron{} deformation on multiquantum mechanics and regenesis}, \href{http://dx.doi.org/10.1103/PhysRevD.106.046002}{{ Phys. Rev. D} {106} no.~4, (2022) 046002}, \href{http://arxiv.org/abs/2104.03852}{{\ttfamily arXiv:2104.03852}}.

\bibitem{Jensen:2016pah}
K.~Jensen, {Chaos in AdS$_2$ Holography}, \href{http://dx.doi.org/10.1103/PhysRevLett.117.111601}{{ Phys. Rev. Lett.} {117} no.~11, (2016) 111601}, \href{http://arxiv.org/abs/1605.06098}{{\ttfamily arXiv:1605.06098}}.

\bibitem{Maldacena:2016upp}
J.~Maldacena, D.~Stanford, and Z.~Yang, {Conformal symmetry and its breaking in two dimensional Nearly Anti-de-Sitter space}, \href{http://dx.doi.org/10.1093/ptep/ptw124}{{ PTEP} {2016} no.~12, (2016) 12C104}, \href{http://arxiv.org/abs/1606.01857}{{\ttfamily arXiv:1606.01857}}.

\bibitem{Engelsoy:2016xyb}
J.~Engels\"oy, T.~G. Mertens, and H.~Verlinde, {An investigation of AdS$_{2}$ backreaction and holography}, \href{http://dx.doi.org/10.1007/JHEP07(2016)139}{{ JHEP} {07} (2016) 139}, \href{http://arxiv.org/abs/1606.03438}{{\ttfamily arXiv:1606.03438}}.

\bibitem{Saad:2019lba}
P.~Saad, S.~H. Shenker, and D.~Stanford, {JT gravity as a matrix integral}, \href{http://arxiv.org/abs/1903.11115}{{\ttfamily arXiv:1903.11115}}.

\bibitem{Rosso:2020wir}
F.~Rosso, {$T\bar{T}$ deformation of random matrices}, \href{http://dx.doi.org/10.1103/PhysRevD.103.126017}{{ Phys. Rev. D} {103} no.~12, (2021) 126017}, \href{http://arxiv.org/abs/2012.11714}{{\ttfamily arXiv:2012.11714}}.

\bibitem{Ebert:2022gyn}
S.~Ebert, H.-Y. Sun, and Z.~Sun, {$ T\overline{T} $-deformed free energy of the Airy model}, \href{http://dx.doi.org/10.1007/JHEP08(2022)026}{{ JHEP} {08} (2022) 026}, \href{http://arxiv.org/abs/2202.03454}{{\ttfamily arXiv:2202.03454}}.

\bibitem{Guica:2017lia}
M.~Guica, {An integrable Lorentz-breaking deformation of two-dimensional CFTs}, \href{http://dx.doi.org/10.21468/SciPostPhys.5.5.048}{{ SciPost Phys.} {5} no.~5, (2018) 048}, \href{http://arxiv.org/abs/1710.08415}{{\ttfamily arXiv:1710.08415}}.

\bibitem{Apolo:2018qpq}
L.~Apolo and W.~Song, {Strings on warped AdS$_{3}$ via $ \mathrm{T}\bar{\mathrm{J}} $ deformations}, \href{http://dx.doi.org/10.1007/JHEP10(2018)165}{{ JHEP} {10} (2018) 165}, \href{http://arxiv.org/abs/1806.10127}{{\ttfamily arXiv:1806.10127}}.

\bibitem{Chakraborty:2018vja}
S.~Chakraborty, A.~Giveon, and D.~Kutasov, {$ J\overline{T} $ deformed CFT$_{2}$ and string theory}, \href{http://dx.doi.org/10.1007/JHEP10(2018)057}{{ JHEP} {10} (2018) 057}, \href{http://arxiv.org/abs/1806.09667}{{\ttfamily arXiv:1806.09667}}.

\bibitem{Aharony:2018ics}
O.~Aharony, S.~Datta, A.~Giveon, Y.~Jiang, and D.~Kutasov, {Modular covariance and uniqueness of $J\bar{T}$ deformed CFTs}, \href{http://dx.doi.org/10.1007/JHEP01(2019)085}{{ JHEP} {01} (2019) 085}, \href{http://arxiv.org/abs/1808.08978}{{\ttfamily arXiv:1808.08978}}.

\bibitem{LeFloch:2019rut}
B.~Le~Floch and M.~Mezei, {Solving a family of $T\bar{T}$-like theories}, \href{http://arxiv.org/abs/1903.07606}{{\ttfamily arXiv:1903.07606}}.

\bibitem{Hernandez-Chifflet:2019sua}
G.~Hern\'andez-Chifflet, S.~Negro, and A.~Sfondrini, {Flow Equations for Generalized $T\overline{T}$ Deformations}, \href{http://dx.doi.org/10.1103/PhysRevLett.124.200601}{{ Phys. Rev. Lett.} {124} no.~20, (2020) 200601}, \href{http://arxiv.org/abs/1911.12233}{{\ttfamily arXiv:1911.12233}}.

\bibitem{Rodriguez:2021tcz}
P.~Rodr\'\i{}guez, D.~Tempo, and R.~Troncoso, {Mapping relativistic to ultra/non-relativistic conformal symmetries in 2D and finite $ \sqrt{T\overline{T}} $ deformations}, \href{http://dx.doi.org/10.1007/JHEP11(2021)133}{{ JHEP} {11} (2021) 133}, \href{http://arxiv.org/abs/2106.09750}{{\ttfamily arXiv:2106.09750}}.

\bibitem{He:2024yzx}
S.~He and X.-C. Mao, {Irrelevant and marginal deformed BMS field theories}, \href{http://dx.doi.org/10.1007/JHEP04(2024)138}{{ JHEP} {04} (2024) 138}, \href{http://arxiv.org/abs/2401.09991}{{\ttfamily arXiv:2401.09991}}.

\bibitem{Tempo:2022ndz}
D.~Tempo and R.~Troncoso, {Nonlinear automorphism of the conformal algebra in 2D and continuous $ \sqrt{T\overline{T}} $ deformations}, \href{http://dx.doi.org/10.1007/JHEP12(2022)129}{{ JHEP} {12} (2022) 129}, \href{http://arxiv.org/abs/2210.00059}{{\ttfamily arXiv:2210.00059}}.

\bibitem{Ebert:2023tih}
S.~Ebert, C.~Ferko, and Z.~Sun, {Root-TT\textasciimacron{} deformed boundary conditions in holography}, \href{http://dx.doi.org/10.1103/PhysRevD.107.126022}{{ Phys. Rev. D} {107} no.~12, (2023) 126022}, \href{http://arxiv.org/abs/2304.08723}{{\ttfamily arXiv:2304.08723}}.

\bibitem{Ebert:2024zwv}
S.~Ebert, C.~Ferko, C.~L. Martin, and G.~Tartaglino-Mazzucchelli, {Flows in the space of interacting chiral boson theories}, \href{http://dx.doi.org/10.1103/PhysRevD.110.046005}{{ Phys. Rev. D} {110} no.~4, (2024) 046005}, \href{http://arxiv.org/abs/2403.18242}{{\ttfamily arXiv:2403.18242}}.

\bibitem{Ferko:2023sps}
C.~Ferko, Y.~Hu, Z.~Huang, K.~Koutrolikos, and G.~Tartaglino-Mazzucchelli, {$T \overline{T}$-like flows and $3d$ nonlinear supersymmetry}, \href{http://dx.doi.org/10.21468/SciPostPhys.16.1.038}{{ SciPost Phys.} {16} no.~1, (2024) 038}, \href{http://arxiv.org/abs/2302.10410}{{\ttfamily arXiv:2302.10410}}.

\bibitem{Ferko:2024yua}
C.~Ferko, J.~Hou, T.~Morone, G.~Tartaglino-Mazzucchelli, and R.~Tateo, {${T\overline{T}}$-like Flows of Yang-Mills Theories}, \href{http://arxiv.org/abs/2409.18740}{{\ttfamily arXiv:2409.18740}}.

\bibitem{Bandos:2020hgy}
I.~Bandos, K.~Lechner, D.~Sorokin, and P.~K. Townsend, {On p-form gauge theories and their conformal limits}, \href{http://dx.doi.org/10.1007/JHEP03(2021)022}{{ JHEP} {03} (2021) 022}, \href{http://arxiv.org/abs/2012.09286}{{\ttfamily arXiv:2012.09286}}.

\bibitem{Ferko:2024zth}
C.~Ferko, S.~M. Kuzenko, K.~Lechner, D.~P. Sorokin, and G.~Tartaglino-Mazzucchelli, {Interacting chiral form field theories and $ T\overline{T} $-like flows in six and higher dimensions}, \href{http://dx.doi.org/10.1007/JHEP05(2024)320}{{ JHEP} {05} (2024) 320}, \href{http://arxiv.org/abs/2402.06947}{{\ttfamily arXiv:2402.06947}}.

\bibitem{Brennan:2019azg}
T.~D. Brennan, C.~Ferko, and S.~Sethi, {A Non-Abelian Analogue of DBI from $T \overline{T}$}, \href{http://dx.doi.org/10.21468/SciPostPhys.8.4.052}{{ SciPost Phys.} {8} no.~4, (2020) 052}, \href{http://arxiv.org/abs/1912.12389}{{\ttfamily arXiv:1912.12389}}.

\bibitem{Ferko:2024ali}
C.~Ferko and L.~Smith, {An Infinite Family of Integrable Sigma Models Using Auxiliary Fields}, \href{http://arxiv.org/abs/2405.05899}{{\ttfamily arXiv:2405.05899}}.

\bibitem{Bielli:2024khq}
D.~Bielli, C.~Ferko, L.~Smith, and G.~Tartaglino-Mazzucchelli, {T-Duality and $T \overline{T}$-like Deformations of Sigma Models}, \href{http://arxiv.org/abs/2407.11636}{{\ttfamily arXiv:2407.11636}}.

\bibitem{Fukushima:2024nxm}
O.~Fukushima and K.~Yoshida, {4D Chern-Simons theory with auxiliary fields}, \href{http://arxiv.org/abs/2407.02204}{{\ttfamily arXiv:2407.02204}}.

\bibitem{Bielli:2024ach}
D.~Bielli, C.~Ferko, L.~Smith, and G.~Tartaglino-Mazzucchelli, {Integrable Higher-Spin Deformations of Sigma Models from Auxiliary Fields}, \href{http://arxiv.org/abs/2407.16338}{{\ttfamily arXiv:2407.16338}}.

\bibitem{Bielli:2024fnp}
D.~Bielli, C.~Ferko, L.~Smith, and G.~Tartaglino-Mazzucchelli, {Auxiliary Field Sigma Models and Yang-Baxter Deformations}, \href{http://arxiv.org/abs/2408.09714}{{\ttfamily arXiv:2408.09714}}.

\bibitem{Babaei-Aghbolagh:2022itg}
H.~Babaei-Aghbolagh, K.~Babaei~Velni, D.~M. Yekta, and H.~Mohammadzadeh, {Manifestly SL(2, R) Duality-Symmetric Forms in ModMax Theory}, \href{http://dx.doi.org/10.1007/JHEP12(2022)147}{{ JHEP} {12} (2022) 147}, \href{http://arxiv.org/abs/2210.13196}{{\ttfamily arXiv:2210.13196}}.

\bibitem{Ferko:2023wyi}
C.~Ferko, S.~M. Kuzenko, L.~Smith, and G.~Tartaglino-Mazzucchelli, {Duality-invariant nonlinear electrodynamics and stress tensor flows}, \href{http://dx.doi.org/10.1103/PhysRevD.108.106021}{{ Phys. Rev. D} {108} no.~10, (2023) 106021}, \href{http://arxiv.org/abs/2309.04253}{{\ttfamily arXiv:2309.04253}}.

\bibitem{Floss:2023nod}
T.~Fl\"oss, D.~Roest, and T.~Westerdijk, {Non-linear electrodynamics from massive gravity}, \href{http://dx.doi.org/10.1007/JHEP02(2024)194}{{ JHEP} {02} (2024) 194}, \href{http://arxiv.org/abs/2308.04349}{{\ttfamily arXiv:2308.04349}}.

\bibitem{Morone:2024ffm}
T.~Morone, S.~Negro, and R.~Tateo, {Gravity and $T \bar{T}$ flows in higher dimensions}, \href{http://dx.doi.org/10.1016/j.nuclphysb.2024.116605}{{ Nucl. Phys. B} {1005} (2024) 116605}, \href{http://arxiv.org/abs/2401.16400}{{\ttfamily arXiv:2401.16400}}.

\bibitem{Morone:2024sdg}
T.~Morone and R.~Tateo, {Solutions to the Ricci Flow via Einstein Field Equations}, \href{http://arxiv.org/abs/2411.10265}{{\ttfamily arXiv:2411.10265}}.

\bibitem{Hao:2024stt}
F.~Hao, X.-Y. Ran, and M.~Yamada, {Geometric realization via irrelevant deformations induced by the stress-energy tensor}, \href{http://arxiv.org/abs/2410.02537}{{\ttfamily arXiv:2410.02537}}.

\bibitem{Shi:2021nkx}
Y.-H. Shi { et~al.}, {Quantum simulation of Hawking radiation and curved spacetime with a superconducting on-chip black hole}, \href{http://dx.doi.org/10.1038/s41467-023-39064-6}{{ Nature Commun.} {14} no.~1, (2023) 3263}, \href{http://arxiv.org/abs/2111.11092}{{\ttfamily arXiv:2111.11092}}.

\bibitem{Basteiro:2022pyp}
P.~Basteiro, F.~Dusel, J.~Erdmenger, D.~Herdt, H.~Hinrichsen, R.~Meyer, and M.~Schrauth, {Breitenlohner-Freedman Bound on Hyperbolic Tilings}, \href{http://dx.doi.org/10.1103/PhysRevLett.130.091604}{{ Phys. Rev. Lett.} {130} no.~9, (2023) 091604}, \href{http://arxiv.org/abs/2205.05081}{{\ttfamily arXiv:2205.05081}}.

\bibitem{Chen:2023cad}
J.~Chen { et~al.}, {AdS/CFT Correspondence in Hyperbolic Lattices}, \href{http://arxiv.org/abs/2305.04862}{{\ttfamily arXiv:2305.04862}}.

\bibitem{Swingle:2009bg}
B.~Swingle, {Entanglement Renormalization and Holography}, \href{http://dx.doi.org/10.1103/PhysRevD.86.065007}{{ Phys. Rev. D} {86} (2012) 065007}, \href{http://arxiv.org/abs/0905.1317}{{\ttfamily arXiv:0905.1317}}.

\bibitem{Gubser:2016guj}
S.~S. Gubser, J.~Knaute, S.~Parikh, A.~Samberg, and P.~Witaszczyk, {$p$-adic AdS/CFT}, \href{http://dx.doi.org/10.1007/s00220-016-2813-6}{{ Commun. Math. Phys.} {352} no.~3, (2017) 1019--1059}, \href{http://arxiv.org/abs/1605.01061}{{\ttfamily arXiv:1605.01061}}.

\bibitem{Chen:2021ipv}
L.~Chen, X.~Liu, and L.-Y. Hung, {Emergent Einstein Equation in p-adic Conformal Field Theory Tensor Networks}, \href{http://dx.doi.org/10.1103/PhysRevLett.127.221602}{{ Phys. Rev. Lett.} {127} no.~22, (2021) 221602}, \href{http://arxiv.org/abs/2102.12022}{{\ttfamily arXiv:2102.12022}}.

\bibitem{Chen:2022wvy}
L.~Chen, K.~Ji, H.~Zhang, C.~Shen, R.~Wang, X.~Zeng, and L.-Y. Hung, {CFTD from TQFTD+1 via Holographic Tensor Network, and Precision Discretization of CFT2 }, \href{http://dx.doi.org/10.1103/PhysRevX.14.041033}{{ Phys. Rev. X} {14} no.~4, (2024) 041033}, \href{http://arxiv.org/abs/2210.12127}{{\ttfamily arXiv:2210.12127}}.

\bibitem{Cheng:2023kxh}
G.~Cheng, L.~Chen, Z.-C. Gu, and L.-Y. Hung, {Precision reconstruction of rational CFT from exact fixed point tensor network}, \href{http://arxiv.org/abs/2311.18005}{{\ttfamily arXiv:2311.18005}}.

\bibitem{Hung:2024gma}
L.-Y. Hung and Y.~Jiang, {Building up quantum spacetimes with BCFT Legos}, \href{http://arxiv.org/abs/2404.00877}{{\ttfamily arXiv:2404.00877}}.

\bibitem{Babaei-Aghbolagh:2024uqp}
H.~Babaei-Aghbolagh, S.~He, and H.~Ouyang, {Generalized $ T\overline{T} $-like deformations in duality-invariant nonlinear electrodynamic theories}, \href{http://dx.doi.org/10.1007/JHEP09(2024)137}{{ JHEP} {09} (2024) 137}, \href{http://arxiv.org/abs/2407.03698}{{\ttfamily arXiv:2407.03698}}.

\bibitem{Bialynicki-Birula:1981}
I.~Bialynicki-Birula, Nonlinear electrodynamics: Variations on a theme by born and infeld, in { Quantum Theory of Particles and Fields: Birthday Volume Dedicated to Jan Lopuszanski}, B.~Jancewicz and J.~Lukierski, eds., pp.~31--48.
\newblock World Scientific Publishing Co Pte Ltd, 1983.

\bibitem{Bialynicki-Birula:1992rcm}
I.~Bialynicki-Birula, Field theory of photon dust, { Acta Phys. Polon. B} {23} (1992) 553--559.

\bibitem{Gibbons:1995ap}
G.~W. Gibbons and D.~A. Rasheed, Sl(2,r) invariance of nonlinear electrodynamics coupled to an axion and a dilaton, \href{http://dx.doi.org/10.1016/0370-2693(95)01272-9}{{ Phys. Lett. B} {365} (1996) 46--50}. [arXiv:hep-th/9509141 [hep-th]].

\bibitem{Deser:1997mz}
S.~Deser, A.~Gomberoff, M.~Henneaux, and C.~Teitelboim, Duality, selfduality, sources and charge quantization in abelian n form theories, \href{http://dx.doi.org/10.1016/S0370-2693(97)00338-9}{{ Phys. Lett. B} {400} (1997) 80--86}. [arXiv:hep-th/9702184 [hep-th]].

\bibitem{Deser:1997se}
S.~Deser, A.~Gomberoff, M.~Henneaux, and C.~Teitelboim, P-brane dyons and electric magnetic duality, \href{http://dx.doi.org/10.1016/S0550-3213(98)00179-5}{{ Nucl. Phys. B} {520} (1998) 179--204}. [arXiv:hep-th/9712189 [hep-th]].

\end{thebibliography}
\end{document}